%% file: MainThesis.tex
\documentclass[instructions]{uqthesis}
\input{LaTexPackages.tex}
\renewcommand{\hbar}{\mathchar'26\mkern-9mu h}
\title{Quantum thermodynamics of integrable and near-integrable atomic systems }

\author{Raymon Watson}
\currentdegrees{MPhys}

\orcid{\href{https://orcid.org/0000-0002-3243-2110}{0000-0002-3243-2110}}

\date{2024}
\submittedfor{Doctor}

\school{School of Mathematics and Physics}

\begin{document}

\frontmatter
\maketitle
\clearpage

\section{Abstract}
\normalfont
\input{./Abstract/Abstract.tex}

\clearpage
\section*{Declaration by author}
\input{./Authordeclaration.tex}

\clearpage
\input{./PreliminaryAndBackPages/Preliminary.tex}

\mainmatter


\input{./Chapter1/Introduction.tex}

\input{./Chapter2/Chapter2.tex}
\input{./Chapter3/Chapter3.tex}

\input{./Chapter4/Chapter4.tex}

\input{./Chapter5/Chapter5.tex}

\input{./Chapter6/Chapter6.tex}

\input{./Chapter7/Chapter7.tex}


\input{./Conclusion/Conclusion.tex}




\bibliographystyle{uqthesis}

\bibliography{./References/Bibliography}




\input{./PreliminaryAndBackPages/Back.tex}

\end{document}

%% file: LaTexPackages.tex
\usepackage[normalem]{ulem}

\usepackage{pdfpages}			 
\usepackage{wrapfig}			 
\usepackage{framed}			     
\usepackage{microtype}			 
\usepackage{color}				 
\usepackage{transparent}	     
\usepackage{placeins}			 
\usepackage{mdframed,mdwlist}    
\usepackage{graphicx}            
\usepackage{float}               
\usepackage{longtable}           
\usepackage{array}               
\usepackage{enumerate}           
\usepackage[all]{xy}             
\usepackage{amsmath}           
\usepackage{amssymb}        

\usepackage{fancyhdr}            
\usepackage{blindtext}           
\usepackage{tikz}                
\usepackage[figuresright]{rotating}	
\usetikzlibrary{shapes.geometric, arrows}

\usepackage{subcaption}
\usepackage{bbold}
\usepackage{times}
\usepackage{xcolor}

\usepackage{setspace}

\usepackage{breakurl}
\usepackage{seqsplit}

\makeatletter

\usepackage{enumitem}
\newlist{todolist}{itemize}{2}
\setlist[todolist]{label=$\square$}
\usepackage{pifont}

\usepackage[normalem]{ulem}

%% file: Abstract/Abstract.tex






The field of quantum thermodynamics has flourished in the past decade, due in large part to the rapid advancement in experimental control over, and theoretical simulation of, ultracold atomic systems. Notably, this has led to the realization of single-body quantum heat engines in the laboratory, which represents the ultimate limit in the creation of infinitesimal machines.
Beyond this single-particle limit, however, many-body interacting systems offer a plethora of quantum resources, such as quantum coherence, entanglement, and correlations, which may be utilized to enhance performance of quantum thermal machines, or even perform tasks not possible classically. To characterise and understand such systems in the context of quantum thermodynamics requires theoretical or computational treatment of their quantum many-body dynamics at nonzero temperatures. However, real-time, finite-temperature dynamics of many-body interacting quantum systems can be difficult to evaluate, often becoming entirely intractable in highly nonequilibrium scenarios. An important exception to this is in simulation of integrable systems, exactly solvable through the Bethe ansatz, and near-integrable systems, which has seen much progress since the landmark experimental realization of the near-integrable quantum Newton's cradle in 2006. In particular, the recently developed theory of generalized hydrodynamics (GHD) has been shown to be capable of accurate modeling of the large-scale nonequilibrium dynamics of these systems.

In this thesis, we explore various aspects of equilibrium and nonequilibrium thermodynamics for ultracold atomic gases, with a focus on the experimentally realisable one-dimensional (1D) Bose gas. This is a paradigmatic example of an interacting many-body system, which is integrable in the uniform limit and near-integrable otherwise.
We first investigate a quantum thermodynamic Otto cycle driven by a quench of interaction strength of a 1D Bose gas. For the case of a sudden quench in a uniform 1D Bose gas, we demonstrate how the performance of this highly nonequilibrium quantum thermal machine may be expressed in terms of atom-atom correlations.
Further, we derive a new Maxwell relation which allows one to express entropy, which is generally difficult to ascertain in quantum systems, in terms of Glauber's local second-order correlation function, which are experimentally measurable.
We next perform benchmarks of GHD in order to ascertain the regimes where such a theory may be applied, and then utilize it to investigate the interaction strength-driven quantum Otto cycle in a  harmonically trapped 1D Bose gas. However, to realise such an engine in an experimentally realistic manner would require modelling thermalisation with external reservoirs, which is currently beyond the scope of GHD.
Yet, the classical $c$-field stochastic-projected Gross-Pitaevskii equation (SPGPE) method is capable of simulating tunnel-coupling pairs of 1D Bose gases in the weakly interacting regime. We therefore utilise this numerical method to simulate the \emph{full} quantum Otto engine cycle, realised through a single working fluid tunnel-coupled to two 1D Bose gases, which constitute the external (hot and cold) reservoirs.


In Chapters \ref{Chap:Intro} and \ref{Chap:2} we introduce the reader to the current state of the field and outline the theoretical descriptions and techniques which are used throughout this thesis.

In Chapter \ref{Chap:3} we introduce and explore a quantum heat engine powered by control over inter-particle interaction strengths. 
In particular, we demonstrate that, through a sudden interaction quench protocol, the net work of an Otto engine cycle may be expressed in terms of a difference of local second-order correlation functions. This brings Glauber's local second-order correlation function, a quantity which has relevance broadly across many fields of physics, into the field of quantum thermodynamics in a clear and direct manner. The net work and efficiency of this quantum  engine cycle are expressed entirely in terms of observables of thermal equilibrium states. Further, we derive similar results for a variety of other quantum thermal machines, such as the refrigerator, thermal accelerator, and heater. 
We note that, even though we have focused on analysing performance of these thermal machines with the 1D Bose gas in mind, the  theory developed in this Chapter is applicable more generally to any atomic system where interactions are mediated via short-range $s$-wave scattering.

In Chapter \ref{Chap:Maxwell}, we derive a new Maxwell relation which links Glauber's local second-order correlation function with the thermodynamic entropy of an ultracold quantum gas.
To demonstrate the utility of this Maxwell relation, we perform a numerical experiment using the SPGPE, and calculate the entropy of a uniform 1D Bose gas, which has not previously accomplished via the SPGPE method. We believe this new Maxwell relation will have direct impact on experimental and theoretical investigation of ultracold atomic gases, where entropy is an important thermodynamic quantity that is often difficult to measure in practice, whereas local correlations are significantly more accessible.

In Chapter \ref{Chap:5}, we benchmark the recently developed theory of GHD, which is capable of simulating the large-scale dynamics of integrable and near-integrable systems, against a variety of alternative theoretical and computational methods. In particular, we focus on a set of highly nonequilibrium protocols, such as expansion from a localised density bump in a 1D Bose gas, which produces a dispersive quantum shock, and expansion from a density dip, which instead sheds a train of grey solitons. We explore these dynamics over the full range of interaction strengths, and compare GHD with a variety of alternative theoretical models such as the SPGPE method, the infinite matrix product states method (iMPS), truncated Wigner approximation, and exact diagonalization. Through this, we characterise the regimes of applicability for GHD, and explain the reason for its breakdown outside of these regions. Further, we benchmark higher-order Navier-Stokes GHD, which is capable of modeling thermalization in models with weakly broken integrability, and compare it against the SPGPE method for a quantum Newton's cradle scenario, reporting excellent agreement between the two methods for the case of a double to single-well quench.

In Chapter \ref{Chap:6}, we utilize GHD to study the nonequilibrium dynamics of a quantum thermochemical Otto engine cycle, including in the strongly interacting regime that is not accessible through the $c$-field SPGPE method. In particular, this quantum thermal machine is driven via control over inter-particle interactions, and fuelled through diffusive contact with the external reservoir, which is mediated by particle transfer. We employ the approach developed in Chapter \ref{Chap:3} for the sudden quench protocol to explore how the finite time performance scales between this idealized limit and that of the infinitely slow isentropic quench protocol. Through this analysis, we observe that the time window over which performance transitions between these two limits is associated with the timescale of longitudinal breathing modes. To the best of our knowledge, these results represent the first application of GHD to study the dynamics of a quantum engine cycle.

In Chapter \ref{Chap:7} we investigate an experimentally realistic quantum Otto engine cycle, modelled via three tunnel-coupled 1D Bose gases. We apply the $c$-field SPGPE method to simulate the entire interaction-driven engine cycle, where thermalisation strokes consist of tunnel-coupling with external 1D Bose gases, and the unitary work strokes consist of a finite time quench of interaction strength. Through this, we are able to calculate the output power, which depends on the total cycle time, and evaluate the trade-off between power and efficiency as a function of the work stroke duration. We find that the sudden quench protocol results in significantly higher power than the near-quasistatic cycle, while at the same time resulting in a reasonably high efficiency, implying a favourable trade-off between power and efficiency.

Lastly, in Chapter \ref{Chap:Conclusion}, we provide a summary of our main results, and highlight possibilities for future work. A selection of these outlooks include: extending the investigation of the sudden quench Otto cycle to other models and cycles, applying the newly derived Maxwell relations to other models and thermodynamic quantities, both of which would be of experimental relevance, and exploring further application of GHD to study nonequilibrium dynamics within other quantum thermodynamics devices.

\noindent



%% file: Authordeclaration.tex
%

\noindent
This thesis is composed of my original work, and contains no material previously published or written by another person except where due reference has been made in the text. I have clearly stated the contribution by others to jointly-authored works that I have included in my thesis.\\

\noindent
I have clearly stated the contribution of others to my thesis as a whole, including statistical assistance, survey design, data analysis, significant technical procedures, professional editorial advice, financial support and any other original research work used or reported in my thesis. The content of my thesis is the result of work I have carried out since the commencement of my higher degree by research candidature and does not include a substantial part of work that has been submitted to qualify for the award of any other degree or diploma in any university or other tertiary institution. I have clearly stated which parts of my thesis, if any, have been submitted to qualify for another award.\\

\noindent
I acknowledge that an electronic copy of my thesis must be lodged with the University Library and, subject to the policy and procedures of The University of Queensland, the thesis be made available for research and study in accordance with the Copyright Act 1968 unless a period of embargo has been approved by the Dean of the Graduate School. \\

\noindent
I acknowledge that copyright of all material contained in my thesis resides with the copyright holder(s) of that material. Where appropriate I have obtained copyright permission from the copyright holder to reproduce material in this thesis and have sought permission from co-authors for any jointly authored works included in the thesis.

%% file: PreliminaryAndBackPages/Preliminary.tex


\section*{Publications included in this thesis}

The following articles have been published during my candidature, and are included in this thesis as Chapters \ref{Chap:5} and \ref{Chap:7}, respectively: 

    \begin{enumerate}

    \item[] \cite{watson2023benchmarks} \textbf{R. S. Watson}, S.A. Simmons, and K. V. Kheruntsyan, \href{https://journals.aps.org/prresearch/abstract/10.1103/PhysRevResearch.5.L022024}{Benchmarks of generalized hydrodynamics for one-dimensional Bose gases}, Phys. Rev. Research, \textbf{5}, L022024 (2023)

    \item[] \cite{nautiyal2024finitetime} V. V. Nautiyal, \textbf{R. S. Watson}, and K. V. Kheruntsyan, \href{https://arxiv.org/abs/2404.16470}{A finite-time quantum Otto engine with tunnel coupled one-dimensional Bose gases}, New Journal of Physics \textbf{26}, 063033 (2024).

    \end{enumerate}

 \noindent The contribution by each named author to the publication of Ref.~\cite{watson2023benchmarks} appears in Table.~\ref{Tab:Chap5}, on the page directly following Chapter \ref{Chap:5}.

 \noindent The contribution by each named author to the publication of Ref.~\cite{nautiyal2024finitetime} appears in Table.~\ref{Tab:Chap7}, on the page directly following Chapter \ref{Chap:7}.



\section*{Submitted manuscripts included in this thesis}

The following articles have been submitted for publication during my candidature, and are included in this thesis as chapters \ref{Chap:3} and \ref{Chap:Maxwell}, respectively: 

    \begin{enumerate}

    \item[] \cite{watson2024interaction} \textbf{R. S. Watson}, and K. V. Kheruntsyan, \href{https://arxiv.org/abs/2308.05266}{Quantum many-body thermal machines enabled by atom-atom correlations}, submitted to SciPost Physics on 11 June 2024.

    \item[] \cite{watson2024maxwell} \textbf{R. S. Watson}, C. Coleman, and K. V. Kheruntsyan, \href{https://arxiv.org/abs/2405.04159}{Maxwell relation between entropy and atom-atom pair correlation}, Phys. Rev. Lett. \textbf{133}, 100403 (2024).

    \end{enumerate}

 \noindent The contribution by each named author to the publication of Ref.~\cite{watson2024interaction} appears in Table.~\ref{Tab:Chap3} on the page directly following Chapter \ref{Chap:3}.

  \noindent The contribution by each named author to the publication of Ref.~\cite{watson2024maxwell} appears in Table.~\ref{Tab:Chap4} on the page directly following Chapter \ref{Chap:Maxwell}.






\section*{Contributions by others to the thesis}

My supervisor Karén Kheruntsyan has made substantial input into the work presented in this thesis through his contribution to: the conception and design of the project, analysis and interpretation of results, the writing of included manuscripts, and proofreading of the thesis itself.

\noindent In chapter \ref{Chap:7}, V. N. Nautiyal provided the results based on simulations of the stochastic-projected Gross-Pitaevskii equation. 



\section*{Statement of parts of the thesis submitted to qualify for the award of another degree}
No work submitted towards another degree have been included in this thesis



\section*{Research involving human or animal subjects}
No animal or human subjects were involved in this research.



\clearpage
\section*{Acknowledgments}

I would like to thank my supervisor Professor Karén Kheruntsyan for the amount of time and effort you have spent helping me throughout my time as a research student. In particular, I would like to thank you for your understanding, your motivating influence, and your ability to always simplify the task at hand, providing a way forward whenever I was stuck.

I would like to thank my family for all the support that you have given me over these four years. Particular thanks go to my sister Michel, who was always free to call when I needed it. Thanks also must go to Justine and Ian, who would always let me stay with them when times got tough (thanks also to Lola for being a big sweetie).
I am also extremely grateful to my parents, who have always supported me and always will. Also thanks to my dog Frankie, for being a big baby but also just the sweetest girl.

Finally, thanks to all my friends in Brisbane and in Perth. In particular, thanks to Liam, Blair, and Tom, who, despite the distance, remain some of my closest friends for well over a decade now. I can't wait to try to explain all of this work to you.


\clearpage
\section*{Financial support}

This research was supported by an Australian Government Research Training Program Scholarship
(includes tuition fee offset and living allowance stipend).

\noindent Part of this research was performed while employed as a research assistant at the University of Queensland under Professor Karén V. Kheruntsyan (Start date: April 2, 2024)


\section*{Keywords}

Quantum thermodynamics, quantum Otto engine, quantum thermal machines, ultra-cold atomic gases, Bose gases, quasi-condensate, one-dimensional systems, Lieb-Liniger model, Tonks-Girardeau gas, integrable and near-integrable systems, thermodynamic Bethe ansatz, thermalisation of isolated quantum systems, generalised hydrodynamics, dispersive quantum shock waves, quantum Newton's cradle, Glauber second-order correlation function, atom-atom correlations, classical field method, stochastic projected Gross-Pitaevskii equation, thermodynamic Maxwell relations. 


\section*{Australian and New Zealand Standard Research Classifications (ANZSRC)}

    \noindent ANZSRC code: 510801, Degenerate quantum gases and atom optics, 70\% \\
    \noindent ANZSRC code: 401207, Fundamental and theoretical fluid dynamics, 20\% \\
    \noindent ANZSRC code: 510401, Condensed matter characterisation technique development, 10\%



\section*{Fields of Research (FoR) Classification}

\noindent FoR code: 5108, Quantum physics, 70\% \\
\noindent FoR code: 5104, Condensed matter physics, 25\% \\
\noindent FoR code: 5102, Atomic, molecular and optical physics, 5\%



\clearpage
\pagestyle{headings}

\tableofcontents
	\clearpage
\listoffigures
	\clearpage
\input{./PreliminaryAndBackPages/Symbols.tex} 


%% file: PreliminaryAndBackPages/Symbols.tex
%

\chapter[List of Abbreviations and Symbols]{List of Abbreviations and Symbols}


\begin{center}
	\small
	\begin{longtable}{ll}
	\toprule
	Abbreviations & {} \\
	\bottomrule
	1D				& One-dimensional \\
	2D				& Two-dimensional \\
	3D				& Three-dimensional \\
        BEC				& Bose-Einstein condensate \\
	CHD				& Conventional hydrodynamics \\
	GGE				& Generalized Gibbs ensemble \\
	GHD				& Generalized Hydrodynamics \\
	GPE			    & Gross-Pitaevskii equation \\
	IBG				& Ideal Bose gas \\
	iMPS			& Infinite matrix product states \\
	LDA				& Local density approximation \\
	QHE				& Quantum heat engine \\
	QTM				& Quantum thermal machine \\
	SPGPE			& Stochastic Projected Gross-Pitaevskii equation \\
	TBA				& Thermodynamic Bethe ansatz \\
	TF				& Thomas-Fermi \\
	TG				& Tonks-Girardeau \\
	TWA				& Truncated Wigner approximation \\
	\hline
	\end{longtable}
\end{center}

\begin{center}
	\small
	\begin{longtable}{ll}
	\toprule
	Symbols  & {} \\
	\bottomrule
	$\hbar$		& Reduced Planck's constant \\
	$k_B$		& Boltzmann's constant \\
	$m$		& Particle mass \\
	$z$		& Longitudinal coordinate \\
	$t$		& Time \\
	$\hat{H}$		& Hamiltonian \\
        $L $		& Total system length  \\ 
	$V$		& External trapping potential \\
	$T$		& Temperature \\
	$\tau$		& Dimensionless temperature \\
	$\mu$		& Chemical potential \\
        $\omega $		&  Harmonic trap frequency  \\ 
	$\hat{\Psi}$		& Bosonic field annihilation operator \\
	$\hat{\Psi}^\dagger$		& Bosonic field creation operator \\
	$\rho$		& Particle number density \\
	$N$		& Total particle number \\
	$a_s$		& 3D s-wave scattering length \\
	$g$		& Inter-particle interaction strength \\
	$\gamma$		& Dimensionless interaction strength, or Lieb-Liniger parameter \\
	$l_h$		& Healing length \\
	$l_\phi$		& Phase coherence length \\
	$l_T, \, \Lambda_T$		& Thermal de-Broglie wavelength \\
	$g^{(2)}(z,z')$		&  Glauber's normalised second-order correlation function\\
	$G^{(2)}(z,z')$		&  Glauber's unnormalised second-order correlation function\\
	$\overline{G^{(2)}}$		&   Glauber's total (integrated) correlation function\\

	$E$		& Total energy \\
	$S$		&  Entropy\\
	$s$		&  Entropy density\\
	$F$		& Helmholtz free energy \\ 
        $Q $		& Heat  \\ 
        $\eta $		&  Engine efficiency \\     
        $P $		& Power  \\ 
        $CoP $		& Coefficient of performance  \\ 
        $\Xi $		& Efficient work  \\


	$\lambda,\theta$		& Quasiparticle rapidity \\
        $\varphi $		& Scattering shift  \\ 
        $\hat{\rho} $		& One-body density matrix  \\ 
        $\alpha,\,\beta^i $		& Lagrange multipliers  \\ 
       $Q_i $		& Conserved charge  \\ 
	$ q_i $		& Charge density \\
	$ j_i $		& Charge current \\
	$f_p$		&  Quasiparticle root density, or particle density \\
	$f_h$		& Quasiparticle hole density  \\
	$f_s$		&  Total quasiparticle density \\
	$n$		&  Fermi filling factor \\
        $\varepsilon $		& Quasiparticle excitation energy \\ 
        $v^\mathrm{eff} $		& Effective velocity  \\ 
        $f^\mathrm{dr} $		& Dressed function \\ 
        $\mathcal{D} $		& Diffusion kernel  \\

	$\Psi_\mathbf{C}$, $\psi^{(\mathbf{C})}$		& $c$-field complex amplitude \\
        $\mathcal{P}^{(\mathbf{C})} $		& Projector for high energy cut-off \\ 
        $\mathcal{L}^{(\mathbf{C})} $		& Gross-Pitaevskii operator \\ 
       $dW_\Gamma $		& Complex white noise term  \\ 
        $\Gamma $		& SPGPE growth rate  \\ 
        $J $		& Tunnel coupling rate  \\

 	\hline
	\end{longtable}
\end{center}


%% file: Chapter1/Introduction.tex
\chapter[Introduction]{Introduction}
\label{Chap:Intro}


\textit{In this chapter we give a brief introduction to the areas of physics that will be relevant throughout this thesis. First, we give an overview of quantum thermodynamics, focusing on the theoretical and experimental development of quantum heat engines. We then provide a short discussion on the history of integrability and integrability breaking in quantum models, highlighting recent developments in the simulation of nonequilibrium dynamics of these systems. This leads into a discussion of the model of interest in this thesis, the 1D Bose gas, which is a paradigmatic example of an integrable model in its uniform configuration. Finally, we outline the motivation and scope of this thesis, presenting a breakdown of the contents of each chapter. }

\section{Quantum thermodynamics}

The field of quantum thermodynamics is broadly interested in the emergence of thermodynamic phenomena in the context of quantum systems \cite{kosfloff2014quantum,vinjanampathy2016quantum,deffner2019quantum,kosloff2013quantum}. The history of the field may be traced back to 1953, to H. E. D. Scovil and E. O. Schulz-Dubois's analysis of the three-level maser as a heat engine \cite{scovil2959three}.
In the following decades, a great deal of foundational work essential to the later development of quantum thermodynamics was performed \cite{RAlicki_1979,spohn2007irreversible}. For example, 
following the development of the Gorini–Kossakowski–Sudarshan–Lindblad equation for modelling open quantum systems \cite{davies1974,gorini1976,gorini1976completely,lindblad1976}, a number of individual expressions for the laws of nonequilibrium thermodynamics in the context of open quantum systems were derived \cite{ALICKI1976249,mcadory1977,spohn2007irreversible,RAlicki_1979,Spohn1978}.
In particular, H. Spohn and J. L. Lebowitz provided an expression for the second law of thermodynamics for nonequilibrium open quantum systems weakly coupled to thermal reservoirs in Ref.~\cite{spohn2007irreversible}, which was soon generalised by R. Alicki in a seminal work studying open quantum systems through the lens of heat engine operation \cite{RAlicki_1979}. 

Later, in a seminal work done by C. Jarzynski in 1997 \cite{CRPHYS_2007__8_5-6_495_0}, an equality was established between the fluctuations of the work distribution and the free energy difference for a classical system. Such a fluctuation relation may be understood as a generalisation of the second law of thermodynamics to nonequilibrium statistical mechanics \cite{deffner2019quantum}. This instigated an important series of works into fluctuation theorems within nonequilibrium statistical mechanics \cite{crooks1998,RevModPhys.81.1665}, which was soon extended to application in quantum systems \cite{kurchan2001quantum,PhysRevE.75.050102}. Study into quantum fluctuation relations is an important example of early work that laid the foundation for quantum thermodynamics, and remains an active area of research for both theoretical \cite{PhysRevE.85.051107,PhysRevA.102.043312,vinjanampathy2016quantum,watanabe2020quantum,RevModPhys.83.771,PhysRevLett.113.250601} and experimental investigation \cite{PhysRevLett.110.230602,PhysRevLett.110.230601,PhysRevLett.113.140601,PhysRevLett.115.190601,cheng2024experimental}.
However, it wasn't until the late 2000's that the field of quantum thermodynamics became well established \cite{vinjanampathy2016quantum,alicki2018introduction}.
This field has expanded rapidly in the past decade as a result of both theoretical advances, allowing for simulation of systems with greater complexity, and experimental progress, allowing for quantum heat engines and other thermal machines, such as quantum refrigerators \cite{PhysRevE.87.042131,Correa2014,10.1063/1.373503,Schmiedmayer_PRXQuantum} or quantum batteries \cite{PhysRevE.87.042123,PhysRevLett.122.210601}, to be realized in the laboratory \cite{rossnagel2016single,Ca-ion-spin-engine,Nitrogen-vacancy-heat-engine,bouton2021quantum,deffner2019quantum}.

Modern quantum thermodynamics is often interested in exploring how one may utilize quantum resources to boost the performance of thermal machines beyond their classical counterparts \cite{vinjanampathy2016quantum,deffner2019quantum}. This particular concept was first expressed in the early 2000's, with the work of  M. O. Scully \textit{et al}. in Refs.~\cite{scully2002afterburner} and \cite{scully2003extracting}, where they demonstrated the ability to boost the performance of an Otto and Carnot engine cycle, respectively, beyond their classical counterparts. In particular, their work on the nonequilibrium quantum Carnot cycle in Ref.~\cite{scully2003extracting} demonstrated how inducing quantum coherence in thermal atoms via interaction with a laser cavity can directly result in efficiency exceeding the traditional Carnot limit.
The capability of quantum thermal machines to outperform comparable classical devices, or even to perform new tasks not possible classically, is a key facet driving interest in the field \cite{vinjanampathy2016quantum,deffner2019quantum,kosloff2017quantum,kosloff2013quantum,kosfloff2014quantum}.

Remarkably, using Alicki's definitions for the first and second laws of thermodynamics in open quantum systems \cite{RAlicki_1979}, it may be shown that the maximum efficiency of a quantum Carnot cycle is unchanged from the classical value of $\eta_\mathrm{Carnot}\!=\!1-\frac{T_c}{T_h}$, where $T_c$ and $T_h$ are the temperatures of the hot and cold thermal baths, respectively \cite{vinjanampathy2016quantum}.
More generally, during the two thermalization strokes of an engine cycle, if the working fluid is allowed to fully equilibrate with external thermal baths at two distinct temperatures, engine performance becomes universally limited to operate within the conventional bounds on efficiency stipulated by classical thermodynamics \cite{SchroederD_ThermalPhysics,vinjanampathy2016quantum,gardas2015thermodynamic,niedenzu2015performance}.
In particular, it was noted by B. Gardas and S. Deffner in Ref.~\cite{gardas2015thermodynamic} that, if complete equilibration with external reservoir is assumed, such cycles are necessarily limited by the Carnot bound on efficiency when one accounts for the energy spent to maintain coherence and correlations between the working fluid and the environment. 
Thus, in order to demonstrate improvement on the limits of classical thermodynamics, it is necessary to operate in scenarios where the working fluid is not allowed to fully equilibrate with the external reservoirs. As an example of this, it was shown in Ref.~\cite{Rossnagel2014nanoscale} that utilizing squeezed thermal baths can have a beneficial impact on the performance of nonequilibrium engine cycles. In this work, they formulated a generalised expression for the efficiency at maximum power, a quantity which dates back to the first exploration of the endoreversible classical Carnot cycle by F. L. Curzon and B. Ahlborn in Ref.~\cite{curzon1975efficiency}, and demonstrated that squeezing the thermal bath could result in efficiency surpassing the standard Carnot limit.

The drive to realize uniquely \textit{quantum} thermal machines, in combination with the rapid advancement in control over various quantum platforms, has led to the realisation of single-particle quantum devices. Examples of these include single ion heat engines \cite{rossnagel2016single,Ca-ion-spin-engine}, engines realized within nitrogen vacancy centers in diamond \cite{Nitrogen-vacancy-heat-engine}, and more recently, single-atom impurities immersed in an ultra-cold atomic bath \cite{bouton2021quantum}. However, in order to fully access the breadth of possible quantum resources, and assess how they may be exploited to enhance current -- or even design new -- quantum thermal machines, one must apply the lens of thermodynamics to quantum many-body interacting systems \cite{mukherjee2024promises}.

Quantum many-body interacting systems are able to employ various quantum phenomena, such as entanglement \cite{brandao2008entanglement,funo2013thermo,alicki2013entanglement,hilt2009system}
, correlations \cite{oppenheim2002thermo,llobet2015extractable,Huber_2015}
, or quantum coherence \cite{narasimhachar2015low,Korzekwa2016extraction,lostaglio2015description,bernardo2020unraveling,kammerlander2016coherence,Cwiklinski2015limitations}, as a resource to enhance the performance of conventional thermodynamic devices. Further, many-body quantum heat engines have been shown to be capable of outperforming an ensemble of single-particle engines operating with the same resources \cite{jaramillo2016quantum}, or even performing entirely new tasks that would be impossible classically \cite{halpern2019quantum}.
Many-body interacting quantum heat engines are therefore at the forefront of both experimental and theoretical advancement, having been recently realized in the laboratory for the first time \cite{koch2022making,simmons2023thermodynamic}. For an excellent overview of the prospects of many-body quantum thermal machines, see the recent review Ref.~\cite{mukherjee2024promises}.

As is the case with heat engines introduced in classical thermodynamics \cite{SchroederD_ThermalPhysics,Callen_book}, the study of the performance of quantum heat engines is chiefly achieved in terms of heat flow and work output. In particular, the classical expressions for net work, efficiency, and power may be equivalently expressed in the context of quantum systems, and further may often, surprisingly, be bound by the same limits as found in the classical case. Indeed, both the classical Carnot efficiency \cite{carnot1890reflections} and the Curzon-Ahlborn efficiency at maximum work of an endoreversible engine cycle \cite{curzon1975efficiency} have been found to be generally applicable in the context of quantum thermodynamics if the systems are bound to operate between two reservoirs at fixed finite temperatures \cite{deffner2019quantum,vinjanampathy2016quantum,kosloff2013quantum}. Further, the plethora of 
engine cycles available in classical thermodynamics and nonequilibrium statistical mechanics are often available to be translated directly into quantum systems \cite{quan2007quantum}, a fact which we make use of when investigating the quantum Otto engine cycle in Chapters \ref{Chap:3}, \ref{Chap:6}, and \ref{Chap:7} of this thesis.

Though this thesis largely focuses on quantum thermal machines, the field of quantum thermodynamics is more broadly interested in how fundamental concepts from thermodynamics and statistical physics apply in the context of quantum systems \cite{kosloff2017quantum}.
Notably, in the late 1990's, two new classical fluctuation theorems were derived: Jarzynski's equality \cite{CRPHYS_2007__8_5-6_495_0}, and the more general Crook's fluctuation theorem, from which the Jarzynski equality can be derived \cite{crooks1998}. These relate the free energy difference between two states to the work done via an arbitrary protocol.
Though these were originally derived in the context of classical nonequilibrium statistical mechanics, the Jarzynski equality in particular may be derived in the context of quantum mechanics. Further, in the past two decades there has been an effort to test these two theorems in quantum systems.
Notably, the Crook's fluctuation theorem was recently experimentally validated for the case of a single nuclear spin using a nitrogen vacancy center in diamond by W. Cheng \textit{et al}. in Ref.~\cite{cheng2024experimental}. There, they tested this fluctuation relation using a two-point measurement protocol via non-demolition measurements, validating the Crook's fluctuation theorem via these two-point measurement protocols in quantum systems for the first time.

Additionally, understanding how closed quantum systems thermalise has been a significant fundamental question within quantum thermodynamics, and quantum physics more broadly \cite{vinjanampathy2016quantum,bloch2008many,polkovnikov2011colloquium}. Over the past two decades, there has been extensive investigation into quantum quench scenarios, which attempt to probe the relaxation of quantum systems following a sudden change in the system Hamiltonian. As time evolution in closed quantum systems is unitary, with the microscopic description being generally time-reversal invariant, this presents conceptual difficulties in explaining how such systems relax and reach a seemingly thermal equilibrium
state, and even what form the relaxed states take. Further, for integrable models, which possess an extensive number of conserved quantities, how such a set of conservation equations constrains thermalisation remains an interesting open question \cite{kollar2011gge,gring2012relaxation}, and is discussed briefly in Chapter \ref{Chap:2}, where we introduce generalised hydrodynamics \cite{castro2016emergent,Bertini_2016_Transport}.
Finally, we note that an important aspect of quantum thermodynamics is its ability to bring together experts from a wide range of disciplines, such as information theory, resource theory, AMO, and condensed matter physics to provide a variety of outlooks on fundamental quantum problems.
For an excellent overview of the breadth of topics included in quantum thermodynamics, see the review by S. Vinjanampathy and J. Anders in Ref.~\cite{vinjanampathy2016quantum}.

\section{Integrable quantum systems}

The history of quantum integrability dates back nearly as far as quantum mechanics itself. As noted in Ref.~\cite{Thacker-RMP}, the first nontrivial example of a many-body quantum integrable system was given by H. Bethe in 1931, who provided an exact solution, now commonly known as the Bethe ansatz, to the isotropic Heisenberg antiferromagnet model \cite{bethe1931theorie}.
The essence of Bethe's solution is its ability to reduce many-body dynamics to a product of two-body elastic interactions \cite{korepin1993}.
The method Bethe developed for constructing the exact eigenvectors of interacting quantum systems is central to quantum integrability, so much so that the term `Bethe ansatz solvable' is often synonymous with quantum integrability \cite{sutherland2004beautiful}, and it was later applied to solve a variety of physical models. Perhaps most famously, it was utilised in the explanation of anomalous behaviour observed in the resistance of metals at low temperatures, known as the Kondo problem \cite{kondo1964resistance,Wiegmann1981kondo}. Further, throughout the 1950's and 1960's this method was applied to formulate exact solutions to a wide variety of lattice models, such as the XXZ spin chain \cite{orbach1958linear,walker1959anti,yang1966one}, the six-vertex model \cite{sutherland1967exact}, the 1D Fermi Hubbard model \cite{Lieb1968Absence,LIEB20031}, and the XY spin chain \cite{LIEB1961407}.

An exact solution for a non-trivial \textit{continuum} model was first demonstrated for the uniform one-dimensional Bose gas with delta-function contact interactions. This was accomplished by M. Girardeau in 1960 \cite{girardeau1960relationship}, where he exploited a correspondence between a system of impenetrable (i.e. infinitely strongly interacting) bosons and a free Fermi gas. Through this, the $N$-body system simplifies to $N$ single body solutions for the wavefunction.
Three years later, E.H. Lieb and W. Liniger utilized the Bethe ansatz to find a solution for the general case of finite strength contact interactions \cite{Lieb-Liniger-I,Lieb-Liniger-II}. In two landmark papers, they completely described the ground state of the 1D Bose gas for a finite system and in the thermodynamic limit \cite{Lieb-Liniger-I}, and evaluated the fundamental excitations of the model, which take the form of particle and hole quasiparticles \cite{Lieb-Liniger-II}. For further details, we provide a brief review of the ground state of the 1D Bose gas, otherwise known as the Lieb-Liniger model, in Chapter \ref{Chap:2}.

Finite temperature solutions for the 1D Bose gas was first achieved in 1969 by C.N. Yang and C.P. Yang \cite{yang1969thermodynamics}. In their seminal paper on finite temperature thermodynamics of the repulsive 1D Bose gas, they developed the theory that is now commonly known as the thermodynamic Bethe ansatz (TBA), and otherwise known as Yang-Yang thermodynamics. Their solution for a thermal equilibrium state at temperature $T$ consisted of first utilizing the particle and hole excitations to formulate an expression for the entropy, $S$. The entropy of finite temperature integrable systems may be calculated from the total number of microstates that correspond to a state of approximately the same energy. Minimizing the Helmholtz free energy, $F\!=\!E\!-\!TS$, where $E$ is the total system energy, results in an integral equation commonly known as the Yang-Yang equation, and which may be numerically solved to describe the thermal equilibrium state. The importance of this theory is in its applicability more broadly to any Bethe ansatz solvable model; we review the mathematical formulation of the TBA in Chapter \ref{Chap:2}.

Despite the existence of exact solutions for the ground state and thermal equilibrium states, the dynamics of integrable models is often extremely nontrivial, or even computationally intractable. Integrable models possess an extensive set of conserved charges, which results in an infinite set of conserved charges in the thermodynamic limit. This is key to the formulation of their exact equilibrium solution. However, each conserved charge naturally implies the presence of a microscopic continuity equation, thus resulting in an infinite set of equations for solving their dynamics. Further, at hydrodynamic scale, if one assumes local relaxation to a thermal equilibrium state provided by the TBA, the presence of this infinite set of microscopic continuity equations results in an infinite set of coupled conservation equations for the large-scale hydrodynamics  \cite{Doyon-lectures}. One solution is truncating this series to a finite set of equations, which can result in conventional hydrodynamics (CHD), or Euler hydrodynamics, when truncated to three continuity equations typical of conventional theory of hydrodynamics for generic (nonintegrable) systems. However, this fails to faithfully model the dynamics of integrable systems, in particular for highly nonequilibrium scenarios as shown in Ref.~\cite{GHD_onatomchip}.

In 2016, two groups independently discovered a method for exactly solving the hydrodynamics of integrable systems \cite{bertini2016transport,Castro-Alvaredo_2016_Emergent}. Utilizing the fact that the TBA encodes the complete set of conserved charges through a continuous quasiparticle density function known as the root density, they provided a set of equations of motion for the hydrodynamic evolution of this root density, now commonly known as generalised hydrodynamics (GHD). In particular, they formulated an effective velocity that models the speed at which the quasiparticle excitations move when interacting with a dense background of other quasiparticles. Similar to how the Bethe ansatz simplifies many-body interactions to a product of two-body interactions \cite{korepin1993}, GHD simplifies the motion of a particle moving through a dense background to the motion of a single quasiparticle renormalized by these interactions through a product of two-body scattering events \cite{bouchoule2022generalized}. As this method is written in the language of the TBA, GHD is applicable to any TBA-solvable model, which includes quantum and classical models. We present an overview of the GHD framework in Chapter \ref{Chap:2}, focusing on the case of a uniform one-dimensional Bose gas.

Extensions and higher-order expansions for GHD were developed in the years following its discovery. In 2017, GHD was extended to systems with broken translational invariance \cite{doyon2017note}, which is notable as it broadens the applicability of GHD to models with weakly broken integrability, otherwise known as near-integrable models \cite{bastianello2021hydrodynamics}. Such systems eventually relax to a thermal equilibrium state described by the canonical Gibbs ensemble, as the infinite set of conservation equations no longer applies at late times. To model  this relaxation process, it was required to extend GHD to higher order in spatial derivative, which results in diffusive corrections to the base GHD equations of motion \cite{DeNardis_Diffusion_2018}. Dynamics in the presence of an external trapping potential was demonstrated by Bastianello \textit{et al}. to inevitably result in thermalisation when this diffusion term is present in Ref.~\cite{bastianello_thermalization_2020}, where they rigorously demonstrated the long-time relaxation of the quantum Newton's cradle to a thermal state. Further investigation into thermalization of a quantum Newton's cradle was performed by K. F. Thomas \textit{et al.} in Ref.~\cite{Thomas2021}, where they utilised the $c$-field SPGPE method to demonstrate thermalization of a more conventional quantum Newton's cradle setup under a Bragg pulse protocol \cite{kinoshita2006quantum}, in a weakly interacting 1D Bose gas in the quasicondensate regime.

The ability to simulate GHD in systems with external trapping potentials allowed for direct comparison with experiment, as 1D systems typically require confinement in their longitudinal direction. This was first done for a weakly interacting one-dimensional Bose gas in Ref.~\cite{GHD_onatomchip}, and soon after for a strongly interacting system in Ref.~\cite{malvania2021generalized}. In particular, in Ref.~\cite{GHD_onatomchip}, Schemmer \textit{et al}. demonstrated that GHD surpasses CHD in the description of highly nonequilibrium phenomena. This was most clearly shown for their experimental realization of a double-well release quench, where CHD failed to model the nonequilibrium dynamics, while GHD remained accurate in describing the system evolution. In Ref.~\cite{malvania2021generalized}, Malvania \textit{et al}. showed that GHD was capable of simulating the long time dynamics of a harmonic trap quench for a strongly interacting 1D Bose gas. Further, they demonstrated that GHD was capable of accurately modelling the dynamical variation of particle statistics under a 100-fold quench of the strength of the harmonic trap. As such, GHD has demonstrated the unique ability to accurately model the dynamics of near-integrable quantum systems across a broader range of parameters---from the weakly to strongly interacting regimes---than any other single theory. For more details on the benchmarking of GHD against alternative theories, see Chapter \ref{Chap:5}.

We further highlight a handful of extensions to GHD that have been developed in the years since its original discovery. Calculation of the dynamical generation of correlations between space-time separated points was first achieved for Euler-scale GHD in Ref.~\cite{Doyon_correlation_2018} (see also \cite{De_Nardis_2022}). Similarly for the zero-temperature fluid, quantum GHD was developed by Ruggiero \textit{et al}. in Ref.~\cite{ruggiero2021quantum} to understand the dynamics of correlations due to quantum fluctuations, which were incorporated via quantized fluctuations on top of the Euler-scale, i.e. non-diffusive, GHD background. Other extensions include incorporating atom losses via a Lindbladian open system formalism in Ref.~\cite{Bouchoule_AtomLoss_2020}, modelling anomalous diffusion in spin chains in Refs.~\cite{Bulchandani_2021}, investigating GHD in the presence of impurities in Ref.~\cite{rylands2023impurity}, and modelling dimensional crossover in a 1D Bose gas in Ref.~\cite{moller2020extension}. For further discussion on these topics, see the recent special issue introduced in Ref.~\cite{bastianello2022introduction}, and the recent perspective articles Ref.~\cite{doyon2023generalized,Kerr2023}.

\section{The one-dimensional Bose gas}

Low-dimensional systems are unique amongst physical models, largely due to the counter-intuitive properties which they often possess \cite{RevModPhys.83.1405,giamarchi2003quantum}. For one-dimensional (1D) systems, this is caused by the inevitability of interactions between particles; as noted in Ref.~\cite{giamarchi2003quantum}, 1D interacting models have no individual motion, the only type of motion is collective, as particles are unable to avoid each other. 
Further, there is a degree of universality amongst one-dimensional models, with the low energy, long-wavelength dyanamics of 1D interacting quantum systems being universally described by the Tomonaga-Luttinger liquid model \cite{tomonaga1950,Mattis2004,luttinger1963,Imambekov_1Dreview_2012}.
1D models are therefore of great interest to study both theoretically and, where one can find a real-world implementation, experimentally. Fortunately, in 1998, M. Olshanii derived a relationship between the theoretical model of a 1D Bose gas with contact interactions, and a 3D ultracold Bose gas with tight harmonic confinement in two radial directions \cite{Olshanii1998}, allowing for realisation of the 1D Bose gas in quasi-1D experimental setups. 

Rapid development of experimental platforms throughout the 1980's and 1990's led to the first realisation of Bose-Einstein condensation in the laboratory in 1995, for which the 2001 Nobel prize in physics was awarded to E. A. Cornell, C. Wieman, and W. Ketterle \cite{Davis1995BEC,anderson1995observation,cornell2002nobel,ketterle2002nobel}. Soon after, experimental realisation of quasi-1D gases were achieved in highly anisotropic traps \cite{gorlitz2001realization,greiner2001phase,schreck2001quasipure}.
Unlike its higher-dimensional counterparts, the 1D Bose gas generally lacks any sharp phase transition, or indeed a transition to a true Bose-Einstein condensate phase in the thermodynamic limit \cite{Pitaevskii_Stringari_book,Pethick_Smith_book,petrov2004low}. Instead, the 1D Bose gas in the weakly interacting regime undergoes a smooth crossover transition to a phase-fluctuating quasicondensate, and more generally it possesses a rich range of different physical regimes---from weak to strong interactions---with smooth crossovers in between as a function of dimensionless interaction strength and temperature \cite{Gangardt2003,kheruntsyan2003pair, kheruntsyan2005finite}. These regimes may be classified through Glauber's local second-order correlation function, which was first accomplished in a series of papers by D. M. Gangardt \textit{et al.} and K. V. Kheruntsyan \textit{et al.} in Refs.~\cite{Gangardt2003,kheruntsyan2003pair, kheruntsyan2005finite} (see also Ref.~\cite{petrov2004low} for a review of low-dimensional trapped gases). We give a brief review of this method  for characterising the parameter regimes of a 1D Bose gas in Chapter \ref{Chap:2} in the context of a uniform system.


Experimentally, there are two common setups capable of implementing the tight transverse trapping required to realise the low dimensionality of this system. The first is in a 2D optical lattice, where the ultracold gas is initially prepared in a 3D trap, and then adiabatically loaded into an optical potential consisting of a 2D array of 1D tubes \cite{Kinoshita2005,Haller2009,PhysRevA.83.031604,PhysRevLett.86.5413,kinoshita2006quantum,Kinoshita1125,malvania2021generalized,kinoshita2005local,Wilson1461}. Here, the intensity gradient of two counter-propagating lasers provides the means of confining the atoms. Measurements are performed by averaging over the array of 1D tubes, with each tube being an individual realisation of a 1D system, where atom numbers vary between tubes. Drawbacks to this method, however, include the fact that the requirement to realise an array of 1D Bose gases means that the possible presence of tunneling grows with increasing atom numbers, restricting this setup to low atom numbers. Additionally, the necessity of averaging over many independent tubes, each with their own atom number and hence each in a slightly different parameter regime, means measurement of fluctuations and correlations becomes challenging \cite{bouchoule2022generalized}. This method is therefore generally used to realise an array of strongly interacting particles in the Tonks-Girardeau regime, as the 1D dimensionless interaction strength, $\gamma$, is inversely dependent on the 1D atom number density, $\rho$ \cite{Lieb-Liniger-I}, which we note is not the case in 3D. It was in this experimental setup that Kinoshita \textit{et al.} first realised the quantum Newton's cradle in 2006 \cite{kinoshita2006quantum}, helping instigate, alongside work of Hofferberth \textit{et al.} in Ref.~\cite{hofferberth2007non}, both experimental and theoretical study into relaxation and thermalization in integragble and near-integrable systems that would flourish over the next two decades  \cite{rigol2008thermalization,PhysRevLett.93.142002,gring2012relaxation,PhysRevX.9.021027}.

The second method utilises what is commonly known as an atom chip setup \cite{PhysRevLett.100.090402,refId0,jacqmin2011sub,Armijo_ThreeBody_2010,PhysRevLett.83.3398,PhysRevLett.84.4749,reichel2011atom,Armijo2011,GHD_onatomchip,PhysRevLett.121.200401}. In this experimental setup, 1D Bose gases are realised through tight confinement provided by a parallel set of microwires which produce strong electric currents, in turn creating strong transverse magnetic fields.
Such systems are typically used to produce 1D Bose gases in the weakly interacting, quasicondensate regime, as they are capable of containing more atoms and achieving higher densities, lowering the dimensionless interaction strength, $\gamma$, mentioned above.
Further, such systems tend to have greater control over the type of longitudinal confinement employed, and are able to shape the longitudinal trapping potential effectively up to the level of a fourth order polynomial \cite{GHD_onatomchip}. As the atom chip setup produces only a single ultracold atomic cloud, one may use them to study the nature of fluctuations and correlations in such systems. The ability to study the nonequilibrium dynamics of coherence in a single 1D Bose gas therefore allowed for greater detail in studying the nature of thermalization within near-integrable systems \cite{schumm2007matterwave,hofferberth2007non,Schmiedmayer_1D_PRL_2010}.


Thanks in part to its integrability, the 1D Bose gas has a range of theoretical tools that are capable of simulating its dynamics. Notably, one can rely on the mean-field Gross-Pitaevskii equation to calculate the ground state and simulate nonequilibrium dynamics in the weakly interacting regime \cite{Gross,Pitaevski,Pitaevskii_Stringari_book,Pethick_Smith_book}. This may be extended to finite temperatures via the classical $c$-field method known as the stochastic projected Gross-Pitaevskii equation (SPGPE) \cite{gardiner2002stochastic,gardiner2003stochastic,rooney2012stochastic,Bradley2015,spgpe}. We make extensive use of this particular method in Chapters \ref{Chap:Maxwell}, \ref{Chap:5}, and \ref{Chap:7} of this thesis, and therefore provide a mathematical introduction to it in Chapter \ref{Chap:2}. Other $c$-field techniques, such as the positive $P$ representation, and the truncated Wigner method which is used in Chapter \ref{Chap:5}, are utilised for describing the low temperature weakly interacting system in the presence of quantum fluctuations. We refer to the supplementary material of  Ref.~\cite{whatisqushock} for further details on these particular methods in the context of out-of-equilibrium dynamics of the 1D Bose gas.

For simulating the dynamics of the 1D Bose gas in the strongly interacting regime, one may utilize the Bose-Fermi mapping, first discovered by M. Girardeau \cite{girardeau1960relationship} as mentioned in the previous subsection. This maps the ground state of the hard-core, or infinitely strongly interacting, 1D Bose gas to a noninteracting, or free, Fermi gas, which itself is exactly solvable in terms of single-particle wavefunctions. From this, one may exactly solve for the dynamics by independently evolving each eigenstate of the non-interacting problem \cite{Girardeau2000,Girardeau-Wright-2000,Yukalov_2005,PhysRevLett.97.100402,GangardtMinguzziExact}. One may extend this method beyond the ground state, as was done by Y. Y. Atas \textit{et al.} in Ref.~\cite{atas_TG_exact}, where finite-temperature dynamics were again reduced to a single-particle basis with occupations dictated by the Fermi-Dirac distribution.

Finally, we highlight the time-dependent density-matrix renormalisation group (tDMRG) approach \cite{PhysRevLett.69.2863,PhysRevLett.93.076401,nakatani2018matrix}, and related NRG-TSA-ABACUS method \cite{PhysRevLett.98.147205,largescale_Doyon}, which may utilise matrix product states (MPS), or infinite matrix product states (iMPS), to accurately simulate the zero-temperature dynamics of the 1D Bose gas through a discretized version of the system Hamiltonian, which maps on to a Bose-Hubbard model. However, although this method is capable of simulating the Lieb-Liniger model across the entire range of interaction strengths, it is currently limited to simulating zero-temperature systems with relatively low particle numbers, being restricted to systems of up to only $\sim\!100$ particles. Again, for details on this method we refer to the supplementary material of Ref.~\cite{whatisqushock}, where they provide an overview of iMPS, TWA, and positive-$P$ methods; for all other numerical methods utilized in this thesis, such as GHD, GPE, and SPGPE, see Chapter \ref{Chap:2}.

\section{Motivation and scope of this thesis}

Research into quantum thermodynamics can largely be broken down into the drive to answer two complimentary questions: (1) how do thermodynamic principles arise within and from quantum mechanics; and (2) how can quantum properties be beneficially exploited within the framework of quantum heat engines? The work presented in this thesis utilizes integrable quantum systems, with a particular focus on the 1D Bose gas, in order to explore aspects of these questions.
In particular, we leverage exact equilibrium solutions and the large variety of numerical tools available for the 1D Bose gas in order to study operation of various quantum thermal machines, including their real-time, finite-temperature dynamics, and provide new means of calculating an important thermodynamic quantity: entropy.

The remaining chapters of this thesis are organised as follows.

In Chapter \ref{Chap:2} we provide the necessary background theory for quantum thermodynamics and integrable quantum systems, focusing on the mathematical tools and numerical techniques used throughout this thesis. We provide a brief introduction to the 1D Bose gas, outlining its exact ground state and thermal equilibrium solutions available through the Bethe ansatz and thermodynamic Bethe ansatz, respectively. We then introduce Glauber's local second-order correlation function, which may be utilised to define the rich parameter space of the 1D Bose gas as a function of interaction strength and temperature. Following this, we provide a short introduction to the $c$-field stochastic projected Gross-Piteavskii equation (SPGPE) method, before giving a detailed description of the recently developed generalised hydrodynamic (GHD) framework. In this description of GHD, we focus on providing an intuitive understanding of how GHD works at the level of quasiparticle excitations, through which GHD may be viewed as free evolution renormalized by moving through a background of other quasiparticles \cite{bouchoule2022generalized}.

In Chapter \ref{Chap:3}, we introduce and explore a quantum thermal device whose performance may be expressed in terms of Glauber's local second-order correlation function. The quantum Otto cycle introduced here is driven by a sudden quench of interparticle interaction strengths for models with $s$-wave interactions, such as the 1D Bose gas. We then utilize the formulae derived to explore engine operation, whose performance is expressed entirely in terms of thermal equilibrium properties of the working fluid. We then extend this analysis to the three other thermodynamic devices, that of a heater, accelerator, and refrigerator, analysing their operation as a function of interaction strength and temperature ratios. As analytic solutions are available in each asymptotic parameter regime of the 1D Bose gas, we use them to exactly derive the limiting behaviour of this quantum thermal machine, in addition to providing analytic expressions for both the net work and efficiency of the engine cycle in each parameter regime.

In Chapter \ref{Chap:Maxwell}, we further our investigation into the role that Glauber's second-order correlation function plays in the thermodynamics of the uniform 1D Bose gas. In particular, we derive a new thermodynamic Maxwell relation which relates the atom-atom correlations of a uniform 1D Bose gas, which are experimentally measurable, to the entropy, which remains challenging to access experimentally.
We then apply this Maxwell relation to calculate the entropy of a finite-temperature 1D Bose gas using the classical $c$-field SPGPE method, which, similar to experiment, had been previously challenging to calculate.
These calculations may be considered as a numerical experiment, and therefore stand as a proof-of-principle demonstration for the utility of this method, which may be employed to deduce the entropy of a quantum gas from the measured atom-atom correlations.


In Chapter \ref{Chap:5}, we perform benchmarks for the recently developed theory of GHD, which is capable of modelling the nonequilibrium dynamics of integrable systems, such as the uniform 1D Bose gas, and near-integrable systems, where the integrability breaking we consider is provided by an external trapping potential.
We investigate a set of highly nonequilibrium scenarios for the case of a 1D Bose gas at various values of both interaction strength and temperature, benchmarking GHD against an array of alternative theoertical methods. In particular, we study how GHD performs under expansion from a localized density perturbation, which results in a dispersive shock wave for the case of a density bump, and a train of grey solitons for a density dip. We observe that, while GHD performs very well at sufficiently high temperatures and strong interactions, it cannot capture the short-range phenomena that arises in the case of a shock wave in low temperature weakly interacting systems. However, we find that GHD agrees well with a coarse-grained approximation of these alternative theoretical methods, where the specific coarse-graining procedure that we propose and apply is based on convolution averaging that approximates finite imaging resolution in ultracold atom experiments.
We further contrast higher-order Navier-Stokes GHD with the classical $c$-field SPGPE method for the case of a quantum Newton's cradle. Here, we find that GHD demonstrates excellent agreement with the highly nonequilibrium dynamics present in SPGPE at short times, and further agrees with both the rate of relaxation, and the final relaxed state at long times.


In Chapter \ref{Chap:6}, we utilize GHD, along with theoretical methods developed in Chap.~\ref{Chap:3}, to study the operation of an experimentally realizable quantum Otto engine cycle, fuelled by diffusive contact realized through particle transfer with external reservoirs.
We numerically evaluate the performance of this engine cycle in both the weakly interacting quasicondensate regime, and in the strongly interacting near-Tonks-Girardeau regime of the 1D Bose gas. We make connections with previous results derived by Keller \textit{et al}. in Ref.~\cite{keller2020feshbach} for the case of an adiabatic engine cycle at zero temperature, which in turn provides an upper bound to the net work and efficiency of this engine cycle at nonzero temperatures.
Further, we apply GHD to study the finite-time operation, demonstrating how the scaling between the two idealised limits of sudden and quasistatic operation is dictated by timescales associated with longitudinal breathing modes excited by the interaction strength quench.

In Chapter \ref{Chap:7} we employ the $c$-field SPGPE method to investigate the same Otto engine cycle introduced in Chapter \ref{Chap:6}, here utilizing the capability of SPGPE to simulate tunnel-coupled 1D Bose gases to model all four strokes of the thermochemical engine cycle, rather than only the unitary work strokes as was done in Chapter \ref{Chap:6}. In particular, we numerically simulate a set of three harmonically trapped 1D Bose gases, defining one as the working fluid and the remaining two as the external reservoirs, which are alternately coupled to the working fluid to simulate the two thermalization strokes. The ability to simulate the entire Otto cycle is often not possible, as exact simulation of the thermalization strokes can be difficult to implement.
Through our analysis, we demonstrate that operation of a sudden quench protocol for the work strokes demonstrates a favourable trade-off in terms of power and efficiency when contrasted with slower, and even quasistatic operation.

Lastly, in Chapter \ref{Chap:Conclusion} we provide a summary of our main results, and highlight outlooks for possible future work.

%% file: Chapter2/Chapter2.tex

\chapter[Theoretical background]{Theoretical background}
\label{Chap:2}	
\pagestyle{headings}
\textit{In this chapter we introduce the background theory, computational tools, and mathematical formalism that are utilised throughout this thesis, with a focus on the uniform 1D Bose gas described by the integrable Lieb-Liniger model. We briefly outline the history of exact solutions for this model, provided through the Bethe ansatz for the ground state, and the thermodynamic Bethe ansatz (TBA) for thermal equilibrium states. We then introduce Glauber's local second-order correlation function, which provides a means to characterize the parameter regimes of the 1D Bose gas. We additionally demonstrate how solutions for the uniform 1D Bose gas may be extended, via a local density approximation, to systems with broken translational invariance through external trapping potentials.}

\textit{Sections \ref{Sec:SPGPE} and\ref{Sec:GHD} focus on describing the two main tools utilized to simulate the thermodynamics and real-time dynamics of the 1D Bose gas. First, in Section \ref{Sec:SPGPE}, we give an overview of the stochastic projected Gross-Pitaevskii equation (SPGPE), a method capable simulating equilibrium states and nonequilibrium dynamics for the weakly interacting finite-temperature 1D Bose gas. Finally, in Section \ref{Sec:GHD}, we give a detailed overview of the recently developed theory of generalized hydrodynamics (GHD), a theory capable of simulating the large-scale nonequilibrium dynamics of integrable and near-integrable systems for arbitrary interactions strengths and temperatures. }






\section{The one-dimensional Bose gas}
\label{Sec:1DBoseGas}

The physical model most often considered throughout this thesis is the ultracold 1D Bose gas with repulsive contact interactions. The uniform case, commonly known as the Lieb-Liniger model, is a paradigmatic integrable quantum model, described by the following second-quantized Hamiltonian \cite{Lieb-Liniger-I,Pitaevskii_Stringari_book}
\begin{equation}\label{eq:LL_Ham}
    \begin{split}
        \hat{H} =- \frac{\hbar^2}{2m} \int \!dz\, \hat{\Psi}^\dagger \frac{\partial^2 \hat{\Psi}}{\partial z^2} 
        + \frac{g}{2} \int \!dz\, \hat{\Psi}^\dagger \hat{\Psi}^\dagger \hat{\Psi} \hat{\Psi}.
    \end{split}
\end{equation}
Here, $\hat{\Psi}^\dagger(z)$ and $\hat{\Psi}(z)$ are the boson creation and annihilation field operators, respectively, and obey the canonical bosonic commutation relations 
\begin{equation}
    \left[\hat{\Psi}(z),\hat{\Psi}(z')\right]=\left[\hat{\Psi}^\dagger(z),\hat{\Psi}^\dagger(z')\right]=0, \quad \left[\hat{\Psi}(z),\hat{\Psi}^\dagger(z')\right]=\delta(z - z').
\end{equation}
Furthermore, $m$ is the bosonic mass, and $g$ is the one-dimensional interaction strength which is taken to be positive ($g\!>\!0$) for repulsive interactions. For an experimentally realistic, quasi-1D system \cite{petrov2004low}, with strong transverse harmonic confinement of frequency $\omega_\perp$, the interaction strength may be calculated as a product of this transverse harmonic frequency and the 3D s-wave scattering length $a_s$, as $g\!=\!2\hbar \omega_\perp a_s$, away from confinement-induced resonances \cite{Olshanii1998}. As this system is translationally invariant, the 1D particle number density is given by $\rho\!=\!N/L$, where $N$ is the total atom number and $L$ is the system length.

\subsection{Ground state solutions}\label{sec:groundstate}
Historically, the first exact solution for the ground state of the 1D Bose gas model was provided by Girardeau, who treated the case of hard-core interactions, corresponding to $g\!\to\!\infty$ in the Hamiltonian \eqref{eq:LL_Ham}. In this limit, there is an exact correspondence for a number of key observables with those of a free fermionic model, as provided by the Bose-Fermi mapping \cite{girardeau1960relationship}. 
This was soon extended by Lieb and Liniger \cite{Lieb-Liniger-I}, who found the exact eigenstates of the model for finite $g$, describing its ground state properties via application of the Bethe ansatz, as mentioned in Chapter \ref{Chap:Intro}.
Through dimensional analysis, they demonstrated that the ground state energy may be written in terms of a single dimensionless interaction strength parameter, commonly known as the Lieb-Liniger parameter, given by
\begin{equation}\label{eq:lieb_parameter}
    \gamma = \frac{mg}{\hbar^2 \rho}. 
\end{equation}
Importantly, at $T\!=\!0$, the ground state solution to the Lieb-Liniger model is defined entirely by this single parameter, a solution that consists of a summation of a set of $N$ plane waves \cite{Lieb-Liniger-I}. This summation extends over all permutations of the plane wave momentum, with each term in the sum weighted by a coefficient that may be computed from a transcendental algebraic equation, which itself arises from the boundary condition provided by the delta function interaction in Eq.~\eqref{eq:LL_Ham}. Notably, this form of solution is general to all Bethe ansatz integrable models, with the particular form of algebraic equation dependent on the model in question \cite{bethe1931theorie,giamarchi2003quantum,korepin1993}.

A key feature of the dynamics of integrable models is the factorization of $N$-body scattering events into a product of two-body scattering processes \cite{sutherland2004beautiful,korepin1993}. In particular, for the model under consideration, the delta-function interaction results in nondiffractive scattering, where the outgoing wavefunctions from the scattering event match those incoming multiplied by a symmetric phase shift \cite{sutherland2004beautiful}. This implies an overall conservation of each \textit{individual} velocity, hereafter referred to as the rapidities. Further, the delta function interaction results in each of these rapidities being unique \cite{korepin1993}, and the single dimensionality of the model allows for simple labelling of the particles \cite{giamarchi2003quantum}. Hence, we may track the particle with rapidity $\lambda$, and refer to this tracer particle as the quasiparticle of the model. This concept is vital to the derivation and exploration of generalized hydrodynamics presented in Section \ref{Sec:GHD}.

As each rapidity is a conserved quantity, their sum is also conserved, implying a conservation of total momentum,
\begin{equation}
    P = m\sum_{\alpha=1}^N \lambda_\alpha,
\end{equation}
where $\lambda_\alpha$ is the rapidity of the particle labelled $\alpha$. Further, as each \textit{individual} rapidity is conserved, one may take any power of these rapidities in the summation above and be ensured that this quantity is also conserved. This implies an extensive set of conserved quantities. Two common examples of this include the energy,
\begin{equation}
    E = \frac{m}{2}\sum_{\alpha=1}^N \lambda_\alpha^2,
\end{equation}
and the total particle number,
\begin{equation}\label{eq:particle_number}
    N = \sum_{\alpha=1}^N \lambda_\alpha^0.
\end{equation}
To simplify notation, we refer to any conserved charge of the Lieb-Liniger model with power $i$ as $Q_i$, where
\begin{equation}
    Q_i = \sum_{\alpha=1}^N \lambda_\alpha^i.
\end{equation}
Here, we have omitted the physical prefactor for simplicity. These may be re-introduced for calculation of physical observables (i.e. $N\!=\!Q_0$, $P\!=\!mQ_1$, $E\!=\!(m/2)Q_2$, ...). Importantly, extensivity of this set of conserved charges directly implies the integrability of the system, which requires $N$ independent conserved charges for an $N$-body system \cite{doi:10.1142/0858}.

In the thermodynamic limit, $N\!\to\!\infty$, $L\!\to\!\infty$, $\rho\!=\!N/L\!=\!\mathrm{const.}$, the density of the quasiparticle rapidities may be approximated by a continuous distribution function, $f_s(\lambda)$ \cite{Lieb-Liniger-I}.
This distribution was first shown in Ref.~\cite{Lieb-Liniger-I} to be obtainable via solution to the following Fredholm integral equation,
\begin{equation}\label{eq:f_s_T0}
    2 \pi f_s(\lambda) = 1 + \int_{-\lambda_F}^{\lambda_F} d\theta \varphi(\lambda\!-\!\theta)f_s(\theta),
\end{equation}
where the kernel of this integral equation, $\varphi$, which is known as the scattering shift, results from the delta-function boundary condition mentioned above, and is expressed as
\begin{equation}\label{eq:scattering_shift}
    \varphi(\lambda) = \frac{2mg}{g^2 + (\hbar \lambda)^2}.
\end{equation}
This scattering shift is directly derived from the phase shift incurred through two-body scattering \cite{sutherland2004beautiful}, and may alternatively be derived from the scattering picture for the elastic scattering of a two-body system. In particular, in the asymptotic limit following a two-body scattering event, this kernel is simply the derivative of the phase shift incurred.

One may utilise this quasiparticle distribution, $f_s$, to calculate the particle density,
\begin{equation}\label{eq:gs_density}
    \rho = \int_{-\lambda_F}^{\lambda_F} d\lambda f_s(\lambda),
\end{equation}
where the integration limits, $\pm \lambda_F$, are determined by the particle number density, which is an externally controlled parameter. This also allows for expression of the total energy as,
\begin{equation}\label{eq:gs_energy}
    E = \frac{m}{2} L\int_{-\lambda_F}^{\lambda_F} d\lambda f_s(\lambda) \lambda^2, 
\end{equation}
and more generally, any conserved charge $Q_i$,
\begin{equation}\label{eq:Q_thermolim}
    Q_i = L \int_{-\lambda_F}^{\lambda_F}d\lambda f_s(\lambda) \lambda^i .
\end{equation}
This set of conserved charges, as it was extensive for the $N$-body system introduced, is infinite in the thermodynamic limit. 
The presence of an infinite set of conserved charges serves to thoroughly constrain the evolution of integrable systems in the thermodynamic limit. This results in a highly non-trivial dynamics, where one must effectively solve an infinite set of coupled conservation equations \cite{Doyon-lectures}. However, the solution to this was recently expressed in 2016 with the discovery of the generalized hydrodynamic method \cite{bertini2016transport,castro2016emergent}. For further discussion see Section \ref{Sec:GHD}.

\begin{figure*}[!tbp]
\begin{center}
   \includegraphics[width=8.5cm]{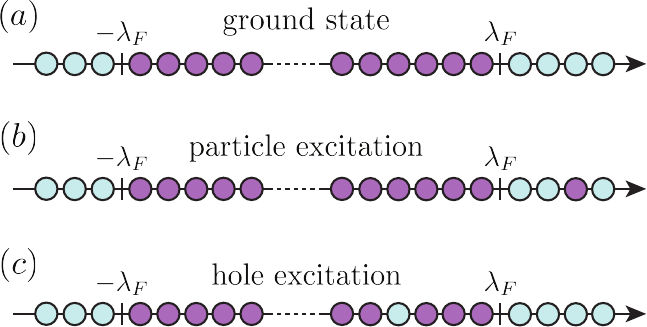}   
   \caption{A cartoon illustration of the Lieb-Liniger ground state, along with the two distinct types of quasiparticle excitation. In panel (a), we illustrate the ground state of the Lieb-Liniger model, where particles occupy rapidities up to the cutoff $\lambda_F$, which may be calculated through the fixed density condition given in Eq.~\eqref{eq:gs_density}. Excitations above the ground state come in two forms: particle excitations, where a single quasiparticle is added to the state with rapidity exceeding the cutoff, and hole excitations, where a single quasiparticle is removed from the ground state.
   }
  \label{fig:Quasiparticle_Excitations}
  \end{center}
\end{figure*}

\subsection{Thermal equilibrium solutions}\label{sec:tba}
For the majority of this thesis, we are interested in finite temperature solutions to the Lieb-Liniger model. This was first provided by Yang and Yang in their seminal 1969 paper \cite{yang1969thermodynamics}, where they developed the thermodynamic Bethe ansatz, otherwise known as Yang-Yang thermodynamics. This method utilises the two types of excitation present above the ground state of the Lieb-Liniger model, shown in Fig.~\ref{fig:Quasiparticle_Excitations}. These are particle excitations, corresponding to adding a quasiparticle with rapidity, $\lambda\!>\lambda_F$, above those contained in the ground state distribution as shown in Fig/~\ref{fig:Quasiparticle_Excitations}(b), and hole excitations, where a quasiparticle is removed from the ground state solution as shown in Fig/~\ref{fig:Quasiparticle_Excitations}(c).

In the thermodynamic limit, these excitations are described by the particle density, or root density, $f_p(\lambda)$, and the hole density, $f_h(\lambda)$. Together, these sum to the \textit{total} density of quasimomenta, $f_s(\lambda) \!=\! f_p(\lambda) \!+\! f_h(\lambda)$, used previously for the ground state solutions.
Through the introduction of these distribution functions, we may introduce the filling factor, 
\begin{equation}\label{eq:filling_function}
n(\lambda) = \frac{f_p(\lambda)}{f_p(\lambda) + f_h(\lambda)}= \frac{f_p(\lambda)}{f_s(\lambda)}.
\end{equation}
This dimensionless function expresses the fraction of possible excited states that are filled with particle excitations. 

One may express the total quasiparticle density, $f_s(\lambda)$, in the finite temperature thermodynamic limit as a smooth function with no rapidity cutoff \cite{yang1969thermodynamics,kheruntsyan2005finite}, achieved through an extension of the Fredholm integral equation introduced previously in Eq.~\eqref{eq:f_s_T0},
\begin{equation}
    2 \pi f_s(\lambda) = 1 + \int_{-\infty}^\infty d \theta\varphi(\lambda - \theta) f_p(\theta) . 
\end{equation}
We may make use of the filling function, defined in Eq.~\eqref{eq:filling_function}, to re-express the above integral relation in an alternate form,
\begin{equation}\label{eq:total_density}
     f_s(\lambda) = \frac{1}{2\pi} + \int_{-\infty}^\infty \frac{d \theta}{2\pi} \varphi(\lambda - \theta) n(\theta) f_s(\theta)  . 
\end{equation}
Though this is simply a re-expression of the above formula, Eq.~\eqref{eq:total_density} is directly relevant to the various formulas and expressions utilized in Section \ref{Sec:GHD}.
As was the case for the ground state thermodynamic limit, the conserved charge $Q_i$ may be expressed through a simple integral relation. However, for the finite temperature thermodynamic state, the integral is over the occupied quasiparticle states, given by the root density,
\begin{equation}\label{eq:Q_observables}
    Q_i = L \int_{-\infty}^{\infty} d\lambda f_p(\lambda) \lambda^i.
\end{equation}

In the thermodynamic limit at finite temperature, if given a state described by fixed $f_s$ and $f_p$, one may find many sets of microscopic states that are consistent with these distributions \cite{yang1969thermodynamics}.
In particular, for a given interval $d\lambda$, the total number of quasiparticle excitations is $L f_p d\lambda$, with $Lf_s d\lambda$ vacancies for these excitations to fill. The total number of microstates that are consistent with a given $d\lambda$ is thus given by
\begin{equation}
    \frac{[f_s(\lambda) L d\lambda]!}{[f_p(\lambda) L d\lambda]! [f_h(\lambda) L d\lambda]!},
\end{equation}
where $L f_h d\lambda$ is the number of unfilled, or `hole', states.
The logarithm of the total number of states gives a contribution to the overall entropy. Thus, integrating gives the entropy density of the uniform system,
\begin{equation}
    s = \frac{S}{L} = k_B \int_{-\infty}^\infty d\lambda f_s \left[ (1-n) \ln (1-n) - n \ln n \right] ,
\end{equation}
where we have utilised the filling factor $n(\lambda)$ for simplicity of expression.

Minimizing the total free energy, $F\!=\!E\!-\!TS$, for a fixed temperature $T$, while constrained by the fixed total particle number, $N$, may be achieved through the method of Lagrange multipliers, where $\mu$ is the multiplier associated with the total particle number. Following Ref.~\cite{yang1969thermodynamics}, one arrives at an additional integral equation,
\begin{equation}
    -\mu + m\lambda^2 + k_B T \ln \frac{f_p(\lambda)}{f_h(\lambda)} - \frac{k_B T}{2\pi} \int_{-\infty}^{\infty} d\theta \varphi(\lambda - \theta) \ln \left(1 + \frac{f_p(\theta)}{f_h(\theta)} \right)= 0
\end{equation}
where the Lagrange multiplier, $\mu$, can be shown to coincide with the chemical potential \cite{yang1969thermodynamics}. This is more commonly expressed as an integral equation for the excitation spectrum, $\varepsilon(\lambda)= k_B T \ln(f_h/f_p)$,
\begin{equation}\label{eq:yangyang_integral}
    \varepsilon(\lambda) = -\mu + \lambda^2 - \frac{k_B T}{2\pi} \int_{-\infty}^\infty d\theta \varphi(\lambda - \theta) \ln \left(1 + e^{-\varepsilon(\theta)/k_B T} \right) ,
\end{equation}
commonly known as the Yang-Yang integral equation. Further, as first proved in their original paper on the thermodynamic Bethe ansatz, this integral equation may be solved by a process of iteration \cite{yang1969thermodynamics}.

The solution to the integral equation Eq.~\eqref{eq:yangyang_integral} fully characterizes the thermal equilibrium state of temperature $T$.
In particular, the excitation energy may be inverted to obtain the filling factor, $n(\lambda) \!=\! \left(1+\exp(\varepsilon(\lambda)/k_B T)\right)^{-1}$. Using the first integral equation for $f_s(\lambda)$, given in Eq.~\eqref{eq:total_density}, the root density $f_p(\lambda)$ is obtained by $f_p(\lambda)\!=\!f_s(\lambda)n(\lambda)$. From this, all observables may be calculated through Eq.~\eqref{eq:Q_observables}. We note that, as there is a bijection between the filling factor, $n(\lambda)$, the root density, $f_p(\lambda)$, the exictation energy, $\varepsilon(\lambda)$, and even the hole density, $f_h(\lambda)$, the TBA treatment may be expressed entirely in terms of any of these functions, a fact which will be utilized in Section \ref{Sec:GHD}.

Finally, after obtaining the exact solution for the quasiparticle distribution function, one may express the Helmholtz free energy via,
\begin{equation}
    \frac{F}{N} = \frac{E - k_BTS}{N} = \mu - \frac{k_B T}{2 \pi \rho}  \int_{-\infty}^\infty \ln \left( 1 + e^{-\varepsilon(\lambda)/k_B T} \right) d\lambda,
\end{equation}
which is employed in the next Section on calculating Glauber's local second-order correlation function for thermal equilibrium states.

\subsection{Glauber's local second-order correlation function}\label{sec:g2_glauber}

The normalized two-point particle-particle correlation  is  defined in terms of the field operators as the expectation value of a normally-ordered product of two density operators, $\hat{\rho}(z)=\hat{\Psi}^\dagger(z)\hat{\Psi}(z)$ and $\hat{\rho}(z')=\hat{\Psi}^\dagger(z')\hat{\Psi}(z')$: 
\begin{equation}
    g^{(2)}(z,z') = \frac{\langle \hat{\Psi}^\dagger(z)\hat{\Psi}^\dagger(z')\hat{\Psi}(z')\hat{\Psi}(z) \rangle}{\rho(z)\rho(z')}.
\end{equation}
In other words, the pair correlation $g^{(2)}(z,z')$ is a normalized and normally-ordered density-density correlation function. It is normalized to the product of mean densities $\rho(z)=\langle \hat{\rho}(z)\rangle$ and $\rho(z')=\langle \hat{\rho}(z')\rangle$ at points $x$ and $x'$ so that for uncorrelated particles (with $\langle \hat{\Psi}^\dagger(z)\hat{\Psi}^\dagger(z')\hat{\Psi}(z')\hat{\Psi}(z) \rangle=\langle \hat{\Psi}^\dagger(z)\hat{\Psi}(z) \rangle \langle \hat{\Psi}^\dagger(z')\hat{\Psi}(z')\rangle$), one has $g^{(2)}(z,z')=1$. For values of $g^{(2)}(z,z')\neq 1$, the pair correlation characterizes an enhanced ($g^{(2)}(z,z')> 1$) or suppressed ($g^{(2)}(z,z')< 1$) probability of finding two particles at positions $z$ and $z'$, respectively, compared to uncorrelated particles.

Due to the translational invariance of the uniform system that we are considering, where $\rho(z')\!=\!\rho(z)\!=\!\rho$, the above pair correlation $g^{(2)}(z,z')$ can only depend on the relative distance $|z-z'|$ between the two particles, i.e., $g^{(2)}(z,z')\!=\!g^{(2)}(|z-z'|)$.  The local or the same-point ($z\!=\!z'$) correlation then corresponds to
\begin{equation}\label{eq:g2-definition}
    g^{(2)}(0)= \frac{\langle \hat{\Psi}^{\dagger}(z)\hat{\Psi}^{\dagger}(z) \hat{\Psi}(z)\hat{\Psi}(z)\rangle }{\rho^2}.
\end{equation}  
The thermodynamic solution for the ground state energy, given in Eq.~\eqref{eq:gs_energy}, together with the Hellmann-Feynman theorem, allows for calculation of the normalized local second-order correlation function at $T\!=\!0$,
\begin{equation}
    \frac{d E}{dg} = \bigg\langle \frac{d\hat{H}}{dg} \bigg\rangle = \frac{L}{2}\langle \hat{\psi}^\dagger(z) \hat{\psi}^\dagger(z) \hat{\psi}(z) \hat{\psi}(z) \rangle,
\end{equation}
meaning
\begin{equation}
    g^{(2)}(0) = \frac{d E}{dg}.
\end{equation}

\begin{figure*}[!tbp]
\begin{center}
   \includegraphics[width=10cm]{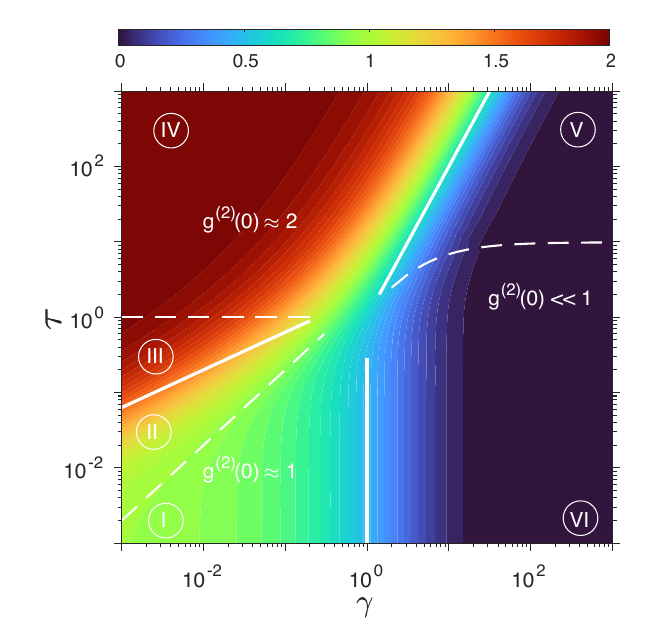}   
   \caption{Atom-atom correlations, described by Glauber's $g^{(2)}(0)$ correlation function, for the uniform 1D Bose gas evaluated using the exact Yang-Yang TBA \cite{kheruntsyan2003pair,yang1969thermodynamics}. The correlation function, which is dependent on dimensionless interaction strength, $\gamma$, and dimensionless temperature, $\tau$, is shown as a contour diagram, in which we also show the crossover boundaries (white solid and dashed lines) between the different asymptotic analytic regimes \cite{kheruntsyan2003pair}.
   }
  \label{fig:g2_parameter_space_simple}
  \end{center}
\end{figure*}

Calculation of the local second-order correlation function from thermal equilibrium states at $T\!\neq \!0$, provided by solution to the Yang-Yang equation given in Eq.~\eqref{eq:yangyang_integral}, may likewise be achieved through application of the Hellmann-Feynman theorem applied to the Hemholtz free energy $F$. In the canonical formalism, the partition function $Z(T,N,L,g)$ can be written in terms of either the Helmholtz free energy $F$ or the Hamiltonian $\hat{H}$ via $Z\!=\! \exp(-F/k_BT) = \mathrm{Tr}\exp(-\hat{H}/k_BT)$. By differentiating the Helmholtz free energy $F(T,N,L,g)\!=\!-k_BT\ln Z$ with respect to the interaction strength $g$, at constant $N$, $L$, and $T$, one finds that \cite{Kheruntsyan2003}
\begin{equation}
 \left(\frac{\partial F}{\partial g}\right)_{T,N,L} \!= \frac{1}{Z} \mathrm{Tr}\left(e^{-\hat{H}/k_BT} \frac{\partial \hat{H}}{\partial g}\right) = \frac{L}{2} \bigg\langle \hat{\Psi}^{\dagger} \hat{\Psi}^{\dagger} \hat{\Psi} \hat{\Psi}\bigg\rangle,
\end{equation}
and hence
\begin{equation}\label{eq:g2-canonical}
    g^{(2)}(0) = \frac{2}{L\rho^2} \left(\frac{\partial F}{\partial g}\right)_{T,N,L}. 
\end{equation}
This relationship between the local pair correlation and the Helmholtz free energy is what was used in Ref. \cite{Kheruntsyan2003} to calculate the $g^{(2)}(0)$ function using the exact Yang-Yang TBA \cite{Yang-Yang} solution for $F$, as a function of the dimensionless interaction strength $\gamma$, defined in Eq.~\eqref{eq:lieb_parameter}, and the dimensionless temperature $\tau$, defined via
\begin{equation}
    \tau = \frac{2mk_BT}{\hbar^2 \rho^2}.
\end{equation}
We note here that these two dimensionless parameters completely characterize the thermodynamic properties of a uniform 1D Bose gas \cite{Yang-Yang, Kheruntsyan2003}.


The 1D Bose gas, which notable contains no phase transition to a Bose-Einstein condensate \cite{Pitaevskii_Stringari_book,Pethick_Smith_book}, can be characterized by six distinct asymptotic regimes defined through the same-point correlation function \cite{kheruntsyan2003pair}, as shown in Fig.~\ref{fig:g2_parameter_space_simple}. 
The weakly interacting ($\gamma\ll 1$), low temperature quasicondensate regime can be treated using the Bogoliubov theory for quasicondensates  \cite{Mora-Castin-2003}, and is characterised by suppressed density fluctuations, but fluctuating phase. This may be subdivided into regions dominated by quantum (I) and thermal (II) fluctuations \cite{kheruntsyan2003pair}. At higher temperatures, the gas becomes nearly ideal, and can be treated using perturbation theory with respect to $\gamma$ \cite{kheruntsyan2003pair,Sykes_2008}. This asymptotic region may in turn be subdivided into quantum degenerate (III) and non-degenerate (IV) regimes. Finally, in the strongly interacting regime ($\gamma \gg 1$), where the Fermi-Bose gas mapping applies, the 1D Bose gas can be well approximated by a nearly ideal Fermi gas, and can be treated using pertubation theory with respect to $1/\gamma$ \cite{kheruntsyan2003pair,Sykes_2008}.
This regime can be further subdivided into two regions corresponding to high-temperature (V) and low-temperature (VI) fermionization.

In each of these asymptotic regimes, the pair correlation function $g^{(2)}(0)$ can be derived in closed approximate analytic form, where we additionally define the boundary of these regimes in terms of $\gamma$ and $\tau$
Further, we may express the total energy and entropy of the thermal equilibrium states in each asymptotic regime (see Ref.~\cite{kerr2024analytic} for further details). We make extensive use of these analytic functions in Chapters \ref{Chap:3} and \ref{Chap:6}, for calculating the performance of various quantum thermodynamic devices, and in Chapter \ref{Chap:Maxwell}, where they are utilized in the derivation and evaluation of a new Maxwell relation.

\subsection{External trapping potential}\label{sec:LDA}
In the presence of an external trapping potential, $V(z)$, the Hamiltonian acquires an additional term,
\begin{equation}\label{eq:LL_Ham_V}
    \begin{split}
        \hat{H} =- \frac{\hbar^2}{2m} \int \!dz\, \hat{\Psi}^\dagger \frac{\partial^2 \hat{\Psi}}{\partial z^2} 
        + \frac{g}{2} \int \!dz\, \hat{\Psi}^\dagger \hat{\Psi}^\dagger \hat{\Psi} \hat{\Psi} + \int\! dz\, V(z) \hat{\Psi}^\dagger \hat{\Psi}.
    \end{split}
\end{equation}
Through this, the model becomes generically non-integrable outside of the ideal (noninteracting) Bose gas limit or the Tonks-Girardeau limit of infinitely strong interactions. This breaking of integrability due to breaking of translational invariance can be intuitively understood as, in the presence of an external trap, the individual rapidities are no longer conserved, meaning we no longer have the extensive set of conserved quantities described in Section \ref{sec:groundstate}.

However, exact solutions derived for the uniform gas can be still utilized for finding thermodynamic properties of inhomogeneous gases via the local density approximation (LDA) \cite{kheruntsyan2005finite}. In the LDA framework, the Yang-Yang equations are solved using a local chemical potential $\mu(z)\!=\!\mu_0\!-\!V(z)$, where $\mu_0$ is the global chemical potential of the system and effectively defines the total atom number; as a result, the local dimensionless interaction strength $\gamma(z)\!=\!mg/\hbar^2\rho(z)$ acquires position-dependence through the inhomogeneity of the density profile $\rho(z)$.

For the LDA to be valid, the correlation length, $l_c(z)$, is required to be much smaller than the characteristic inhomogeneity length, $l_\mathrm{inh}(z)$, associated with the trapping potential,
\begin{equation}
    l_c(z) \ll l_\mathrm{inh}(z) = \frac{\rho(z)}{|d\rho(z)/dz|}.
\end{equation}
The form of this correlation length is dependent on which of the regimes, introduced in Section \ref{sec:g2_glauber}, that the gas inhabits. In regimes I and II, this correlation length coincides with the healing length of the low temperature Bose gas, $l_c(z)\!=\!\sqrt{\hbar^2 / m g \rho(z)}$ \cite{Pitaevskii_Stringari_book,Pethick_Smith_book,kheruntsyan2005finite}. For the strongly interacting regime VI, this lengthscale is given by the average inter-particle separation, i.e. $l_c(z)\!=\!1/\rho(z)$ \cite{kheruntsyan2005finite}. Finally, for high temperature regimes IV and V, the correlation length is on the order of the thermal de Broglie wavelength, $\Lambda_T\!=\!\sqrt{2 \pi \hbar^2 / m k_B T}$ \cite{kheruntsyan2005finite}.

\begin{figure*}[!tbp]
\begin{center}
   \includegraphics[width=9cm]{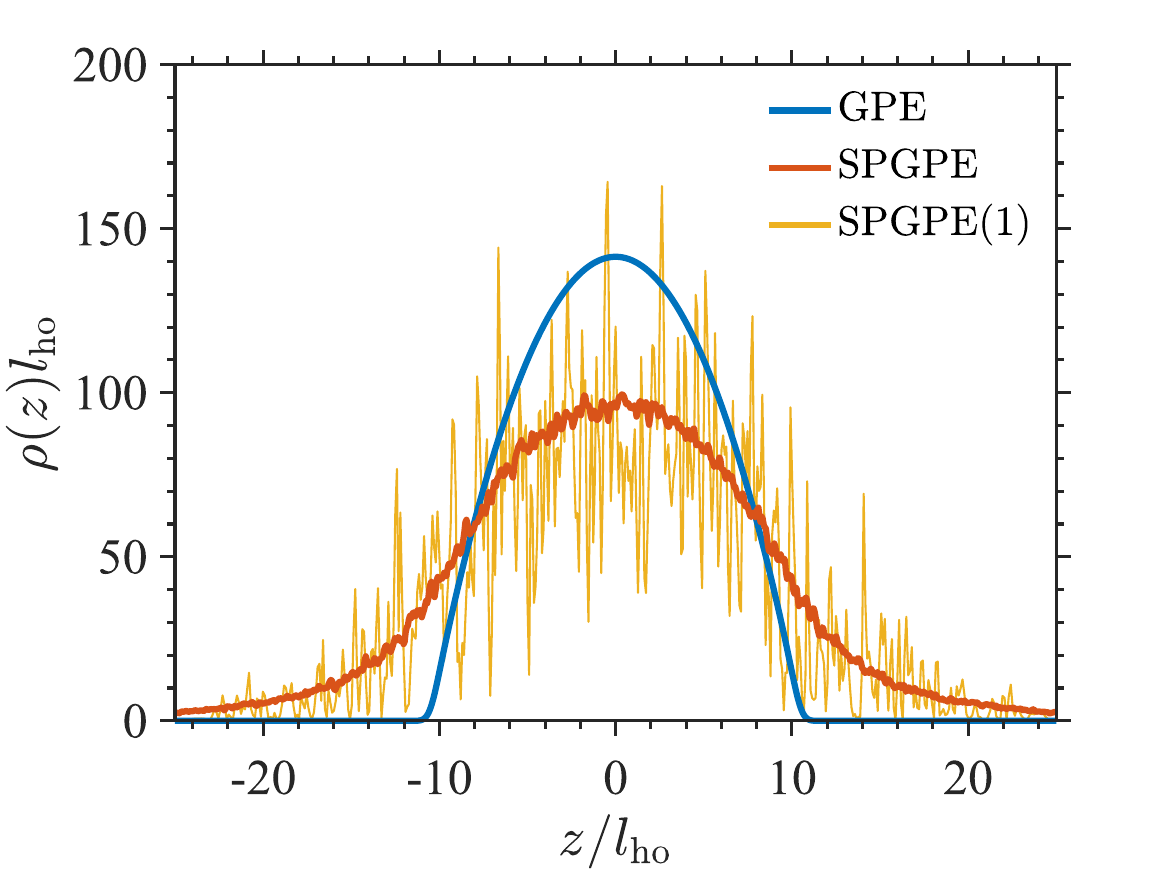}   
   \caption{An example of equilibrium states at zero and finite temperature for a 1D Bose gas consisting of $N\!=\!2000$ particles in a harmonic trap of frequency $\omega$. The ground state (i.e. $T\!=\!0$) of the harmonically trapped 1D Bose gas is generated via imaginary time evolution of the Gross-Pitaevskii equation (GPE), and is shown in blue. Shown in red is the thermal density profile generated by the classical $c$-field stochastic projected GPE (SPGPE) method, averaged over 1000 individual samples. A \textit{single} realisation from the SPGPE method is shown in yellow, and effectively corresponds to single-shot measurements in experiment \cite{whatisqushock}. All observables are given in natural units of the harmonic oscillator of frequency $\omega$, where $l_\mathrm{ho}\!=\!\sqrt{\hbar/m\omega}$ is the harmonic oscillator length.
   }
  \label{fig:SPGPE_Intro}
  \end{center}
\end{figure*}

For the case of an external harmonic trap of frequency $\omega$, $V(z)\!=\!0.5 m \omega^2 z^2$, one may utilize the Thomas-Fermi approximation for the weakly interacting ground state, which results in the familiar parabolic density profile shown as the blue solid line in Fig.~\ref{fig:SPGPE_Intro} \cite{Pethick_Smith_book,Pitaevskii_Stringari_book}. This may be expressed analytically as
\begin{equation}\label{eq:TF_T0}
    \rho(z) = \rho(0) \left( 1 - \frac{z^2}{R^2}\right),
\end{equation}
where
\begin{equation}
    \rho(0) = \left( \frac{9 m N^2 \omega^2}{32 g} \right)^{1/3} , \quad R = \left( \frac{3 N g}{2 m \omega^2} \right)^{1/3},
\end{equation}
are the peak density and Thomas-Fermi radius, respectively. Despite the fact that this approximation is strictly applicable to the ground state gas, it often remains approximately valid for the density profile in regimes I and II \cite{kheruntsyan2005finite}.

Likewise, the ground state density profile of the strongly interacting system may be given by \cite{kheruntsyan2005finite},
\begin{equation}
    \rho(z) = \rho(0) 
    \sqrt{1 - \frac{z^2}{R^2}},
\end{equation}
where
\begin{equation}
    \rho(0) = \sqrt{\frac{2 m N \omega}{\pi^2 \hbar}}, \quad R = \sqrt{\frac{2 \hbar N}{m \omega}}.
\end{equation}
are the peak density and density profile radius, respectively. This approximation may be utilized in the low temperature, strongly interacting regime VI, as it agrees well with the density profile obtained via the LDA \cite{kheruntsyan2005finite}.

Finally, at high temperatures, the 1D Bose gas becomes dominated by thermal fluctuations, and becomes well described by the Boltzmann density profile for a classical gas,
\begin{equation}
    \rho(z) = \rho(0) 
    \exp\left(-\frac{z^2}{R^2}\right),
\end{equation}
where
\begin{equation}\label{eq:rho_highT}
    \rho(0) = N \sqrt{\frac{m \omega^2}{2 \pi k_B T }}, \quad R = \sqrt{\frac{2 k_B  T}{m \omega^2}}.
\end{equation}
are the peak density and characteristic radius, respectively \cite{kheruntsyan2005finite}. The density profiles for the 1D Bose gas introduced in Eqs.~\eqref{eq:TF_T0}--\eqref{eq:rho_highT} are utilized extensively in Chapter \ref{Chap:6}, where we investigate the performance of a quantum heat engine for a harmonically trapped system.

\subsection{The stochastic projected Gross-Pitaevskii equation}
\label{Sec:SPGPE}
In the weakly interacting limit, $\gamma\!\ll\!1$ (see Section \ref{sec:groundstate}), at $T\!=\!0$, one may study the nonequilibrium dynamics of the 1D Bose gas through the mean field approximation \cite{Pethick_Smith_book,Pitaevskii_Stringari_book}. This consists of replacing the bosonic field operator with a classical field, i.e. $\hat{\Psi}(z,t)\!\to\!\langle \hat{\Psi}(z,t)\rangle \!=\!\Psi(z,t)$, and deriving the equation of motion from the Heisenberg representation for Eq.~\eqref{eq:LL_Ham_V}, resulting in
\begin{equation}
    i \hbar \frac{\partial}{\partial t} \Psi(z,t) = \left(-\frac{\hbar^2}{2m} \frac{\partial^2}{\partial z^2} + V(z) + g |\Psi(z,t)|^2 \right) \Psi(z,t).
\end{equation}
This method is commonly known as the Gross-Pitaevskii equation (GPE) \cite{Gross,Pitaevski,Pethick_Smith_book,Pitaevskii_Stringari_book}, and is utilized for simulating the nonequilibrium dynamics of a quantum shockwave at $T\!=\!0$ in Chapter \ref{Chap:5}.

Extending the mean-field method to $T\!>\!0$ may be achieved through application of the stochastic projected GPE (SPGPE) \cite{castin2000,blakie2008dynamics,spgpe}.
The SPGPE is a classical or $c$-field method for computing thermal equilibrium and dynamical properties of degenerate Bose gases at finite temperatures \cite{castin2000,Sinatra_PRL_2001,Bradley_2005,blakie2008dynamics,spgpe,stimming2010,Grisins2011,stimming2011,Bradley2015,Bouchoule2016,Deuar2018,Thomas2021,Bayocboc2022,Bayocboc2023}.
In this approach, the quantum field operator $\hat{\Psi}(z,t)$ is decomposed into two regions,
a $c$-field region and an incoherent thermal region. The $c$-field region
contains highly occupied low-energy modes and is described
by a single complex-valued classical field $\Psi_{\mathbf{C}}(z,t)$. The incoherent region, on the other hand, 
contains sparsely occupied high-energy modes that act as
an effective thermal bath, treated as static, with temperature
$T$ and chemical potential $\mu$ that governs the thermal average
number of particles in the system (in the $c$-field region). The boundary between these two regions is defined by an appropriately chosen energy cutoff $\epsilon_{\text{cut}}$.

In this approach, the thermal equilibrium state of the system is prepared by evolving the simple
growth stochastic projected Gross-Pitaevskii equation
(SPGPE) for the complex $c$-field $\Psi_{\mathbf{C}}(z,t)$ \cite{blakie2008dynamics,spgpe},
\begin{align}
	\label{eq:breathing_SPGPE}
	\mathrm{d}\Psi_{\mathbf{C}}&(z,t)=\mathcal{P}^{(\mathbf{C})}\!\left\{ - \frac{i}{\hbar}\mathcal{L}_{0}^{(\mathbf{C})}\Psi_{\mathbf{C}}(z,t)\,\mathrm{d}t 
	\right. \nonumber\\
	&\left.
	+ \frac{\Gamma}{k_BT}(\mu-\mathcal{L}_{0}^{(\mathbf{C})})\Psi_{\mathbf{C}}(z,t)\, \mathrm{d}t + dW_{\Gamma}(z,t)\!\right\}.
\end{align}
Here, the projection operator $\mathcal{P}^{(\mathbf{C})}\{\cdot\}$ sets up the high-energy cutoff 
$\epsilon_{\mathrm{cut}}$, whereas $\Gamma$ is the so-called growth rate responsible for the coupling between the $c$-field and the effective reservoir (served by the incoherent region). In addition, $\mathcal{L}_{0}^{(\mathbf{C})}$ is the Gross-Pitaevskii operator defined by
\begin{equation}
	\mathcal{L}_{0}^{(\mathbf{C})} = -\frac{\hbar^{2}}{2m}\frac{\partial^{2}}{\partial x^{2}} + V(z) + g|\Psi_{\mathbf{C}}(z,t)|^{2}, 
\end{equation}
where $V(z,t)$ is the external trapping potential, if any. The last term, $dW_{\Gamma}(z,t)$, in Eq.~\eqref{eq:breathing_SPGPE} is a complex-valued stochastic white noise term with the following nonzero correlation:
\begin{equation}
	\langle dW_{\Gamma}^{*}(z,t)dW_{\Gamma}(z',t) \rangle = 2\Gamma\delta(z-x')dt.
\end{equation}

The stochastic realisations of the $c$-field $\Psi_{\mathbf{C}}(z,t)$ prepared via the SPGPE after a sufficiently long evolution time sample the grand-canonical ensemble of thermal equilibrium states of the system. These stochastic realisations can then be evolved in real time according to the mean-field projected Gross-Pitaevskii equation \cite{blakie2008dynamics}, following a certain out-of-equilibrium protocol. This represents real-time dynamical evolution of the system starting from an initial thermal equilibrium state. 
Further, the thermal equilibrium values of physical observables may be calculated in terms of expectation values of products of $\Psi_{\mathbf{C}}(z)$ and its complex conjugate $\Psi_{\mathbf{C}}^{*}(z)$ 
averaged over a large number of stochastic realizations. This is much in the same way as calculating the same observables in terms of expectation values over normally-ordered products of quantum field operators $\hat{\Psi}(z)$ and $\hat{\Psi}^{\dagger}(z)$, except that their non-commuting nature is ignored.
As an example, the particle number density $\rho(z) = \langle \hat{\Psi}^{\dagger}(z) \hat{\Psi}(z) \rangle$ in the SPGPE approach is calculated as  $\rho(z)  = \langle \Psi_{\mathbf{C}}^{*}(z) \Psi_{\mathbf{C}}(z) \rangle$, where the brackets $\langle{...}\rangle$ refer to ensemble averaging over a large number of stochastic trajectories; similarly the pair correlation function $g^{(2)}(0)$ can be computed according to:
\begin{equation}
g^{(2)}=\frac{\langle \Psi_{\mathbf{C}}^{*}(z)\Psi_{\mathbf{C}}^{*}(z) \Psi_{\mathbf{C}}(z) \Psi_{\mathbf{C}}(z)\rangle}{\langle \Psi_{\mathbf{C}}^{*}(z) \Psi_{\mathbf{C}}(z) \rangle^2}.
\label{g2-SPGPE}
\end{equation}

\section{Generalized hydrodynamics}
\label{Sec:GHD}

\begin{figure*}[!tbp]
\begin{center}
   \includegraphics[width=8.5cm]{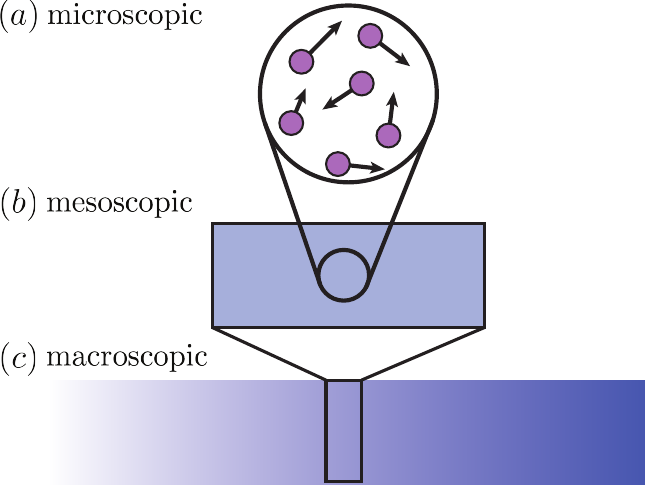}   
   \caption{A cartoon illustration of the three scales involved in the hydrodynamic description. The microscopic scale, shown in panel (a), is governed by the individual atomic trajectories and collisions between particles. On the larger mesoscopic scale, shown in (b), these collisions cause the system to rapidly approach a local equilibrium state, which coincides with a generalized Gibbs ensemble (GGE) for integrable systems (see text). This rapid local equilibration to a GGE is the hydrodynamic assumption underlying the derivation of generalized hydrodynamics, which in turn governs the macroscopic dynamics of the system, shown in panel (c). 
   }
  \label{fig:hydroscales}
  \end{center}
\end{figure*}

Integrable models, as detailed in Sections \ref{sec:groundstate} and \ref{sec:tba}, have exact ground state and thermal equilibrium solutions in the thermodynamic limit thanks to possession of an infinite set of conserved charges.
However, as briefly mentioned in the paragraph below Eq.~\eqref{eq:Q_thermolim}, this infinite set of conserved quantities often results in nontrivial dynamics.
In particular, at the macroscopic scale shown in Fig.~\ref{fig:hydroscales}(c), application of a hydrodynamic assumption, where the gas is assumed to locally (i.e. at the mesoscopic scale) relax to a thermal equilibrium state, results in an infinite set of coupled continuity equations \cite{Doyon-lectures}. In the following, we give a brief overview of this hydrodynamic assumption and its application to integrable models, providing a useful background to the discovery of generalized hydrodynamics, which is explored throughout the remainder of this chapter.

By definition, local equilibrium states will maximise (or extremise) the entropy, $S = -k_B\mathrm{Tr}\left[ \hat{\rho} \log \hat{\rho} \right]$, with respect to the conserved charges, $Q_i$, and where $\hat{\rho}$ is the one-body density matrix \cite{Jaynes_I,Jaynes_II,Doyon-lectures}.
Hence, one may utilize the method of Lagrange multipliers, with a multiplier $\beta^i$ for each charge $Q_i$, along with an additional parameter $\alpha$ that encodes the normalisation condition for the distribution. Hence, under entropy maximisation, $\delta S \!=\! 0$, one finds
\begin{equation}
    \delta \mathrm{Tr} \left[ \hat{\rho} \left( \log\hat{\rho} + \sum_{i=-1}^M \beta^i Q_i + \alpha \right) \right] = 0,
\end{equation}
where $M$ is the number of known conservation laws. This may be rewritten as,
\begin{equation}\label{eq:S_delta_rho}
    \mathrm{Tr} \left[ \delta\hat{\rho} \left( \log\hat{\rho} + \sum_{i=-1}^M \beta^i Q_i + \alpha + 1 \right) \right] = 0.
\end{equation}
Hence, the one-body density matrix, $\hat{\rho}$, has the form \cite{Doyon-lectures}
\begin{equation}
    \hat{\rho} =\frac{ \exp \left(  -\sum_{i=1}^\infty \beta^i Q_i  \right)}{\mathrm{Tr}\left[\exp \left(  -\sum_{i=1}^\infty \beta^i Q_i  \right) \right]}.
\end{equation}
In particular, for conventional hydrodynamics, where there are typically at most three conserved charges, that of total particle number, momentum, and energy, the maximum entropy state consists of a Gibbs ensemble (GE) \cite{Jaynes_I,Jaynes_II},
\begin{equation}\label{eq:GE}
    \hat{\rho}' = \frac{\exp\left( -\sum_{i=1}^3 \beta^i Q_i \right)}{\mathrm{Tr}\left[\exp\left( -\sum_{i=1}^3 \beta^i Q_i \right)\right]}.
\end{equation}
It is important to emphasize that this is a \textit{local} GE, this means that the trace is averaging over the microscopic states in a localized region of space, commonly known as a mesoscopic `fluid cell' (see Fig.~\ref{fig:hydroscales}) \cite{Doyon-lectures}.

Extending this to an \textit{infinite} set of conserved charges consists of removing the upper limit on the summations in Eq.~\eqref{eq:GE}, resulting in the generalized Gibbs ensemble (GGE),
\begin{equation}\label{eq:GGE}
    \hat{\rho} = \frac{\exp\left( -\sum_{i=1}^\infty \beta^i Q_i \right)}{\mathrm{Tr}\left[\exp\left( -\sum_{i=1}^\infty \beta^i Q_i \right)\right]}.
\end{equation}
At the mesoscopic level, the integrable system consists of \textit{local} conserved densities, $\langle q_i \rangle(z,t)$, which are derived from the related conserved charge on the macroscopic scale \cite{Doyon-lectures},
\begin{equation}\label{eq:conservedcharge_GHD}
    Q_i(t) = \int dz \langle q_i \rangle(z,t) = \int dz \int d\lambda f_p(\lambda;z,t) \lambda^i,
\end{equation}
where we have employed the definition of total charge $Q_i$, given in Eq.~\eqref{eq:Q_observables} for the TBA, which corresponds here to the total charge over a possibly non-uniform system constituted of uniform mesoscopic fluid cells of length $L$. Over these mesoscopic fluid cells, the local conserved density may be expressed as \cite{Castro-Alvaredo_2016_Emergent,Doyon-lectures}
\begin{equation}\label{eq:conserved_density}
    \langle q_i \rangle(z,t) = \int_{-\infty}^\infty d\lambda f_p(\lambda;z,t) \lambda^i,
\end{equation}
where the angled brackets, $\langle \cdot \rangle$, in Eq.~\eqref{eq:conservedcharge_GHD} represent averages over local GGE,
\begin{equation}
    \langle \hat{o} \rangle = \mathrm{Tr}\left\{ \hat{\rho} \hat{o} \right\},
\end{equation}
for any system observable $\hat{o}$. 
Here, we have used the fact that the local thermal equilibrium state of an integrable system may be expressed through the Yang-Yang thermodynamic Bethe ansatz as detailed in the Section prior. In particular, we utilize Eq.\eqref{eq:Q_observables} to express the local conserved densities, which themselves depend on the space-time coordinates, in terms of the quasiparticle density $f_p$. In doing this, we have promoted the quasiparticle density, or root density $f_p$, to a space-time dependent function, $f_p(\lambda;z,t)$ \cite{Doyon-lectures}.

Conservation of the global charge $Q_i$ implies a local conservation of the charge density, $q_i$. This, in combination with the hydrodynamic assumption of local relaxation to a thermal equilibrium state, results in $\langle q_i \rangle(z,t)$ satisfying a local continuity equation \textit{between} mesoscopic fluid cells, i.e. at the macroscopic scale \cite{Doyon-lectures},
\begin{equation}\label{eq:continuity_eqs}
    \partial_t \langle q_i \rangle(z,t) + \partial_z \langle j_i \rangle(z,t) = 0, \quad i=1,2,3,\dots \, .
\end{equation}
The conserved currents, $j_i$, for charge $i$ are known exactly for a Galilean fluid with $3$ conserved charges, and result in the typical Euler hydrodynamic equations under the hydrodynamic assumption of local relaxation to a Gibbs state (e.g. $\hat{\rho}'$ in Eq.~\eqref{eq:GE}) \cite{Doyon-lectures,landau2013fluid}. 
For an integrable system however, this results in an infinite set of continuity equations at the hydrodynamics scale, making the large-scale dynamics computationally intractable if one were to attempt to apply Eq.~\eqref{eq:continuity_eqs} directly.

However, in 2016 two separate groups postulated a new theory of hydrodynamics for any integrable model solvable by the thermodynamic Bethe ansatz \cite{bertini2016transport,Castro-Alvaredo_2016_Emergent}.
The first of these papers focused on the XXZ spin chain, and utilized a kinetic theory to demonstrate that the late-time steady-state dynamics followed an equation of hydrodynamic type \cite{bertini2016transport}. Shortly after, a second group derived a similar set of equations more generally for all models that were integrable through the assumption of local relaxation to a GGE (see Eq.~\eqref{eq:GGE}) \cite{Castro-Alvaredo_2016_Emergent}. This was achieved by postulating a form for the conserved current $j_i$ mentioned above. Rigorous proofs of this current formula were later provided for individual models in an important series of papers \cite{10.21468/SciPostPhys.6.2.023,Yoshimura_CollisionRate,10.21468/SciPostPhys.8.2.016} (see Refs.~\cite{Borsi_2021} and \cite{Cubero_2021} for recent reviews on rigorous proofs of current operator relations).

In the following subsections, we give an overview of the aspects of GHD theory which we utilize in this thesis. In Sec.~\ref{Chap:2_Sec:GHD}, we provide a kinetic picture of GHD, focusing on how the effective velocity may be understood in terms of summing elastic collisions between quasiparticles. Through this, we hope to give an intuitive picture of how GHD operates at the level of quasiparticles in all applicable integrable systems. Next, in Sec.~\ref{Chap:2_Sec:T0}, we detail the zero-entropy, or $T\!=\!0$, methods for GHD. Here, we demonstrate how the zero temperature limit of the thermodynamic Bethe ansatz can be utilized to simplify the evolution equations of GHD, using a method first detailed in Ref.~\cite{largescale_Doyon}. Finally, we detail developments in GHD that have occurred more recently, focusing on the aspect of integrability breaking and the higher-order Navier-Stokes expansion to GHD, which incorporates diffusive effects and describes thermalization at long times.

\subsection{The generalized hydrodynamic equations}\label{Chap:2_Sec:GHD}

Generalized hydrodynamics is most commonly expressed using the language of Yang and Yang's thermodynamic Bethe ansatz (TBA) \cite{yang1969thermodynamics,mossel2012generalized,Bertini_2016_Transport,Castro-Alvaredo_2016_Emergent}. Importantly, the current operation relation that forms the foundation for the theory of GHD, and which is mentioned in the paragraphs below Eq.~\eqref{eq:continuity_eqs}, may be formulated without use of the TBA, for systems of finite size \cite{PhysRevX.10.011054,PhysRevLett.125.070602,10.21468/SciPostPhys.8.2.016}. Indeed, GHD in the thermodynamic limit may effectively be seen as a consequence of this finite-size theory \cite{bouchoule2022generalized}. Interestingly, the thermodynamic theory of GHD was actually discovered before its finite-size counterpart \cite{Bertini_2016_Transport,Castro-Alvaredo_2016_Emergent}. Here we present a brief formulation of first order (or Euler-scale) GHD in the finite-temperature thermodynamic limit in terms of the density of quasiparticles, $f_p(\lambda;z,t)\!\equiv\!f_p(\lambda)$, and the filling factor, $n(\lambda;z,t)\!\equiv\!n(\lambda)$ given in Eq.~\eqref{eq:filling_function}, for the TBA. We suppress space-time dependence from here onwards for simplicity of notation. 

The theory of GHD may be expressed in terms of an effective `change of basis' from the infinite set $\langle q_i\rangle$, to the root density, $f_p$, which are related via Eq.~\eqref{eq:conservedcharge_GHD} \cite{Doyon-lectures}. This was achieved by postulating the form of the corresponding conserved current, $\langle j_i\rangle$, in terms of an effective velocity, $v^\mathrm{eff}$ \cite{Bertini_2016_Transport,Castro-Alvaredo_2016_Emergent,bouchoule2022generalized},
\begin{equation}
    \langle j_i \rangle = \int_{-\infty}^\infty d\lambda v^\mathrm{eff}[f_p(\lambda)] f_p(\lambda) \lambda^i,
\end{equation}
which had previously appeared for quantum integrable systems in Ref.~\cite{PhysRevLett.113.187203}. Importantly, this effective velocity is dependent on the space-time coordinates, as it is a function of the quasiparticle density, $f_p$, which is itself space-time dependent.
Utilizing this postulated form for the conserved current in conjunction with an expression for the local conserved density, given in Eq.~\eqref{eq:conserved_density},
one may express the infinite set of continuity equations given in Eq.~\eqref{eq:continuity_eqs} in terms of the quasiparticle density, $f_p(\lambda)$, as,
\begin{equation}
    \partial_t \left[ \int_{-\infty}^\infty d\lambda f_p(\lambda) \lambda^i  \right] + \partial_z \left[ \int_{-\infty}^\infty d\lambda v^\mathrm{eff}[f_p(\lambda)] f_p(\lambda) \lambda^i  \right] = 0
\end{equation}
which in turn may be rearranged as
\begin{equation}
    \int_{-\infty}^\infty d\lambda \lambda^i\left[ \partial_t  f_p(\lambda)   + \partial_z \left( v^\mathrm{eff}[f_p(\lambda)] f_p(\lambda) \right)  \right] = 0.
\end{equation}
From here, one arrives at the core `Bethe-Boltzmann' equation of Euler-scale GHD \cite{Bertini_2016_Transport,Castro-Alvaredo_2016_Emergent,bulchandani2018bethe},
\begin{equation}\label{eq:GHD}
    \partial_t f_p(\lambda) + \partial_z \big( v^\mathrm{eff}[f_p(\lambda)] f_p(\lambda) \big) 
    =   0.
\end{equation}
Expression of this effective velocity, $v^\mathrm{eff}[f_p(\lambda)]$, which is a functional of the root density, was the most important breakthrough in the original formulation of GHD, and is key to understanding how this hydrodynamic model works. As such, we dedicate Section \ref{Chap:2_Sec:veff} to an intuitive derivation of the effective velocity for the case of a 1D Bose gas.
Importantly, this hydrodynamic equation is applicable to \textit{any} integrable model, quantum or classical, that is solvable via the thermodynamic Bethe ansatz.


As was the case for the Yang-Yang thermodynamic treatment, GHD may be expressed in terms of a variety of quasiparticle densities.
In particular, utilizing the bijection between the root density, $f_p(\lambda)$, and the filling function, $n(\lambda)$, we may perform a change of basis for Eq.~\eqref{eq:GHD}, transforming the GHD equation of motion to an \textit{advective} form \cite{Castro-Alvaredo_2016_Emergent}
\begin{equation}
    \partial_t n(\lambda) + v^\mathrm{eff}[n(\lambda)] \partial_z n(\lambda)=0.
\end{equation}
This allows for the identification of the filling factor as the \textit{normal modes} of GHD \cite{Doyon-lectures,EL201611}, meaning they are conserved throughout dynamics. We mention this as it is the most common method for simulating GHD, and is utilized for the majority of simulations presented in Chapters \ref{Chap:5} and \ref{Chap:6} of this thesis.
In our GHD simulations of systems evolving from finite-temperature thermal equilibrium states, we used the \emph{iFluid} software package \cite{moller2020introducing}, which is an efficient, easily expandable open-source numerical framework based in Matlab. Simulations of systems evolving from zero-temperature ground states, on the other hand, were carried out using zero-entropy subspace methods found in Ref.~\cite{largescale_Doyon} and detailed in Section \ref{Chap:2_Sec:T0} below.

\begin{figure*}[!tbp]
\begin{center}
   \includegraphics[width=4cm]{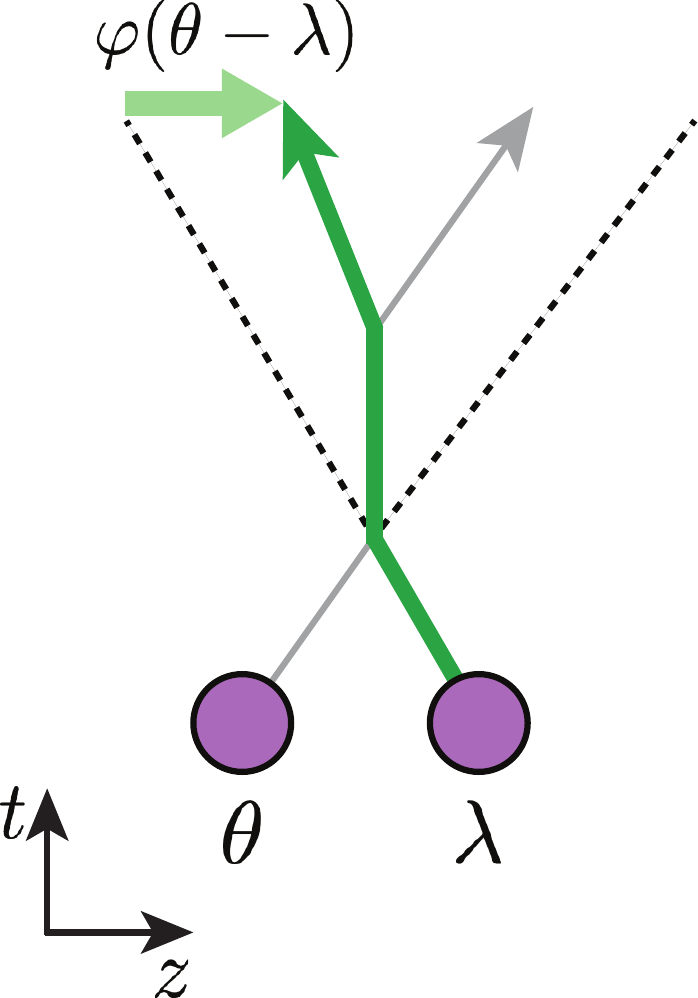}   
   \caption{An illustration of the physical displacement incurred via elastic collision between quasiparticles. Here, the tracer particle of rapidity $\lambda$ scatters with another quasiparticle of rapidity $\theta$. The tracer particle's trajectory, highlighted in green, following the scattering event has the same rapidity, $\lambda$, as the initial state due to the elasticity of the collision. The collision instead encodes a physical displacement in the position coordinate of the particle through the scattering shift $\varphi(\theta\!-\!\lambda)$.
   }
  \label{fig:Scattering_Shift}
  \end{center}
\end{figure*}

\subsubsection{The effective velocity}\label{Chap:2_Sec:veff}

The formulation of the effective velocity, $v^\mathrm{eff}[f_p]$, in terms of quasiparticle excitations is central to original development of GHD at Euler scale, i.e. at first order in spatial derivative \cite{Doyon-lectures,DeNardis_Diffusion_2018}.
In this subsection, we focus on providing intuition for the mechanisms behind the core GHD equation given in Eq.~\eqref{eq:GHD} by providing a derivation for this effective velocity. In particular, we focus on the fact that, by expressing this effective velocity in terms of a sum over quasiparticle collisions, the GHD evolution may be interpreted as free transport of a single quasiparticle renormalized by moving through a dense background of other quasiparticles \cite{Doyon-lectures,bouchoule2022generalized}.

\begin{figure}
    \centering
    \includegraphics[scale=0.2]{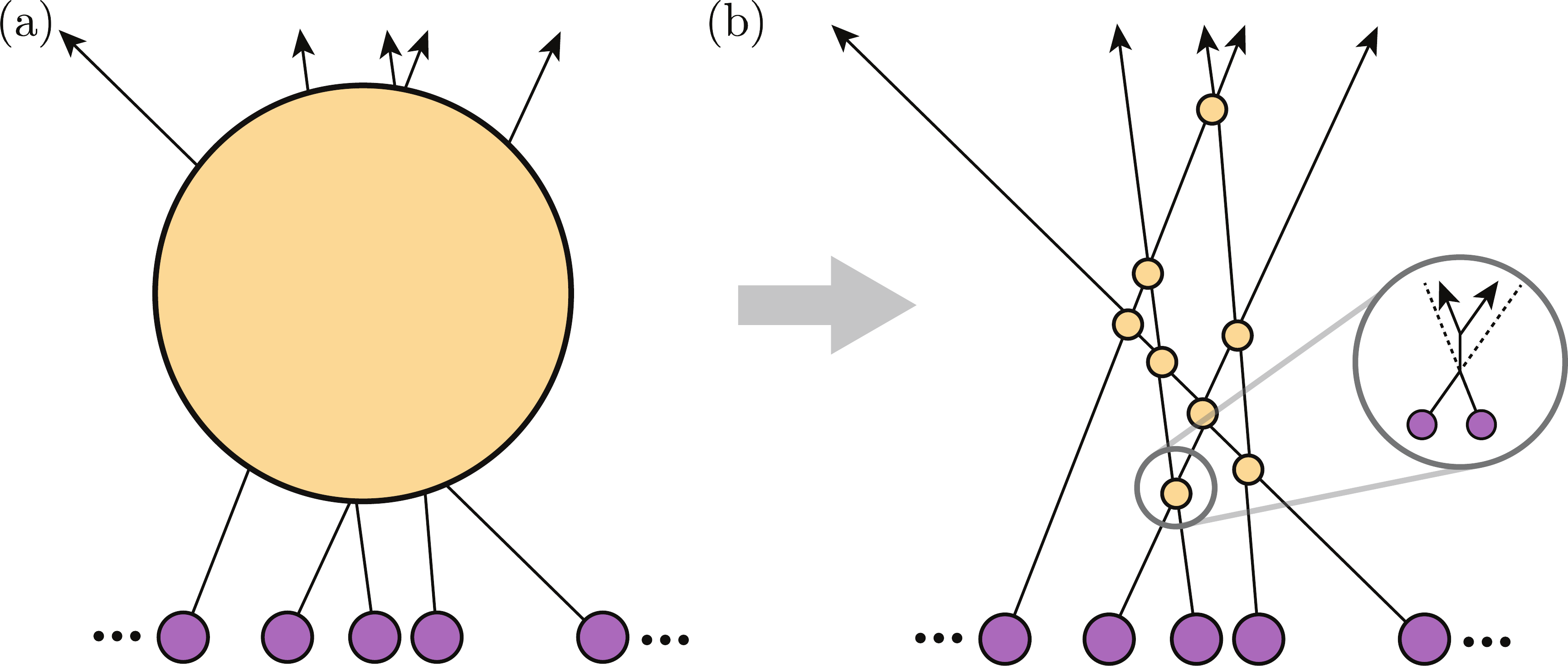}
    \caption{A key feature of integrable models is that $N$-body scattering processes, a schematic of which is shown in panel (a), may always be factorized into a combination of two-body elastic scattering processes, shown in (b), each of which may be described in terms of the scattering shift (see Fig.~\ref{fig:Scattering_Shift} for further details on the scattering shift).}
    \label{fig:scattering_factorization}
\end{figure}

To derive this effective velocity, we must therefore try to understand the average speed that a single quasiparticle moves when subject to collisions with this dense background of other quasiparticles.
In the thermodynamic limit, we are therefore interested in the average distance, $\langle \Delta z \rangle$, that our tracer particle (introduced for the Lieb-Liniger model in Section \ref{sec:groundstate}) moves over time $\Delta t$. Over this time, the particle undergoes a large number of collisions with other quasiparticles, with each collision incurring a physical displacement due to a phase shift. This phase shift results in a physical displacement, or scattering shift $\varphi(\lambda,\theta)$, where $\lambda$ is the tracer particle's rapidity, and $\theta$ is the rapidity of the quasiparticle collided with, as shown in Fig.~\ref{fig:Scattering_Shift} and introduced in terms of two-body scattering for the Lieb-Liniger model in Eq.~\eqref{eq:scattering_shift}.
We may thus express the distance travelled via a sum of the distance the quasiparticle would move without collisions, given by the free evolution at rapidity (i.e. quasiparticle velocity) $\lambda$, resulting in a displacement $\lambda \Delta t$, and add the cumulative effect of these collisions,
\begin{equation}\label{eq:cumulative_shift}
    \langle \Delta z \rangle = \lambda \Delta t + \mathcal{C}_{\Delta t}[f_p](\lambda),
\end{equation}
where we have defined $\mathcal{C}_{\Delta t}[f_p](\lambda)$ to be the cumulative displacement from collisions with a uniform background of quasiparticles with root density $f_p$ over a time $\Delta t$.
To evaluate this, we utilize the fact that any $N$-body scattering process may be factorized into a product of two-body scattering events, as illustrated in Fig.~\ref{fig:scattering_factorization}, and introduced in terms of the Lieb-Liniger model in Section \ref{sec:groundstate}.

\begin{figure}
    \centering
    \includegraphics[width=15cm]{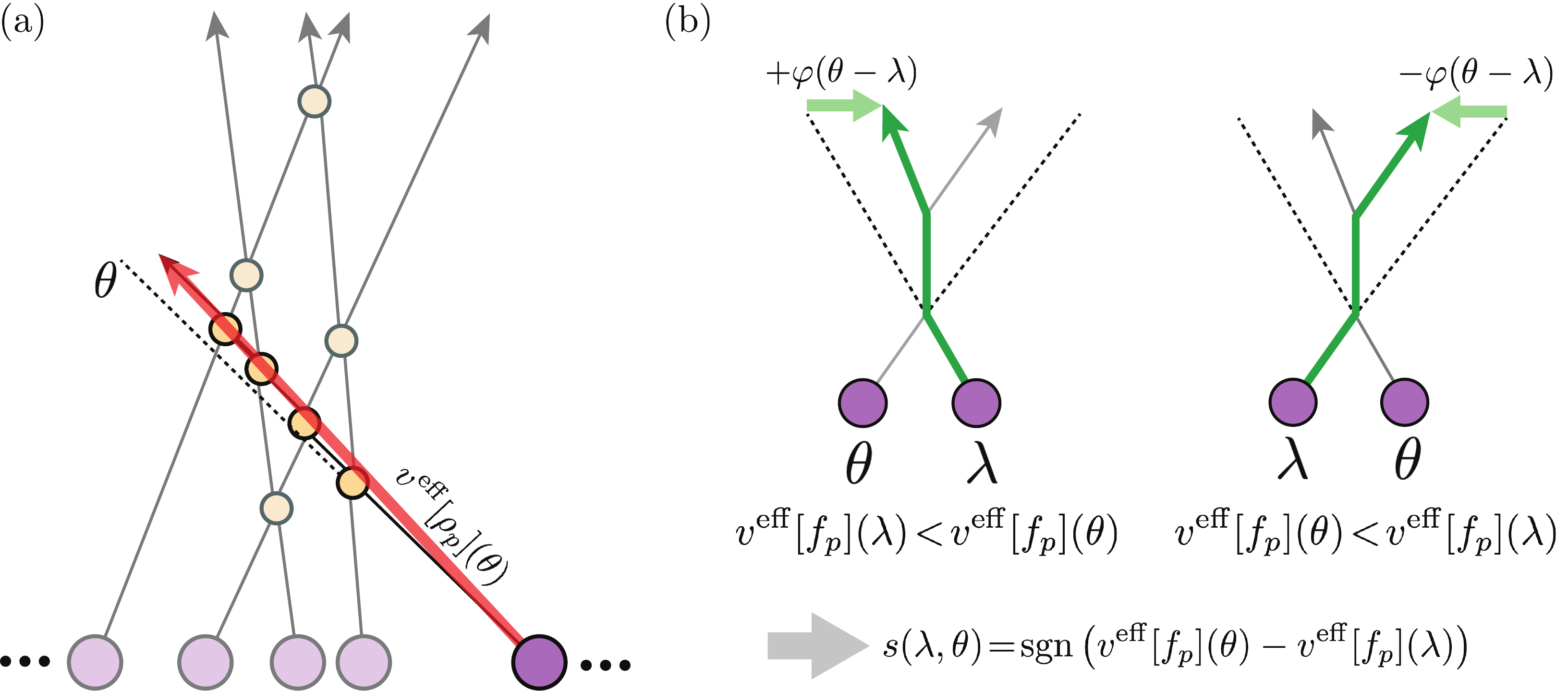}
    \caption{The effective velocity, which may be formulated as a dressing of the group velocity by moving through a dense background of quasiparticles, is demonstrated in panel (a) as a red line, distinct from the line of constant rapidity, $\theta$, shown as a dashed black line. In panel (b) we demonstrate the effect of the phase shift, $\varphi(\theta - \lambda)$, on a quasiparticle of rapidity $\lambda$ undergoing a collision with one of rapidity $\theta$, where the trajectory of the quasiparticle labelled by $\lambda$ is highlighted in green. This demonstrates that the direction of the scattering shift, denoted $s(\theta,\lambda)$, away from the unperturbed free evolution, shown as the black dashed lines, is dependent \textit{only} on the sign of the difference of the effective velocities.}
    \label{fig:effective_velocity}
\end{figure}

To find an expression for this cumulative scattering shift, which is schematically illustrated in Fig.~\ref{fig:effective_velocity}(a), we focus on the motion within a locally uniform mesoscopic `fluid cell', shown schematically in Fig.~\ref{fig:hydroscales}(b). Thus, we first use the fact that the average distance between particles in a uniform gas, $d$, may be approximated through the inverse of the particle number density, i.e.
\begin{equation}
    d = \frac{1}{\rho} 
\end{equation}
where $N$ is the number of particles in our fluid cell of length $L$.
Further, assuming that each quasiparticle is moving at its own effective velocity \cite{Doyon-lectures}, we may define the \textit{average magnitude} of the effective velocity per quasiparticle as
\begin{equation}\label{eq:abs_veff}
    \overline{v^\mathrm{eff}}[f_p] = \frac{L}{N} \int d \theta f_p(\theta) |v^\mathrm{eff}[f_p](\theta)|.
\end{equation}
Thus, for any given position coordinate $z$, the average time between quasiparticles crossing this point may be approximated by
\begin{equation}\label{eq:t_avg_collision}
    \overline{t} = \frac{d}{\overline{v^\mathrm{eff}}[f_p]},
\end{equation}
where the absolute value in Eq.~\eqref{eq:abs_veff} accounts for left-moving and right-moving particles.

However, if we are instead interested in the average time between \textit{collisions} of the quasiparticle moving at effective velocity $v^\mathrm{eff}[f_p](\lambda)$ and all other quasiparticles, a quantity which we here denote $\overline{t}_\lambda$, we may simply provide a Galilean boost of velocity $v^\mathrm{eff}[f_p](\lambda)$ to the velocity term in Eq.~\eqref{eq:t_avg_collision}, to find
\begin{equation}
    \overline{t}_\lambda = \frac{d}{\frac{L}{N}\int d \theta f_p(\theta) |v^\mathrm{eff}[f_p](\theta) - v^\mathrm{eff}[f_p](\lambda)|}.
\end{equation}
As we have assumed a locally uniform fluid, we use $d = 1/\rho = L/N$ to simplify this approximation to
\begin{equation}
    \overline{t}_\lambda = \frac{1}{\int d \theta f_p(\theta) |v^\mathrm{eff}[f_p](\theta) - v^\mathrm{eff}[f_p](\lambda)|}.
\end{equation}
We are interested in the cumulative effect of these scattering shifts over time $\Delta t$. We may therefore estimate the \textit{total} number of scattering events that our tracer particle, $\lambda$, incurs in this time, $N^\mathrm{scatt.}$, as
\begin{equation}\label{eq:Nscatt}
    N^\mathrm{scatt.}(\Delta t; \lambda) = \frac{\Delta t}{\overline{t}_\lambda} = \Delta t \int d \theta f_p(\theta) |v^\mathrm{eff}[f_p](\theta) - v^\mathrm{eff}[f_p](\lambda)|.
\end{equation}
Finally, we use the fact that each scattering event between two particles with rapidities $\lambda$ and $\theta$ results in a physical displacement by an amount $\varphi(\theta-\lambda)$, where the direction of this displacement depends \textit{only} on whether the quasiparticle colliding with our tracer quasiparticle, labelled by $\lambda$, is coming from the left, resulting in a positive displacement, or from the right, resulting in a negative displacement, as illustrated in Fig.~\ref{fig:effective_velocity}(b). This results in simply replacing the absolute values around the effective velocity terms in Eq.~\eqref{eq:Nscatt} with ordinary brackets, as $x = \mathrm{sgn}(x) |x|$.

Thus, we may incorporate the cumulative effect of these scattering terms into this integral by multiplying each scattering event by the resulting displacement, or scattering shift $\varphi(\lambda\! -\! \theta)$,
\begin{equation}
    \mathcal{C}[f_p](\lambda) =  \int d\theta f_p(\theta) (v^\mathrm{eff}[f_p](\theta) - v^\mathrm{eff}[f_p](\lambda)) \varphi(\lambda - \theta) \Delta t.
\end{equation}
Combining this with Eq.~\ref{eq:cumulative_shift}, we find that the average distance, $\langle \Delta z \rangle$, moved by a quasiparticle of rapidity $\theta$ over a time $\Delta t$, may be expressed as an integral relation,
\begin{equation}
    \langle \Delta z \rangle = \lambda \Delta t +\int d\theta f_p(\theta) (v^\mathrm{eff}[f_p](\theta) - v^\mathrm{eff}[f_p](\lambda)) \varphi(\lambda - \theta) \Delta t
\end{equation}
As we have defined the effective velocity as the effective speed at which a quasiparticle moves when accounting for the cumulative scatering with the background, we may identify this effective velocity as the average speed at which a quasiparticle moves, i.e. $v^\mathrm{eff} = \langle \Delta z \rangle / \Delta t$, meaning
\begin{equation}
    v^\mathrm{eff}[f_p](\lambda) = \lambda + \int d \theta \varphi(\lambda - \theta) \left( v^\mathrm{eff}[f_p](\theta) -  v^\mathrm{eff}[f_p](\lambda)\right) f_p(\theta).
\end{equation}
This is the second core equation for GHD, and, through this, one has all that is required to calculate the dynamics under Euler-scale GHD.

However, we may utilize the TBA once more in order to express the effective velocity in a final, and more suggestive form. To do this, we note that the total quasiparticle density, which is expressed in terms of the filling factor in Eq.~\eqref{eq:total_density}, may be written as the solution to a `dressing equation',
\begin{equation}\label{eq:dressing_operation}
    g^\mathrm{dr}(\lambda) = g(\lambda) + \int \frac{d \theta}{2\pi} \varphi(\lambda - \theta) n(\theta) g^\mathrm{dr}(\theta).
\end{equation}
By comparison with Eq.~\eqref{eq:total_density}, we observe that the total quasiparticle density may be expressed as $f_s(\lambda)\!=\!(1/2\pi)^\mathrm{dr}[f_p](\lambda)$, where $1(\lambda)\!=\!1$. One may then rearrange the effective velocity formula by placing all terms multiplying $v^\mathrm{eff}[f_p](\lambda)$ to the left hand side, resulting in
\begin{equation}
    v^\mathrm{eff}[f_p](\lambda) \left( 1 + \int d \theta \varphi(\lambda - \theta) f_p(\theta) \right) =\left( \lambda + \int d \theta \varphi(\lambda - \theta)  v^\mathrm{eff}[f_p](\theta) f_p(\theta) \right).
\end{equation}
Substituting in the definition for the filling factor, $n(\theta)\!=\! f_p(\theta)/f_s(\theta)$, and dividing both sides by a factor of $2\pi$, we find
\begin{equation}
    v^\mathrm{eff}[f_p](\lambda) \left( \frac{1}{2\pi} + \int \frac{d \theta}{2\pi} \varphi(\lambda - \theta) n(\theta)f_s(\theta) \right) =\left( \frac{\lambda}{2\pi} + \int \frac{d \theta}{2\pi} \varphi(\lambda - \theta)  (\theta) n(\theta) v^\mathrm{eff}[f_p] f_s(\theta) \right)
\end{equation}
By comparison of both bracketed terms with the dressing operation, and utilizing the fact that $(\lambda/2\pi)^\mathrm{dr}(\theta)\!=\!\lambda^\mathrm{dr}(\theta) / 2 \pi$, we finally arrive at
\begin{equation}\label{eq:effective_velocity}
    v^\mathrm{eff}[f_p](\lambda) = \frac{(\lambda)^\mathrm{dr}[f_p]}{(1)^\mathrm{dr}[f_p]}.
\end{equation}
This formula may be thought of as a `dressed' version of the group velocity found in free evolution under a collisionless Boltzmann equation \cite{bouchoule2022generalized}, as illustrated schematically in Fig.~\ref{fig:effective_velocity}(a).

In the limit of a non-interacting model, which occurs either in the limit of $c\to\infty$, where the gas is equivalent to free fermions, or the $c\to0$ limit of free bosons, the dressing formula reduces to the identity operation, $f^\mathrm{dr}[f_p](\theta) = f(\theta)$, meaning
\begin{equation}
        v^\mathrm{eff}[f_p](\lambda) = \frac{(\lambda)^\mathrm{dr}[f_p]}{(1)^\mathrm{dr}[f_p]}=\lambda,
\end{equation}
recovering their free evolution.

\subsection{Zero-entropy subspace methods and integrability breaking}\label{Chap:2_Sec:T0}

\begin{figure}
    \centering
    \includegraphics[width=9cm]{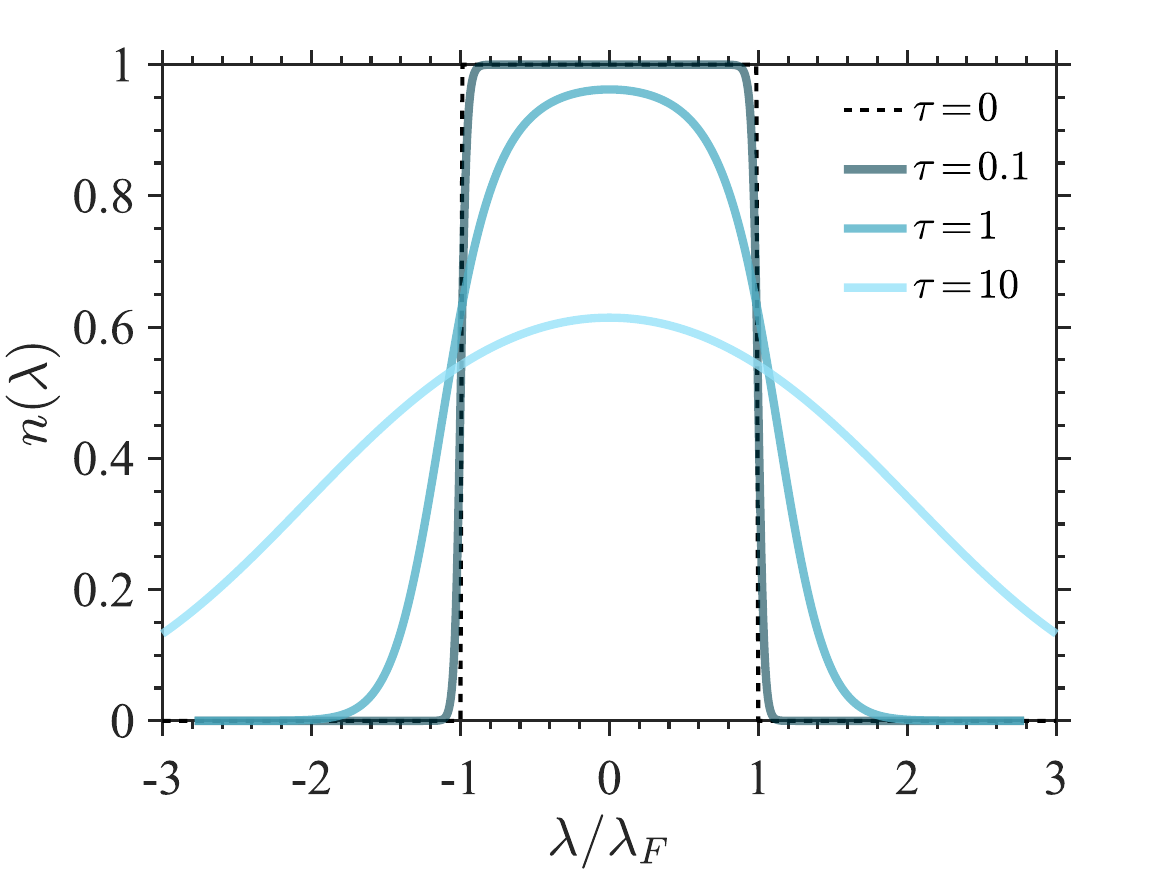}
    \caption{The zero-temperature limit of the filling function, $n(\lambda)$. As temperature is decreased from the degenerate regime, $\tau\!>\!1$, to $\tau\!=\!0$, the shape of the filling function becomes sharper, eventually being well described by an indicator function for the region $\lambda\!\in\![-\lambda_F,\lambda_F]$, where $\lambda_F$ is set either by the local effective chemical potential (see text), or the local atom number through Eq.~\eqref{eq:gs_density}. Importantly, this limiting behaviour is independent of dimensionless interaction strength $\gamma$. }
    \label{fig:T0_theta}
\end{figure}

In the limit of zero temperature, i.e. $T\!\to\!0$, the filling function, $n(\lambda)$, converges to an indicator function $n_{T=0}(\lambda) = \mathbf{1}_{[- \lambda_F, \lambda_F]}(\lambda)$, as illustrated in Fig.~\ref{fig:T0_theta} \cite{largescale_Doyon}.
This function is equal to 1 in the interval $[- \lambda_F, \lambda_F]$, and uniformly zero outside of this region, where $\lambda_F$ is the Fermi rapidity and is generally dependent on the local effective chemical potential via the local density approximation as described in Section \ref{sec:LDA}. If the local atom number is known, rather than the local chemical potential, $\lambda_F$ is instead constrained by Eq.~\eqref{eq:gs_density}.

\begin{figure}
    \centering
    \includegraphics[width=9cm]{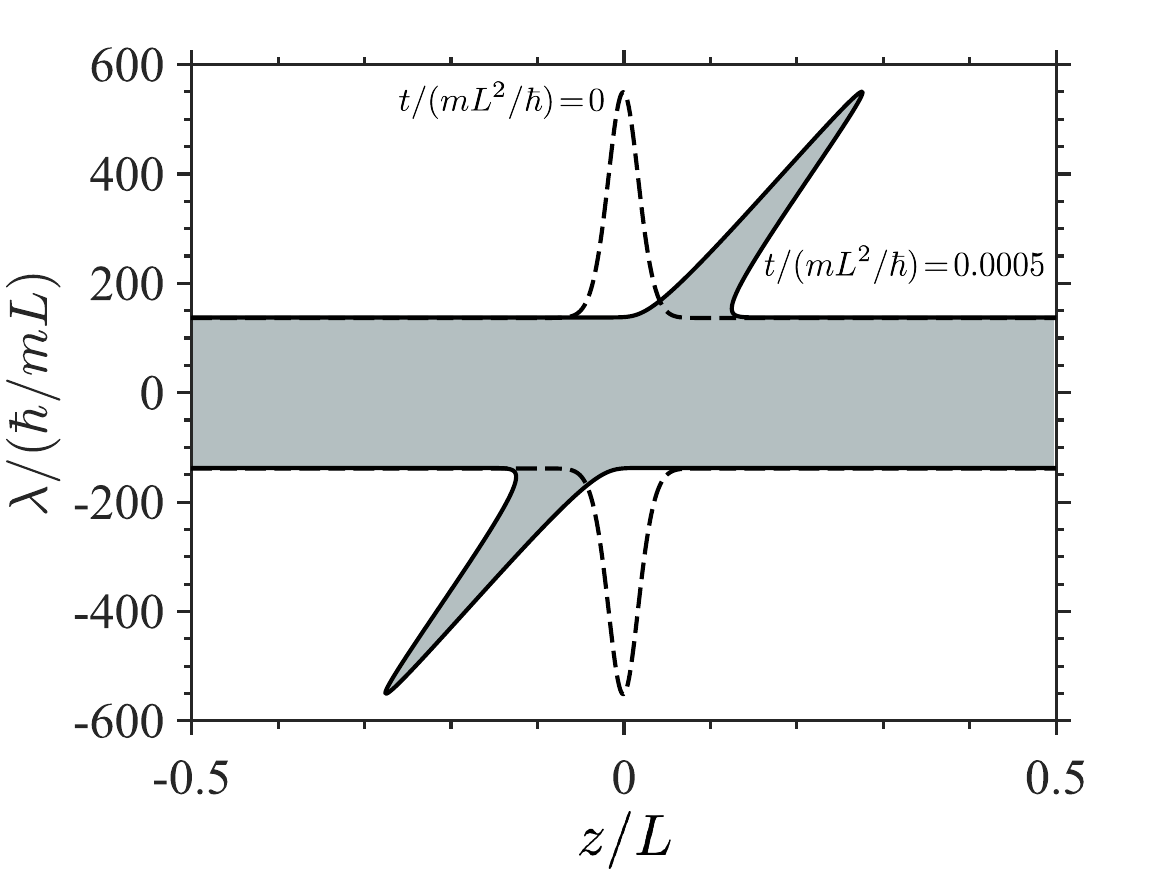}
    \caption{An illustration of evolution of the filling function at zero temperature for a shockwave emanating from an initial density bump, where the initial profile is shown in the dashed line. 
    The filling function takes on a value of 1 between the two curves, and zero everywhere else. Importantly, evolution of the edge of the filling function (i.e. the Fermi contour, see text) leads to wave breaking, where it becomes multi-valued for a single value of position, occuring as the higher rapidity points move faster than those lower on the density bump. This demonstrates the multiple rapidity points $\theta_j^+$ for the upper curve, and $\theta_j^-$ for the lower curve (see Eq.~\eqref{eq:T0_dressing}). For further details, see Chapter \ref{Chap:5}, which deals extensively with this shockwave scenario. }
    \label{fig:SW_T0_fillingfactor}
\end{figure}

As the GHD evolution equations conserve the value of the local filling factor, evolution may thus be described entirely through modelling the dynamics of the edges of this region, i.e. the points defined by $\lambda_F$. These edges are often referred to as the `Fermi contour' of the gas, in particular for systems of finite size \cite{malvania2021generalized,ruggiero2020quantum}. Numerically solving the GHD equations of motion for these systems may thus be accomplished through a zero-entropy algorithm, first detailed in Ref.~\cite{largescale_Doyon}. 
This algorithm works with the finite set of curves defined by the edges of the zero-temperature occupation number, with each point on these curves given by a position and rapidity coordinate $(z_i,\lambda_i)$. For a single time-step, $\delta t$, the position $x_i$ is shifted by an amount $v^\mathrm{eff}_{\{ \lambda \}}(\lambda_i) \delta t$, where calculating the effective velocity given in Eq.~\eqref{eq:effective_velocity} through the dressing operation in Eq.~\eqref{eq:dressing_operation} is simplified due to the form of the zero-temperature occupation number described above,
\begin{equation}\label{eq:T0_dressing}
    f^\mathrm{dr}(\lambda) = f(\lambda) + \sum_{i=1}^{k} \int_{\theta_j^-}^{\theta_j^+} d\theta \varphi(\lambda-\theta)f^\mathrm{dr}(\theta).
\end{equation}
Here, the filling factor, $n(\lambda)$, defines only the edges of the integration region $\theta_j^-$ and $\theta_j^+$, being exactly equal to $1$ within these bounds. Evolution of the GHD equation thus simplifies to the evolution of the Fermi contour, and may be written as
\begin{equation}
    \partial_t \lambda_j^\pm + v^\mathrm{eff}(\lambda_j^\pm)\partial_z\lambda_j^\pm = 0.
\end{equation}
Here, the effective velocity is calculated by utilizing Eq.~\eqref{eq:T0_dressing} in Eq.~\eqref{eq:effective_velocity}.
Notably, the GHD evolution may result in wave breaking, illustrated in Fig.~\ref{fig:SW_T0_fillingfactor}, where the Fermi contour becomes multi-valued for a single value of the position coordinate $z$. In this scenario, one utilizes the multiple $\lambda_j^{\pm}$ points seen in Eq.~\eqref{eq:T0_dressing}, and is explored in detail in Chapter \ref{Chap:5}.

\subsubsection{Integrability breaking}

Only a year after the first two papers introducing the GHD framework, the hydrodynamic equations of motion for the theory were extended to include the effects of inhomogeneous fields in Ref.~\cite{doyon2017note}. Accounting for the presence of an external trapping potential, $V(z)$, causes the Euler-scale evolution equation to take the form,
\begin{equation}
    \partial_t f_p(\lambda) + \partial_z \big( v^\mathrm{eff}[f_p(\lambda)] f_p(\lambda) \big) 
    =    \frac{\partial_z V(z)}{m}  \partial_\lambda f_p(\lambda) ,
\end{equation}
which mirrors the extension of conventional hydrodynamics to systems with broken translational invariance \cite{landau2013fluid}. Importantly, the breaking of translational invariance results in the rapidity distribution no longer being conserved, thus breaking the integrability of the model as described in Section \ref{sec:groundstate}.

In 2018, higher order corrections to the Euler-scale hydrodynamics were formulated, extending GHD to the Navier-Stokes scale and incorporating diffusive effects \cite{DeNardis_Diffusion_2018,gopalakrishnan2018hydrodynamics,bastianello_thermalization_2020,durnin2021diffusive,bastianello2021hydrodynamics,Bulchandani_2021}. At this level, the hydrodynamic equation of motion is modified by the incorporation of a diffusion operator, $\mathcal{D}$, arising through two-body scattering processes among quasiparticles \cite{de2019diffusion}. Subsequently, it was shown in Ref.~\cite{bastianello_thermalization_2020}, and later fully justified in Ref.~\cite{durnin2021diffusive}, that the diffusive hydrodynamic equation for the quasiparticle density, $f_p$, in the presence of an external trapping potential is given by
\begin{equation}\label{eq:DiffusiveGHD}
\begin{split}
\begin{aligned}
\partial_t & f_p + \partial_z(v^\mathrm{eff} f_p) = \frac{\partial_z V(z)}{m} \partial_\lambda f_p + \partial_z(\mathcal{D} \, \partial_z f_p),
\end{aligned}
\end{split}
\end{equation}
where $(\mathcal{D} \partial_z f_p)(\lambda) = \int d\lambda' \mathcal{D}(\lambda,\lambda'; z,t) \partial_z f_p(\lambda')$, and $\mathcal{D}(\lambda,\lambda')$ is the diffusion kernel \cite{DeNardis_Diffusion_2018,de2019diffusion}. Importantly, all elements required to simulate the higher-order evolution equation are already present in the Euler-scale theory of GHD, requiring no additional assumptions about the form of the diffusion operator.

It was subsequently shown that, when an external trapping potential breaks the integrability of a gas, diffusive dynamics inevitably lead the system to thermalize at late times \cite{bastianello_thermalization_2020,bastianello2021hydrodynamics}. This allowed for a complete simulation of the dynamics of thermalization for the quantum Newton's cradle experiment, first observed in experiment in 2006 \cite{kinoshita2006quantum,bastianello_thermalization_2020}.
Diffusive dynamics are soluble through a second-order backwards implicit algorithm, first demonstrated in Ref.~\cite{bulchandani2018bethe}, available within the \textit{iFluid} package \cite{moller2020introducing}.

%% file: Chapter3/Chapter3.tex
\chapter[An interaction-driven quantum many-body engine enabled by atom-atom correlations]{An interaction-driven quantum many-body engine enabled by atom-atom correlations.}
\label{Chap:3}	
\pagestyle{headings}

\textit{Particle-particle correlations, characterized by the second-order Glauber correlation function, play an important role in the understanding of various phenomena in radio and optical astronomy, quantum and atom optics, particle physics, condensed matter physics, and quantum many-body theory. However, the relevance of such correlations to quantum thermodynamics has so far remained illusive. Here, we propose and investigate a class of quantum many-body thermal machines whose operation is directly enabled by second-order atom-atom correlations in an ultracold atomic gas. More specifically, we study quantum thermal machines that operate in a sudden interaction-quench Otto cycle and utilize a one-dimensional Lieb-Liniger gas of repulsively interacting bosons as the working fluid. 
The atom-atom correlations in such a gas are different to those of a classical ideal gas, and are a result of the interplay between interparticle interactions, quantum statistics, and thermal fluctuations. 
We show that operating these thermal machines in the intended regimes, such as a heat engine, refrigerator, thermal accelerator, or  heater, would be impossible without such atom-atom correlations. Our results constitute a step forward in the design of conceptually new quantum thermodynamic devices which take advantage of uniquely quantum resources such as quantum coherence, correlations, and entanglement. }


\section{Introduction}
\label{Sec:g2_Engine_Intro}	






\noindent \noindent Quantum thermal machines (QTM), such as quantum heat engines (QHE), refrigerators, and quantum batteries, are central in the theoretical and experimental development of the emerging field of quantum thermodynamics \cite{vinjanampathy2016quantum,kosfloff2014quantum}. Their primary utility is to explore the fundamental laws of thermodynamics in the quantum realm and to  demonstrate possible advantages gained by utilising quantum resources. Accordingly, understanding QTM's are expected to play a similar role in the development of quantum technologies as classical heat engines played in fostering scientific advances during
the Industrial Revolution. In the past decade, progress in the control over quantum platforms, such as single ions \cite{rossnagel2016single,Ca-ion-spin-engine}, nitrogen vacancy centers \cite{Nitrogen-vacancy-heat-engine}, and single-atom impurities immersed in an ultra-cold atomic bath \cite{bouton2021quantum}, have led to the realization of single-particle QHE's. Such single-particle QHE's represent the ultimate limit in the realization of an `infinitesimal machine' \cite{feynman2018there}.

However, in order to utilize the breadth of quantum resources available, one must move beyond single-particle systems---to engines that utilize interacting many-particle systems. 
Such QHE's are uniquely positioned to take advantage of quantum resources, such as entanglement \cite{brandao2008entanglement,funo2013thermo}
, correlations \cite{oppenheim2002thermo,llobet2015extractable}
, or quantum coherence \cite{narasimhachar2015low,Korzekwa2016extraction},
to enhance the performance of classical heat engines \cite{jaramillo2016quantum} or perform entirely new tasks that would be impossible classically \cite{halpern2019quantum}.
In particular, control over inter-particle interactions allows for the creation of  strictly many-body QHE's, \cite{jaramillo2016quantum,mathieu2016scaling,li2018efficient,chen2019interaction,keller2020feshbach}
which have recently been realized in the laboratory \cite{koch2022making,simmons2023thermodynamic}. 
These very recent experimental developments underscore the need for further studies of thermodynamic processes in the context of interacting quantum many-body thermal machines.

Here, we propose a quantum many-body Otto heat engine---as well as related thermal machines such as Otto refrigerator, thermal accelerator, and heater---using a uniform one-dimensional (1D) Bose gas with repulsive contact interactions as the working fluid. In the proposed Otto cycles, the unitary work strokes are driven by a sudden quench of the interaction strength, and we demonstrate how the thermodynamic performance, in particular net work and efficiency, of these many-body QTM's can be calculated through the experimentally measurable atom-atom local pair correlation \cite{kheruntsyan2003pair,kheruntsyan2005finite,kinoshita2005local}.

The atom-atom local correlation, $g^{(2)}(0)$, is described by the second-order Glauber correlation function $g^{(2)}(r)$ at zero interparticle separation (i.e., when $r=0$, where $r=|z-z'|$ is the distance between the two particles with positions $z$ and $z'$ ) and characterises the probability of pairs of atoms to be found at the same point, relative to uncorrelated atoms. As such, the correlation function may be measured experimentally through, amongst other methods, monitoring the photoassociation rate to a bound molecular state, which is directly proportional to the probability of two atoms being at the same point. Such a measurement was originally accomplished for a strongly interacting 1D Bose gas by Kinoshita \textit{et al.} in 2005 \cite{kinoshita2005local}, and by Partridge \textit{et al.} for a 3D Fermi gas, also in 2005 \cite{partridge2005molecular}.

The benefits of using the 1D Bose gas as the working fluid in the proposed Otto cycles is that the underlying theoretical model---the Lieb-Liniger model---is exactly solvable in the uniform limit via the Yang-Yang thermodynamic Bethe ansatz (TBA) \cite{Lieb-Liniger-I,yang1969thermodynamics,korepin1993}, in addition to being experimentally realizable using ultracold atomic gases confined to highly anisotropic traps 
\cite{gorlitz2001realization,greiner2001phase,schreck2001quasipure,moritz2003exciting,tolra2004observation,paredes2004tonks,greiner2001bose,richard2003momentum,paredes2004tonks,stoferle2004exciting,meyrath2005box,Kinoshita1125,kinoshita2005local,kinoshita2006quantum,trebbia2006experimental,esteve2006observation,hofferberth2007non,Karen_Yang_2008,Schmiedmayer_1D_PRL_2010,Schmiedmayer_1D_PRL_2011,Armijo_ThreeBody_2010,jacqmin2011sub,Bouchoule_dimensional_crossover,Atom_chips,Schmiedmayer_1D_PRL_2013,1D_entropy_2017,Schmiedmayer_Universal_2018,Schmiedmayer_1D_Optics_2019}.
This offers unique opportunities for gaining physical insights into the performance of such Otto QTM's as a tractable and testable quantum many-body problem.
Additionally, the Lieb-Liniger gas has a rich phase diagram spanning several nontrivial regimes \cite{kheruntsyan2003pair,kheruntsyan2005finite}, from the weakly interacting quasicondensate through to the strongly interacting Tonks-Girardeau regime of fermionization \cite{Girardeau_1960,Kinoshita1125,Girardeau2000,Girardeau_Bose_Fermi,Mora-Castin-2003,castin2000,Gangardt2003,kheruntsyan2003pair,kheruntsyan2005finite,Yukalov_2005,GangardtMinguzziExact}.
The atom-atom correlation within these regimes takes on a range of values between $0<g^{(2)}(0)<2$, depending on the temperature and interaction strength, which aids the operation of the proposed Otto cycles under a variety of conditions.
We evaluate the performance of the 1D Bose gas Otto QTM's, but we emphasise that the broad conclusions arrived at here are not limited to the Lieb-Liniger model.

~

\section{Interaction-driven Otto cycle}

\noindent We start by considering an interaction-driven Otto heat engine cycle with a uniform 1D Bose gas as the working fluid. In a uniform 1D Bose gas, described by the integrable Lieb-Liniger model \cite{Lieb-Liniger-I}, the interatomic interaction strength $g$ can be expressed in terms of the harmonic trap frequency $\omega_{\perp}$ in the tightly confined (transverse) direction and the 3D $s$-wave scattering length $a_s$ as $g \simeq 2 \hbar \omega_\perp a_s$ \cite{Olshanii1:998}. Accordingly, changing the interaction strength $g$ may be achieved by either tuning the external trapping potential that controls the transverse confinement $\omega_{\perp}$ or by changing the scattering length $a_s$ by means of a magnetic Feshbach resonance \cite{chin2010feshbach}. The former option leads to a volume change of the gas (i.e. transverse expansion or compression), and hence can be thought of as analogous to mechanical work in the conventional Otto cycle, however, changing $g$ via a change of the scattering length leads to identical results, which then justifies our referral to the engine cycle as the Otto cycle regardless of the means of tuning the interaction strength.
We emphasize, however, that the dynamics of the quantum Otto cycle explored here are strictly longitudinal, with the gas always remaining in its transverse ground state.

\begin{figure}[!tbp]
\begin{center}
   \includegraphics[width=12cm]{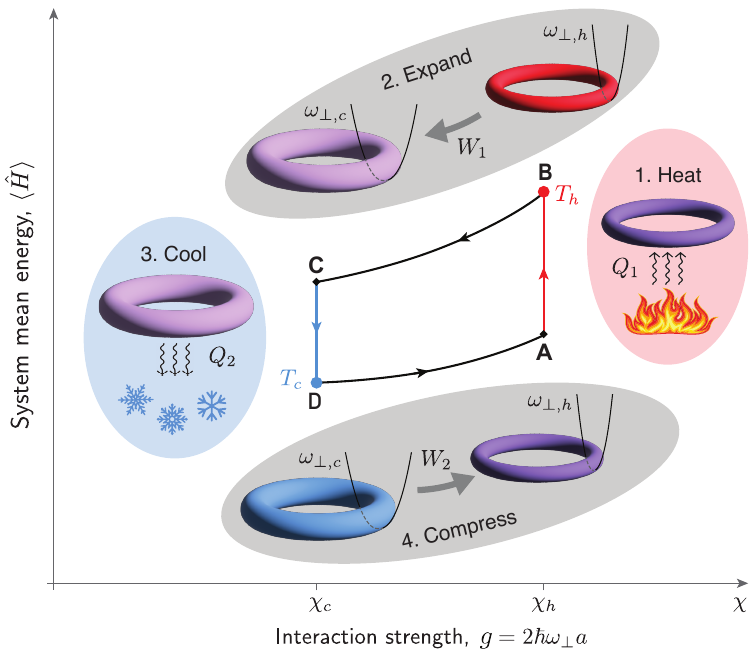}  
   \caption{An interaction-driven quantum many-body Otto cycle, operating between two interaction strengths, $g_{c}$ and $g_{h}$. Unitary work strokes (\textbf{BC} and \textbf{DA}) are shown in black, while non-unitary thermalization strokes (\textbf{AB} and \textbf{CD}) are color-coded to the cold (blue) and hot (red) reservoirs at temperatures $T_{c}$ and $T_{h}$, respectively.}
  \label{fig:Engine_Diagram}
\end{center}
\end{figure}

The uniform 1D Bose gas with repulsive contact interactions  \cite{Lieb-Liniger-I} is described by the second-quantized Hamiltonian
\begin{equation}\label{eq:Hamiltonian_LL}
\begin{split}
\begin{aligned}
    \hat{H} &= \hat{H}^{kin} + \hat{H}^{int} 
    \\&=- \frac{\hbar^2}{2m}\int dz \hat{\Psi}^\dagger \frac{\partial^2}{\partial z^2} \hat{\Psi}
    + \frac{g}{2} \int dz  \hat{\Psi}^\dagger \hat{\Psi}^\dagger \hat{\Psi} \hat{\Psi},
\end{aligned}
\end{split}
\end{equation}
where $m$ is the atomic mass, $g$ is the strength of the contact interactions, and $\hat{\Psi}^\dagger(z)$ and $\hat{\Psi}(z)$ are the bosonic field creation and annihilation operators, respectively. We here highlight the separation of the Lieb-Liniger Hamiltonian into its kinetic energy, $\hat{H}^{kin}$, and interaction energy, $\hat{H}^{int}$, components, to be referred to later.
Ground state solutions to this integrable model are dependent only on a single dimensionless interaction strength, $\gamma\!=\!m g/\hbar^2 \rho$, where $\rho\!=\!N/L$ is the linear density for $N$ particles in a system of size $L$. Finite temperature solutions, on the other hand, can be obtained using the Yang-Yang thermodynamic Bethe ansatz (TBA) \cite{yang1969thermodynamics}, and can be parametrised by an additional dimensional parameter, the dimensionless temperature $\tau \!=\! 2 m k_B T/\hbar^2 \rho^2$ \cite{kheruntsyan2003pair}.

The interaction-driven Otto engine cycle, which we thus consider, consists of four strokes (see Fig.~\ref{fig:Engine_Diagram}): 
\begin{itemize}
\item[(1)] \textit{Thermalization with hot reservoir}, \textbf{A}$\to$\textbf B: the working fluid, consisting of $N$ total atoms at interaction strength $g_{h}$, generally beginning in a nonequilibrium state, is connected to a hot ($h$) reservoir at temperature $T_{h}$, where it is left to equilibrate, taking in heat $Q_1\!=\!\langle \hat{H}\rangle_{\textbf{B}} \!-\!\langle \hat{H}\rangle_{\textbf{A}}\!>\!0$, which is to be partially converted into beneficial work in the subsequent stroke. Here $\hat{H}$ is the system Hamiltonian, and $\langle \hat{H}\rangle_{\textbf{j}}$ is its expectation value, i.e., the total energy of the system, in state $\textbf{j}=\{\textbf{A,B,C,D}\}$.
\item[(2)] \textit{Unitary expansion, \textbf{B}$\to$\textbf{C}}: the working fluid, now in a thermal equilibrium state at temperature $T_{h}$, is decoupled from the hot reservoir and has its interaction strength quenched from $g_{h}$ to $g_{c}\!<\!g_{h}$, generating beneficial 
work $W_1\!=\!\langle \hat{H}\rangle_{\textbf{C}} \!-\!\langle \hat{H}\rangle_{\textbf{B}}\!<\!0$ done by the fluid. 
\item[(3)] \textit{Thermalization with cold reservoir}, \textbf{C}$\to$\textbf{D}: the working fluid is connected to a cold ($c$) reservoir at temperature $T_{c}<T_{h}$ and allowed to equilibrate at constant interaction strength $g_{c}$ while dumping energy in the form of heat $Q_2\!=\!\langle \hat{H}\rangle_{\textbf{D}} \!-\!\langle \hat{H}\rangle_{\textbf{C}}\!<\!0$ into the cold reservoir. 
\item[(4)] \textit{Unitary compression}, \textbf{D}$\to$\textbf{A}: disconnected from the cold reservoir, the working fluid has its interaction strength quenched from $g_{c}\!\to\!g_{h}$, with work $W_2\!=\!\langle \hat{H}\rangle_{\textbf{A}} \!-\!\langle \hat{H}\rangle_{\textbf{D}}\!>\!0$ done on the fluid, and the system returning to the initial state of the overall cycle.
\end{itemize}
Such an engine cycle generates net beneficial work (done by the fluid) if the total work $W \!=\! W_1 \!+\! W_2\!<\!0$, i.e., if $|W_1|>W_2$ (or $Q_1>|Q_2|$), with efficiency $\eta \!=\! -W/Q_1 \!=\! 1 \!-\! |Q_2|/Q_1$, where we used the conservation of energy $W+Q=0$, with $Q=Q_1 + Q_2$ being the total heat \cite{CallenHerbertB1985Taai}.

We note that, though each pair of consecutive nodes ($\{\textbf{A,B,C,D}\}$) in the engine cycle diagram shown in Fig.~\ref{fig:Engine_Diagram} are connected via continuous lines, which is typically indicative of quasistatic processes \cite{Callen_book,SchroederD_ThermalPhysics}, at no point along \emph{any} of these connecting lines is the system in question in an equilibrium state. Indeed, the only equilibrium states contained in Fig.~\ref{fig:Engine_Diagram} are the thermal equilibrium states $\textbf{B}$ and $\textbf{D}$. Here, continuous lines connecting the various nodes are solely employed in service of simplifying the description of our chosen engine cycle.

~

\noindent \textbf{\large{Work from second-order Glauber correlations}}

In this chapter, we specifically consider a sudden or instantaneous quench of the interaction strength $g$ in the unitary strokes (2) and (4). 
Under a sudden interaction quench, the initial ($i$) and final ($f$) expectation values over field operators in the system Hamiltonian, i.e., the expectation values before and after the quench,
remain unchanged as the system did not have sufficient time to evolve into a new state. Hence, the only contribution to the difference in total energy before and after the quench, $ \langle\hat{H} \rangle_f - \langle \hat{H} \rangle_i$, comes from the difference between the interaction terms, $\frac{1}{2}g_f\int dz \langle \hat{\Psi}^{\dagger}\hat{\Psi}^{\dagger} \hat{\Psi}\hat{\Psi} \rangle_f-\frac{1}{2}g_i\int dz \langle \hat{\Psi}^{\dagger}\hat{\Psi}^{\dagger} \hat{\Psi}\hat{\Psi} \rangle_i$, where $\langle \hat{\Psi}^{\dagger}\hat{\Psi}^{\dagger} \hat{\Psi}\hat{\Psi} \rangle_f=\langle \hat{\Psi}^{\dagger}\hat{\Psi}^{\dagger} \hat{\Psi}\hat{\Psi} \rangle_i$ in a sudden quench, and $\hat{\Psi}^{\dagger}(z)$ and $\hat{\Psi}(z)$ represent the field creation and annihilation operators. Accordingly, the energy difference can be expressed as $ \langle\hat{H} \rangle_f - \langle \hat{H} \rangle_i\!=\!  \frac{1}{2}(g_f \!-\! g_i)\overline{G^{(2)}_i}$,
where we have defined the total (integrated) second-order correlation of the thermal equilibrium state $ \overline{G^{(2)}_i} \!=\! \int dz \langle \hat{\Psi}^\dagger \hat{\Psi}^\dagger \hat{\Psi} \hat{\Psi} \rangle_i$ \cite{kheruntsyan2005finite}.

Identifying the $i$ and $f$ states as points \textbf{B} (hot, $h$) and \textbf{C}, or as \textbf{D} (cold, $c$) and \textbf{A} in the digram of Fig. \ref{fig:Engine_Diagram}, the net work of the Otto engine can be expressed as 
\begin{equation}\label{eq:Work_G2}
    W = -\frac{1}{2}(g_{h} - g_{c})\left( \overline{G^{(2)}_{h}} - \overline{G^{(2)}_{c}}\right).
\end{equation}
Likewise, the efficiency of the engine may be expressed as
\begin{equation}\label{eq:Efficiency_G2}
    \eta = 1 - \frac{ \langle\hat{H} \rangle_{h} - \langle \hat{H} \rangle_{c} - \frac{1}{2}\left(g_{h} - g_{c} \right) \overline{G^{(2)}_{h}}}{ \langle\hat{H} \rangle_{h} - \langle \hat{H} \rangle_{c}   - \frac{1}{2}\left(g_{h} - g_{c} \right)  \overline{G^{(2)}_{c}}}.
\end{equation}
These equations allow for investigation of the interaction-driven Otto engine  under a sudden quench protocol through solely the equilibrium properties of the gas, as all expectation values involved are with respect to $h$ (\textbf{B}) or $c$ (\textbf{D}) states.

\begin{figure*}[!tbp]
\begin{center}
   \includegraphics[width=16cm]{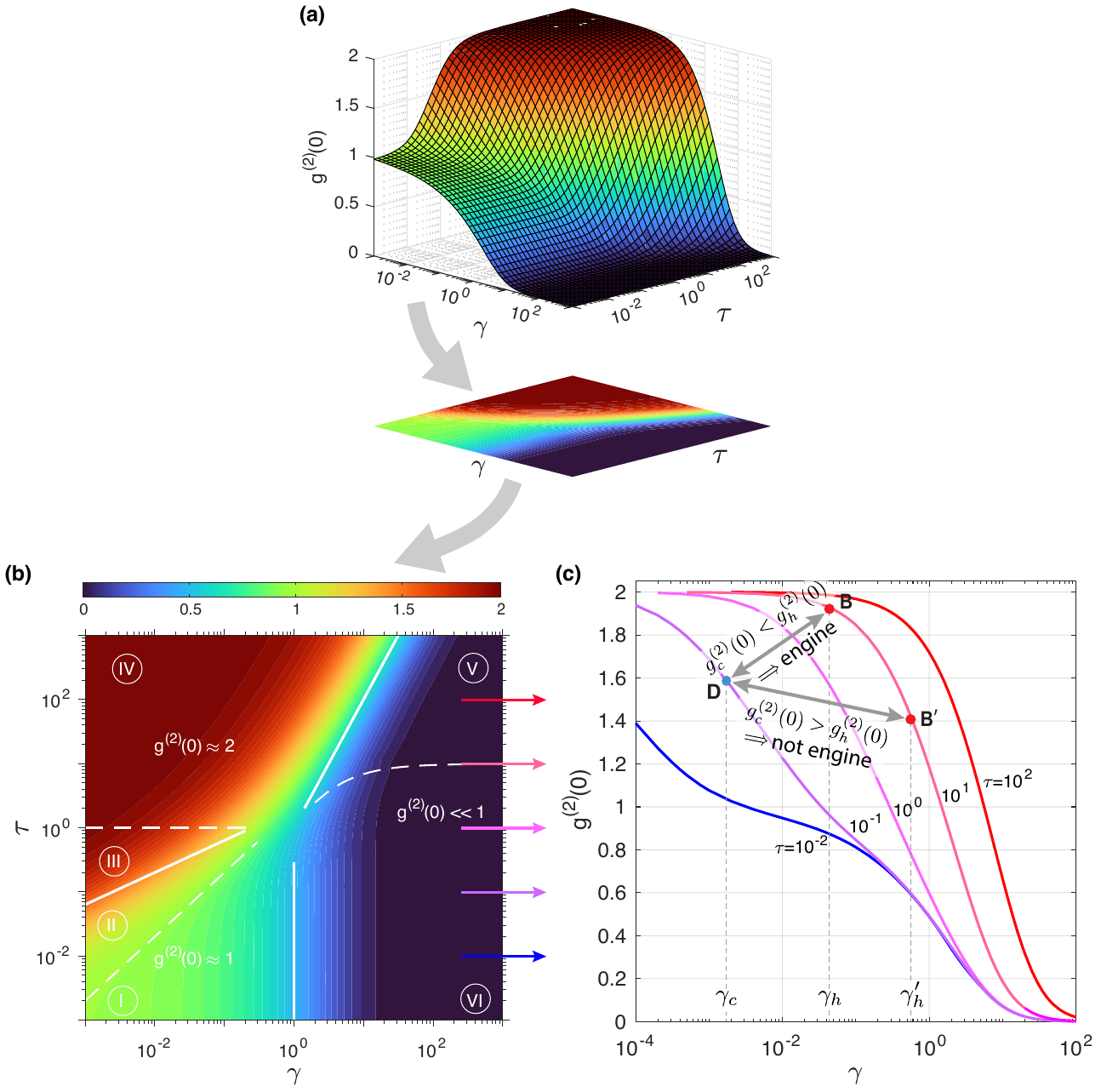}   
   \caption{Atom-atom correlations, described by Glauber's $g^{(2)}(0)$ correlation function, for the uniform 1D Bose gas evaluated using the exact Yang-Yang TBA \cite{kheruntsyan2003pair,yang1969thermodynamics}. Panel \textbf{a} shows $g^{(2)}(0)$ as a function of the dimensionless interaction strength, $\gamma$, and temperature, $\tau$. In panel \textbf{b}, this is translated into a contour diagram, in which we also show the crossover boundaries (white solid and dashed lines) between the different asymptotic analytic regimes \cite{kheruntsyan2003pair}.
   Panel ($\textbf{c}$) shows line plots of $g^{(2)}(0)$ vs $\gamma$, at different fixed values of $\tau$, together with two possible choices, $\textbf{D-B}$ or $\textbf{D-B}'$, of the thermal equilibrium operating points of the Otto cycle from Fig.~\ref{fig:Engine_Diagram}; as we see, according to Eq.~\eqref{eq:Work_Uniform}, operating the Otto cycle as an engine (with $W<0$) can be achieved between the points $\textbf{D-B}$ ($\gamma_c\longleftrightarrow \gamma_h$), where $g^{(2)}_c(0)\!<\!g^{(2)}_h(0)$, but not between $\textbf{D-B}'$ ($\gamma_c\longleftrightarrow \gamma'_h$), where $g^{(2)}_c(0)\!>\!g^{(2)}_h(0)$ due to the stronger interaction quench, even though the temperature at $\textbf{D}$ is still lower than at $\textbf{B}'$.
   }
  \label{fig:g2_parameter_space}
  \end{center}
\end{figure*}

Realistically, a sudden quench of interaction strength from $g_{h(c)}$ to $g_{c(h)}$ would still occur over a finite duration $\Delta t$. The ``instantaneity'' of the quench refers to the assumption that $\Delta t$ is much shorter than the characteristic time scale for longitudinal dynamics, i.e. that $\Delta t \!\ll\! m l_{\text{cor}}^2/\hbar$, where $l_{\text{cor}}$ is the characteristic short-range correlation length in the system. This may be given in each regime, defined in Section \ref{sec:g2_glauber}, by: the healing length $l_h\!=\!\hbar/\sqrt{m g \rho}$ in regimes I and II;  thermal phase coherence length $l_\phi \!=\! \hbar^2 \rho / m k_B T$ in regime III; thermal de Broglie wavelength $\lambda_T \!=\! \sqrt{2 \pi \hbar^2 / m k_B T}$ in regime IV; absolute value of the 1D scattering length $|a_{1D}|=2 \hbar^2/mg$ in the regime of high-temperature fermionization V; and the Fermi wavelength $\lambda_F \!=\! 2 / \rho$ in the Tonks-Girardeau regime of low-temperature fermionization VI. 
Thus, it is with respect to the \emph{longitudinal} dynamics that we refer to our quench as sudden. With respect to the \emph{transverse} dynamics, on the other hand, we are assuming that
$\Delta t$ is sufficiently long ($\Delta t\gg 2\pi/\omega_{\perp}$) compared to the characteristic transverse timescale, $2\pi/\omega_{\perp}$, governed by the transverse harmonic trap frequency $\omega_{\perp}$ \cite{Olshanii1:998}.
As a result, the quench would retain the system in the transverse ground state, and hence would not compromise the 1D character of the system. 
As such, the work done on (or by) the system during the unitary strokes can be regarded as transversely quasistatic.

For a magnetically trapped ultracold 1D Bose gas, the work done via transverse compression and expansion is ultimately magnetic: it is done by the magnetic field on the atomic dipole moments when $\omega_{\perp}$ is increased, or vice versa -- by the atomic dipole moments on the magnetic field when  $\omega_{\perp}$ is decreased. Alternatively, the change in the interaction strength $g$ is implemented through control over the $s$-wave scattering length $a_s$ via a magnetic Feshbach resonance \cite{chin2010feshbach}, in which case the nature of the work is still magnetic.
We use the term Otto cycle in the same sense as used to describe, e.g., a harmonic oscillator Otto engine \cite{Kosloff1984quantum,li2018efficient,chen2019interaction,koch2022making,bouton2021quantum}, wherein the harmonic oscillator frequency (rather than the volume of the system) is fixed as an external parameter during the thermalization strokes. In our case, it is the interaction strength that is fixed, which itself is proportional to the transverse harmonic confinement frequency of the 1D Bose gas.

\section{Quantum Otto engine cycle}

Finite-temperature uniform 1D Bose gases have no phase transition to a true Bose-Einstein condensate in the thermodynamic limit, unlike Bose-Einstein condensation in three dimensions \cite{Pethick_Smith_book}.
However, there still exists a rich crossover phase diagram of different regimes that can be characterised by the normalized local (same-point) atom-atom pair correlation function $g^{(2)}(0)$ \cite{kheruntsyan2003pair}. The pair correlation function is a thermodynamic quantity that can be calculated from the exact TBA, as well as using approximate analytic methods \cite{kheruntsyan2003pair}, and is shown in Figs.~\ref{fig:g2_parameter_space} \textbf{a} and \textbf{b}. 
This function, and hence the different regimes of the uniform 1D Bose gas, can be parameterized by dimensionless interaction strength, $\gamma \!=\! m g/\hbar^2 \rho$, and dimensionless temperature, $\tau \!=\! 2 m k_B T/\hbar^2 \rho^2$, where $m$ is the boson mass and $\rho\!=\!N/L$ is the 1D density for $N$ atoms in a system of length $L$.

\begin{figure}[!tbp]
\begin{center}
   \includegraphics[width=12cm]{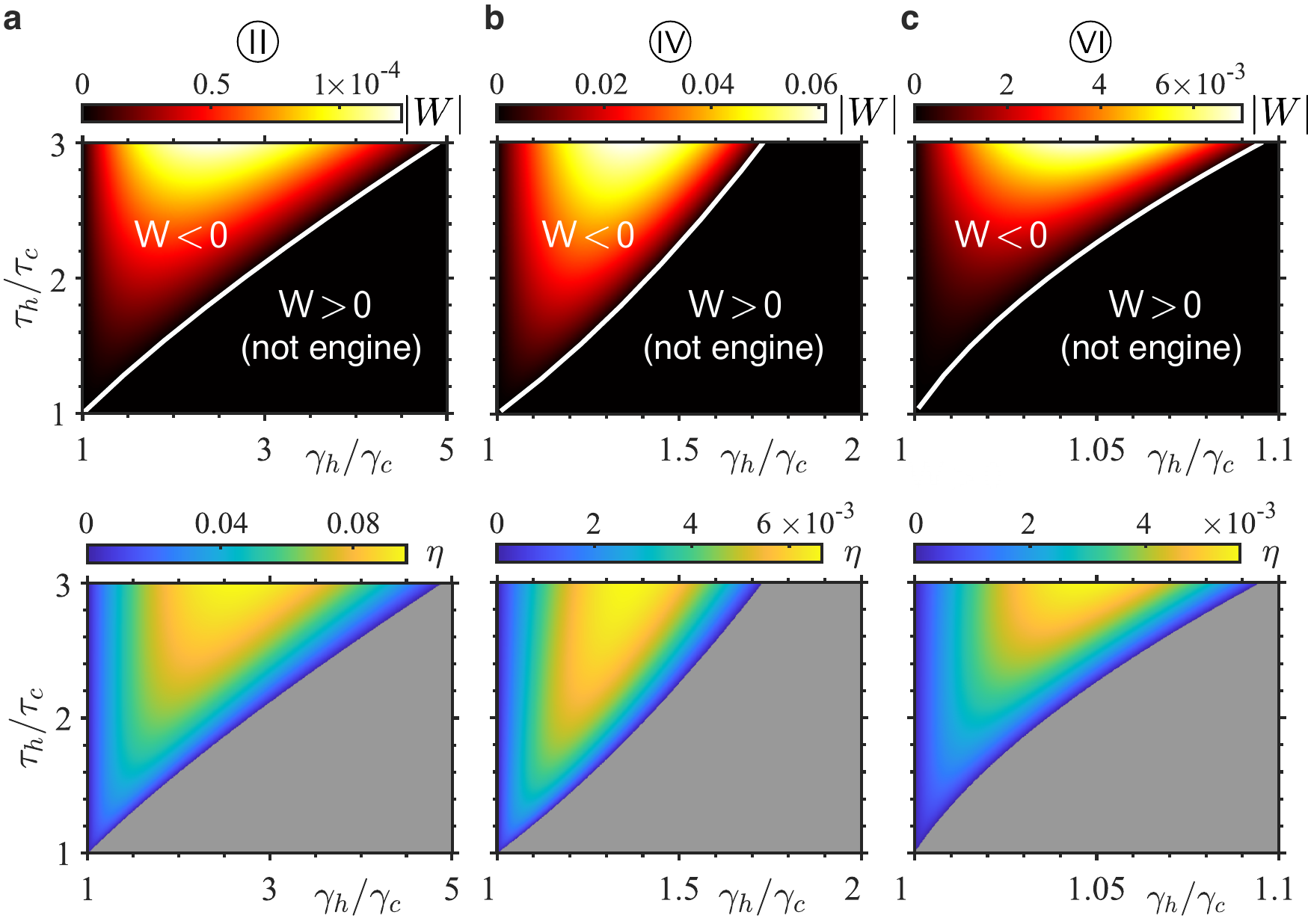}   
   \caption{Performance of the interaction-driven quantum Otto engine. Columns, \textbf{a}, \textbf{b}, and \textbf{c},  demonstrate net work, $W$ (in units of $\hbar^2 \rho^2 / 2m$), and efficiency, $\eta$, as a function of the ratio of interaction strength, $\gamma_h/\gamma_c$, and temperature, $\tau_h/\tau_c$, between the hot ($h$) and cold ($c$) thermal equilibrium states. The example of panel  \textbf{a} is for $\gamma_c\!=\!10^{-3}$ and $\tau_c\!=\!10^{-2}$, where $\gamma_h/\gamma_c$ and $\tau_h/\tau_c$ explores the parameter range within the region II of the equilibrium regimes diagram of Fig.~\ref{fig:g2_parameter_space}\textbf{b}. Similarly, panel \textbf{b} explores region IV, with $\gamma_c\!=\!1$ and $\tau\!=\!10$, whereas panel \textbf{c} explores region VI, with $\gamma_c\!=\!10$ and $\tau\!=\!1$.
   }
  \label{fig:engine_performance}
\end{center}
\end{figure}


For a uniform 1D Bose gas, the total correlation in the hot  or cold  thermal equilibrium state may be expressed as $\overline{G^{(2)}_{h\textbf{c}}} \!=\!N \rho g^{(2)}_{h\textbf{c}}(0)$.    Combining this with Eq.~\eqref{eq:Work_G2}, the net work per particle can be expressed as
\begin{equation}\label{eq:Work_Uniform}
    \frac{W}{N} = -\frac{\hbar^2 \rho^2}{ 2 m}(\gamma_{h} - \gamma_{c})\left( g^{(2)}_{h}(0) - g^{(2)}_{c}(0) \right),
\end{equation} 
i.e., the net work is directly proportional to the difference between atom-atom correlations of the 1D Bose gas in the hot and cold thermal equilibrium states. This simple relationship between thermodynamic work and Glauber second-order correlation function represents one of the key results of this chapter.

From Eq.~\eqref{eq:Work_Uniform}, and given that $\gamma_{h}$ is always larger than $\gamma_{c}$, we see that if the local pair correlations did not depend on the respective interaction strengths and temperatures, i.e. if they were the same, $ g^{(2)}_{h}(0)\! =\! g^{(2)}_{c}(0)$, then the net work per particle would vanish. 
We therefore conclude that extracting net work ($W\!<\!0$) from this Otto cycle, and hence operating it as a heat engine, can only be enabled by atom-atom correlations; more specifically, the only way to extract net work is to have $ g^{(2)}_{h}(0)\!>\!g^{(2)}_{c}(0)$ (see Figs.~\ref{fig:g2_parameter_space} \textbf{c}). We note that this implies a saturation of net work as the temperature of both hot and cold thermal states are raised such that they both occur deep within regime IV of Fig.~\ref{fig:g2_parameter_space}(b). This is in contrast to the conventional Otto cycle, where work strokes are performed through control over the system volume, with performance remaining linear in temperature ratio in this limit of high temperatures.

Net work, Eq.~\eqref{eq:Work_Uniform}, and efficiency, Eq.~\eqref{eq:Efficiency_G2}, of this quantum Otto engine, calculated for simplicity using analytic approximations to the atom-atom correlation function and total energy \cite{kheruntsyan2003pair}, are shown in Fig.~\ref{fig:engine_performance} as a function of the ratio of interaction strengths, $\gamma_{h}/\gamma_{c}$, and temperatures, $\tau_{h}/\tau_{c}$, for three of the six asymptotic regimes. Notably, in this Otto engine cycle, for any fixed value of the temperature ratio, the interaction strength quench corresponding to maximum net work is approximately the same as that providing maximum efficiency; this occurs as, to first order, the heat intake $Q_1$ varies slowly with $\gamma_h/\gamma_c$, meaning $\eta \!\propto\! W$.
The observed increase of net work and efficiency in all regimes under large temperature ratio may be attributed 
to the fact that the local correlation of the hot thermal state in Eq.~\eqref{eq:Work_Uniform} is always a monotonically increasing function of $\tau$.
However, as the correlation function is also monotonically decreasing under $\gamma_h$, this results in no extractable net work under sufficiently large interaction strength ratios for any given temperature ratio.

At a glance, one may conclude that the net work, which is enabled through the $g^{(2)}(0)$ correlation function, is maximized under the largest possible difference in correlation function, i.e. $g^{(2)}_h(0)\!-\!g^{(2)}_c(0)\!\simeq\! 2$.
However, to achieve this, while also guaranteeing that $\gamma_h \!>\! \gamma_c$, would require an unrealistically high (from practical point of view) temperature ratio to operate between regimes VI and IV, shown in Fig.~\ref{fig:g2_parameter_space}. Rather, we observe that, while the $g^{(2)}(0)$ correlation function is responsible for enabling operation as a heat engine, the \emph{magnitude} of net work is governed more strongly by the difference in the interaction strengths, $\gamma_h-\gamma_c$, which is unrestricted. Consequently, it is in the weakly interacting ($\gamma\ll1$) region II that we observe the lowest magnitude of net work (see Fig.~\ref{fig:engine_performance}\textbf{a}), where $\gamma_h-\gamma_c$ is very small.

In comparison, the magnitude of net work is largest in regime IV, shown in Fig.~\ref{fig:engine_performance}\textbf{b}, where $\gamma\!\sim\!1$ and hence the difference $\gamma_h\!-\!\gamma_c$ can also be on the order of one. The same considerations apply to the strongly interacting ($\gamma\!\gg\!1$) regime VI, where one can operate under the largest magnitudes of interaction strengths, however, in this regime the net work is diminished due to the vanishing of correlation itself ($g^{(2)}\!\ll\!1$) due to the effect of fermionization. In contrast to these observations, the efficiency of the engine, Eq.~\eqref{eq:Efficiency_G2}, is inversely dependent on the total energy of the thermal states, which is minimal in the weakly interacting low temperature regime II, which thus has the largest efficiency. 
Further, we note that the efficiencies presented in Fig.~\ref{fig:engine_performance}, while of low magnitude, are not significantly reduced when compared to the corresponding reversible engine cycle results.


As shown in Eq.~\eqref{eq:Hamiltonian_LL}, the total energy may be separated into its kinetic energy, which scales predominately with temperature, and interaction energy, which scales predominately with interaction strength.
Thus, for a fixed ratio of temperatures, $\tau_h/\tau_c$, the difference between the total energies of the hot and cold thermal state may be given as a sum of two terms: the first is the kinetic energy difference, determined by the temperature ratio and therefore approximately constant, the second given by the interaction energy difference, which scales with the interaction strengths, $\gamma_h$ and $\gamma_c$, of the hot and cold thermal states as
\begin{equation}\label{eq:interaction_energy_diff}
    \langle \hat{H}^{int} \rangle_h \!-\! \langle \hat{H}^{int} \rangle_c \!=\! N \frac{\hbar^2 \rho^2}{2m}\left(\gamma_h g^{(2)}_h(0) \!-\!\gamma_c g^{(2)}_c(0)\right).
\end{equation}

However, when operating within a single asymptotic regime under a moderate quench of interaction strength, the $g^{(2)}(0)$ correlation function is only slowly varying with $\gamma$. This means, to first approximation, $g^{(2)}_h(0) \!\simeq\!g^{(2)}_c(0)$, which in turn transforms the interaction energy difference to
\begin{equation}
    \langle \hat{H}^{int} \rangle_h \!-\! \langle \hat{H}^{int} \rangle_c \!\simeq\! N \frac{\hbar^2 \rho^2}{2m}\left(\gamma_h  \!-\!\gamma_c \right) g^{(2)}_c(0).
\end{equation}
The heat intake, which is given by Eq.~\eqref{eq:Q1}, is therefore well approximated by
\begin{align*}
    Q_1 &= \langle \hat{H} \rangle_h - \langle \hat{H} \rangle_c - \frac{N \hbar^2 \rho^2}{2m}(\gamma_h - \gamma_c) g^{(2)}_c(0) \\ 
    &\simeq \langle\hat{H}^{kin} \rangle_h - \langle \hat{H}^{kin} \rangle_c,
\end{align*}
which is approximately constant, as detailed above. Therefore, under a fixed ratio of temperatures, the efficiency, which is given by $\eta \!=\!W/Q_1$, scales predominately with $W$, hence $\eta\!\propto\!W$.

\begin{figure*}[!tbp]
\begin{center}
    \includegraphics[width=12cm]{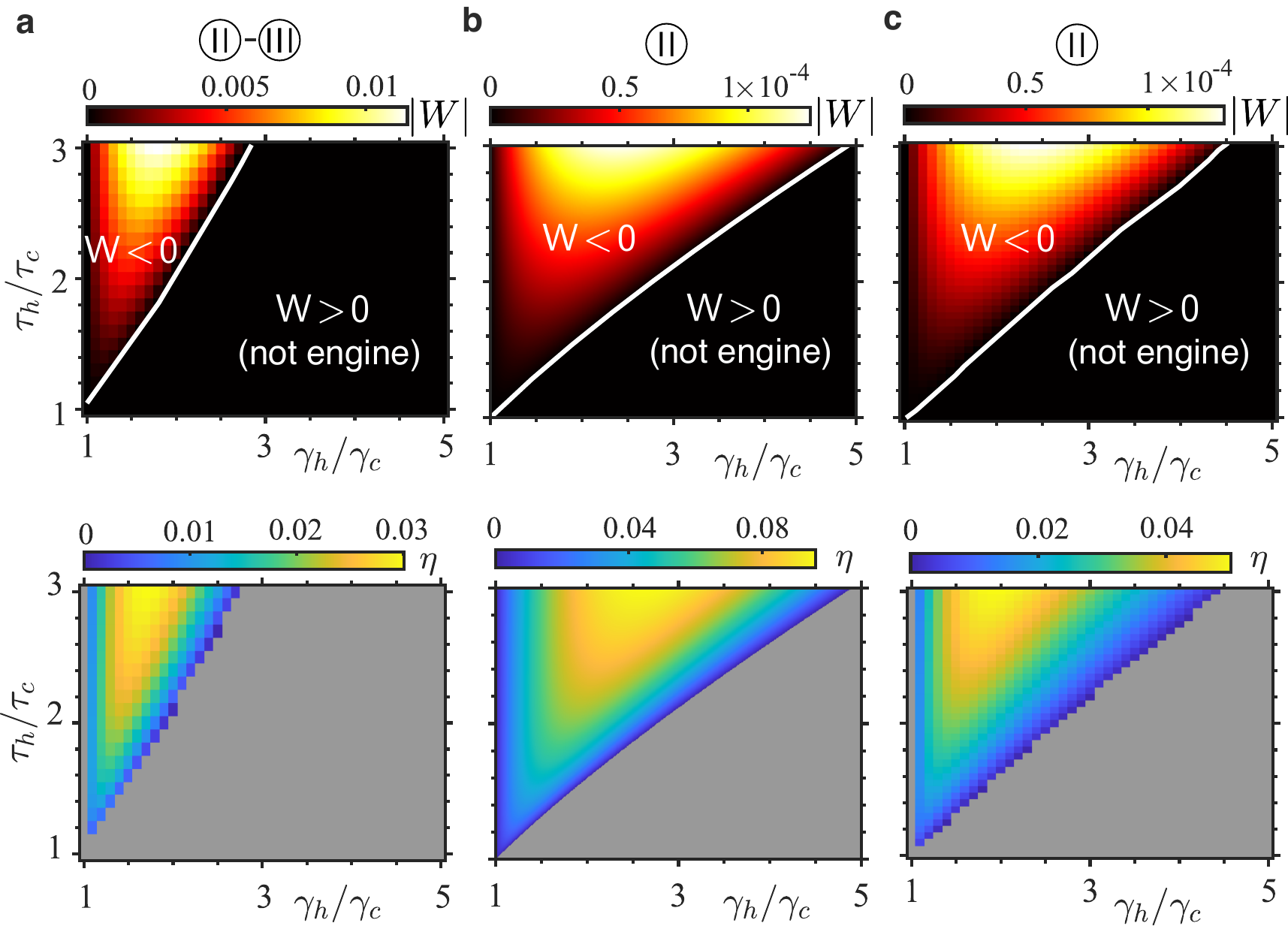}
    \caption{Performance of the sudden interaction quench quantum Otto cycle, numerically evaluated via the thermodynamic Bethe ansatz. Panel \textbf{a} demonstrates numerically evaluated net work and efficiency for a system with a cold thermal state defined by $\gamma_c \!=\! 0.1$, $\tau_c \!=\! 0.5$, lying on the border of regimes II and III (see Fig.~\ref{fig:g2_parameter_space}), and thus lying outside the range of the analytic approximations utilized previously. Panel \textbf{b} is a copy of Fig.~\ref{fig:engine_performance}\textbf{a} for comparison with the numerical evaluation of the same cycle in panel \textbf{c} using the TBA. Here, there is excellent agreement in the net work between panels \textbf{b} and \textbf{c}, with small disagreement under large interaction strength and temperature ratios, as the hot thermal state is approaching the edge of the asymptotic regime where it is applicable.}
\label{fig:Numeric_Comparison}
\end{center}
\end{figure*}

\subsubsection{Comparison with TBA numerics}

Experimental realization of a 1D Bose gas often falls outside the asymptotic regimes where analytic approximations are applicable. In such situations, we may utilize the exact Yang-Yang thermodynamic Bethe ansatz \cite{yang1969thermodynamics,korepin1993} to evaluate the equilibrium properties of the gas required for calculating net work and efficiency via Eqs.~\eqref{eq:Work} and \eqref{eq:Efficiency}, respectively. This is presented in Fig.~\ref{fig:Numeric_Comparison}\textbf{a} for experimentally realistic set of system parameters that inhabit the boundary between asymptotic parameter regimes II and III (see Fig.~\ref{fig:g2_parameter_space}). 
This highlights the broader applicability of the engine cycle introduced in this chapter, and suggests that an interesting avenue of future research would be to further investigate engine performance between or even on the border of the many parameter regimes.

Additionally, one may utilize the exact TBA to confirm the results derived via approximate analytics. This is illustrated in Figs.~\ref{fig:Numeric_Comparison} \textbf{b} and \textbf{c}, where we see excellent agreement between these results when the parameters $\gamma$ and $\tau$ are sufficiently deep into the analytic asymptotic regimes.

\subsubsection{Isentropic engine cycle}\label{Chap:3_Sec:isentropic_g2}

The work strokes of the conventional Otto engine cycle are, by definition, fully reversible, and as such the overall engine cycle obtains the maximum possible values for both net work and efficiency \cite{SchroederD_ThermalPhysics}. The isentropic interaction-driven Otto cycle for the uniform 1D Bose gas was introduced in Ref.~\cite{chen2019interaction}, where they employed the Tomonaga-Luttinger liquid (TLL) theory to derive approximate analytic results for the net work and efficiency. The TLL theory captures the low energy regimes of the uniform 1D Bose gas \cite{1D_entropy_2017}, corresponding to regions I and VI of Fig.~\ref{fig:g2_parameter_space}. Notably, as discussed in the supplementary material of Ref.~\cite{chen2019interaction}, though the isentropic cycle does not guarantee that the final state reached following the work strokes is thermal, the difference in performance between the engine cycle evaluated via an adiabatic work stroke, and that evaluated via entropy conservation between fully thermal states, i.e. the method employed here, was observed to be minimal.

\begin{figure*}[!tbp]
\begin{center}
    \includegraphics[width=12cm]{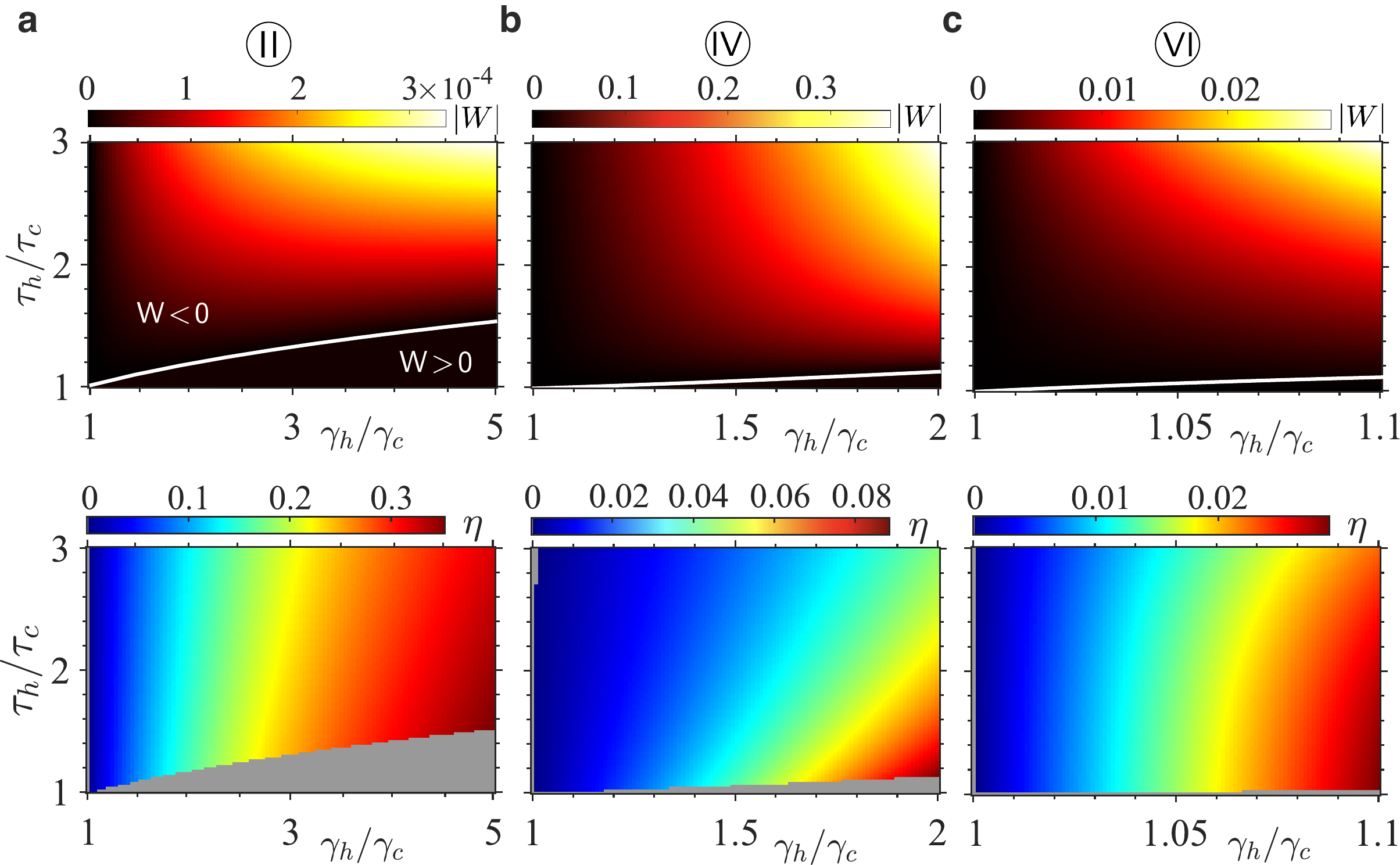}
    \caption{Performance of the isentropic interaction-driven quantum Otto cycle, for the same parameters as in the sudden quench scenarios of Figs.~\ref{fig:engine_performance}\,\textbf{a}--\textbf{c}. As one approaches the boundary of the engine regime, defined by $W\!=\!0$, the efficiency, which is given by the ratio of the net work, $W$, to the heat intake, $Q_1$, is approximately constant, unlike the $W$ itself which approaches 0 near the boundary. However, as the efficiency is only a valid metric in the engine regime, it is set to gray scale outside the engine regime.
    }
\label{fig:SQ_AD}
\end{center}
\end{figure*}

Here, we analyze the performance of the isentropic Otto engine cycle under higher temperatures, i.e., outside of the regime of applicability of the TTL theory. This can be done by 
utilizing the conservation of entropy condition for the work strokes $W_1$ and $W_2$ of Fig.~\ref{fig:Engine_Diagram} \cite{chen2019interaction}, along with the TBA or approximate analytic results for entropy available in the uniform 1D Bose gas (see Ref.~\cite{kerr2024analytic} for full details on entropy in each parameter regime).
In Fig.~\ref{fig:SQ_AD} we demonstrate performance of the isentropic engine cycle (using the analytic results), which may be directly compared against that of the sudden quench engine cycles explored in Fig.~\ref{fig:engine_performance}. Notably, over the range of parameters that the sudden quench Otto cycle operates as an engine, its performance remains relatively close in both net work and efficiency to the comparative isentropic results, despite being highly nonequilibrium and thus irreversible. Further optimization of the sudden quench Otto cycle is left for future work.

~

\section{Interaction-driven Otto accelerator, heater, and refrigerator}

\noindent 


\noindent Glauber's $g^{(2)}(0)$ correlation function for the 1D Bose gas is inherently dependent on the interaction strength and temperature \cite{kheruntsyan2003pair}. This implies that the condition for the Otto cycle to operate as a heat engine, $g^{(2)}_c(0)\!<\!g^{(2)}_h(0)$, where $\gamma_c\!<\!\gamma_h$, cannot hold under large quenches of interaction strength from $\gamma_c$ to $\gamma_h$ as the gas becomes increasingly fermionized (i.e., $g_h^{(2)}(0)\!\to\!0$) in the limit $\gamma \! \to \! \infty$ \cite{Girardeau_1960}.
However, beyond the heat engine operation regime, a further three QTM's are thermodynamically allowed \cite{buffoni2019quantum}: namely, accelerator [A], heater [H], and refrigerator [R].
For these QTM's, one may define a coefficient of performance (CoP) according to the following principle \cite{SchroederD_ThermalPhysics}:
\begin{equation}
    \mathrm{CoP[QTM]} = \frac{\mathrm{benefit\,of\,operation}}{\mathrm{cost\,of\,operation}}.
    \label{eq:CoP}
\end{equation}

\begin{figure*}[!tp]
    \includegraphics[width=17cm]{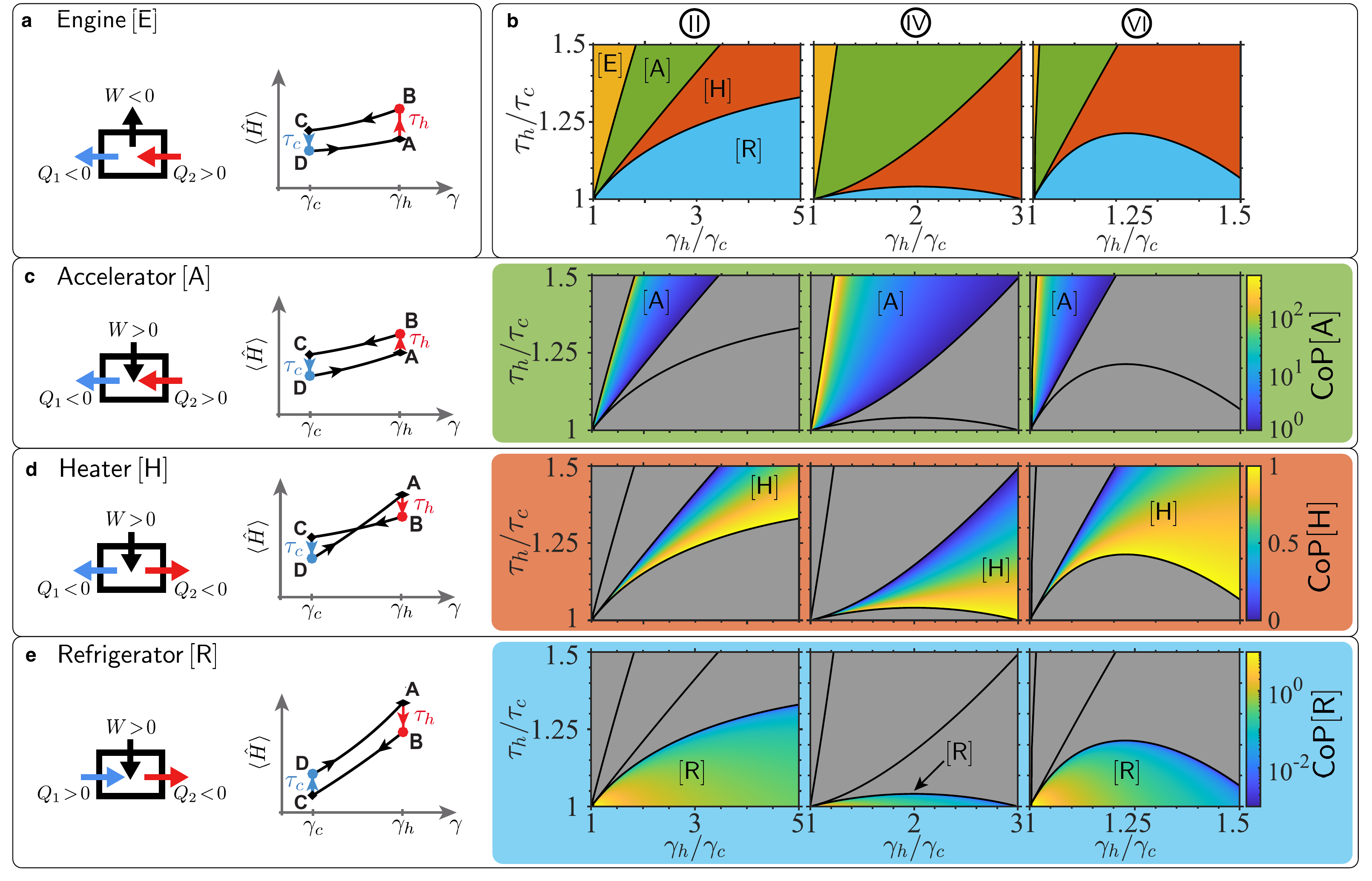}
    \caption{Performance of the interaction-driven quantum Otto cycles in the accelerator ([A]), heater ([H]), and refrigerator ([R]) regimes. An energy flow diagram (left) for the interaction-driven Otto engine ([E]) is shown for comparison in panel \textbf{a} alongside a simplified version of the cycle diagram (right), originally depicted in Fig.~\ref{fig:Engine_Diagram}. Panel \textbf{b} shows a simplified layout of how the operating regimes of different QTM's depend on the ratio of interaction strengths, $\gamma_h/\gamma_c$, and temperatures, $\tau_h/\tau_c$, of the hot and cold thermal states in same the three asymptotic regimes as in Fig.~\ref{fig:engine_performance}. The energy flow and cycle diagrams for the accelerator QTM are illustrated in panel \textbf{c}, where we additionally plot its coefficient of performance, CoP[A] (see text), in the chosen asymptotic parameter regimes. Likewise, in panels \textbf{d} and \textbf{e} we show the energy flow, cycle diagrams, and coefficients of performance for the heater and refrigerator regimes, respectively.}
\label{fig:operational_regimes}
\end{figure*}

In the left two columns of  Fig.~\ref{fig:operational_regimes}, we show the simplified schematics of these additional QTM's compared to the heat engine from Fig.~\ref{fig:Engine_Diagram}, which we repeat here in \textbf{a} for comparison; in panel \textbf{b} we show the operating boundaries of these different QTM's. Finally, in the color coded right panels in \textbf{c}--\textbf{e}, we show the magnitudes of the respective CoP's, which we now defined and discuss in greater detail.

\subsection{Accelerator}
The conditions of operating the Otto cycle as a thermal accelerator ([A])
are given by: $W\!>\!0, \, Q_1 \!>\! 0 , \, Q_2\! <\! 0$, where $W\!=\!0$ defines the border between the heat engine and the accelerator.
This QTM \cite{buffoni2019quantum} enhances the natural flow 
of heat, taken into the working fluid from the hot reservoir, $Q_1$, and transferred to the cold reservoir, $Q_2$, by investing net mechanical work, $W$, in the process. According to Eq.~\eqref{eq:CoP}, its CoP is given by:
    \begin{equation}
        \mathrm{CoP[A]} = -\frac{Q_2}{W} = 1 + \frac{Q_1}{W} > 1,
    \end{equation}
where,
\begin{align}
    Q_2&= -\langle \hat{H} \rangle_h + \langle \hat{H} \rangle_c + \frac{N \hbar^2 \rho^2}{2m}(\gamma_h - \gamma_c) g^{(2)}_h(0),
\label{eq:Q2}\\
    Q_1 &= \langle \hat{H} \rangle_h - \langle \hat{H} \rangle_c - \frac{N \hbar^2 \rho^2}{2m}(\gamma_h - \gamma_c) g^{(2)}_c(0).
\label{eq:Q1}
\end{align}
The magnitude of CoP[A] is shown Fig.~\ref{fig:operational_regimes}\,\textbf{c}, where we note that, at the border between [E] and [A], the coefficient of performance diverges due to its inverse dependence on $W$.
Operation of this QTM, enabled through $g^{(2)}_h(0)\!<\!g^{(2)}_c(0)$, additionally requires that the work in and out, which are proportional to $g^{(2)}_c(0)$ and $g^{(2)}_h(0)$, respectively, remain small in comparison to the energy gap between the hot and cold thermal states, as shown in the cycle diagram in Fig.~\ref{fig:operational_regimes}\,\textbf{c}. This is in direct contrast with the two further QTM's described below.

\subsection{Heater}
Operating the Otto cycle in the heater ([H]) regime requires: $W > 0, Q_1 < 0, \, Q_2 < 0$, and is shown schematically in Fig.~\ref{fig:operational_regimes}\,\textbf{d}. This QTM utilizes mechanical work to heat up both hot and cold thermal reservoirs. The border with the accelerator regime is defined by $Q_1\!=\!0$.
This condition depends on the competition between $\langle \hat{H} \rangle_h$ and $\gamma_h g^{(2)}_c(0)$, with all other terms fixed by $\gamma_c$ and $\tau_c$, defining the cold thermal state.
However, we note that for a fixed temperature ratio, $\tau_h/\tau_c$, an arbitrarily large interaction strength quench inevitably incurs operation as a heater. This may be intuitively explained by the fact that such large quenches incur large irreversible excitations in the working fluid, the additional energy generated by these strokes must then be dumped into both baths at the end of the respective interaction strokes, heating both baths.

The CoP for the heater regime, if one considers the benefit of operation to be heating of \textit{both} reservoirs then, due to the conservation of energy, is trivially $\mathrm{CoP[H]}\!=\!-Q/W \!=\! 1$. Instead, in Fig.~\ref{fig:operational_regimes}\,\textbf{d}, we define the benefit of operation to be the heating of the hot reservoir
, thus 
    \begin{equation}\label{eq:cop_H}
         \mathrm{CoP[H]} = -\frac{Q_1}{W} = 1 - \frac{|Q_2|}{W} < 1.
    \end{equation}
One may alternatively define the benefit of the heater as heating the cold reservoir,
    \begin{equation}
         \mathrm{CoP[H]}' = -\frac{Q_2}{W} = 1 - \mathrm{CoP[H]} < 1.
    \end{equation}
Both definitions of $\mathrm{CoP[H]}$, however, are limited by energy conservation to be less than or equal to $1$.

\subsection{Refrigerator}
The conditions of operating the Otto cycle
as a refrigerator ([R]) are: $W>0, \, Q_1 < 0, \, Q_2 > 0$. The purpose of this thermal machine is to cool down the cold reservoir by extracting heat and dumping it into the hot reservoir, with the aid mechanical work done by the working fluid. The boundary between [H] and [R] is defined by $Q_2\!=\!0$.
The CoP for the refrigerator is given by \cite{SchroederD_ThermalPhysics}
    \begin{equation}
         \mathrm{CoP[R]} = \frac{Q_2}{W}=\frac{|Q_1|}{W}-1,
    \end{equation}
and is shown in Fig.~\ref{fig:operational_regimes}\,\textbf{e}; it diverges in the limit of infinitesimal quenches in both interaction strength and temperature due to the rapid convergence of $W\!\to\!0$, which is faster than that of $Q_1 \!\to\! 0$.

The conditions on $g^{(2)}_h(0)$ and $g^{(2)}_c(0)$ for operating in this regime run directly counter to operation as an accelerator. For refrigeration, the correlation functions must remain large enough that the work in and out exceed the energy gap between the two thermal states, as shown in the cycle diagram in Fig.~\ref{fig:operational_regimes}\,\textbf{e}.
However, as noted above, large interaction strength quenches inevitably reduce the work out as $g^{(2)}_h(0)$ reduces rapidly for large $\gamma_h$, thus incurring operation as a heater for any given temperature ratio. This implies that the refrigerator occurs only over a finite parameter region, most clearly visible in regimes IV and VI of Fig.~\ref{fig:operational_regimes}\,\textbf{e}.

Under large interaction strength quenches, for fixed temperatures $\tau_c$ and $\tau_h$, it was noted previously that the heater is the inevitable mode of operation. This scenario requires the fulfilment of two conditions: the work in, $W_1$, exceeds the energy gap between the hot and cold thermal states, whereas the work out, $W_2$, remains less than this same gap (see Fig.~\ref{fig:operational_regimes}\,\textbf{d}). As detailed above, in the section on maximum efficiency at maximum work, the total energy difference between the hot and cold thermal states for fixed temperature ratios, $\tau_h/\tau_c$, is given by a sum of the kinetic energy difference, which is approximately constant, and the interaction energy difference, given by Eq.~\eqref{eq:interaction_energy_diff}.

Here, for a \textit{large} quench in interaction strength, the correlation function is no longer approximately constant, and $g^{(2)}_h(0)$ is strongly monotonically decreasing as a function of $\gamma_h$, i.e. $g^{(2)}_h(0) \!<\!g^{(2)}_c(0)$. We therefore find that the work input, $W_2$, exceeds the interaction energy difference given in Eq.~\eqref{eq:interaction_energy_diff},
\begin{equation}
    W_2 \propto (\gamma_h - \gamma_c) g^{(2)}_c(0) > \gamma_h g^{(2)}_h(0) \!-\!\gamma_c g^{(2)}_c(0).
\end{equation}
Further, as the kinetic energy term is approximately constant, $W_2$ inevitably exceeds the difference in total energy between the hot and cold thermal states due to its linear dependence on $\gamma_h$.

Similarly, since $g^{(2)}_h(0)$ monotonically decreases with $\gamma_h$ for large quenches of interaction strength, the magnitude of the work output, $|W_1| \propto (\gamma_h - \gamma_c) g^{(2)}_h(0)$, remains less than the energy gap between the hot and cold thermal states:
\begin{equation}
    |W_1| \propto (\gamma_h - \gamma_c) g^{(2)}_h(0) < \gamma_h g^{(2)}_h(0) \!-\!\gamma_c g^{(2)}_c(0).
\end{equation}
These two conditions, taken together, imply operation as a heater under large interaction quenches.

In contrast, for any fixed value of $\gamma_c$ and $\gamma_h$, an increasingly higher temperature of the hot thermal state, $\tau_h$, means that the corresponding correlation function, $g^{(2)}_h(0)$, monotonically increases towards its maximum value of $g^{(2)}_h(0) \!\simeq\!2$, which is achieved in regime IV. Thus, there is always a value of $\tau_h$ such that $g^{(2)}_h(0) \!>\!g^{(2)}_c(0)$, turning the Otto cycle into the engine operation regime.

Finally, in the refrigerator thermal operation regime, for $\tau_h\!=\!\tau_c$ and an infinitesimal quench of interaction strength, $\gamma_h\!-\!\gamma_c\!=\!\delta \gamma$, the net work vanishes as $W\!\propto  (\gamma_h \!-\!\gamma_c)(g^{(2)}_h(0)\!-\!g^{(2)}_c(0)) \!\propto\!\delta \gamma^2$. This occurs as the zeroth order terms in the correlation function cancel when taking their difference in a single asymptotic regime. In contrast, the heat intake, $Q_1$, depends on the difference in total energy, which to first order vanishes as $Q_1\!\propto\!\delta \gamma$. This results in $\mathrm{CoP[R]}\!=\!|Q_1|/W\!-\!1\!\propto\delta\gamma^{-1}$, which diverges as $\delta \gamma \!\to\! 0$, as noted previously.

~

\section{Conclusions}

In this chapter, we proposed and investigated a new quantum Otto engine cycle, where work strokes are performed via a quench of interparticle interaction strengths. Extracting net work from such an engine was shown, for the case of a sudden quench, to be enabled by atom-atom correlations, as demonstrated in Eqs.~\eqref{eq:Work_G2} \& \eqref{eq:Work_Uniform}. 
Such correlations are characterized by Glauber's local second-order correlation function, $g^{(2)}(0)$, which, for the system investigated, is a thermodynamic quantity that can be calculated from the exact thermodynamic Bethe ansatz solution through application of the Hellmann-Feynman theorem to the Helmholtz free energy \cite{kheruntsyan2003pair}. This fact may be further leveraged to calculate various other thermodynamic properties in relevant system, a property which we utilize to derive a new Maxwell relation for evaluation of entropy in ultracold atomic gases in Chap.~\ref{Chap:Maxwell}.

Here, we investigated operation of the introduced Otto cycle applied to the case of a repulsive 1D Bose gas as a working fluid. This system contains a variety of asymptotic parameter regimes which may be characterized by the local second-order correlation function, and depend on the dimensionless interaction strength, $\gamma$, and temperature, $\tau$. We utilized analytic formulas for the correlation function, total energy, and entropy, in these regimes to investigate the performance of the sudden quench engine cycle in a handful of these regimes, in particular focusing on regimes of experimental relevance. We expressed both net work and efficiency using these analytic formulas, which in turn allowed us to evaluate the limiting factors to engine performance as a function of the interaction strength ratio.

Further, utilizing the analytic results available for this model, were were able to explain why the respective maxima of net work and efficiency approximately align.
Comparison of the analytic results with exact TBA numerics was shown to generally agree, as expected, within the regions of their applicability. We additionally made use of these numerics to explore engine operation in a regime outside of asymptotic regions where the analytic approximations are valid. These results demonstrated the general applicability of the formulas introduced to regimes outside of analytic approximation.

Maximum net work and efficiency of thermodynamic engine cycles are achieved in the quasistatic, or isentropic, limit. Here, the work strokes are performed slowly enough such that there are no irreversible excitations of the working fluid, thus creating no irreversible entropy \cite{Varizi2020quantum}. We employed recently formulated expressions for the entropy of a uniform 1D Bose gas in order to evaluate the operation of this interaction-driven engine in the isentropic limit. Importantly, we found that the sudden quench cycle, where it provides beneficial net work, operates with performance in close proximity to this idealized limit of operation.

As the formulas introduced are applicable to quantum thermal machines beyond that of an engine cycle, we investigated performance of the sudden quench Otto cycle as a thermal accelerator, heater, and refrigerator cycles, defining and examining their coefficients of performance. Through this exploration, we developed an understanding of the limiting regimes of operation due to the limiting behaviours of the correlation function. In particular, we showed that under arbitrarily large temperature ratios, engine operation is inevitable due to the monotonic scaling of the correlation function with temperature in all parameter regimes. Likewise, for arbitrarily large quenches of interaction strength, heater operation was shown to be inevitable in all parameter regimes. This in turn implied that operation as a refrigerator, which is a more desirable operation than that of a heater due to its ability to cool an already ultracold atomic gas, occurs \textit{only} over a limited range of interaction strength quenches.

Though our specific results for the net work and the efficiency were calculated for a uniform 1D Bose as an example, the broad conclusions arrived at here are applicable to any other many-body system with short-range contact interactions, in particular those describable by $s$-wave scattering. These results should therefore aid the tests of quantum thermodynamic concepts and realization of novel quantum thermal machines in laboratory settings.


~



~

\newpage

\noindent
The work presented in Chapter \ref{Chap:3} was adapted from the submitted publication of Ref.~\cite{watson2024interaction}, and the contribution of each named author to that work is presented below in Table.~\ref{Tab:Chap3}. 

\noindent
\cite{watson2024interaction} \textbf{R. S. Watson}, and K. V. Kheruntsyan, \href{https://arxiv.org/abs/2308.05266}{Quantum many-body thermal machines enabled by atom-atom correlations}, submitted to Phys. Rev. Lett. on 1 March 2024.

\begin{table}[h]
	\begin{center}
	\begin{tabular}{|c|l|l|}
		\hline
		Contributor & Statement of contribution & \% \\
		\hline
		\textbf{R. S. Watson}				& writing of text 					& 80\\
															& proof-reading							& 50 \\
															& analytic derivations 	& 25\\
															& numerical calculations 		& 100\\
															& preparation of figures 		& 60 \\
		\hline
		K. V. Kheruntsyan								& writing of text 					& 20\\
															& proof-reading							& 50 \\
															& supervision, guidance 		& 20\\
															& analytic derivations  	& 75\\
															& preparation of figures 		& 40 \\
															& initial concept						& 100 \\
		\hline
	\end{tabular}
	\end{center}
 \caption{}\label{Tab:Chap3}
\end{table}

%

%% file: Chapter4/Chapter4.tex

\chapter[Maxwell relation between entropy and atom-atom pair correlation]{Maxwell relation between entropy and atom-atom pair correlation}
\label{Chap:Maxwell}	
\pagestyle{headings}

\textit{For many-particle systems with short range interactions the local (same point) particle-particle pair correlation function represents a thermodynamic quantity that can be calculated using the Hellmann-Feynman theorem. 
Here we exploit this property to derive a thermodynamic Maxwell relation between the local pair correlation and the entropy of an ultracold Bose gas in one dimension (1D). 
To demonstrate the utility of this Maxwell relation, we apply it to the computational formalism of the stochastic projected Gross-Pitaevski equation (SPGPE) to determine the entropy of a finite-temperature 1D Bose gas from its atom-atom pair correlation function. Such a correlation function is easy to compute numerically within the SPGPE and other formalisms, which is unlike computing the entropy itself. Our calculations can be viewed as a numerical experiment that serves as a proof-of-principle demonstration of an experimental method to deduce the entropy of a quantum gas from the measured atom-atom correlations. }


\section{Introduction}
\label{Sec:label}	


Entropy plays a fundamental role in thermodynamics, statistical mechanics, and quantum information theory. However, measuring it directly or calculating it from the defining multiplicity function or the density matrix of an interacting many-body system often represents a formidable challenge. Instead, the entropy is often deduced from other thermodynamic quantities (such as the heat capacity or the free energy) using the relevant thermodynamic relations \cite{Huang_book,Callen_book}. Here, we derive and discuss a thermodynamic Maxwell relation by which the entropy of a quantum many-body system with short-range interactions can instead be related to, and hence deduced from, the local particle-particle correlation function. Such a pair correlation function characterizes the probability of two particles to be found in the same position compared to uncorrelated particles and can often be computed 
using methods of many-body and quantum field theory either analytically or numerically \cite{Gangardt2003,Kheruntsyan2003}.
It can also be measured experimentally in, e.g., ultracold quantum gas experiments using photoassociation \cite{Weiss_g2_2005}.

The surprising aspect of the Maxwell relation between the pair correlation 
and the entropy that we discuss here is that the pair correlation function is usually viewed and treated as a typical two-body observable, whereas the entropy is a thermodynamic quantity. However, what promotes the pair correlation into a thermodynamic quantity as well is the fact that we are only considering many-body systems with short-range interactions that can be characterized by the $s$-wave scattering length \cite{Pethick_Smith_book,Pitaevskii_Stringari_book}. In this case, the inter-particle interactions can be approximated by a simple contact interaction, meaning that the two-body correlation function at zero inter-particle separation indeed becomes a thermodynamic quantity. This was first demonstrated by Lieb and Liniger in their seminal work on the exact Bethe ansatz treatment of a uniform one-dimensional (1D) Bose gas with repulsive contact (delta-function) interactions \cite{Lieb-Liniger-I,Lieb-Liniger-II}. By using the Hellmann-Feynman theorem and differentiating the total ground state (zero-temperature, $T\!=\!0$) energy of the gas with respect to the interaction strength, Lieb and Liniger were able to calculate the mean interaction energy component, which itself is proportional to the unnormalized pair correlation function (see below).

The extension of the Hellmann-Feynman theorem to finite temperature systems \cite{Kheruntsyan2003,kheruntsyan2005finite}, together with the exact Yang-Yang thermodynamic Bethe ansatz (TBA) solution for the 1D Bose gas at finite temperature \cite{Yang-Yang}, was later utilized to calculate the local pair correlation at any temperature and interaction strength. In this case, the pair correlation function is related to the partial derivative of the Helmholtz free energy with respect to the interaction strength \cite{Kheruntsyan2003,kheruntsyan2005finite,Kerr2024F}. In this chapter, we take this relationship a step further by combining it with the fact that the partial derivative of the same Helmholtz free energy with respect to the temperature, on the other hand, gives the entropy of the system according to the canonical ensemble formalism of statistical mechanics. Therefore, by using the commutative property of mixed second derivatives of the Helmholtz free energy (with respect to the interaction strength and temperature) one obtains the Maxwell relation between the pair correlation and the entropy that we discuss here.

As a practical application of this Maxwell relation, we utilize it  for computing the entropy of a weakly interacting 1D Bose gas in the quasicondensate regime in the context of the classical $c$-field approach of the stochastic projected Gross-Pitaevskii equation (SPGPE) \cite{Bradley_2005,spgpe,gardiner2002stochastic,gardiner2003stochastic}. The SPGPE is a well established and widely used numerical approach for computing thermal equilibrium and dynamical properties of finite temperature Bose gases, such as partially condensed Bose-Einstein condensates in 2D and 3D , or phase-fluctuating quasicondensates in 1D \cite{Bradley2015}. Despite its wide applicability to ultracold quantum gas systems, computing the entropy of such systems within the SPGPE has not been accomplished previously. Here, we compute the entropy of a 1D quasicondensate within the SPGPE approach; we restrict ourselves to the 1D Bose gas because of the availability of the exact TBA solution for both the entropy and the pair correlation function, to which we compare, and hence validate, our numerical SPGPE results. However, we point out that the Maxwell relation derived and discussed here is equally applicable to 2D and 3D systems, as well as to Fermi gas systems with similar contact interactions.


\section{Maxwell relation}

We now recall that the partial derivative of the same Helmholtz free energy with respect to temperature $T$ gives the entropy $S$ of the system:
\begin{equation}
S=-\left(\frac{\partial F}{\partial T}\right)_{g,L,N}.
\label{eq:entropy}
\end{equation}

Combining  Eq.~\eqref{eq:entropy} with Eq.~\eqref{eq:g2-canonical} introduced in Chapter \ref{Chap:2}, and using the commutative property of mixed second derivatives of $F$, i.e. $\frac{\partial}{\partial g}\left( \frac{\partial F}{\partial T}\right)_{L,N} = \frac{\partial}{\partial T}\left( \frac{\partial F}{\partial g}\right)_{L,N} $, leads to the following Maxwell relation: 
\begin{equation}
\boxed{\left(\frac{\partial S}{\partial g} \right)_{T,L,N}=- \frac{Ln^2}{2}\left( \frac{\partial g^{(2)}}{\partial T}\right)_{g,L,N}.}
\label{eq:Maxwell}
\end{equation}
Here, for simplicity of notation, we define $g^{(2)}\!\equiv\!g^{(2)}(0)$.
Equation \eqref{eq:Maxwell} is one of this chapter's key results and implies that the entropy of the gas can be calculated by integrating the partial derivative of the pair correlation function $g^{(2)}$ with respect to $T$ over the interaction strength $g$. Importantly, as this formula makes no specific reference to the parameter regime of operation, it remains valid over the entire space of dimensionless parameters which define the finite temperature 1D Bose gas.
Further, the entropy of the uniform 1D Bose at a specific value of $g$ (and some fixed values of $T$, $L$, and $N$) can be evaluated via
\begin{equation}\label{eq:entropy_from_integral}
S(g,T,L,N)=S(g_0,T,L,N) 
-\frac{Ln^2}{2}\int_{g_0}^{g}\left( \frac{\partial g^{(2)}(g',T,L,N)}{\partial T}\right)_{g',L,N}dg',
\end{equation}
where $S(g_0,T,L,N)$ serves the role of the integration constant and is assumed to be known for the method to work. In practice, $S(g_0,T,L,N)$ can be chosen to correspond to the entropy of an ideal ($g_0=0$) 1D Bose gas, $S(0,T,L,N)$, which can indeed be calculated for any $T$ using standard methods of statistical mechanics \cite{Huang_book,kerr2024analytic}.

As a simple analytic illustration of the utility of Eq.~\eqref{eq:entropy_from_integral}, we calculate the entropy of a 1D Bose gas in highly degenerate, nearly ideal Bose gas regime that can be treated using perturbation theory with respect to $\gamma$ (see the results for the so-called decoherent quantum regime in Ref. \cite{Kheruntsyan2003,kerr2024analytic}, valid in the region $2\sqrt{\gamma}\ll \tau \ll 1$). In this regime, the pair correlation function $g^{(2)}$, which was calculated in Ref. \cite{Kheruntsyan2003} without resorting to the Helmholtz free energy, is given by
\begin{equation}
g^{(2)}=2-4\gamma/\tau^2.
\label{g2_III}
\end{equation}
Therefore, Eq.~\eqref{eq:entropy_from_integral} yields the following result for the corresponding entropy:
\begin{equation}
S=S_{\text{IBG}}-4k_BN\gamma^2/\tau^3.
\label{S_III}
\end{equation}
where $S_{\text{IBG}}=S(0,T,L,N)$ is entropy of the ideal 1D Bose gas at the same temperature \cite{kerr2024analytic}.

\subsection{Thermodynamic potential behind the Maxwell relation for the atom-atom correlation function.}
We outline 
the fundamental thermodynamic identities behind the Maxwell relation given in Eq.~\eqref{eq:Maxwell}. First, we note that in the $G^{(2)}$ function, defined in Sec.~\ref{sec:g2_glauber}, which follows from the Hellmann-Feynman theorem, 
the integrated correlation function $\overline{G^{(2)}}=\int dx \langle \hat{\Psi}^{\dagger}(x)\hat{\Psi}^{\dagger}(x)\hat{\Psi}(x)\hat{\Psi}(x)\rangle$ (times a factor of $1/2$) can be viewed as an \emph{extensive} thermodynamic parameter
that characterizes the variation of the system's internal energy $U=\langle \hat{H} \rangle$ with the interaction strength $g$ which itself is an \emph{intensive} parameter conjugate to $\overline{G^{(2)}}/2$.  (The factor of $1/2$ is an artifact of the conventional definition of the interaction part of the Hamiltonian, with $\langle \hat{H}_{\mathrm{int}}\rangle =\frac{g}{2}\overline{G^{(2)}}$, where $1/2$ can be either absorbed into the re-definition of the coupling constant, $g/2\to g$, or  kept as a multiplier in front of $\overline{G^{(2)}}$ whenever we talk about the integrated pair correlation as an extensive parameter.)

The extensive nature of $\overline{G^{(2)}}$ follows immediately from the fact that $g \overline{G^{(2)}}/2$ (where $g$ is intensive) is simply the expectation value of the interaction part of the Hamiltonian, $\langle \hat{H}_{\mathrm{int}}\rangle=g \overline{G^{(2)}}/2$, which itself must be extensive because the overall Hamiltonian energy and the kinetic energy part alone are also extensive. For a uniform system of length $L$, this can also be seen explicitly through $\overline{G^{(2)}}=L\langle \hat{\Psi}^{\dagger}\hat{\Psi}^{\dagger}\hat{\Psi}\hat{\Psi}\rangle=Ln^2g^{(2)}$, where $L$ is extensive whereas $n$ and $g^{(2)}$ are intensive.

The variation of the generalized Helmholtz free energy in the canonical formalism, $F=F(T,L,N,g)$, where for the 1D system the role of the volume $V$ is played by the length $L$, can therefore be written as  
\begin{equation}
dF=-SdT-PdL+\mu dN+(\overline{G^{(2)}}\!/2)dg,
\label{eq:dF}
\end{equation}
where $S=-(\partial F/\partial T)_{L,N,g}$ is the entropy, $P=-(\partial F/\partial L)_{T,N,g}$ is the pressure, $\mu =(\partial F/\partial N)_{T,L,g}$ is the chemical potential, and $\overline{G^{(2)}}/2=(\partial F/\partial g)_{T,L,N}$. From this, one can derive a set of Maxwell relations as usual, including the one between $S$ and $\overline{G^{(2)}}$, i.e., Eq.~\eqref{eq:Maxwell}. 

According to the standard formalism of thermodynamics (see, e.g., \cite{Callen_book}), the fundamental equation \eqref{eq:dF} can be obtained via a Legendre transform,
\begin{equation}
F=U-TS+g (\overline{G^{(2)}}/2),
\end{equation}
from the Euler equation for the generalized internal energy of the system,
\begin{equation}
U=TS-PL+\mu N-g (\overline{G^{(2)}}\!/2),
\end{equation}
which is a function of only extensive parameters, $U=U(S,L,N,\overline{G^{(2)}}\!/2)$. The differential  of $U$ is given by  
\begin{equation}
dU=TdS-PdL+\mu dN-g d(\overline{G^{(2)}}\!/2),
\label{eq:dU}
\end{equation}
where $T\!=\!(\partial U/\partial S)_{L,N,\overline{G^{(2)}}}$, $P\!=\!-(\partial U /\partial L)_{S,N,\overline{G^{(2)}}}$, $\mu\!=\!(\partial U/\partial N)_{S,L,\overline{G^{(2)}}}$, and $g \!=\! -\left(\partial U/\partial(\overline{G^{(2)}}\!/2)\right)_{S,L,N}$. We note that one can also show within the standard microcanonical formalism of statistical mechanics that $g$ can be alternatively calculated from $S$ using
$g =\!T\left(\partial S/\partial(\overline{G^{(2)}}\!/2)\right)_{U,L,N}$, which is analogous to showing that
$P=-(\partial U/\partial L)_{S,N}=T(\partial S/ \partial L)_{U,N}$.
Further, the negative sign in $g \!=\! -\left(\partial U/\partial(\overline{G^{(2)}}\!/2)\right)_{S,L,N}$ makes physical sense for positive $g$ (repulsive interactions) as the internal energy of the system increases when the pair correlation is decreased when approaching the `fermionized' regime of particle-particle antibunching where $g^{(2)}(0)\! \to\! 0$, as opposed to the weakly interacting regime where the gas displays bosonic bunching  $g^{(2)}(0) \!\to\! 2$ \cite{Kheruntsyan2003}.

Furthermore, an equation similar to Eq.~\eqref{eq:dF} can be written down for the grand-canonical thermodynamic potential $\Omega=F-\mu N=U-TS+g (\overline{G^{(2)}}/2)-\mu N$, 
\begin{equation}
d\Omega=-SdT-PdL-Nd\mu+(\overline{G^{(2)}}\!/2)dg,
\label{eq:dO}
\end{equation}
with $\Omega=\Omega(T,L,\mu,g)$. Using additionally $\Omega=-PL$ for homogeneous systems, Eq.~\eqref{eq:dO} can be further rewritten as
\begin{equation}
LdP-SdT-Nd\mu+(\overline{G^{(2)}}\!/2)dg=0,
\label{eq:Gibb-Duhem}
\end{equation}
which takes the role of the generalized Gibbs-Duhem relation and implies, in particular, that among the four intensive parameters $\{P,T,\mu,g\}$ only three are independent; this in turn implies that the functional dependence of the fourth parameter on the other three takes the role of the thermodynamic equation of state, such as $P=P(T,\mu,g)$. For explicit examples of such equations of state for the uniform 1D Bose gas, see a recent review in Ref.~\cite{kerr2024analytic}.

We emphasize that all these generalizations of the thermodynamic relations, accounting for the changes of the interaction strength $g$, are applicable only to ultracold atomic gases in which the interactions are short-ranged and can be accounted for via a single parameter ($g$), which itself can be varied via the $s$-wave scattering length $a$. These generalized thermodynamic relations can be adopted to describe short-range interacting Fermi gases \cite{Tan_contact_2008,Zwerger_Tan_2011,Braaten_Tan_review,Pitaevskii_Stringari_book,Cherny2021}, Fermi and Bose gas mixtures 
\cite{Pitaevskii_Stringari_book,Pethick_Smith_book}, as well as lattice models, such as Bose and Fermi Hubbard models \cite{Pitaevskii_Stringari_book,Pethick_Smith_book}, and Heisenberg-like models of interacting spins \cite{Sachdev_2011} 
where the role of the correlation function $\overline{G^{(2)}}$ is taken by the neighbouring spin-spin correlation function. We also note that these thermodynamic relations are similar to those that have been derived in the context of Tan's contact parameter and the related Tan's thermodynamic relations  \cite{Tan_contact_2008,Werner2009,Zwerger_Tan_2011,Braaten_Tan_review}; see, e.g., the treatments summarized in chapter 18.3 of Ref. \cite{Pitaevskii_Stringari_book} and in Ref. \cite{Cherny2021}, which we closely followed here. In retrospect this is not surprising, because Tan's contact,  which characterizes the strength of the tails of the momentum distribution of an ultracold atomic gas, is known to be directly proportional to the (spatial) local atom-atom pair correlation function $g^{(2)}$ \cite{Zwerger_Tan_2011,Minguzzi_2013,De_Rosi_2023,kerr2024analytic}. We emphasize, however, that deriving the thermodynamic and Maxwell relations presented in this chapter does not rely on, and does not require the knowledge of, Tan's contact and Tan's thermodynamic relations.


\section{Application to the SPGPE approach}

\begin{figure}[t] 
\begin{center}
    \includegraphics[width=9cm]{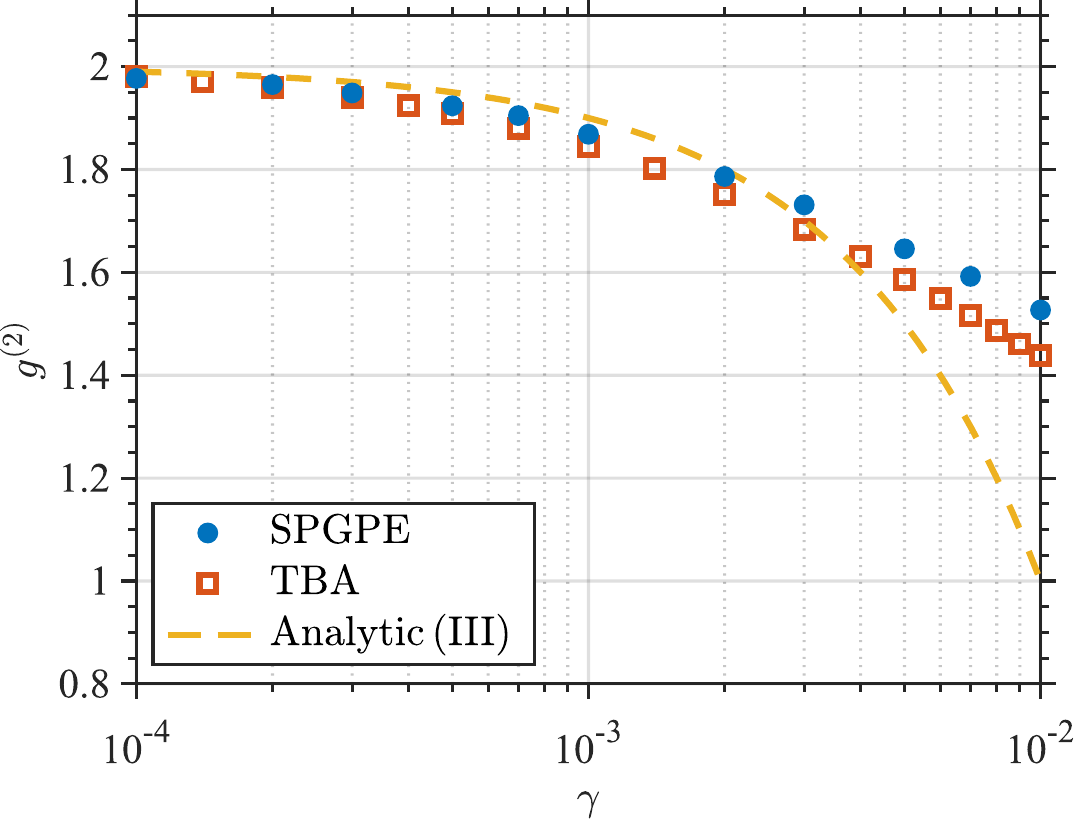}
    \caption{Local pair correlation $g^{(2)}$ for a 1D quasicondensate as a function of the dimensionless interaction strength $\gamma$, for a fixed dimensionless temperature $\tau=0.2$. The numerically computed data from the SPGPE simulations are shown as circles and are compared to the exact TBA result (squares), and the analytic approximation of Eq.~\eqref{g2_III} (dashes).}
    \label{fig:g2}
\end{center}
\end{figure}

We now illustrate the utility of Eq.~\eqref{eq:entropy_from_integral} using a numerically computed pair correlation function within the SPGPE approach, introduced in Chapter \ref{Chap:2}. This itself can be viewed as a numerical experiment demonstrating how one can deduce the entropy of a quantum gas from the measured atom-atom correlations. 
We are interested in the pair correlation function, $g^{(2)}$, of a 1D quasicondensate at thermal equilibrium, in the parameter range  
$\sqrt{\gamma}\ll \tau \ll 1$, where the $c$-field approach is valid \cite{castin2000,Bayocboc2023}. In Fig.~\ref{fig:g2} we show the dependence of this correlation function over a range of the dimensionless interaction strength $\gamma\in [10^{-4}, 10^{-2}]$, for a fixed value of the dimensionless temperature $\tau=0.2$, obtained from the SPGPE approach. For comparison, we also show the exact TBA result (squares) and the approximate analytic result of Eq.~\eqref{g2_III} (dashed line).

As we see, in the limit of an ideal Bose gas ($\gamma \to 0$) at finite temperature, the pair correlation approaches the value of $g^{(2)}=2$, which is the Hanbury Brown--Twiss effect of bosonic bunching ($g^{(2)}>1$) first observed for photons from a chaotic (thermal) light source \cite{HBT_1956_Stellar,HBT_1956_two_coherent,HBT_1956_question} and more recently for an ultracold atomic gas above the transition to a Bose-Einstein condensate \cite{HBT_Esslinger_2005,HBT_BEC_Aspect_2005,Jeltes2007}. It corresponds to large density fluctuations and an enhanced probability of detecting two indistinguishable bosons in the same position due to the constructive interference of the respective probability amplitudes. As the strength of the repulsive interaction increases, the said probability decreases and manifests itself in the reduction of the value of $g^{(2)}$ below $2$ \cite{Kheruntsyan2003,Weiss_g2_2005}. At some finite, but still weak ($\gamma\! \ll\! 1$) interaction strength, the pair correlation crosses the coherent level of $g^{(2)}\!=\!1$ characteristic of a phase-fluctuating quasicondensate with suppressed density fluctuations, which itself shares the properties of a weakly interacting Bose-Einstein condensate in the mean-field description \cite{HBT_Esslinger_2005,HBT_BEC_Aspect_2005}. 

As the interaction strength increases further and approaches the regime of very strong or hard-core repulsion ($\gamma \to \infty$), also known as the Tonks-Girardeau limit of fermionization, the pair correlation reduces further down to $g^{(2)}\to 0$ (see Ref. \cite{Kheruntsyan2003,Weiss_g2_2005}). This reduction reflects the fact that the bosons are now strongly (anti)correlated and behave effectively as fermions, wherein the bosonic hard-core repulsion mimics the fermionic Pauli blocking. In the pair correlation function, such repulsion manifests itself as antibunching ($g^{(2)}<1$), which itself is due to the destructive interference of probability amplitudes for detecting two indistinguishable fermions in the same position \cite{Jeltes2007}. This regime, however, is beyond the applicability of the SPGPE approximation ($\gamma\ll \!1$, $2\gamma\!\ll\! \tau\!\ll \!1$; see, e.g., Ref.~\cite{Bayocboc2023,kerr2024analytic}, and references therein), and this is why in Fig.~\ref{fig:g2} we do not show the behaviour of the $g^{(2)}$ beyond the weakly interacting regime of $\gamma \!\ll \!1$. Because of the same approximate nature of the SPGPE approach, we see that the SPGPE data for $g^{(2)}$, while agreeing well with the exact TBA results at small $\gamma$, starts to deviate from the TBA results as $\gamma$ increases and approaches its upper bound of $\gamma=0.01$, where the condition $2\gamma\ll \tau$ is not well satisfied. Similarly, the analytic result of Eq.~\eqref{g2_III} deviates from TBA to a larger extend as $\gamma$ is increased, because it is applicable in an even more restricted region of $2\sqrt{\gamma}\!\ll\! \tau\ll \!1$ \cite{Kheruntsyan2003,kerr2024analytic}.

\begin{figure}[t] 
\begin{center}
    \includegraphics[width=9cm]{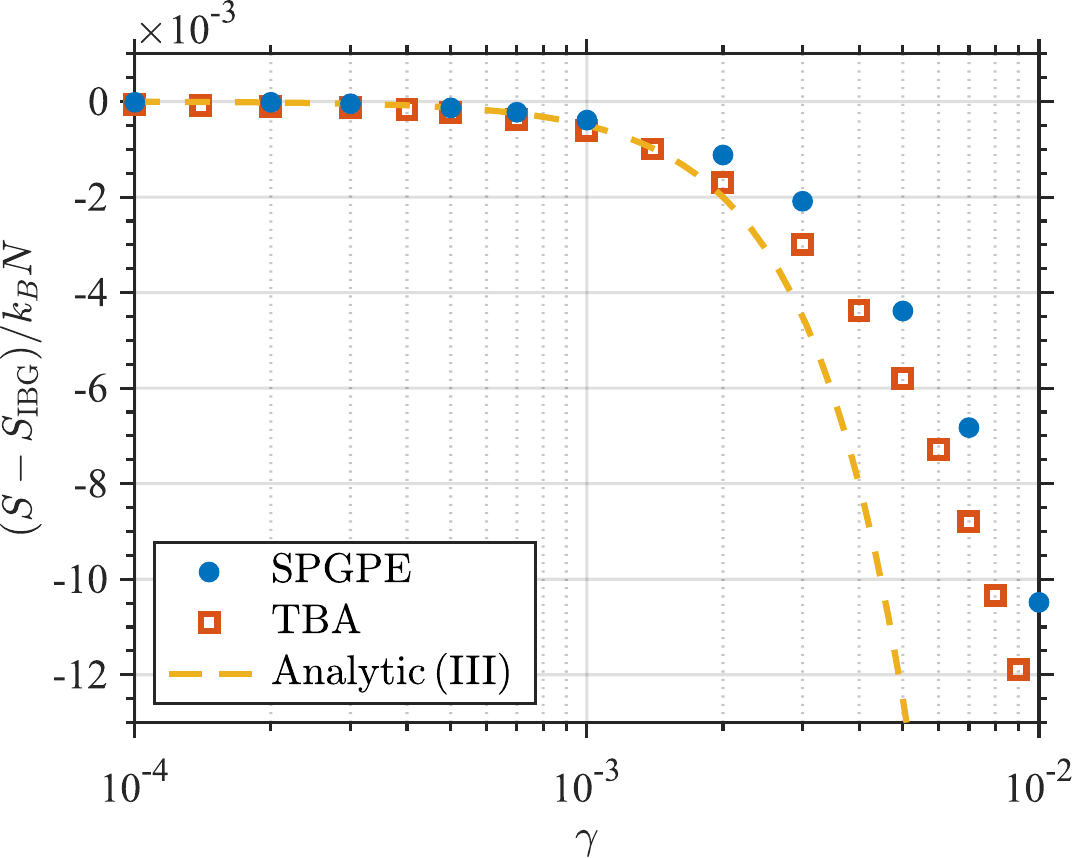}
    \caption{Entropy of a 1D quasicondensate as a function of the dimensionless interaction strength $\gamma$, relative to that of an ideal Bose gas at the same temperature, $S(\gamma,\tau)-S_{\text{IBG}}(\tau)$, for $\tau=0.2$. The SPGPE data computed using Eq.~\eqref{eq:entropy_from_integral} is shown as circles and is compared to the exact TBA data (squares), and the analytic approximation of Eq.~\eqref{S_III} (dashes).}
    \label{fig:S}
\end{center}
\end{figure}

We next deduce the entropy of the 1D quasicondensate using Eq.~\eqref{eq:entropy_from_integral} and the SPGPE data for $g^{(2)}$, except that now the integration in Eq.~\eqref{eq:entropy_from_integral}  is done numerically. To obtain the dependence of $S(g,T,L,N)$ on $g$ at a fixed $T$, or rather on the dimensionless $\gamma$ at a fixed $\tau$, we convert Eq.~\eqref{eq:entropy_from_integral} to the dimensionless units and first evaluate the derivative $\left(\partial g^{(2)}(\gamma',\tau)/\partial \tau\right)_{\gamma'}$ using the central difference scheme, for a range of values of $\gamma'$. We next evaluate the integral over $\gamma'$ numerically, as a function of the upper bound. The upper bound is scanned within $\gamma\in [10^{-4}, 10^{-2}]$, while fixing the lower bound at $\gamma_0=10^{-6}$, which is sufficiently low for the SPGPE results to be nearly identical to the IBG results at finite $T$, for which $g^{(2)}=2$ and $S(\gamma_0,T,L,N)\simeq S_{\text{IBG}}(T,L,N)$.
In Fig.~\ref{fig:S} we show the SPGPE result for the entropy difference per particle, $(S-S_{\text{IBG}})/k_BN$ obtained from the SPGPE approach as a function of $\gamma$, for a fixed value of the dimensionless temperature $\tau$. We again compare these data with the exact TBA result (squares) and the analytic result of Eq.~\eqref{S_III} (dashed line). As we see, the entropy is maximal in the ideal Bose gas limit ($\gamma \to 0$), where $g^{(2)}=2$ is also maximal, reflecting the large density fluctuations and excess randomness (bunching) in the probability of finding two indistinguishable bosons in the same position. As the strength of repulsive interactions increases, the random density fluctuations become more and more suppressed, which is also evident in the decrease of the entropy of the gas, as expected. Overall, we see a good agreement between the SPGPE and TBA results, particularly at small values of $\gamma$, where the condition of validity of the SPGPE approximation is better satisfied; the agreement becomes worse as $\gamma$ is increased, for the same reason as the discrepancy in the behaviour of $g^{(2)}$ discussed earlier.

\section{Conclusions}

In this chapter, we derived and discussed a new Maxwell relation by which the entropy of a quantum many-body system with contact two-body interactions can be related to, and deduced from, the local two-particle correlation function. We have validated this method though a numerical experiment based on the $c$-field SPGPE simulations and computed---for the first time (to the best of our knowledge) within the SPGPE formalism---the thermodynamic entropy of a weakly interacting 1D Bose gas in the quasicondensate regime.

The Maxwell relation derived here may find immediate applications, such as measuring the entropy and deducing the thermodynamic equation of state, in quantum gas experiments that take advantage of tuneable interparticle interactions and measurements of atom-atom correlations using photoassociation or in-situ imaging techniques in quantum gas microscope setups. It can also be applied to other computational approaches, such as DMRG and phase space stochastic gauge methods \cite{Drummond2004,Deuar2009}, that are capable of computing particle-particle correlations from the many-body wave-function or density matrix formalisms, but struggle to compute the entropy from, e.g., the multiplicity or the free energy. 

Beyond quantum gas systems, the main results discussed here, Eqs.~\eqref{eq:Maxwell} and \eqref{eq:entropy_from_integral}, can aid the study of traditional condensed matter systems that are often characterized through the static structure factor that describes scattering experiments. Indeed, the static structure factor $S(\mathbf{k})$ is related to the nonlocal pair correlation $g^{(2)}(\mathbf{r})$ via a Fourier transform, $S(\mathbf{k})\!=\!1\!+\!n\int\! d\mathbf{r} g^{(2)}(\mathbf{r}) e^{-i\mathbf{k}\cdot \mathbf{r}}$. Therefore, measurements or theoretical knowledge of $S(\mathbf{k})$ can be used to deduce the local pair correlation $g^{(2)}(0)\equiv g^{(2)}$ by an inverse Fourier transform, which can then be used for determining the thermodynamic entropy from Eq.~\eqref{eq:entropy_from_integral}. Finally, one can envisage derivation of related Maxwell relations for a large class of spin Hamiltonians in which: (a) the spin-spin interaction term can be similarly calculated using the Hellmann-Feynman theorem; and (b) the relevant spin-spin correlation function can be experimentally measured.

\newpage
 \noindent
The work presented in Chapter \ref{Chap:Maxwell} was adapted from the submitted publication of Ref.~\cite{watson2024maxwell}, and the contribution of each named author to that work is presented below in Table.~\ref{Tab:Chap4}.

\noindent
\cite{watson2024maxwell} \textbf{R. S. Watson}, C. Coleman, and K. V. Kheruntsyan, \href{https://arxiv.org/abs/2405.04159}{Maxwell relation between entropy and atom-atom pair correlation}, Phys. Rev. Lett. \textbf{133}, 100403 (2024).

\begin{table}[h]
	\begin{center}
	\begin{tabular}{|c|l|l|}
		\hline
		Contributor & Statement of contribution & \% \\
		\hline
		\textbf{R. S. Watson}				& writing of text 					& 20 \\
															& numerical calculations 		& 90\\
															& preparation of figures 		& 100 \\
		\hline
		C. Coleman				& writing of text 					& 5\\
															& proof-reading 							& 5 \\
               															& numerical calculations 		& 10\\

		\hline
		K. V. Kheruntsyan								& writing of text 					& 75 \\
															& proof-reading							& 75 \\
															& analytic derivations 	& 100\\
															& initial concept						& 100 \\
		\hline
	\end{tabular}
	\end{center}
 \caption{}\label{Tab:Chap4}
\end{table}


%% file: Chapter5/Chapter5.tex

\chapter[Benchmarks of generalized hydrodynamics for one-dimensional Bose gases]{Benchmarks of generalized hydrodynamics for one-dimensional Bose gases}
\label{Chap:5}	
\pagestyle{headings}

\textit{Generalized hydrodynamics (GHD) is a recent theoretical approach that is becoming a go-to tool for characterizing out-of-equilibrium phenomena in integrable and near-integrable quantum many-body systems. Here, we benchmark its performance against an array of alternative theoretical methods, for an interacting one-dimensional Bose gas described by the Lieb-Liniger model. In particular, we study various quantum shockwave scenarios, along with a quantum Newton's cradle setup, for various interaction strengths and initial temperatures. We find that GHD generally performs very well at sufficiently high temperatures or strong interactions. For low temperatures and weak interactions, we highlight situations where GHD, while not capturing interference phenomena on short lengthscales, can describe a coarse-grained behaviour based on convolution averaging that mimics finite imaging resolution in ultracold atom experiments. In a quantum Newton's cradle setup based on a double-well to single-well trap quench, we find that GHD with diffusive corrections demonstrates excellent agreement with the predictions of a classical field approach.}


\section{Introduction}
\label{Sec:label}	


The study of dynamics of integrable and near-integrable quantum many-body systems has been a thriving area of research for more than a decade 
since the landmark experiments on relaxation in the quantum Newton's cradle setup \cite{kinoshita2006quantum} and in coherently split one-dimensional (1D) Bose gases \cite{hofferberth2007non}. During this time, an in-depth understanding of the mechanisms of thermalization and emergent out-of-equilibrium phenomena within these systems has been developed \cite{rigol2007relaxation,rigol2008thermalization,cr10,pssv11,ge16,Langen207}. A recent breakthrough in this area 
has been the discovery of the theory of generalized hydrodynamics (GHD) \cite{Bertini_2016_Transport,Castro-Alvaredo_2016_Emergent} (for recent reviews, see \cite{Doyon-lectures,bouchoule2022generalized,essler2022short}). This new theory is capable of simulating large-scale dynamics of integrable and near-integrable systems across a significantly broader range of particle numbers and interaction strengths than those accessible using previous approaches \cite{largescale_Doyon,GHD_onatomchip,malvania2021generalized}. Because of its broad applicability, GHD is currently regarded as well on its way to becoming ``a standard tool in the description of strongly interacting 1D quantum dynamics close to integrable points" \cite{malvania2021generalized}.

In the years since its discovery, GHD has been rapidly developed to include diffusive terms \cite{DeNardis_Diffusion_2018,gopalakrishnan2018hydrodynamics,bastianello_thermalization_2020,durnin2021diffusive,bastianello2021hydrodynamics,Bulchandani_2021}, particle loss \cite{Bouchoule_AtomLoss_2020},
calculations of quantum and Euler-scale correlations \cite{ruggiero2020quantum,ruggiero2021quantum,Doyon_correlation_2018,moller2020euler,De_Nardis_2022,Alba_2021}, as well as the incorporation of numerous beyond-Euler scale effects \cite{Fagotti_higherorder_2017,Panfil_Linearized_2019,moller2020extension,Bastianello_dephasing_2020} (see also \cite{Bu_a_2021,Borsi_2021,El_2021,Cubero_2021} in a special issue). Recently, GHD applied to a 1D Bose gas has been experimentally verified in a variant of the quantum Newton's cradle setup in the weakly interacting regime \cite{GHD_onatomchip}, and in a harmonic trap quench in the strongly interacting regime \cite{malvania2021generalized}. In both cases, GHD provided an accurate coarse-grained model of the dynamics, exceeding conventional (classical) hydrodynamics. In addition to comparisons with experiments, GHD was benchmarked against other established theoretical approaches---most prominently for the 1D Bose gas and $XXZ$ spin chain \cite{Castro-Alvaredo_2016_Emergent,Bertini_2016_Transport,largescale_Doyon,bastianello2019generalized,moller2020euler,GHD_onatomchip,Bastianello_dephasing_2020,ruggiero2020quantum,UniversalShock,bulchandani2018bethe,urichuk2019spinDrude,doyon2018geometric}. As the purpose of these initial benchmarks was to validate GHD, the typical dynamical scenarios considered were in regimes where GHD was expected to be a valid theory. In all such cases GHD demonstrated very good agreement with the alternative approaches. On the other hand, in scenarios involving, for example, short wavelength density oscillations due to interference phenomena (which are not captured by GHD), it was conjectured that GHD would nevertheless adequately describe spatial coarse-grained averages of the more accurate theories \cite{largescale_Doyon,GHD_onatomchip,moller2020extension}. More generally, it is of significant interest to scrutinize the performance of GHD by extending its benchmarks to a more challenging set of dynamical scenarios. This is important for understanding exactly how GHD breaks down when it is pushed towards and beyond the limits of its applicability.

In this chapter, we systematically benchmark the performance of GHD for the 1D Bose gas in several paradigmatic out-of-equilibrium scenarios. In particular, we focus on the regime of dispersive quantum shockwaves emanating from a localized density bump of the type explored recently in Ref.~\cite{whatisqushock}. We use an array of theoretical approaches, including finite temperature $c$-field methods, the truncated Wigner approximation, and the numerically exact infinite matrix product state (iMPS) method, spanning the entire range of interaction strengths, from the nearly ideal Bose gas to the strongly interacting Tonks-Girardeau (TG) regime. We also analyse the dynamics of a localized density dip which sheds grey solitons, hence benchmarking GHD in scenarios not previously considered. In doing so we address the question of how well GHD predictions agree with coarse-grained averaging of the results of the more accurate theoretical approaches. Additionally, we explore the dynamics of a thermal quasicondensate in a quantum Newton's cradle setup \cite{Thomas2021,GHD_onatomchip,GHD_newtonscradle} using Navier-Stokes type diffusive GHD \cite{de2019diffusion,DeNardis_Diffusion_2018,durnin2021diffusive}, and address the question of characteristic thermalization rates \cite{Thomas2021,bastianello_thermalization_2020}.

\section{Quantum shockwaves}

We begin our analysis by considering dispersive quantum shockwaves of the type studied recently in Ref.~\cite{whatisqushock}. More specifically, we first focus on the weakly interacting regime of the 1D Bose gas of $N$ particles, and consider the dynamics of the oscillatory shockwave train generated through a trap quench from an initially localized perturbation on top of a flat background to free propagation in a uniform box of length $L$ with periodic boundary conditions \cite{Damski_2004,Damski_2006}. The weakly interacting regime is characterized by the Lieb-Liniger \cite{Lieb-Liniger-I,kheruntsyan2005finite} dimensionless interaction parameter $\gamma_\mathrm{bg}=mg/\hbar^2\rho_\mathrm{bg}\ll1$, defined with respect to the background particle number density, $\rho_\mathrm{bg}$, where $g>0$ is the strength of repulsive contact interaction and $m$ is the mass of the particles. 

\subsection{Dynamics of a localized density bump}

\begin{figure}[t!]  
\begin{center}
   \includegraphics[width=10cm]{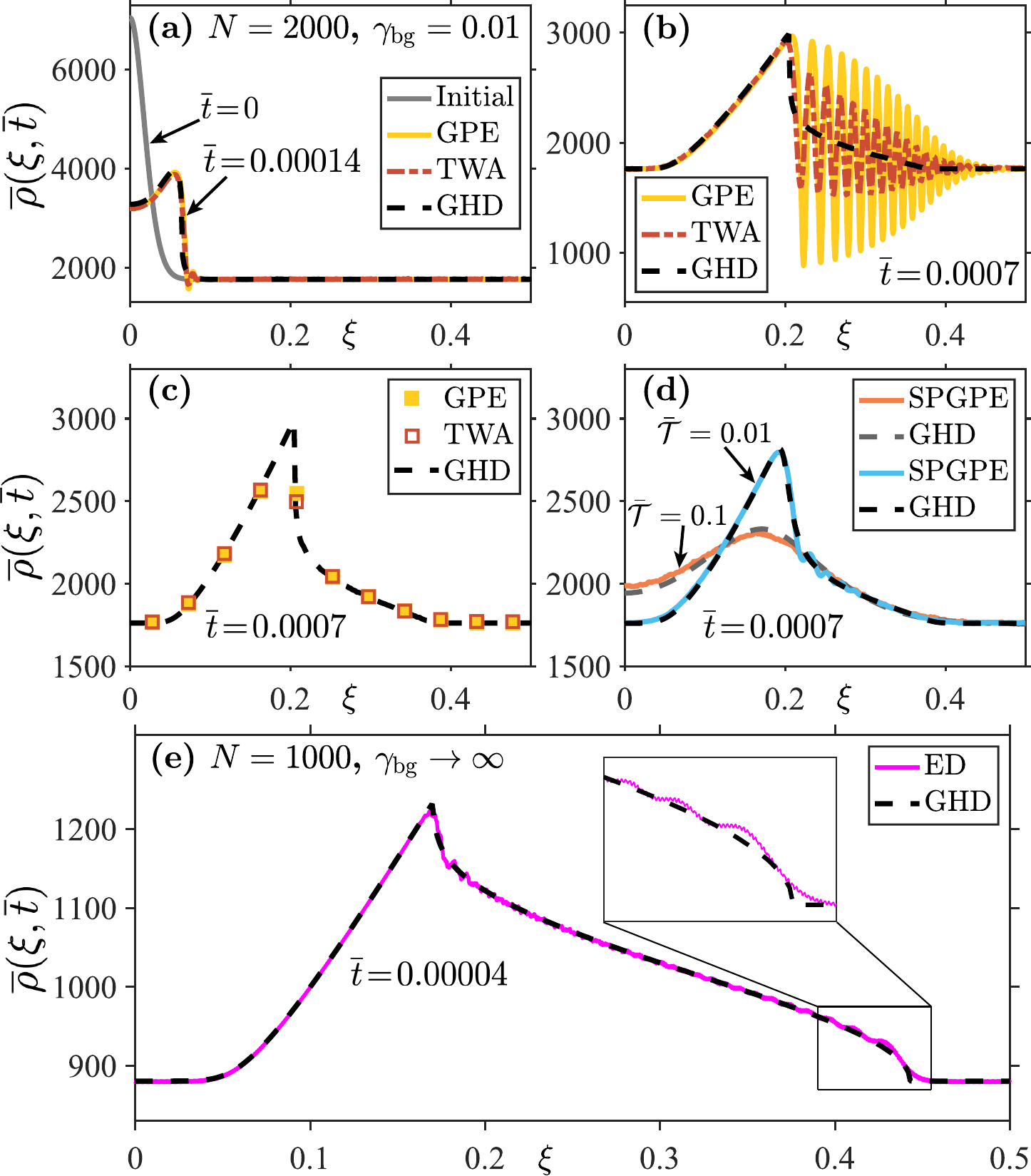}
   \caption{Dimensionless density profiles $\overline{\rho}=\rho L$ of quantum shockwaves in the 1D Bose gas, as a function of the dimensionless coordinate $\xi \!\equiv\! z/L$ at different times $\overline{t} \!\equiv\! \hbar t / m L^2$. In (a) we show the initial ($\overline{t}\!=\!0$) and time-evolved ($\overline{t}\!=\!0.00014$) 
   profiles of a weakly interacting gas 
   at zero temperature, for $\gamma_\mathrm{bg}\!=\! 0.01$ and $N\!=\!2000$ (with $N_\mathrm{bg}\! \simeq \!1761$ the number of particle in the background). Due to the symmetry about the origin, we only show the densities for $\xi\!>\!0$. In (b), the time-evolved profile is shown at $\overline{t} \!=\! 0.0007$. Panel (c) demonstrates the results of finite resolution averaging 
   of both GPE and TWA data from (b) and compares them with the same GHD result. Panel (d) shows the same system as in (b), but at finite temperatures, simulated using the stochastic projected GPE (SPGPE) \cite{whatisqushock}; the dimensionless temperature $\overline{\mathcal{T}}$ here is defined according to $\overline{\mathcal{T}} \!=\! T / T_d$, where $T_d \!= \!\hbar^2 \rho_\mathrm{bg}^2 / 2 m k_B$ \cite{kheruntsyan2005finite}. Panel (e) compares GHD predictions with exact diagonalization (ED) results in the TG regime ($\gamma_\mathrm{bg} \!\to\! \infty$) for $N\!=\!1000$ ($N_\mathrm{bg} \!\simeq \!884$), at $\overline{t}=0.00004$. In all examples, the initial profiles are characterized by the amplitude height $\beta\!=\!1$ and dimensionless width of the bump $\overline{\sigma}\!=\!0.02$. 
   } 
  \label{fig:Bump_largeN}
  \end{center}
\end{figure}

The initial density profile in Fig.~\ref{fig:Bump_largeN}\,(a), in dimensionless units, is set to 
\begin{equation}\label{eq:initialdensity}
   \overline{\rho}(\xi,\overline{t}\!= \!0) \!=\! \overline{\rho}_\mathrm{bg} \big( 1 \!+\! \beta e^{-\xi^2/2 \overline{\sigma}^2} \big)^2,
\end{equation}
where we introduce the dimensionless coordinate, time, and density, respectively, according to $\xi \!\equiv\! z/L$, $\overline{t} \!\equiv\! \hbar t / m L^2$, and $\overline{\rho}(\xi,\overline{t})\!\equiv\!\rho(z,t)L$, with $\overline{\rho}_\mathrm{bg}\!=\!\rho_\mathrm{bg}L\!=\!N_\mathrm{bg}$ being the dimensionless background density equivalent to the total number of particles in the background, $N_\mathrm{bg}\!=\!N/\big( 1+ \frac{\sqrt{\pi} \beta \sigma}{L} [ \beta\, \mathrm{erf}(\frac{L}{2\sigma}) +2 \sqrt{2}  \,\mathrm{erf} (\frac{L}{2\sqrt{2}\sigma}   )  ] \big)$ from the normalization. In addition, the width and amplitude of the bump above the background are characterized by the dimensionless parameters $\bar{\sigma}\equiv\sigma/L$ and $\beta>0$, respectively.

The associated trapping potential that is required for preparation of such a density profile as an initial ground or thermal equilibrium state of the 1D Bose gas in different regimes is discussed in Ref.~\cite{whatisqushock}. Within the mean-field approximation, described by the Gross-Pitaevskii equation, the density profile of Eq. (\ref{eq:initialdensity}) corresponds to the mean field amplitude being initialized as a simple Gaussian bump superimposed on a constant background, $\overline{\Psi}(\xi,\overline{t}\!= \!0) \!=\! \overline{\Psi}_\mathrm{bg} \big( 1 \!+\! \beta e^{-\xi^2/2 \overline{\sigma}^2}\big)$, with $\overline{\rho}_\mathrm{bg}=|\overline{\Psi}_\mathrm{bg}|^2$.

In our first example, 
we consider the case of a large total number of particles, $N\!=\!2000$, and $\gamma_\mathrm{bg}\!=\!0.01$, so that the gas is in the Thomas-Fermi regime where the interaction energy per particle dominates the kinetic energy. We assume that the gas is initialized in the zero-temperature ($T\!=\!0$) ground state of a dimple trap that results in the density profile of Eq.~\eqref{eq:initialdensity}. At time $\overline{t}\!=\!0$, the dimple trap is suddenly switched off, and we follow the evolution of the system in a uniform 1D trap. In Figs.~\ref{fig:Bump_largeN}\,(a) and (b), we show snapshots of the 
density profiles at different times, and compare the GHD results with those obtained using the mean field Gross-Pitaevskii equation (GPE) and the truncated Wigner approximation (TWA) which incorporates the effect of quantum fluctuations ignored in the GPE. 

The snapshot at $\overline{t}\!=\!0.00014$, which corresponds to the onset of a shockwave formation due to a large density gradient, shows excellent agreement between GHD and the more accurate microscopic approaches. Such an agreement  at early times is remarkable given that GHD, which is derived here at Euler scale \cite{spohn2012large}, becomes formally exact only in the limit of infinitely large length and time scales 
\cite{largescale_Doyon,Doyon-lectures,moller2020euler}.

Past this time, the GPE and TWA show the formation of an oscillatory shockwave train, which has been identified in Ref.~\cite{whatisqushock} as a result of self-interference of the expanding density bump with its own background.
The interference contrast in this regime is generally large, even though the quantum fluctuations present in the TWA approach cause a visible reduction in contrast compared with the mean-field GPE result. The GHD prediction, on the other hand, completely fails to capture the oscillations, as these occur on a microscopic lengthscale. The characteristic period of oscillations here (which we note are chirped) 
is given approximately by the healing length $l_h\!=\!\hbar/\sqrt{mg\rho_\mathrm{bg}}$ ($l_h/L = 0.0057$) which is smaller than the width $\sigma$ ($\sigma/L=0.02$) of the initial bump and hence represents the shortest lengthscale of the problem in the bulk of the shockwave train.
Thus, even though the local density approximation (required for GHD to be applicable to an inhomogeneous system in the first place) is valid for the initial Thomas-Fermi density profile, the failure of GHD at later times is expected since it is not supposed to capture phenomena on microscopic lengthscales, which emerge here dynamically.

Despite this failure, GHD clearly captures the average density of the oscillations for the fully formed shockwave train, similar to that shown in \cite{essler2022short}. This is consistent with the analysis of Bettelheim \cite{bettelheim2020whitham}, who showed that the Whitham approach, which allows one to write equations for averaged quantities in the oscillatory shockwave train, is equivalent to GHD in the semiclassical limit ($c \!=\!mg/\hbar^2\!\to\!0$) of the Lieb-Liniger model \cite{whithamBook,kamchatnov2000nonlinear}. This is also consistent with the expectation that GHD in an interfering region would correspond to a coarse-grained average density \cite{GHD_onatomchip,largescale_Doyon}.  To quantitatively assess this expectation, we perform a type of convolution averaging that mimics the finite resolution of \emph{in-situ} imaging systems used in quantum gas experiments. 
As the imaging resolution is usually unable to resolve wavelengths on the order of the healing length (typically in the submicron range), one expects that such averaging will smear out the interference fringes seen in the GPE and TWA data---just as GHD implicitly does. In Fig.~\ref{fig:Bump_largeN}(c) we show the results of convolution-averaged density profiles performed on the GPE and TWA data of Fig.~\ref{fig:Bump_largeN}(b) and compare them with the same GHD curve. The level of agreement between all three curves is now remarkable---a result which was not \textit{a priori} obvious for both GPE and TWA under this model of coarse-graining. This highlights the quantitative success of GHD in describing the dynamics on large scale despite interference or short-wavelength phenomena being present.

In our second set of examples, shown in Fig.~\ref{fig:Bump_largeN}(d), we consider the same shockwave scenario, except now for a phase fluctuating quasicondensate at finite temperatures. Here, the effect of thermal fluctuations is expected to lead to a smearing of the interference contrast due to a reduced thermal phase coherence length in the system, $l_T\!=\!\hbar^2\rho_{\mathrm{bg}}/mk_BT$ \cite{Mora-Castin-2003,Cazalilla_2004,Bouchoule-Arzamasovs-2012}. A well-established theoretical approach to model this is a $c$-field stochastic projected GPE (SPGPE) approach \cite{castin2000,Blakie_cfield_2008}, and we indeed observe such smearing in Fig.~\ref{fig:Bump_largeN}(d), in addition to seeing the expected very good agreement of GHD with these $c$-field results.
For the examples considered in Fig.~1(d), the thermal phase coherence lengths are $l_T/L\!\simeq\!0.1$ for $\overline{\mathcal{T}}\!=\!0.01$ and $l_T/L\!\simeq\!0.01$ for $\overline{\mathcal{T}}\!=\!0.1$. These values, in turn, are comparable or smaller than the width of the initial density bump $\overline{\sigma}=0.02$, implying a reduced phase coherence over the extent of the bump and hence loss of interference contrast upon expansion of the bump into the background \cite{whatisqushock}.

The condition for applicability of the SPGPE is $|\mu| \ll k_B T$ \cite{castin2000}, with the $c$-field region defined in terms of dimensionless interaction and temperature parameters by the condition $\gamma \ll \overline{\mathcal{T}} \ll 1$ \cite{kheruntsyan2005finite,bayocboc2022frequency}, where the dimensionless temeperature $\overline{\mathcal{T}}$ is defined according to $\overline{\mathcal{T}} \!=\! T / T_d$, with $T_d \!= \!\hbar^2 \rho_\mathrm{bg}^2 / 2 m k_B$ being the temperature of quantum degeneracy \cite{kheruntsyan2005finite}. We further focus on the quasicondensate regime, which can be described by the Bogoliubov theory dominated by thermal (rather than vacuum) fluctuations, bounded by the condition $\gamma \ll \overline{\mathcal{T}} \ll \sqrt{\gamma}$ \cite{bayocboc2022frequency}. The truncated Wigner approximation, on the other hand, is valid when the total number of particles in the system $N$ greatly exceeds the number of relevant Bogoliubov modes, $M_B$ \cite{Blakie_cfield_2008}. The systems investigated using this approach in Fig.~1(a)-(c) were reported to follow this criterion in Ref.~\cite{whatisqushock}, where the ratio $N/M_B$ was calculated to be $N/M_B \approx 5$, with variations around this value between $N/M_B \approx 2$ and $N/M_B \approx 10$ producing essentially the same results.

Our third example is shown in Fig.~\ref{fig:Bump_largeN}(e) and lies in the TG regime of infinitely strong interactions, $\gamma_\mathrm{bg}\!\to \!\infty$. It further illustrates
the same observation---that the performance of GHD improves with the loss of phase coherence in the system, wherein interference phenomena are suppressed. Here, we compare the predictions of GHD for the shockwave scenario at $T\!=\!0$ with the results of exact diagonalization. In the TG regime, the system does not posses phase coherence beyond the mean interparticle separation $1/\rho_{bg}$, 
hence the absence of interference fringes in the evolution of a density bump whose initial width is larger than $1/\rho_{bg}$ \cite{whatisqushock}. Accordingly, we see very good agreement of GHD with exact diagonalization, ignoring the small-amplitude density ripples that can be seen in the exact result. Such density ripples (which we note have different origin to Friedel oscillations) have been predicted to occur in the ideal Fermi gas by Bettelheim and Glazman~\cite{quripples}. In particular, utilizing standard bosonization techniques (see, e.g., Refs.~\cite{stone1994bosonization,Cazalilla_2004}), they demonstrated that semiclassical corrections to the Wigner function, which may classically be approximated as a filled Fermi sea for the Tonks-Girardeau gas, appeared as `ripples' for the case of a shockwave emanating from an initial density perturbation.

By the Fermi-Bose mapping \cite{Girardeau_1960,Girardeau-Wright-2000}, these same ripples should emerge in the TG gas, which we confirm here through exact diagonalization. However, their description lies beyond the scope of GHD as a large-scale theory.
In particular, for the example of Fig.~\ref{fig:Bump_largeN}\,(e), which is for $N=1000$ particles, we have been able to discriminate between the Friedel oscillations, which have a period of 1/1000, and the Bettelheim-Glazman density ripples which have a larger oscillation period. This was not possible in the prior work of Ref.~\cite{whatisqushock}, which treated a much smaller number of particles ($N=50$) in the Tonks-Girardeau regime.

In the Tonks-Girardeau limit, there is an exact equivalence between GHD and the semiclassical Wigner function model of a free Fermi gas \cite{quripples,fateofqushock,largescale_Doyon}. This model predicts a smooth density profile of the semiclassical shock, whereas quantum corrections, derived in Refs.~\cite{universalfermi,quripples,protopopov}, predict small-amplitude oscillations (`density ripples') on top of the smooth density profile. These oscillations persist within a finite extent of the shockwave front for times well beyond shockwave formation \cite{fateofqushock}, and have a frequency chirp that is in the opposite direction to the chirp of the interference fringes seen in GPE simulations of dispersive shockwaves in the weakly interacting regime.

\begin{figure*}[tp]
    \includegraphics[width=17cm]{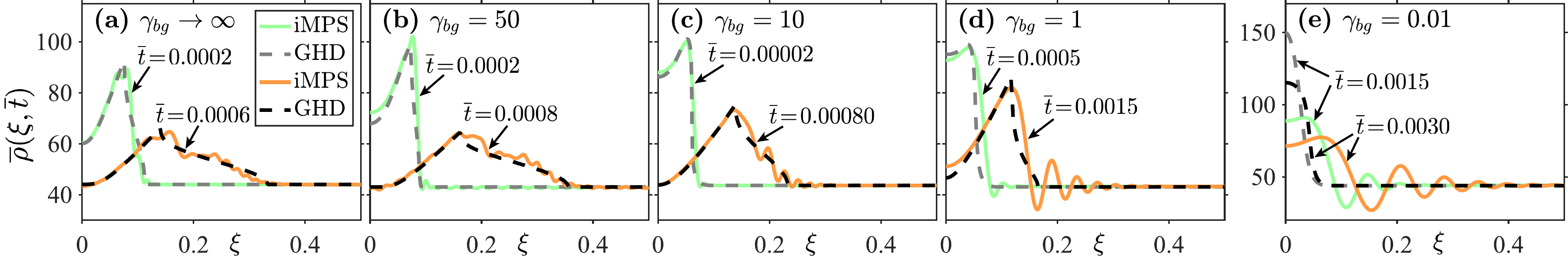}
    \caption{Quantum shockwaves at zero temperature for $N=50$ particles ($N_\mathrm{bg} \simeq 44.03$), over the entire range of interaction strengths. In all examples, the initial density profiles (not shown) closely match Eq.~\eqref{eq:initialdensity}, with $\beta=1$ and $\overline{\sigma}=0.02$. In all panels, we show the GHD (dashed lines) and iMPS (full lines) results for the evolved density profiles at two time instances. In (a) there is no phase coherence beyond the mean interparticle separation ($1/\rho_{bg} L \simeq 0.0227$), whereas in (e) the shortest lengthscale that determines the characteristic period of oscillations is given by the width of the initial Gaussian bump $\sigma$ ($\sigma/L=0.02$), which is much smaller than the healing length $l_h$ ($l_h/L=0.227$).}
\label{fig:Bump_lowN}
\end{figure*}

The final set of examples for the evolution of a density bump is shown in Fig.~\ref{fig:Bump_lowN}. Here, we consider a range of interaction strengths, starting from very strong and going back [from (a) to (e)] to weak interactions, all at zero temperature and $N\!=\!50$. We compare the GHD results with iMPS simulations, which are numerically exact at all interaction strengths \cite{whatisqushock}. At this relatively low particle number, the strongly interacting regime displays Friedel oscillations which appear in the iMPS result and are, as expected, absent from the prediction of GHD. However, there is generally good agreement between GHD and iMPS at large scale. As the interaction strength is reduced, and hence the phase coherence of the gas increases,
the Friedel oscillations disappear and interference fringes return, which now have period $\sim\!\sigma$ (with $\!\sigma<\!l_h$) since the gas is no longer in the Thomas-Fermi regime. The worst performance of GHD is observed for $\gamma_{\mathrm{bg}}\!=\!0.01$, which lies in the nearly ideal (noninteracting) Bose gas regime for $N\!=\!50$. 
In this regime, the local density approximation, intrinsic to GHD \cite{largescale_Doyon,malvania2021generalized,GHD_onatomchip,kheruntsyan2005finite}, is no longer valid even for the initial density profile, and we see that Euler-scale GHD breaks down both spatially and temporally, explaining the failure of GHD to agree with iMPS results even in the coarse-grained sense.


\subsubsection{Finite resolution averaging}

Finite resolution averaging procedure implemented in Fig.~1(c) 
emulates the finite spatial resolution of experimental absorption imaging systems. Following 
Ref.~\cite{Armijo_ThreeBody_2010}, we denote the impulse response function of the imaging system by $\mathcal{A}(z)$, which we here assume to be a normalized Gaussian. 
The impulse response for a pixel of width $\Delta$ centered at $z_p$ is then,
\begin{equation}
    \mathcal{F}(z) = \int_{z_p - \Delta/2}^{z_p+\Delta/2} dz' \mathcal{A}(z'-z).
\end{equation}
The \emph{measured} atom number in the given pixel is then given by
\begin{equation}
    N_m = \mathcal{N} \int_{-\infty}^{+\infty} dz \mathcal{F}(z) \rho(z),
\end{equation}
where $\mathcal{N}$ provides the correct normalization for the total particle number in the limit of zero pixel width.

In our particular example of such averaging, the density profile $\rho(z)$ (at any given time step, with the time argument $t$ being omitted here for notational simplicity) is convoluted with a Gaussian resolution function of width $w = 1$ $\mu$m and then averaged over a finite pixel size $\Delta= 4.5$ $\mu$m, as in Ref.~ \cite{Armijo_ThreeBody_2010}. These absolute values translate to dimensionless values of $w/L = 0.01$ and $\Delta/L = 0.045$, assuming $L\sim 100$~$\mu$m, with results being generally insensitive to the exact values of these parameters around these typical values.
For comparison, the healing length in this example is equal to $l_h/L\!=\!0.0057$. Considering $^{87}$Rb atoms, which have a scattering length of $a\simeq 5.3$~nm, in a system of size $L=100$~$\mu$m, this corresponds to an absolute healing length of $l_h = 0.57$~$\mu$m. These choices of dimensionless parameters, and $\gamma_{\mathrm{bg}}=0.01$, can be realized at a background density of $\rho_\mathrm{bg} \simeq 1.8 \times 10^7$~m$^{-1}$, with an interaction parameter $g\simeq 2 \hbar \omega_\perp a \simeq 1.4\times 10^{-38} \mathrm{J} \! \cdot \! \mathrm{m}$ \cite{Olshanii1:998}, where $\omega_\perp / 2 \pi \simeq 1.9$~kHz is the frequency of the transverse harmonic trapping potential.

\begin{figure}[tbp]
\begin{center}
   \includegraphics[width=10cm]{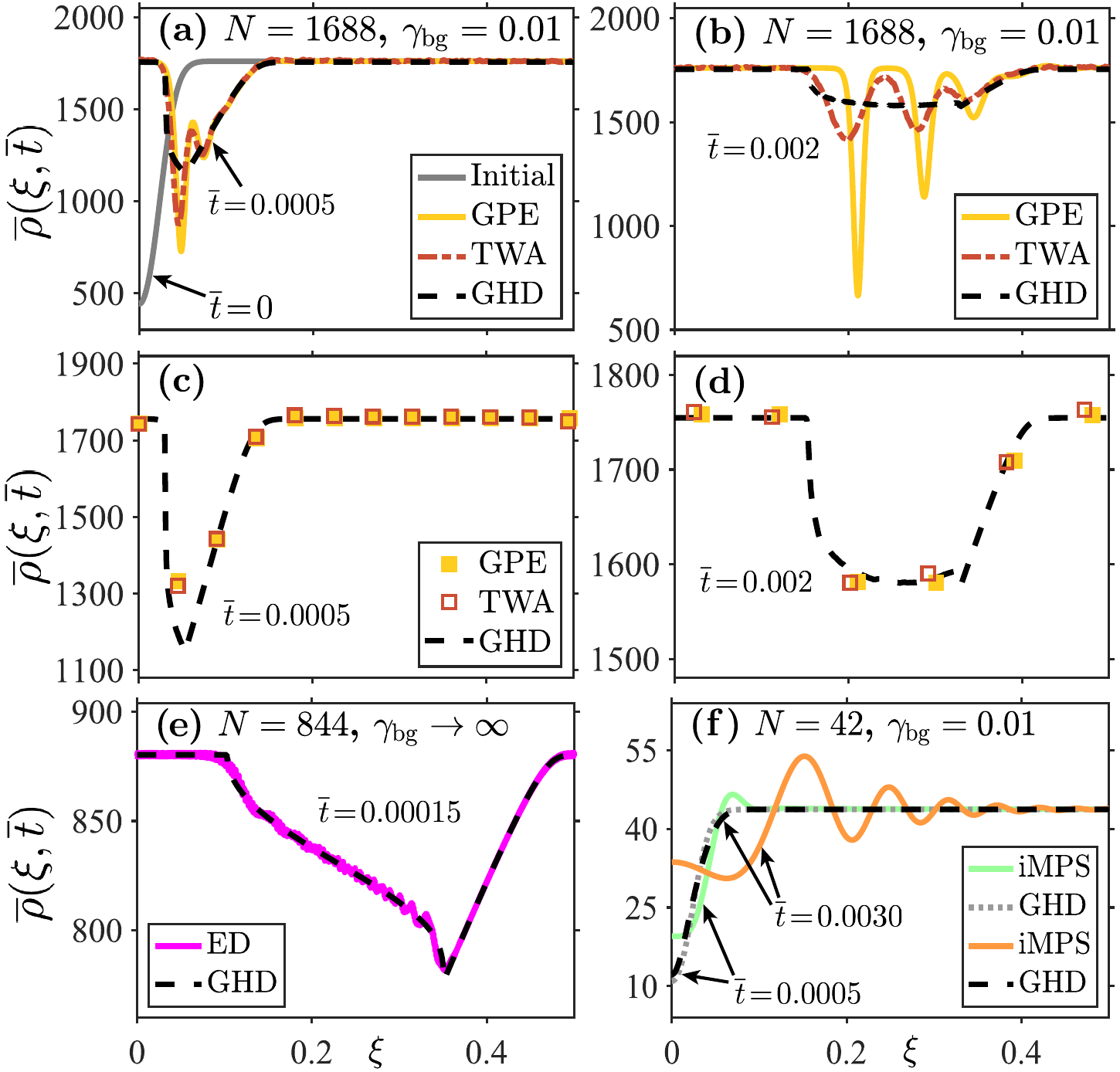}
   \caption{Evolution of a density dip in a 1D Bose gas. Panel (a) shows the initial ($\overline{t}=0$) and time-evolved ($\overline{t}=0.0005$) density profiles from GPE, TWA and GHD simulations, for $\gamma_\mathrm{bg} \!= \!0.01$ and $N \!= \!1688$ ($N_\mathrm{bg} \!\simeq \!1761$); panel (b) shows a time-evolved density profile at a later time ($\overline{t}=0.002$), where we can see a fully formed train of three grey solitons in the mean-field GPE (full yellow) curve.
   Panels (c) and (d) compare the same GHD results (notice the different scale of vertical axis) at $\overline{t}=0.0005$ and $\overline{t}=0.002$ with the outcomes of finite resolution averaging of both GPE and TWA curves. 
   In panel (e), we show a time-evolved snapshot of the density profile in the TG regime ($\gamma_\mathrm{bg} \!\to
   \! \infty$) for $N \!= \!844$ ($N_\mathrm{bg} \!\simeq \!880.5$), and compare the GHD result with that of exact diagonalization (ED). Panel (f) is in the nearly ideal Bose gas regime, with $\gamma_\mathrm{bg} \!= \!0.01$, $N \!= \!42$ ($N_\mathrm{bg} \!\simeq \!44$). In all examples, the initial density profile is given by Eq.~\eqref{eq:initialdensity} with $\beta \!=\! -0.5$ and $\overline{\sigma}\!=\!0.02$.} 
  \label{fig:Dip}
  \end{center}
\end{figure}

\subsection{Dynamics of a localized density dip}
In addition to considering the dynamics of a localized density bump, we also analyse evolution of an initial density dip in a uniform background. This scenario is known to shed a train of grey solitons in the mean-field GPE treatment  \cite{Damski_2004,Damski_2006,Engels_Hoefer_2008}, and the results of comparison of GHD simulations with those of GPE and TWA are presented. The overall conclusions regarding the performance of GHD in this scenario are similar to those for a density bump, including good agreement of GHD with coarse-grained averages of GPA and TWA results in the soliton train region.   

Here, we present the results of evolution of a localized density depression, after quenching (at time $\overline{t}\!=\!0$) the initial trap potential with a localized barrier to uniform. We assume that the initial density profile is given by the same Eq.~\eqref{eq:initialdensity}, except with $\beta$ being negative and satisfying $-1 \!<\! \beta\! <\! 0$. 

In Figs.~\ref{fig:Dip}\,(a) and (b), we consider the weakly interacting regime (with $\gamma_\mathrm{bg} \!= \!0.01$) and show the results of the GPE, TWA, and GHD simulations, 
for a 
gas with $N\!=\!1688$ atoms and the same $N_\mathrm{bg} \!\simeq \!1761$ as in Fig.~\ref{fig:Bump_largeN}\,(a). In this scenario, the steep gradient of the shockwave front forms as the background fluid flows inward and tries to fill the density depression. As a result, one first observes the emergence of large-amplitude structures, forming multiple density troughs, which then evolve into a train of grey solitons propagating away from the origin \cite{Damski_2004,Damski_2006,Engels_Hoefer_2008,el_DSW,Gurevich_Pitaevskii,kamchatnov2000nonlinear}. The differences between the TWA and pure mean-field GPE results, seen in Figs.~\ref{fig:Dip}\,(b), are consistent with previous observations \cite{Dziarmaga_2003,Dziarmaga_2004,Ruostekoski_2010} that quantum fluctuations lower the mean soliton speed and fill in the soliton core. The GHD result, on the other hand, fails to capture the solitonic structures, whose characteristic width (on the order of the microscopic healing length) lies beyond the intended range of applicability of GHD. 

However, GHD still manages to adequately capture the coarse-grained description of the density across the soliton train, which is rather remarkable. This is seen in 
Fig.~\ref{fig:Dip}\,(c) and (d), where we demonstrate the outcomes of finite resolution averaging applied to GPE and TWA results of panels (a) and (b), respectively. 
Similarly to Fig.~1(c), here we used the same normalized Gaussian resolution function of width $1$\,$\mu$m and adopted $^{87}$Rb atoms as an example species for the relevant parameter values. For panel (c) we used the same pixel size ($\Delta = 4.5$\,$\mu$m) as before, whereas for panel (d), due to the presence of fully formed grey solitons whose width is on the order of $(2-4)l_h$, we used a twice larger pixel size ($\Delta = 9.0$\,$\mu$m). A larger pixel size here results in $\Delta/l_h\simeq 16\gg 1$, which is required in order to comply with the large-scale framework of GHD.


The last two examples, shown in Figs.~\ref{fig:Dip}\,(c) and (d), correspond, respectively, to the strongly interacting TG and nearly ideal Bose gas regimes. The overall behaviour and conclusions about the performance of GHD in these examples are the same as in the equivalent scenario of the density bump discussed earlier in Figs.~\ref{fig:Bump_largeN}\,(e) and \ref{fig:Bump_lowN}\,(a).


\section{Quantum Newton's cradle in a thermal quasicondensate}

Our final scenario for benchmarking GHD is in a variant of the quantum Newton's cradle setup for a weakly interacting 1D Bose gas in the quasicondensate regime. Namely, we analyze the release from a symmetric double-well trap to a single-well harmonic trap of frequency $\omega$, similar to the type utilized  in Ref.~\cite{GHD_onatomchip}. Here, we use the SPGPE to simulate collisional dynamics and eventual thermalization, as in Ref.~\cite{Thomas2021}, and for the sake of one-to-one comparison, we also simulate the same system using the Navier-Stokes type of diffusive GHD \cite{DeNardis_Diffusion_2018,de2019diffusion}, solved using a second-order backwards-implicit algorithm 
\cite{gopalakrishnan2018hydrodynamics,bastianello2019generalized,moller2020introducing}.

\begin{figure}[tbp]
\begin{center}
  \includegraphics[width=13cm]{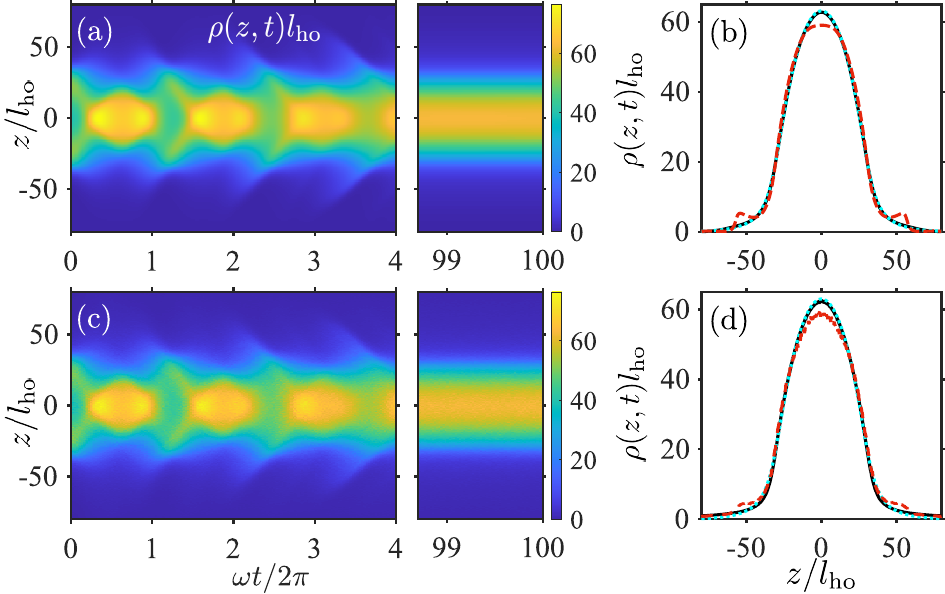}   
   \caption{Evolution and thermalization of the density distribution $\rho(z,t)$ in a quantum Newton's cradle setup initialized from a double-well to single-well trap quench, simulated using (a)-(b) Navier-Stokes GHD, and (c)-(d) SPGPE. The initial cloud of $N=3340$ atoms at temperature $\widetilde{T} = 205$ (in harmonic oscillator units) is prepared in a thermal equilibrium state of a symmetric double-well trap potential. 
    Panel (b) demonstrates the relaxed density profile of the Navier-Stokes GHD evolution at $t\!=\!100/(\omega/2\pi)$ (black solid line), alongside a best-fit thermal equilibrium profile from Yang-Yang thermodynamics at $\widetilde{T}\!\simeq \!213$ (cyan dotted line), and an additional GHD density profile at earlier time $t\!=\!6.79/(\omega/2\pi)$ (red dashed line). Panel (d) is the same as (b), but for the SPGPE, with the relaxed density profile at $t\!\simeq\!100/(\omega/2\pi)$, Yang-Yang thermodyanmic density profile of $\widetilde{T} \!\simeq \!216$, and an additional density profile at $t\!=\!6.81/(\omega/2\pi)$.} 
   
  \label{fig:NewtonsCradle}
  \end{center}
\end{figure}

The initial (pre-quench) double-well trap potential is set to  $\widetilde{V}(\widetilde{z}) \!\simeq\! 2.16 \! \times \! 10^{-3} \widetilde{z}^4 \!-\! 5.27 \! \times \! 10^{-1} \widetilde{z}^2$ in dimensionless form, where $\widetilde{z} \!= \!z/l_\mathrm{ho}$, $\widetilde{V} \!=\! V / \hbar \omega$, and $l_\mathrm{ho} \!=\! \sqrt{\hbar/m \omega}$, where $\omega$ is the post-quench single-well harmonic trap frequency. The initial dimensionless temperature of the cloud, in harmonic oscillator units, is set to  $\widetilde{T} \!=\! T/(\hbar \omega/k_B)\!\simeq\! 205$. In this configuration, the initial density profile for a total of $N=3340$ atoms is double peaked, with the dimensionless interaction strength at either of the peaks given by $\gamma_{\max}\simeq0.0138$.

Comparison of the results using the two methods are shown in Fig.~\ref{fig:NewtonsCradle}, where we illustrate the evolution of the density distribution [(a) --  for diffusive GHD, and (c) -- for SPGPE] over the initial few oscillations, as well as after sufficiently long time, when the system has already thermalized. In Figs.~\ref{fig:NewtonsCradle}\,(b) and (d), we demonstrate the respective relaxed density profiles, along with their corresponding thermal equilibrium profiles from Yang-Yang thermodynamics \cite{yang1969thermodynamics,Karen_Yang_2008, kheruntsyan2005finite}, as well as density profiles at earlier times illustrating their contrast to the relaxed state. The overall conclusion is that GHD demonstrates excellent agreement with SPGPE  in both short- and long-term dynamics, as well as in the characteristic thermalization rate.

It was argued recently, in Ref.~\cite{bastianello_thermalization_2020}, that the stationary states within fluid cells are attained entirely in the time window where diffusive dynamics are the dominant contribution beyond Euler-scale, and that these stationary states are uniquely described by the thermal ensemble. Following this window, higher-order corrections beyond diffusion must be incorporated, however, as thermal states are stationary under exact evolution via the Lieb-Liniger Hamiltonian, the relaxed state will remain thermal beyond this window.

~

\begin{figure}[tb]
\begin{center}
    \includegraphics[width=11cm]{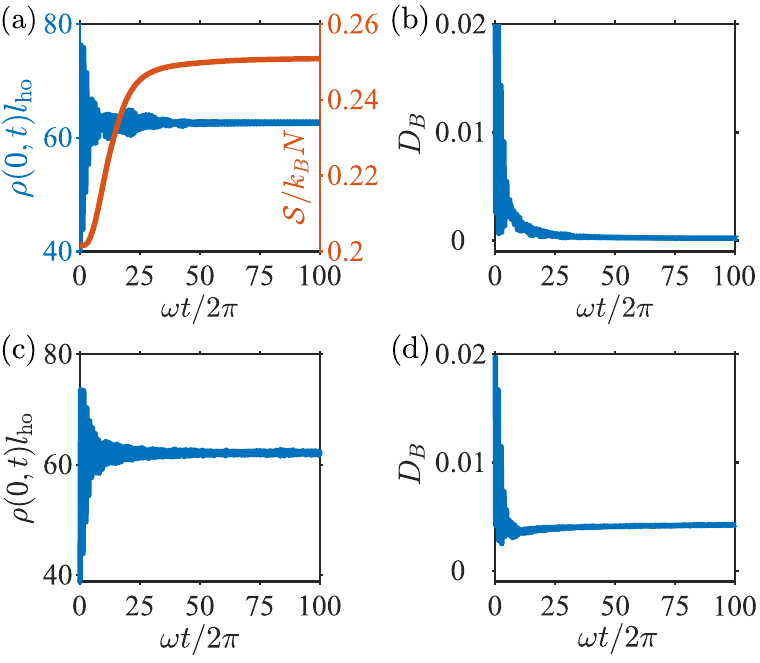}
    \caption{
   Convergence to thermalization for the double-well to single-well trap quench demonstrated in Fig.~\ref{fig:NewtonsCradle}. Panel (a) shows the evolution of the peak density for the Navier-Stokes GHD simulation (averaged over the region $z/l_\mathrm{ho} \in [-2,2]$, where $l_\mathrm{ho} = \sqrt{\hbar / m \omega}$ is the harmonic oscillator length), alongside the respective total entropy per particle, both of which plateau upon reaching the final thermal state. Panel (b) demonstrates the Bhattacharyya statistical distance of the GHD evolving density profile to the corresponding Yang-Yang thermodynamic density profile (fitted to the final relaxed GHD profile) of temperature $\widetilde{T} = T/(\hbar \omega /k_B)\simeq213$ shown in Fig.~\ref{fig:NewtonsCradle}\,(b). The final observed Bhattacharyya distance here is $D_B = 2.23\times10^{-4}$, which is rather small and hence indicates near-perfect overlap of the two distributions. Similarly, panel (c) shows the evolution of the peak density for the SPGPE simulation (again averaged over the region $z/l_\mathrm{ho} \in [-2,2]$) as it approaches thermalization and plateaus; (d) demonstrates the Bhattacharyya distance of the evolving SPGPE density profile to the Yang-Yang thermodynamic density profile of temperature $\widetilde{T}\simeq216$ shown in Fig.~\ref{fig:NewtonsCradle}\,(d), with a final observed distance of $D_B = 4.28\times10^{-3}$.
   }
    \label{fig:Therm_SymDoubleWell}
\end{center}
\end{figure}

Here, we investigate dynamical thermalization of a double-well to single-well trap quench, demonstrated in Fig.~\ref{fig:NewtonsCradle}. To quantify the thermalization rate observed in Navier-Stokes GHD, we have access to the thermodynamic entropy density \cite{yang1969thermodynamics}
\begin{equation}
    s/k_B = -\frac{m}{\hbar}\int d\lambda [ f \ln f - f_p \ln f_p - f_h \ln f_h ],
\end{equation}
which may be integrated for the total entropy per particle, $\mathcal{S}/k_B N = N^{-1} \int dz s/k_B$. The total entropy per particle is known to plateau upon the system reaching a thermalized state \cite{bastianello_thermalization_2020}, and is demonstrated for the Navier-Stokes GHD simulations in Fig.~\ref{fig:Therm_SymDoubleWell}(a), showing thermalization at time $t \simeq 100/(\omega/2\pi)$.

For direct comparison between GHD and SPGPE results (as entropy is not accessible in nonequilibrium SPGPE simulations), we also estimate the rate of thermalization through monitoring the peak density averaged over the region $z/l_\mathrm{ho} \in [-2,2]$, where $l_\mathrm{ho} = \sqrt{\hbar / m \omega}$ is the harmonic oscillator length. This quantity is plotted Figs.~\ref{fig:Therm_SymDoubleWell}(a) and (c) for Navier-Stokes GHD and SPGPE simulation, respectively. Under non-equilibrium Newton's cradle dynamics, the peak density undergoes oscillations at twice the longitudinal trap frequency, eventually relaxing to a final thermal state at time $t \simeq 100/(\omega/2\pi)$ for both simulation methods. This thermalization time agrees with that extracted from the plateauing of GHD entropy per particle. As noted in Ref.~\cite{bastianello_thermalization_2020}, observed thermalization times are generically observable-dependent, however the results presented here demonstrate that, for this system, the thermalization times estimated through relaxation of peak and entropy agree with each other.

The temperature, $\widetilde{T} = T/(\hbar \omega /k_B)$, and global chemical potential, $\mu_0$, of the relaxed state for the Navier-Stokes GHD result are fixed by the initial total energy and number of particles \cite{bastianello_thermalization_2020}. This determines the final temperature of the system to be $\widetilde{T} \simeq 213$, with the corresponding Yang-Yang thermodynamic density profile shown as the cyan dotted line in Fig.~\ref{fig:NewtonsCradle}(b). Temperature estimation of the relaxed state for the SPGPE result is achieved via Yang-Yang thermometry of the relaxed density profile \cite{Karen_Yang_2008,davis2012yang}, and is shown as the cyan dotted line in Fig.~\ref{fig:NewtonsCradle}(d) at a temperature of $\widetilde{T} \simeq 216$. The small difference in the temperatures extracted for the relaxed states in GHD and SPGPE simulations comes from a small difference between the respective final density profiles, so that the best-fit density profiles from Yang-Yang thermodynamics return slightly different values of the respective temperatures.

Additionally, we calculate the proximity of evolving density distributions in both GHD and SPGPE simulations to the respective Yang-Yang thermal density distributions   through the Bhattacharyya statistical distance \cite{bhattacharyya1943measure}
\begin{equation}
    D_B = -\ln [ B(P,P') ],
\end{equation}
where $B(P,P')$ is the Bhattacharyya coefficient of two normalized probability density functions $P(z_i,t)$ and $P'(z_i)$ of the same discrete variable $z_i$, 
\begin{equation}
    B(P,P') = \sum_{i} \sqrt{P(z_i,t) P'(z_i)}.
\end{equation}
Here, $P(z_i,t) = \rho(z_i,t)/\sum_{i} \rho(z_i,t)$ is the normalized evolving density profile of either GHD, shown in Fig.~
3\,(a), or SPGPE, shown in Fig.~\ref{fig:NewtonsCradle}\,(c), with $z_i$ being the position on our computational lattice. $P'(z_i) = \rho_\mathrm{YY}\!(z_i)/\sum_{i}\rho_\mathrm{YY}\!(z_i)$, on the other hand, is taken to be the normalized density profile of the Yang-Yang thermal state, $\rho_\mathrm{YY}\!(z_i)$, fitted to the final relaxed density profile of GHD or SPGPE evolution.

At any given time $t$, the Bhattacharyya distance serves as a measure of similarity between the instantaneous distribution $P(z_i,t)$ and $P'(z_i)$. For $P\to P'$, the Bhattacharyya coefficient becomes $B(P,P')\to 1$ (due to the normalization condition) and hence $D_B\to 0$, implying complete overlap of the two distributions. As we see from Figs.~\ref{fig:Therm_SymDoubleWell}\,(b) and (d), as the evolving density profile approaches the relaxed state, the Bhattacharyya distance, $D_B$, tends to a vanishingly small constant, indicating near perfect overlap between the relaxed and Yang-Yang thermal states, with a slightly higher offset from zero for the SPGPE results, whose tails are truncated due to the dependence on a high energy cut-off (see below), as they are measured against a thermal Yang-Yang density profile.

We further point out here that as the SPGPE is a classical field method \cite{castin2000,Blakie_cfield_2008}, it has an inherent dependency on the high-energy cutoff $\epsilon_{cut}$ separating out the classical field region of relatively highly occupied modes from the sparsely occupied modes treated as an effective reservoir. More specifically, the results of SPGPE simulations can be cutoff dependent subject to the observable in question. In the simulations performed here, the high-energy cutoff was chosen according to the prescription of Ref.~\cite{bayocboc2022frequency}, namely, such that the mode occupancy of the highest energy mode in the harmonic oscillator basis (of Hermite-Gauss polynomials) was on the order of $\sim 0.3$. Such a cutoff mode occupancy, which is somewhat lower than a more conventional cutoff choice at mode occupancy of the order of $\sim 1$ (see, e.g., \cite{Pietraszeqicz2015classical,Pietraszewicz2018classical,Pietraszewicz2018complex} and references therein), must be taken in order to faithfully reproduce the tails of the initial thermal equilibrium density distribution. In doing so, we have also observed that the simulation of the quantum Newton's cradle system under variation of the high-energy cutoff shows a dependence of the thermalization time on this cutoff mode occupancy. In particular, as the cutoff mode occupancy is reduced (corresponding to an increase in the high-energy cutoff $\epsilon_{cut}$), not only does the initial density profile match better to that of Yang-Yang thermodynamics, but the thermalization times are additionally shortened. As increasing the high-energy cutoff corresponds to including a larger number of thermally occupied modes, which speed up collisional relaxation, such shortening   of thermalization times is expected. Ultimately, as we have seen from the results presented in Fig.~\ref{fig:Therm_SymDoubleWell}, the characteristic thermalization times extracted from GHD and SPGPE agreed well with each other under the cutoff mode occupancy of $0.3$. Accordingly, all SPGPE simulations reported here, including the ones corresponding to quantum Newton's cradle with a Bragg pulse (see next Section) were carried out with $\epsilon_{cut}$ corresponding to this optimal cutoff mode occupancy of $0.3$.

\subsection{Bragg pulse quantum Newton's cradle}

Collisional dynamics and thermalization of a quantum Newton's cradle under a Bragg pulse protocol in the finite-temperature quasicondensate regime has recently been studied in Ref.~\cite{Thomas2021}. Here, we provide an additional point of comparison between GHD and SPGPE dynamics through simulation of this system, studying the rate of convergence to thermalization within the two simulation methods. Dynamics are instigated from a thermal state in a harmonic trap of frequency $\omega$, characterized by dimensionless interaction parameter $\gamma_0 = 0.01$ in the trap centre, with temperature $\widetilde{T}\simeq  152$, and total atom number $N\simeq1960$. A Bragg pulse is then applied to initiate Newton's cradle dynamics; this splits the initial atomic wavepacket at rest into two counter-propagating halves corresponding to $\pm 2\hbar k_0$ diffraction orders of Bragg scattering ~\cite{Bragg,Thomas2021}, where $k_0$ is the Bragg momentum in wave-number units.

Considering $^{87}$Rb atoms, with a scattering length of $a\simeq5.3$nm, for an experimentally realistic trapping potential with a longitudinal frequency $\omega / 2 \pi = 3$ Hz, and a dimensionless initial peak density $\rho(0,0) \simeq 1.58 \! \times \! 10^7$m$^{-1}$, this system can be realized with $g\simeq 2 \hbar \omega_\perp a \simeq 1.23\times 10^{-38} \mathrm{J} \! \cdot \! \mathrm{m}$, where $\omega_\perp / 2 \pi \simeq 1.75$ kHz is the frequency of the transverse harmonic trapping potential.

\begin{figure*}[tbp]
    \includegraphics[width=16.7cm]{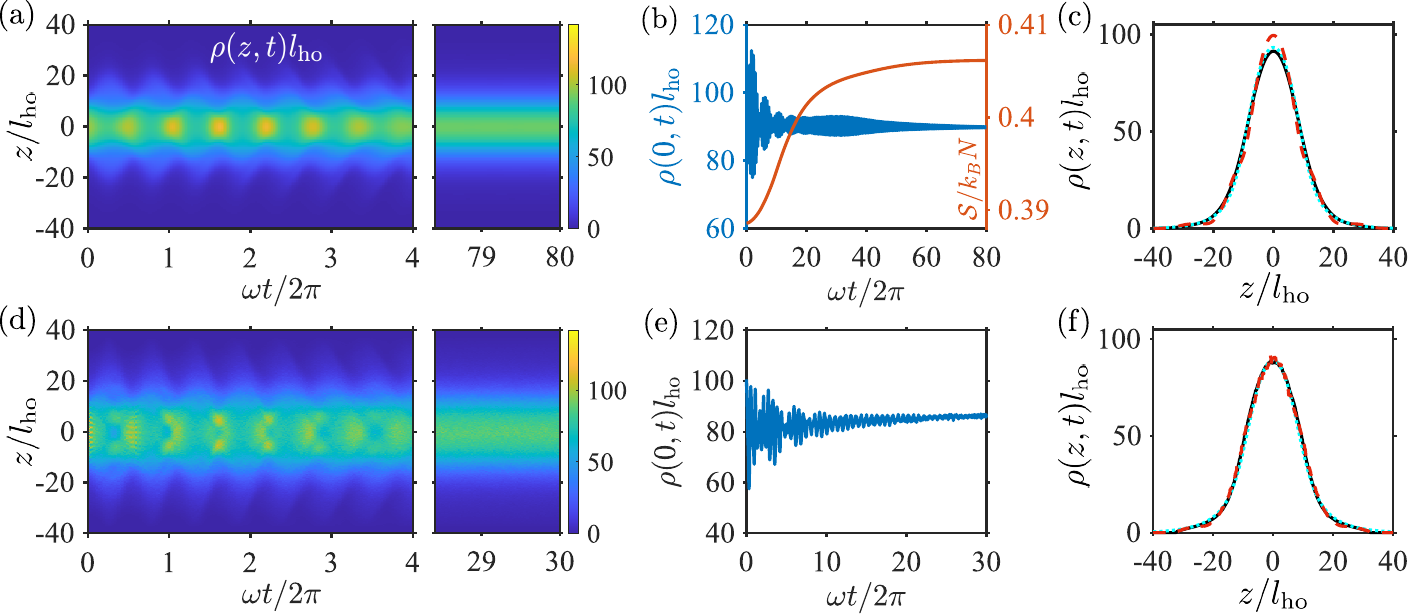}
    \caption{Dynamics of thermalization of a harmonically trapped quasiondensate for GHD and SPGPE evolution in a quantum Newton's cradle setup under a Bragg pulse of momentum $q_0 = k_0 l_\mathrm{ho} = 1$. The corresponding value of Bragg quasimomentum for this system is $\overline{\lambda}_{\mathrm{Bragg}} = 3.19$. Initial thermal states are characterised by a dimensionless interaction parameter at the trap centre of $\gamma_0 = 0.01$, and dimensionless temperature $\widetilde{T}\simeq 152$, with a total atom number of $N\simeq1960$.
    Panel (a) shows Navier-Stokes GHD evolution of the density profile, $\rho(z,t) l_\mathrm{ho}$; (b) demonstrates the evolution of the respective peak density, averaged over the region $z/l_\mathrm{ho} \in [-2,2]$ (blue, fast-oscillating curve), alongside the total entropy per particle of the GHD simulation (red, smooth curve), which both plateau upon reaching the final thermal state; (c) shows the density profile of the relaxed GHD state after dynamics at time $t = 80/(\omega/2\pi)$ (black solid line), as well as the Yang-Yang thermodynamic density profile (cyan dotted line) of temperature $\widetilde{T} \simeq 214$ that fits best to that of the relaxed GHD state. Also shown in (c) is a GHD density profile at an earlier time $t \!\simeq\! 6.56 / (\omega / 2 \pi)$ (red dashed line), chosen to illustrate its deviation from the density profile of the final relaxed state. Density profile evolution under SPGPE simulation is shown in (d); panel (e) demonstrates the evolution of the respective peak density (averaged over the region $z/l_\mathrm{ho} \in [-2,2]$) as it approaches thermalization and plateaus; in (f) we show the relaxed density profile after dynamics at time $t \!=\! 30/(\omega/2\pi)$ (black solid line), as well a best fit Yang-yang thermodynamic density profile of temperature $\widetilde{T} \simeq 261$ (cyan dotted line), and the SPGPE density profile at an earlier time $t\! \simeq\! 6.54 / (\omega / 2 \pi)$ (red dashed line), which is already much closer to the final relaxed density profile compared to that of GHD simulation around the same time.}
\label{fig:q0_1}
\end{figure*}

Implementing the Bragg pulse within the SPGPE method consists of replacing each stochastic realisation of the initial thermal equilibrium state, $\psi(z,t=0^-)$ by a coherent superposition, $\psi(z,t=0^+) = \frac{1}{\sqrt{2}}(e^{i 2 k_0 x} + e^{-i 2 k_0 x}) \psi(z,t=0^-)$, which subsequently evolve in time according to the projected GPE \cite{Bragg,Thomas2021}. Such a superposition of wavefunctions with positive and negative momentum boosts of $\pm2k_0$ is known to be an excellent approximation to experimental implementation of the Bragg pulse \cite{Bragg,Thomas2021,atas_TG_exact} via an external periodic lattice (Bragg) potential formed by two counter-propagating laser beams \cite{kinoshita2006quantum}. The microscopic effect of the realistic Bragg pulse in the SPGPE simulation can be seen as the fast oscillating interference fringes in the spatial density profile at early times, during the periods when the two counter-propagating halves of the density profile overlap spatially.

Simulating a Bragg pulse protocol in GHD, by contrast, is done through the quasiparticle density distribution, $f_p(\lambda;z,t)$, which does not directly depend on the bare atomic momenta, instead relying on the rapidity of quasiparticles, and is generally not equivalent to the momentum distribution, except in free models \cite{Wilson1461, GHD_newtonscradle, malvania2021generalized}. Implementing a Bragg pulse in GHD requires one to work with the quasiparticle density distribution of the initial thermal state, $f^{t=0^-}_p(\lambda;z,t)$, adding positive and negative momenta to the quasiparticles with equal probability. Correspondingly, the post-pulse quasiparticle distribution may be modeled as $f_p^{t=0^+}(\lambda;z,t)\!=\![f^{t=0^-}_p\!(\lambda \!+\! 2 \lambda_\mathrm{Bragg};z,t) \!+\! f^{t =0^-}_p\!(\lambda \!-\! 2\lambda_\mathrm{Bragg};z,t)]/2$ \cite{GHD_newtonscradle}.

\begin{figure*}[tbp]    \includegraphics[width=16.7cm]{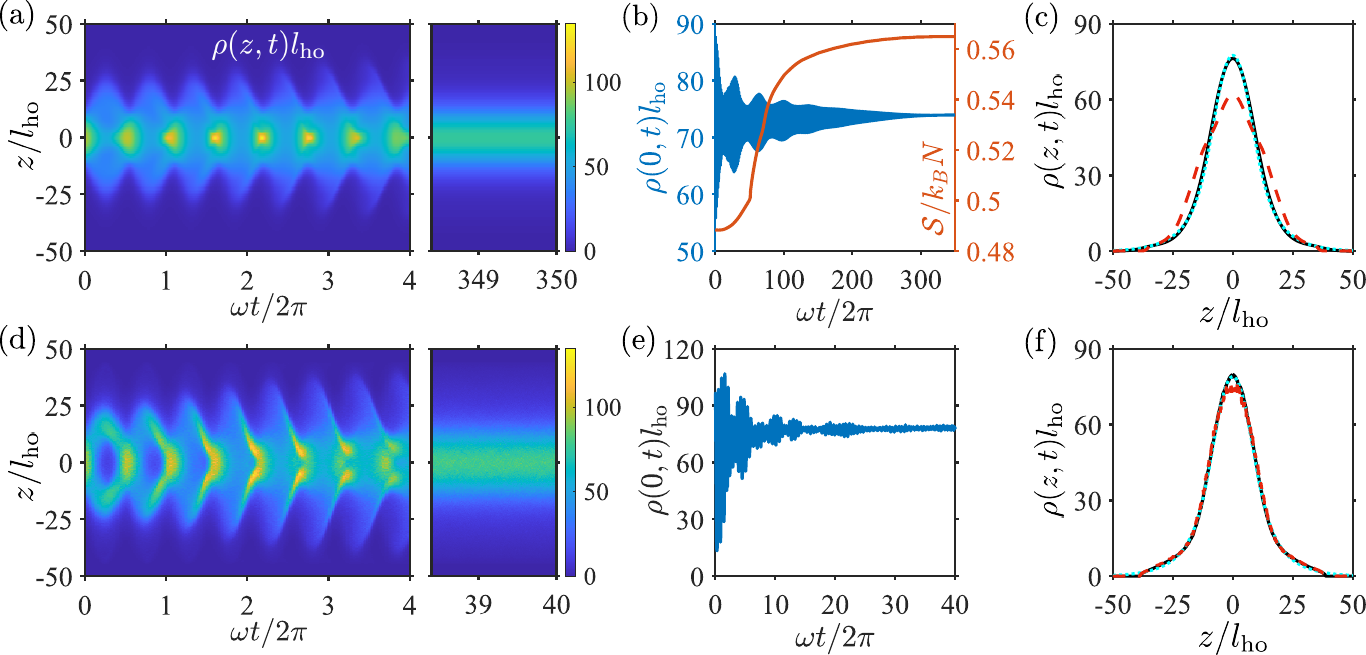}
    \caption{Same as in Fig.~\ref{fig:q0_1}, but for $q_0 = 5$. The corresponding Bragg quasimomentum for this system is $\overline{\lambda}_\mathrm{Bragg} = 5.84$, whereas the total atom number, and the initial values of $\gamma_0$ and $\widetilde{T}$ are the same. The final relaxed density profile from GHD in (c) is at time $t = 350/(\omega/2\pi)$ (black solid line), and is compared with a corresponding Yang-Yang thermodynamic density profile at temperature $\widetilde{T} = 336$ (cyan dotted line), and a respective density profile taken at time $t \simeq 21.5/(\omega/2\pi)$ (red dashed line). The relaxed SPGPE state in (f) is given at a time $t = 40/(\omega/2\pi)$ (black solid line), with a respective Yang-Yang thermodynamic density profile of temperature $\widetilde{T} = 348$ (cyan dotted line), and an additional density profile demonstrated at a time of $t \simeq 22.1/(\omega/2\pi)$ (red dashed line).}
\label{fig:q0_5}
\end{figure*}

Thus, in contrast to the quantum Newton's cradle presented above, which is instigated in exactly the same way---through a sudden quench of the actual external trap potential $V(z)$ from double to single well---in both the SPGPE and GHD simulation, 
the implementation of the Bragg pulse quantum Newton's cradle is achieved in SPGPE and GHD via two different approximations of the post-Bragg pulse state. Additionally, as the GHD is a large-scale theory, its post-Bragg pulse state fails to capture the fast-oscillating interference fringes observed in the SPGPE at early times.

In GHD implementation, one typically chooses the dimensionless quasimomentum of the Bragg pulse, $\overline{\lambda}_\mathrm{Bragg} \!=\! (m/\hbar) \lambda_\mathrm{Bragg} l_\mathrm{ho}$ (given here in  harmonic oscillator units), to be equal to the Bragg momentum, $q_0\equiv k_0 l_{\mathrm{ho}}$, however this is an approximation that assumes a large momentum difference between the clouds \cite{GHD_newtonscradle}. Here, we instead choose the Bragg pulse quasimomentum such that the total energy imparted in the GHD simulation is equal to the energy difference between the initial thermal state and the post-Bragg pulse state calculated in the SPGPE. Using this method, we observe a large difference between the physical Bragg momentum and the corresponding GHD quasimomentum under a low momentum Bragg pulse ($\sim\!\!219 \%$ difference at a Bragg momentum of $q_0 \!=\! 1$), with this difference decreasing for larger momentum Bragg pulses ($\sim\!16.8 \%$ difference at a Bragg momentum of $q_0\! = \!5$).

We first investigate a low momentum Bragg pulse of $q_0 = 1$ and illustrate the dynamics of the density profile for GHD and SPGPE simulations in Fig.~\ref{fig:q0_1}(a) and (d), respectively. We observe a qualitative similarity in the oscillations of collisional dynamics between the two simulation methods, however, there is a clear disagreement in the rate of thermalization. The observed thermalization times of the Navier-Stokes GHD simulation is observed to be $t_f \simeq 80/(\omega/2\pi)$, after approximately 160 oscillation periods, relaxing to a state of temperature $\widetilde{T} \simeq 214$. In comparison, the SPGPE simulation of this $q_0 = 1$ Bragg pulse Newton's cradle system thermalizes at an earlier time of $t_f \simeq 30/(\omega/2\pi)$, after approximately 60 oscillation periods, to a relaxed state at $\widetilde{T} \simeq 261$. The observed difference in thermalization rates between Navier-Stokes GHD and the SPGPE methods of simulating this system likely stems from the approximate method utilized for simulating the Bragg pulse within GHD, described above.

Simulation of the same system, but with a larger Bragg momentum of $q_0 = 5$, is illustrated in Fig.~\ref{fig:q0_5}(a)--(c) for GHD, and in  Fig.~\ref{fig:q0_5}(d)--(f)  for SPGPE, respectively. Here, we observe a greater disparity in the thermalization rates with SPGPE simulations thermalizing at around $t_f \simeq 40/(\omega/2\pi)$ (approximately 80 oscillation periods), and GHD at $t_f \simeq 350/(\omega/2\pi)$ (approximately 700 oscillation periods). This disparity is confirmed when comparing the average peak density between GHD and SPGPE, shown in Fig.~\ref{fig:q0_5}(b) and (e), respectively.

The density profile of the final relaxed state for the GHD simulation is shown in Fig.~\ref{fig:q0_5}\,(c), with its matching thermal density profile of temperature $\widetilde{T} \!\simeq 
\!336$. Likewise, the relaxed density profile for the SPGPE simulation is shown in Fig.~\ref{fig:q0_5}\,(f), with the density profile of its matching Yang-Yang thermodynamic state of temperature $\widetilde{T} \!\simeq \!348$. We thus observe that, for the Bragg pulse quantum Newton's cradle, the discrepancy in thermalization rates is present regardless of the momentum of the simulated Bragg pulse. However, short and intermediate time dynamics, along with the final relaxed states of both simulation methods, are seen to be qualitatively similar.

We emphasize that the quantum Newton's cradle in double-well to single-well quench is implemented in exactly the same way in both the GHD and SPGPE approaches---via a sudden quench of the external trapping potential at time $t=0$. Accordingly, as we discussed previously, the GHD and SPGPE thermalization times agree with each other in this setup. In contrast to this, the original Bragg pulse quantum Newton's cradle scenario of Ref.~\cite{kinoshita2006quantum}, which employs a splitting of the initial quasicondensate wavefunction into two counter-propagating halves, is implemented in GHD in an approximate and qualitatively different way to the SPGPE \cite{Thomas2021,GHD_newtonscradle}. Because of this, we observe significantly different thermalization rates (especially for large Bragg momenta) using GHD and SPGPE simulations, even though the overall dynamics of collisional oscillations are still qualitatively very similar. 

\section{Conclusions}

In conclusion, we have performed benchmarks of generalized hydrodynamics (GHD) in a variety of out-of-equilibrium scenarios in a 1D Bose gas. This was performed against a set of alternative theoretical approaches which are not limited to long-wavelength excitations.
The first part of our analysis focused on systems supporting dispersive quantum shockwaves emanating from a localised density bump. In this highly nonequilibrium scenario, GHD was shown to generally agree with the predictions of the alternative approaches when compared for systems of sufficiently high temperatures and for systems with strong interactions.
For both scenarios, the good agreement stems from a reduced phase coherence length of the gas. In particular, for high temperatures the effect of thermal fluctuations leads to a smearing of interference contrast due to the reduced thermal phase coherence length of the system, $l_T\!=\!\hbar^2 \rho_\mathrm{bg} / m k_B T$. This in turn leads to a suppression of interference phenomena and therefore an absence of high-contrast short-wavelength interference fringes in the density, as clearly demonstrated in the $c$-field SPGPE results. For strong interactions, the coherence length is given by the mean particle separation $1/\rho_\mathrm{bg}$, hence we observe an absence of interference fringes in the evolution density for a bump with initial width larger than $1/\rho_\mathrm{bg}$.

At low temperatures and weak interactions, where interference phenomena are more pronounced, the predictions of GHD only agree with a coarse-grained convolution averaging approximation. The effect of such averaging is similar to having finite imaging resolution in quantum gas experiments, and explains why GHD may perform well when compared to experiments, whilst departing from the predictions of theoretical approaches that are valid at short wavelengths.

At zero temperature, and for low atom numbers, we utilized the infinite matrix product state (iMPS) method, which is capable of simulating dynamics of the 1D Bose gas across the entire range of interaction strengths. Through this, we demonstrate how this shockwave scenario depends on the interaction strength, and how GHD breaks down in the limiting case of weak interactions. In detail, for strong interactions, GHD again accurately models the evolution of the quantum shock, but with the additional presence of Friedel oscillations. As the interaction strength is decreased towards the regime of a near-ideal Bose gas, we observe the emergence of large scale phase coherence in the iMPS data. Here, GHD fails to agree with the iMPS results, as the local density approximation, which is an intrinsic assumption to GHD, is no longer applicable.

We further analysed the scenario of evolution from a density dip. Here, rather than producing a dispersive shockwave, the dynamics instead result in a train of grey solitons. Again, we find that GHD tends to agree with a coarse-grained version of the more microscopic theories. However, due to the larger separation between the localised density dips of the soliton train, this coarse graining requires much more careful application at long times, eventually breaking down when the solitons fully separate due to their density-dependent velocity.

Following the discovery of GHD, a number of extensions to the framework have been developed. In particular, Navier-Stokes diffusive GHD, first postulated in Ref.~\cite{DeNardis_Diffusion_2018}, operates at a higher order in the position gradient expansion, giving more reliable and accurate results, in particular for long time evolutions. Further, this model of GHD is capable of simulating thermalization at long times, a feature which we here utilize to explore and benchmark the thermalization of a quantum Newton's cradle for a harmonically trapped 1D Bose gas in the weakly interacting quasicondensate regime.

In particular, we focus on two protocols for the quantum Newton's cradle. The first being a quench from a double-well to a harmonic trap, of the type explored experimentally in Ref.~\cite{GHD_onatomchip}. Benchmarking is performed against the SPGPE method, which was recently utilized to model thermalization of such a system under a Bragg pulse protocol in Ref.~\cite{Thomas2021}. We observe excellent agreement between these two models, with GHD relaxing to a similar final thermal state over the same timescale. This was evaluated through inspection of the peak density, and the Bhattacharyya distance from the final relaxed state, which both demonstrate an overall agreement in this timescale.

The second protocol utilized a Bragg pulse to separate the atom cloud, which is initially in a thermal equilibrium state in a harmonic trap, into two equal counter-propagating packets. Results were obtained for two sets of Bragg pulse momenta, however we observed an overall lack of agreement in the timescale of relaxation, with SPGPE taking a much greater time to relax to equilibrium after the initial pulse. We believe that this descrepancy may be attributed to the method by which the Bragg pulse was applied within GHD, which was modelled after the method used in Ref.~\cite{GHD_newtonscradle}, and may not be applicable to clouds with significant overall in their post-pulse state, as was the case here.

Overall, we have performed a thorough benchmarking of the GHD method for a 1D Bose gas in a handful of highly nonequilibrium scenarios. Further, we have analyzed where GHD breaks down, and gained an understanding as to why this occurs. GHD is fast becoming a useful tool for simulating ultracold atomic gases, meaning that understanding its regimes of validity is vital to future application.

\newpage

The work presented in Chapter \ref{Chap:5} was adapted from the submitted publication of Ref.~\cite{watson2023benchmarks}, and the contribution of each named author to that work is presented below in Table.~\ref{Tab:Chap5}.

\noindent
\cite{watson2023benchmarks} \textbf{R. S. Watson}, and K. V. Kheruntsyan, \href{https://journals.aps.org/prresearch/abstract/10.1103/PhysRevResearch.5.L022024}{Benchmarks of generalized hydrodynamics for one-dimensional Bose gases}, \textit{Phys. Rev. Research}, \textbf{5}, L022024 (2023)

\begin{table}[h]
	\begin{center}
	\begin{tabular}{|c|l|l|}
		\hline
		Contributor & Statement of contribution & \% \\
		\hline
		\textbf{R. S. Watson}				& writing of text 					& 80\\
															& proof-reading							& 40 \\
															& numerical calculations 		& 80\\
															& preparation of figures 		& 80 \\
               															& initial concept						& 10 \\

		\hline
		S. A. Simmons													& proof-reading							& 20 \\
															& numerical calculations 		& 20\\
		\hline
		K. V. Kheruntsyan								& writing of text 					& 20\\
															& proof-reading							& 40 \\
															& preparation of figures 		& 20 \\
															& initial concept						& 90 \\
		\hline
	\end{tabular}
	\end{center}
 \caption{}\label{Tab:Chap5}
\end{table}

%% file: Chapter6/Chapter6.tex

\chapter[A nonequilibrium quantum Otto cycle fuelled by chemical work	]{A nonequilibrium quantum Otto cycle fuelled by chemical work	}
\label{Chap:6}	
\pagestyle{headings}


\textit{The study of nonequilibrium thermodynamics in many-body interacting systems is often restricted by the complexity of simulating the real-time dynamics of such systems, starting from a finite-temperature thermal equilibrium state.
Integrable systems are one of the few exceptions to this, with the recently developed theory of generalized hydrodynamics (GHD) capable of capturing the  finite-temperature dynamics of integrable and nearly intergrable systems on large scale, and in parameter regimes not accessible by other microscopic theories.
Here, we demonstrate the utility GHD to the field of nonequilibrium quantum thermodynamics through its application in an Otto engine cycle for an experimentally realistic harmonically trapped one-dimensional (1D) Bose gas.
In particular, we analyse  
a thermochemical Otto engine cycle where the unitary work strokes are driven by a quench of interatomic interactions, while the equilibration strokes with hot and cold reservoirs are facilitated via diffusive contact that allows for particle exchange.
We numerically evaluate the performance of this engine cycle in both the weakly interacting quasicondensate regime and the strongly interacting near-Tonks-Girardeau regime of the 1D Bose gas. We further make connections with two previous studies: one in the limit of a sudden quench where approximate analytic results are known at finite temperatures, 
and the other in the limit of an adiabatic engine cycle at zero temperature, which in turn provides an upper bound to the net work and efficiency of this engine cycle at nonzero temperatures.}


\section{Introduction}
\label{Sec:label}	



Interacting quantum many-body systems exhibit a plethora of phenomena, such as quantum coherence or entanglement, that single-particle systems are, by definition, incapable of exhibiting. In the early 2000's, such resources were demonstrated to be capable of boosting the performance of engines realized on the quantum scale, laying the foundation for the field that would later become quantum thermodyanmics. 
Over two decades later, experimental realization of many-body quantum heat engines has been achieved, with recent demonstration of an engine operating at high efficiency and fuelled by control over the quantum statistics of the working fluid, making it notably distinct from a conventional thermal engine cycle \cite{koch2022making}.  

Quantum many-body systems are generally challenging to simulate even approximately. However, there is a special class of integrable systems where exact thermal equilibrium results are obtainable. Further, thanks to the recent development of generalized hydrodynamics (GHD), finite time simulation of integrable and integrability-broken systems are now accessible. This is a unique opportunity for quantum thermodynamics, allowing for finite time simulation of a wide range of quantum many-body interacting systems, which remain at the forefront of progress in the field.

In this Chapter, we analyse an interaction-driven quantum Otto cycle in a harmonically trapped 1D Bose gas, where thermal contact between the reservoirs and the working fluid is replaced by purely diffusive contact, facilitated by particle flow between the reservoirs and working fluid. Such a thermochemical Otto engine is experimentally realizable and first introduced as a `feshbach engine' in Ref.~\cite{keller2020feshbach}, for a zero-temperature weakly interacting gas.
Here, we extend the analysis of this system to finite temperatures and arbitrary interaction strengths, possible through use of the thermodynamics Bethe ansatz (TBA), first developed for this system by Yang and Yang in 1969 \cite{yang1969thermodynamics}. Further, we present finite-time analysis of the cycle using GHD, connecting to the sudden-quench Otto engine introduced in Chapter \ref{Chap:3}.

\section{Interaction-driven Otto cycle}

\begin{figure}[!tbp]\begin{center}\includegraphics[width=10cm]{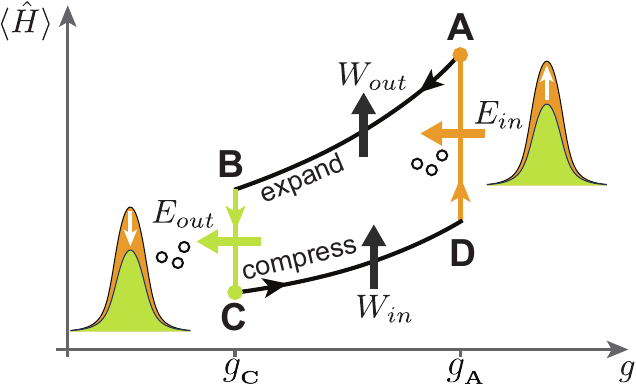}   
   \caption{An interaction-driven thermochemical Otto engine cycle for a harmonically trapped 1D Bose gas operating between two interaction strengths, $g_\textbf{C}$ and $g_\textbf{A}$, at fixed temperature $T$. Unitary work strokes, \textbf{AB} and \textbf{CD}, correspond to mechanical work $W_{out}$ and $W_{in}$, respectively. Non-unitary equilibration strokes, \textbf{BC} and \textbf{DA}, correspond to chemical work via the change in energies denoted $E_{out}$ and $E_{in}$, respectively. We demonstrate this chemical work, accomplished via diffusive contact with two external reservoirs, via the cartoons of the system density profiles that either grow or decrease, depending on whether particles flow in or out of the working fluid.}
   \label{fig:Otto_cycle} 
\end{center}
\end{figure}

We define the interaction-driven thermochemical Otto cycle (see Fig.~\ref{fig:Otto_cycle}) as an engine cycle, operating between two baths both at a fixed temperature $T$,
with equilibration strokes ($\mathbf{DA}$ and $\mathbf{BC}$ in Fig.~\ref{fig:Otto_cycle}) corresponding to \textit{diffusive} rather than \textit{thermal} contact with external reservoirs. This involves the transfer of chemical work, rather than heat, between the working fluid and the external reservoirs, which may be accomplished via control over their respective chemical potentials (see Chapter \ref{Chap:7} for further details).
The working fluid under consideration here is a harmonically trapped 1D Bose gas, with work strokes ($\mathbf{AB}$ and $\mathbf{CD}$ in Fig.~\ref{fig:Otto_cycle}) consisting of unitary quenches of the interaction strength, which may be achieved experimentally via control over either the transverse trapping potential, or via Feshbach resonance as discussed in Chapter \ref{Chap:3}. The duration of time, $\Delta t$, over which the interaction quench is performed may itself be varied, with two important limits being that of a sudden quench ($\Delta t \! \to \! 0$) and adiabatic (or quasistatic) cycle ($\Delta t \! \to \! \infty$). 


In detail, this Otto engine cycle consists of four strokes: (1) \textit{Unitary expansion, \textbf{A}$\to$\textbf{B}}: the working fluid, initially at equilibrium and consisting of $N_\textbf{A}$ total atoms at interaction strength $g_\textbf{A}$, is decoupled from all reservoirs and has its interaction strength quenched to $g_\textbf{C}\!<\!g_\textbf{A}$ over time $\Delta t$, generating beneficial 
work out $W_\mathrm{out}\!=\!\langle \hat{H}\rangle_{\textbf{B}} \!-\!\langle \hat{H}\rangle_{\textbf{A}}\!<\!0$, where $\hat{H}$ is the system Hamiltonian, and $\langle \hat{H}\rangle_{\textbf{i}}$ is its expectation value for the total energy of the system in state $\textbf{i}=\{\textbf{A,B,C,D}\}$. (2) \textit{Equilibration and particle expulsion}, \textbf{B}$\to$\textbf{C}: the working fluid is connected to the external reservoir and allowed to equilibrate at constant interaction strength $g_\textbf{C}$, transferring $\Delta N$ total particles to the reservoir and thus losing energy $E_\mathrm{out}\!=\!\langle \hat{H}\rangle_{\textbf{C}} \!-\!\langle \hat{H}\rangle_{\textbf{B}}\!<\!0$. (3) \textit{Unitary compression}, \textbf{C}$\to$\textbf{D}: again disconnected from all reservoirs, the working fluid has its interaction strength quenched from $g_\textbf{C}\!\to\!g_\textbf{A}$, again over time $\Delta t$, with work $W_\mathrm{in}\!=\!\langle \hat{H}\rangle_{\textbf{D}} \!-\!\langle \hat{H}\rangle_{\textbf{C}}\!>\!0$ done on the fluid. (4) \textit{Equilibration and particle intake}, \textbf{D}$\to$\textbf{A}: the working fluid is connected to a second external resesrvoir, where it is left to equilibrate, taking in $\Delta N$ total particles from the reservoir, and thus intaking energy $E_\mathrm{in}\!=\!\langle \hat{H}\rangle_{\textbf{A}} \!-\!\langle \hat{H}\rangle_{\textbf{D}}\!>\!0$, and returning to the initial equilibrium state of the overall cycle.

Such an engine cycle generates beneficial net work $W \!=\! W_\mathrm{out} \!+\! W_\mathrm{in}\!<\!0$, if $\langle \hat{H} \rangle_{\textbf{A}} \!-\!\langle \hat{H} \rangle_{\textbf{D}} \!>\! \langle \hat{H} \rangle_{\textbf{B}} \!-\! \langle \hat{H} \rangle_{\textbf{C}}$, with efficiency $\eta \!=\! -W/E_\mathrm{in} \!=\! 1 \!-\! E_\mathrm{out}/E_\mathrm{in}$ \cite{SchroederD_ThermalPhysics,CallenHerbertB1985Taai}. 
We note that the definition of efficiency here is taken to be the ratio of the beneficial net work done by the system, $-W$, to the energy intake in the form of chemical work, $E_{in}$. This is notably distinct from the conventional Otto cycle, and as a result is not restricted to the same inequalities as conventional heat engines.

\subsection{Work stroke duration}\label{sec:WorkStrokeDuration}
Any realization of a quantum thermal machine must operate in finite time, meaning the duration of the work strokes, which is generally externally controllable, becomes a key factor in determining engine performance. Both net work and efficiency depend non-trivially on the work stroke duration, and achieve their theoretical maximum value in the limit of an isentropic work stroke. In this limit, the work strokes are performed quasi-statically, meaning the power output vanishes as $P \!=\! -W/t_\mathrm{cycle}\!\to\! 0$, where the total cycle time, $t_\mathrm{cycle}$, is the sum of the work stroke durations, and the time taken for equilibration, which is typically not externally controllable (for further details on this, see Chapter \ref{Chap:7}).


The sudden quench interaction-driven Otto cycle, where the `sudden' nature of the quench is in reference to longitudinal time-scales, was explored for an interaction-driven engine in Chapter \ref{Chap:3}, where expressions were found for the performance in terms of equilibrium observables. In particular, the net work
\begin{equation}\label{eq:Work}
    W^\mathrm{S} = -\frac{1}{2}(g_\mathbf{A} - g_\mathbf{C})\left( \overline{G^{(2)}_\mathbf{A}} - \overline{G^{(2)}_\mathbf{C}}\right),
\end{equation}
where $\overline{G^{(2)}_{\textbf{A}(\textbf{C})}} \! = \! \int dz G^{(2)}(z,z) \! = \! \int dz g^{(2)}(z,z) \rho(z)^2$ is the total integrated correlation function of the thermal equilibrium state $\mathbf{A}(\mathbf{C})$ written in terms of the local atomic density, $\rho(z)$, and Glauber's second-order correlation function,
\begin{equation}\label{eq:g2}
    g^{(2)}(z,z') = \frac{\langle \hat{\Psi}^\dagger(z)\hat{\Psi}^\dagger(z')\hat{\Psi}(z')\hat{\Psi}(z) \rangle}{\rho(z)\rho(z')},
\end{equation}
here taken at the same point $z\!=\!z'$, i.e. $g^{(2)}(z,z)$. The condition $\overline{G^{(2)}_\mathbf{A}} > \overline{G^{(2)}_\mathbf{C}}$ is thus required to ensure that the sudden quench Otto engine cycle provides beneficial net work.

It was shown also that the efficiency of the sudden-quench quantum Otto cycle may be evaluated via
\begin{equation}\label{eq:Efficiency}
    \eta^\mathrm{S} = 1 - \frac{ \langle\hat{H} \rangle_\mathbf{A} - \langle \hat{H} \rangle_\mathbf{C} - \frac{1}{2}\left(g_\mathbf{A} - g_\mathbf{C} \right) \overline{G^{(2)}_\mathbf{A}}}{ \langle\hat{H} \rangle_\mathbf{A} - \langle \hat{H} \rangle_\mathbf{C}   - \frac{1}{2}\left(g_\mathbf{A} - g_\mathbf{C} \right)  \overline{G^{(2)}_\mathbf{C}}},
\end{equation}
where $ \langle\hat{H} \rangle_{\mathbf{A}(\mathbf{C})}$ is the total energy of the equilibrium state $\mathbf{A}(\mathbf{C})$.
These formulas were utilized to evaluate the performance of a uniform 1D Bose gas in Chapter \ref{Chap:3}, where the net work was shown to directly depend on Glauber's local second-order correlation function, $g^{(2)}(0)$.
In this chapter, we make use of these formulas in combination with the numerically exact thermodynamic Bethe ansatz (TBA) \cite{yang1969thermodynamics}, combined with a local density approximation \cite{kheruntsyan2005finite}, in order to evaluate the performance of the sudden quench and isentropic cycles in the context of a harmonically trapped 1D Bose gas.


Operation of this quantum Otto cycle in finite time, i.e. outside the sudden and isentropic limits, typically requires various simulation methods, each capable of operating only over a limited region of the rich parameter space which the 1D Bose gas possesses.
Here, we instead utilize the recently developed generalized hydrodynamic (GHD) framework, which is uniquely capable of simulating the large-scale dynamics of integrable systems across the full breadth of the parameter regimes and their crossover regions. Here, the work strokes are linear quenches of the interaction strength over a time $t_q$.


\section{Performance of the thermochemical Otto engine}

In this Section, we analyze the performance of the quantum thermochemical engine, in terms of the net work and efficiency, with a working fluid consisting of a harmonically trapped 1D Bose gas in the experimentally relevant quasicondensate (weakly interacting) and near-Tonks-Girardeau (strongly interacting) regimes. We focus on the system's dependence on the temperature, which controls the parameter regime which the working fluid inhabits.


\subsection{Weakly interacting working fluid}\label{chap:6_sec:weak}

Despite not possessing any phase transitions, the 1D Bose gas has a rich parameter space described by smooth crossovers between several distinct regimes. These parameter regimes may be identified via Glauber's local second-order correlation function \cite{kheruntsyan2003pair}. In particular, for a harmonically trapped 1D Bose gas, these may be defined by the local second-order correlation function at the trap centre, i.e. $g^{(2)}(0,0)$, as was first identified in Ref.~\cite{kheruntsyan2005finite}. Each parameter regime may thus be characterized via the dimensionless interaction strength, or Lieb parameter, at the trap centre, given by $\gamma_0 \!=\! m g / \hbar^2 \rho(0)$; and, at finite temperature, by an additional dimensionless temperature parameter, $\tau_0 = 2 m k_B T / \hbar^2 \rho(0)^2$.

Weakly interacting systems, $\gamma_0 \!\ll\!1$, at low temperatures, $\tau_0 \!\ll\!1$, possess three distinct regimes as a function of temperature: for $\tau_0 \! \ll \! \sqrt{\gamma_0}$, the system is described by a quasicondensate \cite{Mora-Castin-2003}, where the local correlation function, $g^{(2)}(0,0)\!\simeq\! 1$, reflects the system's coherent nature. This may be further broken down into region I, which is dominated by quantum fluctuations, $\gamma_0 \! \ll \! \tau_0$, and region II, which is where thermal fluctuations are more prominent, $\gamma_0 \!\ll\! \tau_0 \! \ll \! \sqrt{\gamma_0}$. At higher temperatures, $\sqrt{\gamma_0} \! \ll \! \tau_0 \! \ll \! 1$, the system inhabits region III, and is described by a decoherent quantum gas, with correlations approaching that of a classical gas, i.e. $g^{(2)}(0)\!\simeq\!2$.


Here, we explore the operation of the interaction-driven thermochemical Otto cycle, described in the Section above, using a harmonically trapped weakly interacting 1D Bose gas as a working fluid. The interparticle interaction strength at point $\mathbf{C}$ (see Fig.~\ref{fig:Otto_cycle}) is kept constant at $\overline{g}_\mathbf{C}\!=\!g_\mathbf{C}/l_\mathrm{ho} \hbar \omega \!=\!0.4$, written in natural units of the longitudinal harmonic oscillator of frequency $\omega$, with $l_\mathrm{ho}\!=\!\sqrt{\hbar/m \omega}$ the harmonic oscillator length. We demonstrate engine performance, for various values of the temperature $\overline{T}\!=\!k_B T/\hbar \omega$, as a function of the ratio of the interaction strengths at points $\mathbf{A}$ and $\mathbf{C}$. Further, the working fluid consists of $N_\mathbf{C}\!=\!2000$ total particles at point $\mathbf{C}$, with $\Delta N\!=\!300$ particles exchanged between the working fluid and reservoirs when in contact.
The bath temperatures are chosen specifically such that the gas inhabits each of the regimes introduced above, as well as their smooth crossover regions.

\subsubsection{Sudden quench}

\begin{figure}[!tbp] \begin{center}\includegraphics[width=16cm]{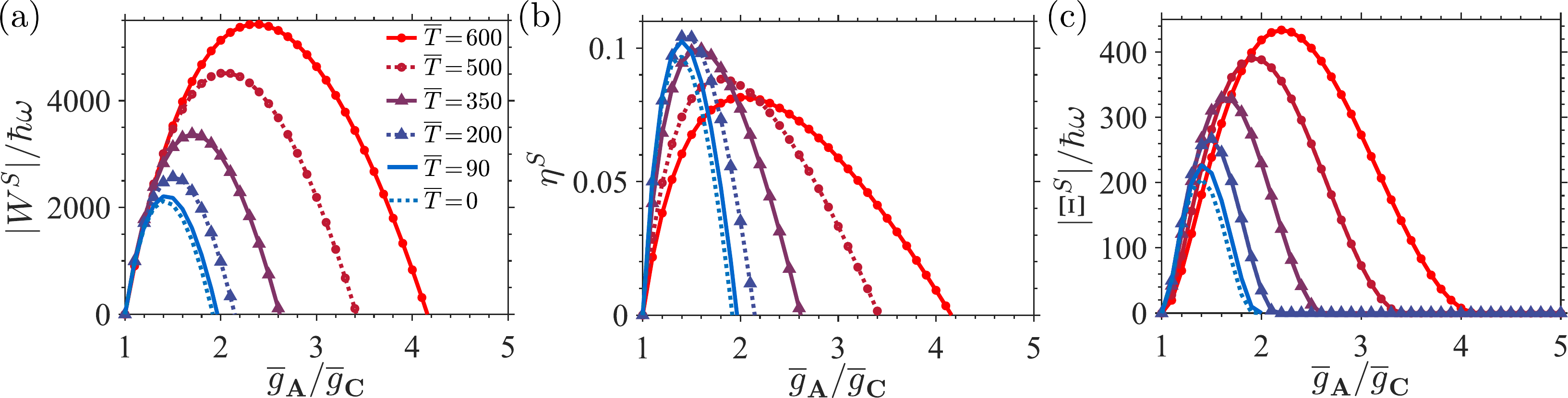}   
   \caption{Performance of the interaction-driven thermochemical Otto cycle with a weakly interacting harmonically trapped 1D Bose gas under a sudden quench protocol. The bath temperature, $\overline{T}$, of the working fluid is varied such that the system inhabits regions II ($\overline{T}\!\leq \!200$), and III ($\overline{T}\!\geq \!500$), as well as their crossover ($\overline{T}\!=\!350$). The system at point $\mathbf{C}$ (see Fig.~\ref{fig:Otto_cycle}) consists of $N_\mathbf{C}\!=\!2000$, with an interaction strength $\overline{g}_\mathbf{C}\!=\!0.4$, in natural units of the harmonic oscillator. The diffusive strokes of this engine cycle consist of transferring $\Delta N \!=\! 300$ particles between the working fluid and the external reservoirs.
   Panels (a) and (b) demonstrate the net work and efficiency, respectively, whereas panel (c) shows the efficient work criterion, $\Xi^S\!=\!W^S \times \eta^S$, introduced in the text.}
   \label{fig:GP_SQ} 
\end{center}
\end{figure}

Performance, in terms of net work, $W^\mathrm{S}$, and efficiency, $\eta^\mathrm{S}$, of the sudden quench quantum Otto cycle in the weakly interaction regime is demonstrated in Figs.~\ref{fig:GP_SQ}(a) and (b), respectively. Operation as an engine, which corresponds to the ability to extract beneficial net work from operation of the Otto cycle, occurs only over a limited range of the ratio of interaction strengths, $\overline{g}_\mathbf{A}/\overline{g}_\mathbf{C}$. This range of operation, for the weakly interacting gas presented here, generally increases for an increase in temperature, which additionally results in a greater maximum value of net work. 

To explain this improvement of the engine cycle in terms of net work for increasing temperatures, we utilize an approximation for the total correlation function: $\overline{G^{(2)}} \! \simeq \! g^{(2)}(0,0) \int dz \rho(z)^2$, first introduced and validated against exact TBA numerics in Ref.~\cite{kheruntsyan2005finite}. This approximation relies on the slow variation of the correlation function, $g^{(2)}(z,z)$, along the longitudinal length of the trap, meaning it is well approximated by the value at the trap centre, i.e. $g^{(2)}(0,0)$. We thus approximate the sudden quench net work, given in Eq.~\eqref{eq:Work}, as
\begin{equation}\label{eq:Work_SQ_Approx}
\begin{split}
    W^\mathrm{S} \simeq -b \frac{g_\mathbf{A} - g_\mathbf{C}}{2} \bigg(g_\mathbf{A}^{(2)}(0,0) \rho_\mathbf{A}(0) N_\mathbf{A}   -g_\mathbf{C}^{(2)}(0,0) \rho_\mathbf{C}(0) N_\mathbf{C} \bigg),
\end{split}
\end{equation}
where $b\!=\!\int dz \rho(z)^2 / N \rho(0)$ is a slowly varying dimensionless parameter of order $1$, dependent only on the density profile. 
Importantly, the formula derived here for the net work of the sudden quench engine cycle is applicable across the entire parameter space of the 1D Bose gas. 

At low temperatures, e.g. $\overline{T}\!=\!90$ in Fig.~\ref{fig:GP_SQ}(a), the working fluid is well approximated by a weakly interacting system in its ground state, where the local correlation function of the ground state working fluid may be well approximated by $g^{(2)}(0,0)\!\simeq\!1$. The ground state density profile may be approximated by the Thomas-Fermi parabola, with peak density $\overline{\rho}(0) \!=\! \rho(0) l_\mathrm{ho}\!=\!(9 N^2/32 \overline{g})^{1/3}$, and constant $b\!=\!4/5$. 
The net work of the ground state system, evaluated via Eq.~\eqref{eq:Work_SQ_Approx}, becomes
\begin{equation}\label{eq:Work_GS_GP}
    W^\mathrm{S}_{T\!=\!0} = \frac{2 (g_\mathbf{A} - g_\mathbf{C})}{5} \left( \rho_\mathbf{A}(0) N_\mathbf{A}  \\ -\rho_\mathbf{C}(0) N_\mathbf{C} \right)
\end{equation}
and is demonstrated in Fig.~\ref{fig:GP_SQ}(a).
Importantly, this formula clearly exhibits its dependence on the peak density, which is inversely proportional to interaction strength as $\overline{\rho}(0) \!\propto\!\overline{g}^{-1/3}$. This dependence is the key factor limiting performance at low temperatures, causing a restriction on beneficial net work, i.e. $W\!<\!0$, which may be expressed as $\overline{\rho}_\mathbf{A}(0) N_\mathbf{A}\!> \overline{\rho}_\mathbf{C}(0) N_\mathbf{C}$. For the results presented in Fig.~\ref{fig:GP_SQ}, this results in $\overline{g}_\mathbf{A}/\overline{g}_\mathbf{C}\!<\!(N_\mathbf{A}/N_\mathbf{C})^5 \!\simeq\!2$, agreeing with the range of operation for low temperatures. Though the results shown here are for regime II, this agreement with the ground state results demonstrates that one may expect limited variation upon decreasing the temperature any further.

In contrast to the results presented for low temperatures, under increasing temperature the density profile becomes less dependent on the interaction strength. Indeed, at high enough temperatures, the density becomes well approximated by a classical Boltzmann density profile, with peak density $\overline{\rho}(0)\!=\!N/\sqrt{2 \pi \overline{T}}$ entirely independent of interaction strength, with the correlation function being still, if weakly, monotonically increasing with temperature.
For intermediate temperatures, around the temperature of quantum degeneracy (i.e. $\tau_0\!\sim\!1$), the correlation function remains strongly monotonically increasing with temperature, while the peak density's dependence on temperature is weakened.
The net work, which again is approximated by Eq.~\eqref{eq:Work_SQ_Approx}, thus remains positive over a significantly broader range as the poor scaling of the density profile is effectively relieved. This results in a greater maximum net work, as the net work, which grows with $\overline{g}_\mathbf{A}\!-\!\overline{g}_\mathbf{C}$, is allowed to increase linearly over a broader range.


The efficiency of the sudden quench cycle, shown in Fig.~\ref{fig:GP_SQ}(b), generally decreases as a function of temperature. This phenomenon occurs over the entire parameter space for both sudden and isentropic quenches, and stems from the fact that, for larger temperatures, the particle intake stroke ($\mathbf{DA}$ in Fig.~\ref{fig:Otto_cycle}) is adding $\Delta N$ particles at temperature $\overline{T}$. The energy transfer, $E_\mathrm{in}$, is therefore greater for higher temperatures, reducing the efficiency which is inversely proportional to this factor.
We examine the efficiency of the ground state engine cycle by utilizing the Thomas-Fermi energy, $\langle \hat{H} \rangle/\hbar \omega \!=\!3 N \overline{g} \overline{\rho}(0) / 5$. The efficiency, calculated through application of Eq.~\eqref{eq:Efficiency}, is presented in Fig.~\ref{fig:GP_SQ}(b), and again broadly agrees with that of the low temperature system.
Notably, the results derived here for the sudden quench Otto cycle in the ground state represent the extension of those first derived in Ref.~\cite{keller2020feshbach} to the operation of a sudden quench.

In order to further evaluate operation and performance of the sudden quench engine cycle, we introduce the concept of \textit{efficient work}, and define it as,
\begin{equation}\label{eq:efficient_work}
    \frac{\Xi}{\hbar \omega} = \frac{W}{\hbar \omega} \times \eta.
\end{equation}
This is a variation of the `efficient power' quantity, first introduced in Ref.~\cite{yilmaz2006new} in order to provide a new criterion for optimal operation of heat engines. In particular, one can consider this function as a means to examine the trade-off between work and efficiency across the entire parameter space of both interaction ratio and temperature ratio \cite{myers2020bosons}, where efficiency at maximum work commonly examined for quantum heat engines simplifies the analysis to focus on the interaction strength ratio which gives maximum work, with temperature ratio being the independent variable. This is an important criterion for the engine cycle under investigation as, for the sudden quench cycle in particular, there is a regime-dependent trade-off between net work and efficiency that suggests the role of both interaction strength and temperature ratios remain essential to the analysis.
Here, as we do not model the full operation cycle, and thus have no exact value for the duration of the equilibration strokes $\mathbf{DA}$ and $\mathbf{BC}$ in Fig.~\ref{fig:Otto_cycle}, we rely only on the net work, rather than the power, of this engine cycle.
Results for the weakly interacting sudden quench engine cycle are demonstrated in Fig.~\ref{fig:GP_SQ}(c), demonstrating an overall improvement of performance under increasing temperature. This is directly related to the large increase in net work with higher temperatures, and is again directly related to the behaviour of the peak density in the weakly interacting regime.

\subsubsection{Isentropic quench}

The isentropic thermochemical Otto cycle with a working fluid consisting of a weakly interacting harmonically trapped 1D Bose gas was first investigated in Ref.~\cite{keller2020feshbach}. There, the Thomas-Fermi approximation was utilized in order to provide a simple expression for efficiency,
\begin{equation}\label{eq:T0_GP_Efficiency}
    \eta^\mathrm{I}_{T=0} = 1 - \left(\frac{g_\mathbf{C}}{g_\mathbf{A}}\right)^{2/3},
\end{equation}
and net work,
\begin{equation}
    W^\mathrm{I}_{T=0} = - \frac{3}{5}\left( \frac{9}{32}\right)^{1/3} \left(\overline{g}_\mathbf{A}^{2/3} - \overline{g}_\mathbf{C}^{2/3}\right)\left(N_\mathbf{A}^{5/3} - N_\mathbf{C}^{5/3}\right).
\end{equation}
Here, we extend these results to finite temperatures via application of the TBA, as described in Sec.~\ref{sec:WorkStrokeDuration}. 

\begin{figure}[!tbp] \begin{center}\includegraphics[width=12cm]{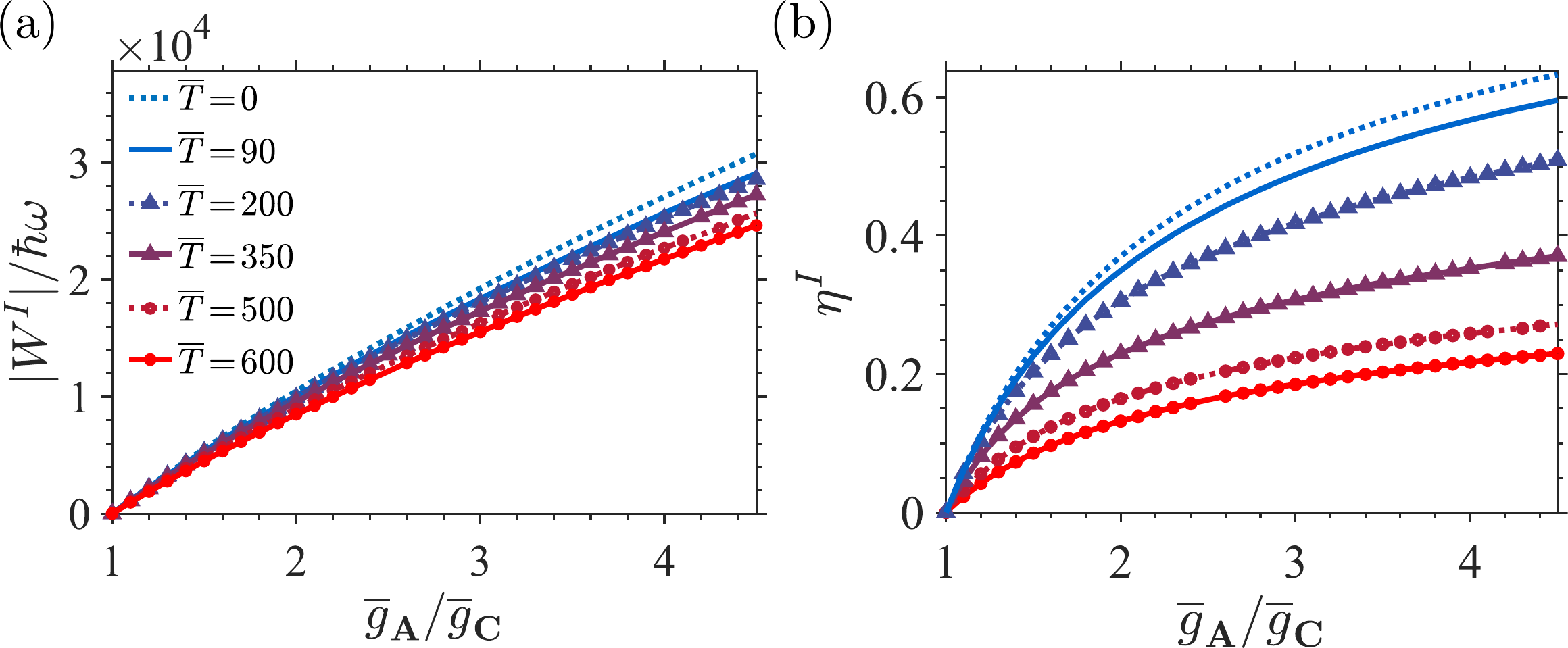}   
   \caption{Performance of the interaction-driven thermochemical Otto cycle with a weakly interacting harmonically trapped 1D Bose gas under an isentropic quench protocol. Net work, $W$, and efficiency, $\eta$, are shown in panels (a) and (b), respectively. System parameters are chosen to be the same as those in Fig.~\ref{fig:GP_SQ}.}
   \label{fig:GP_Isen} 
\end{center}
\end{figure}

The net work and efficiency of the finite temperature isentropic thermochemical engine cycle are shown in Fig.~\ref{fig:GP_Isen}, for the same set of parameters utilized for the sudden quench results. It is clear that, unlike the case of a sudden quench, the overall performance of the isentropic cycle degrades under increasing temperature.
Further, the net work and efficiency of the ground state ($\overline{T}\!=\!0$) engine cycle are shown in Fig.~\ref{fig:GP_Isen}(a) and (b), where they provide an effective upper bound to the finite temperature results of both efficiency and net work. 

This may be explained through an examination of the interaction energy, as the interaction-driven isentropic engine cycle is extracting work from the interparticle interactions. Utilizing the approximation to the integrated correlation function, $\overline{G^{(2)}}$, detailed above, we may express the total interaction energy, as
\begin{equation}\label{eq:H_int}
    \langle \hat{H}^\mathrm{int} \rangle = \frac{g}{2} \overline{G^{(2)}} \simeq  \frac{g}{2} \rho(0) N g^{(2)}(0,0).
\end{equation}
Thus, although the local correlation function increases slowly, from $g^{(2)}(0,0)\!\simeq\!1$ to $g^{(2)}(0,0)\!\simeq\!2$ as a function of temperature, the dependence of the interaction energy on $\rho(0)$, which decreases as a function of temperature due to broadening of the density profile, results in an overall decrease in the total interaction energy. In addition to reducing the net work, this further results in a significant decrease in efficiency under increasing temperature, due to the increased energy intake described above in reference to the sudden quench. Here, unlike the case of the sudden quench, we do not demonstrate the efficient work criterion, $\Xi$, as it is already unambiguous that low temperatures achieve the highest performance in both net work and efficiency.

\subsubsection{Nonequilibrium dynamics}

\begin{figure}[!tbp] \begin{center}\includegraphics[width=14cm]{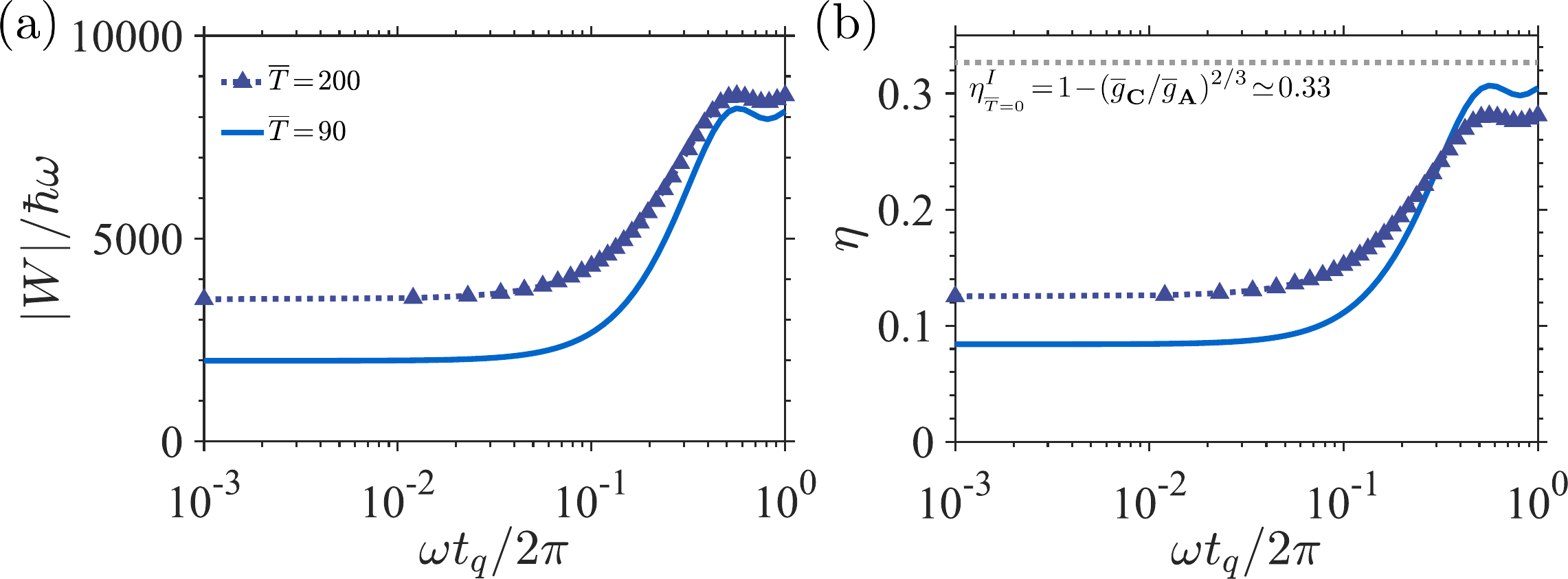}   
   \caption{Finite time engine performance for the weakly interacting thermochemical Otto cycle. Net work and efficiency are demonstrated as a function of work stroke duration, $t_q$, in panels (a) and (b), respectively. The system parameters are chosen to be the same as that investigated in Figs.~\ref{fig:GP_SQ} and \ref{fig:GP_Isen}, with a fixed interaction strength ratio of $\overline{g}_\mathbf{A}/\overline{g}_\mathbf{C}\!=\!1.8$. }
   \label{fig:GHD_GP}
\end{center}
\end{figure}

We employ GHD in order to investigate the crossover between the idealized limits of the sudden quench and isentropic cycles, explored above. In detail, we utilize Navier-Stokes GHD, which incorporates diffusive corrections and is therefore capable of dynamically generating entropy in a time evolved system, and simulating the interaction-driven work strokes of our quantum Otto engine cycle. Results for the net work and efficiency, as a function of the work stroke duration $t_q$, are presented as solid lines in Fig.~\ref{fig:GHD_GP}(a) and (b), respectively. Here, we have chosen a fixed ratio of interaction strength, $\overline{g}_\mathbf{A}/\overline{g}_\mathbf{C}\!=\!1.8$ as it provides a clear separation between the low temperature results. Further, isentropic value for efficiency of the ground state working fluid, given by Eq.~\eqref{eq:T0_GP_Efficiency}, is shown in Fig.~\ref{fig:GHD_GP}(b) as a grey dashed line.

In the limit of rapid work strokes, i.e. $\omega t_q / 2 \pi \!\lesssim\!10^{-1}$, both the net work and efficiency converge near to a constant value of net work and efficiency. However, we find a discrepancy between the analytic sudden quench results and that found from GHD simulation.
The disagreement for the sudden quench protocol is caused by the fact that GHD is applicable at large space-time scales, and is thus not necessarily trustworthy at such short timescales.

For longer work stroke durations, we observe that both net work and efficiency converge to the corresponding isentropic values over a relatively short time window. This window corresponds approximately to the time for a single longitudinal oscillation cycle of the trap, with the first peak of both net work and efficiency corresponding to the first minimum in kinetic energy after the interaction strength ramp. The additional presence of small oscillations in the performance for longer times corresponds to the excitation of longitudinal breathing modes, which would be larger in amplitude for a stronger interaction strength quench. Thus, approximately quasistatic results for engine performance would be obtained only for linear ramps larger than the characteristic oscillation time of the trap.

We note that the time window for convergence to near maximum performance is independent of temperature. Indeed, at higher temperature, these curves tend to flatten. At high temperatures, as noted previously, the density profile becomes independent of interaction strength. Since the convergence to the isentropic results under increasing quench times is caused by the longitudinal dynamics, high temperatures tend to wash these dynamics out, causing the results to be largely independent of work stroke duration.
Finally, we note that this rapid convergence to the isentropic limit is particularly useful for possible experimental realization of such an engine cycle, where work stroke duration is an important parameter which may be tuned to optimize performance.

\subsection{Strongly interacting working fluid} \label{sec:Strong_Interactions}

In the strongly interacting regime, $\gamma_0 \!\gg\!1$, the 1D Bose gas becomes increasingly fermionized, giving rise to the Tonks-Girardeau gas of hard-core bosons in the limit $\gamma_0\!\to\!\infty$, where the correlation function vanishes, i.e. $g^{(2)}(0,0) \!\simeq \! 0$. For strong but finite interaction strengths, the gas is described by corrections to this regime of fermionization, dependent on the system temperature. Here, we investigate engine operation of a strongly interacting working fluid in the near-Tonks-Girardeau regime, $\tau_0\!\ll\!\pi^2/(1\!+\!\gamma_0)^2$ \cite{Kerr2024How}, the regime of high temperature fermionization, $\tau_0\!\gg\!\pi^2/(1\!+\!\gamma_0)^2$, and the smooth crossover between these parameter regimes. 
The dimensionless interaction strength at point $\mathbf{C}$ of Fig.~\ref{fig:Otto_cycle} is fixed at $\overline{g}_\mathbf{C}\!=\!50$, with a working fluid consisting of $N_\mathbf{C}\!=\!20$ particles. Strokes $\mathbf{BC}$ and $\mathbf{DA}$ consist of transferring $\Delta N \!=\!10$ particles between the strongly interacting working fluid and the reservoirs. 

\begin{figure}[!tbp] \begin{center}\includegraphics[width=16cm]{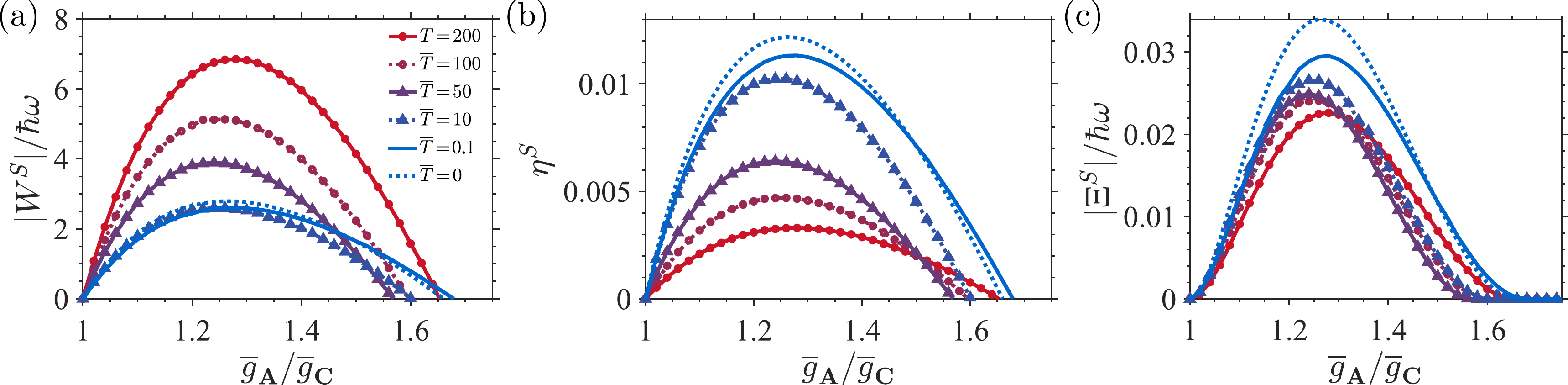}  
   \caption{Performance of the interaction-driven thermochemical Otto engine cycle with a working fluid consisting of a strongly interacting harmonically trapped 1D Bose gas under a sudden quench protocol. The variation of the bath temperature, $\overline{T}$, ensures that the working fluid inhabits the low temperature Tonks-Girardeau regime ($\overline{T}\!\leq\!1$), the regime of high-temperature fermionization ($\overline{T}\!\geq\!50$), and the crossover between these regions ($\overline{T}\!=\!10$). The system at point $\mathbf{C}$ (see Fig.~\ref{fig:Otto_cycle}) consists of $N_\mathbf{C}\!=\!20$, with an interaction strength $\overline{g}_\mathbf{C}\!=\!50$, and transferring $\Delta N \!=\! 10$ particles between the working fluid and the external reservoirs while they are in contact. Panels (a) and (b) demonstrate the net work and efficiency, respectively, as a function of the interaction strength ratio. Panel (c) demonstrates the efficient work criterion introduced in Eq.~\eqref{eq:efficient_work}, and shows that the low temperature regime unambiguously performs better than higher temperatures.}
   \label{fig:TG_SQ} 
   \end{center}
\end{figure}


\subsubsection{Sudden quench}

Performance of the interaction-driven Otto engine cycle in the strongly interacting regime, shown in Fig.~\ref{fig:TG_SQ}, is typically lower than that under operation in the weakly interacting regime. This is due to the combined effect of low atom numbers required to inhabit this regime experimentally, and, more importantly, the suppression of the correlation function due to fermionization.  Net work of the sudden quench is presented in Fig.~\ref{fig:TG_SQ}(a), where we observe a clear separation between the results residing in the two temperature regimes introduced above. In particular, results lying in the low temperature near-Tonks-Girardeau regime, corresponding to temperatures $\overline{T}\!=\!0.1$ and $\overline{T}\!=\!1$, do not significantly deviate from the ground state results, which are obtained via Eq.~\eqref{eq:Work_SQ_Approx} combined with the approximation $g^{(2)}(0,0)\!\simeq\! 4 \pi^2 / 3 \gamma_0^2$ obtained in Ref.~\cite{kheruntsyan2005finite}.
The efficiency, presented in Fig.~\ref{fig:TG_SQ}(b), also agrees with that given for the ground state system, which utilize the total ground state energy of a trapped 1D near-Tonks-Girardeau gas,   $\langle \hat{H} \rangle/\hbar \omega \!=\! N^2/2 - 128\sqrt{2}N^{5/2}/45 \pi^2 \overline{g}$.

In contrast, the performance in the high temperature regime, $\overline{T}\!\geq\!50$, displays a clear pattern of improving performance under increasing temperature, similar to that seen in the weakly interacting gas. This is thanks to the dependence of the net work on the local correlation function, $g^{(2)}(0,0)$, as seen in Eq.~\ref{eq:Work_SQ_Approx}. The local correlation function of the 1D Bose gas monotonically increases as a function of temperature, regardless of regime. The sudden quench engine utilizes this behaviour to boost the beneficial net work, even when competing with the peak density, which strictly decreases as a function of temperature as discussed previously. Thus we observe a clear separation between low and high temperature results in this strongly interacting regime, with results for $\overline{T}\!=\!10$ lying on the crossover between these regimes.

In contrast to the results for efficient work for the weakly interacting gas shown in Fig.~\ref{fig:GP_SQ}(c), efficient work of the strongly interacting gas, shown in Fig.~\ref{fig:TG_SQ}(c), is significantly less dependent on the temperature. This stems from the more subdued scaling of net work in this regime, due largely to the suppression of finite temperature effects in the local second-order correlation function, which is a result of fermionization. The slow scaling of net work with temperature directly results in the strong decrease in efficiency under the same temperature increase. Thus, when multiplied for calculation of efficient work, the combination results in a much weaker overall dependence on temperature. However, there is small improvement to be observed at low temperatures, in terms of peak performance and breadth of operation. This is a reversal of what was seen in the weakly interacting regime, and is entirely due to how the peak density and correlation function differs in these regimes.


\begin{figure}[!tbp] \begin{center}\includegraphics[width=16cm]{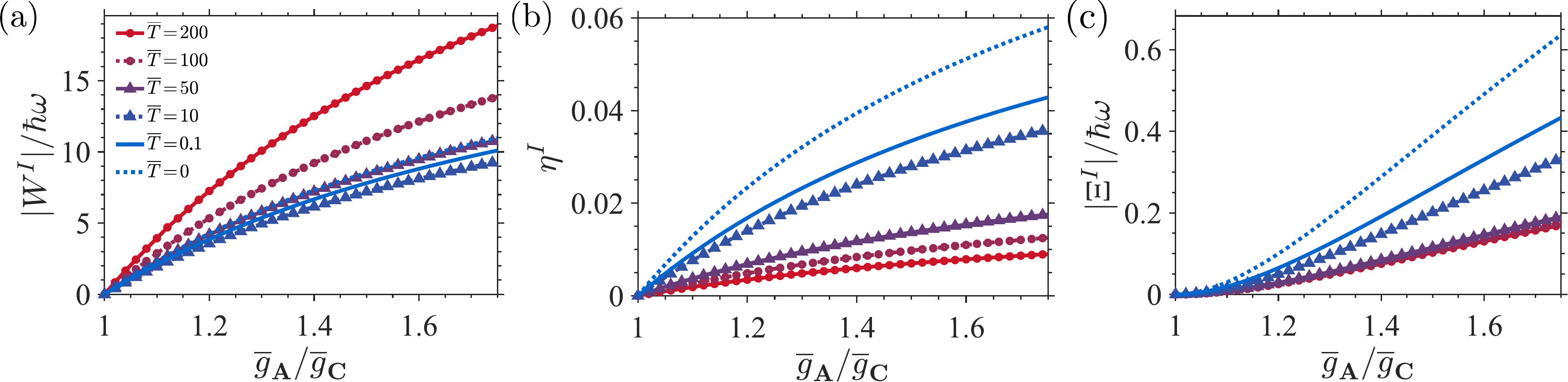}  
   \caption{Performance of the isentropic interaction-driven thermochemical Otto engine cycle with a working fluid consisting of a strongly interacting harmonically trapped 1D Bose gas.  Net work, $W$, and efficiency, $\eta$, are shown in panels (a) and (b), respectively. System parameters are chosen to be the same as those in Fig.~\ref{fig:TG_SQ}. We further demonstrate the efficient work, $\Xi^I$ defined in Eq.~\eqref{eq:efficient_work}, in panel (c), which demonstrates the superior performance in the low temperature regime.}
   \label{fig:TG_Isen} 
   \end{center}
\end{figure}

\subsubsection{Isentropic quench}

The performance of the strongly interacting isentropic thermochemical Otto engine is demonstrated in Figs.~\ref{fig:TG_Isen}. Here, we again observe that the low temperature system does not vary significantly with a change in temperature, as was the case for the sudden quench cycle. Utilizing the total energy of the ground state working fluid, we derive the formula for net work,
\begin{equation}
    W^\mathrm{I}_\mathrm{T\!=\!0} = -\frac{128 \sqrt{2}}{45 \pi^2} \left( N_\mathbf{A}^{5/2} - N_\mathbf{C}^{5/2} \right) \left( \frac{1}{\overline{g}_\mathbf{C}} - \frac{1}{\overline{g}_\mathbf{A}}\right),
\end{equation}
which is demonstrated in Fig.~\ref{fig:TG_Isen}(a). When compared with the results for the weakly interacting regime, given in Eq.~\eqref{eq:Work_GS_GP}, we see that the result for the strongly interacting system demonstrates a trade-off between the higher order dependence on the total atom numbers of states $\mathbf{A}$ and $\mathbf{C}$, which improves the net work under larger $\Delta N$, and a weaker dependence on the interaction strengths, which negatively impacts its scaling with $\overline{g}_\mathbf{A}/\overline{g}_\mathbf{C}$, when compared with the weakly interacting system. Efficiency of the ground state engine cycle may be calculated from the total energy expressed above, and is seen to give an upper bound to the general results for efficiency as a function of temperature, as was seen in the weakly interacting case. These represent an extension to previous results, derived in Ref.~\cite{keller2020feshbach} for the weakly interacting isentropic cycle, to the regime of strong interactions.

For the strongly interacting isentropic cycle, increasing temperature is seen to beneficially impact net work, $W^\mathrm{I}$, which stands in contrast to what was observed in the weakly interacting regime (see Fig.~\ref{fig:GP_SQ}(c)). Here, as was the case for the sudden quench cycle, this is due to the overall increase in interaction energy with increasing temperature, which was suppressed for the weakly interacting gas due to the broadening of the density profile.
Notably, net work for the system with temperature $\overline{T}\!=\!10$, with point $\mathbf{C}$ lying directly on the border between the two temperature regimes, actually \textit{reduces} when compared with the low temperature results. For this system, particle intake causes the working fluid at point $\mathbf{A}$ to lie deeper within the low temperature Tonks-Girardeau regime, since the temperature $\tau_0$ is reduced due to its inverse dependence on the density, $\overline{\rho}(0)$. This causes a reduction in the total interaction energy, which is expressed in Eq.~\eqref{eq:H_int}, due to the strong dependence which the correlation function has on temperature within this crossover region.

We further examine the efficient work criterion, introduced in Eq.~\eqref{eq:efficient_work}, as it is unclear from simply inspecting net work and efficiency, as to whether there is an overall benefit to working at low or high temperatures in this strongly interacting regime. Efficient work of the isentropic quench, $\Xi^I$ is shown in Fig.~\ref{fig:TG_Isen}(c), where we observe a clear increase in efficient work at low temperatures, stemming from the overall increase in efficiency which compensates for the decreased net work. Further, we note that temperatures above $\overline{T}\!=\!50$, i.e. results within regime V, experience a much slower decrease in efficient work. For very high temperatures the 1D Bose gas is well approximated by a classical gas with $g^{(2)}(0,0)\!\simeq\!2$ \cite{kheruntsyan2005finite}. In such a situation, the saturation of the local correlation function implies that interaction energy is approximately unchanging as a function of interaction strength, resulting in a diminished net work. Further, the equipartition theorem implies that the heat intake at high temperatures converges to approximately $Q^I_\mathrm{in}\!\propto\!\Delta N k_B T$. These results, taken together, imply that the efficient work, $\Xi^I\!=\!W^I \times \eta^I \!=\!(W^I)^2 / Q_\mathrm{in}^I$, becomes vanishingly small for high temperatures.




\begin{figure}[!tbp] \begin{center}\includegraphics[width=14cm]{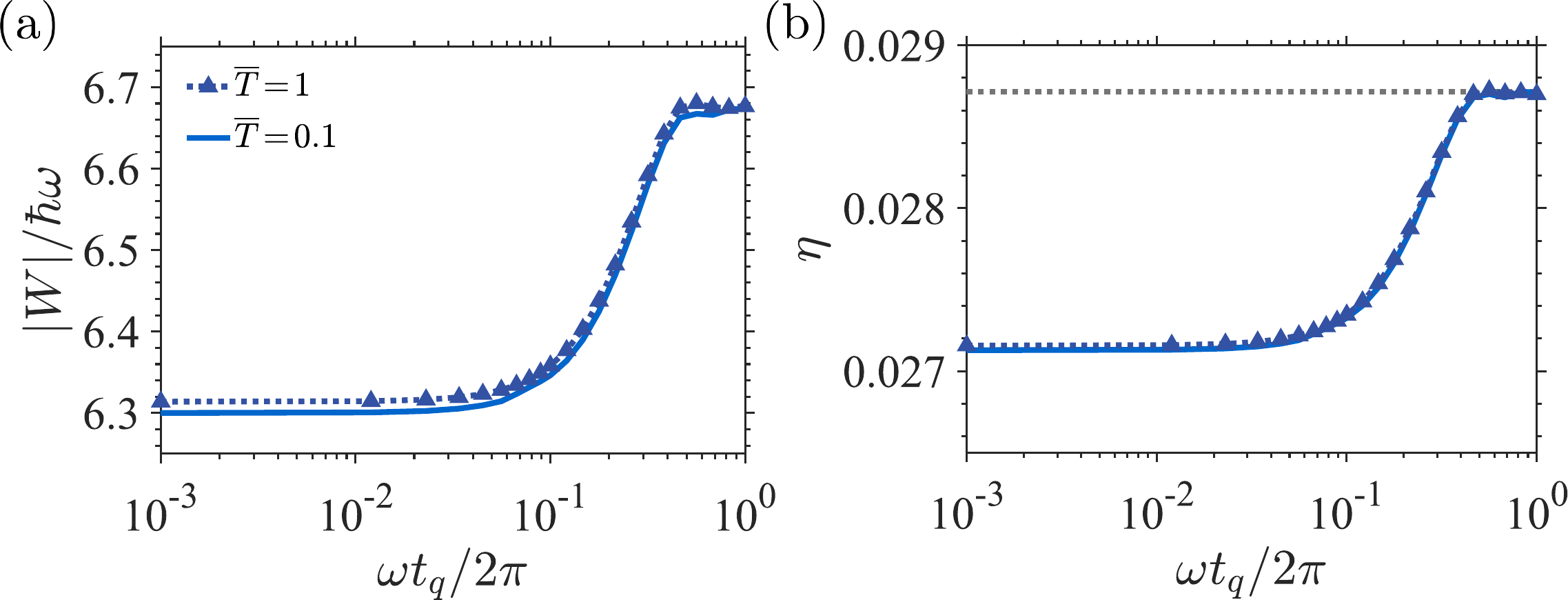}   
   \caption{Finite time engine performance for the strongly interacting thermochemical Otto cycle. Net work and efficiency are demonstrated as a function of work stroke duration, $t_q$, in panels (a) and (b), respectively. The system parameters are chosen to be the same as that investigated in Fig.~\ref{fig:TG_SQ} and \ref{fig:TG_Isen}, with a fixed interaction strength ratio of $\overline{g}_\mathbf{A}/\overline{g}_\mathbf{C}\!=\!1.4$.}
   \label{fig:TG_GHD} 
\end{center}
\end{figure}

\subsubsection{Nonequilibrium dynamics}

We again utilize Navier-Stoke GHD in order to explore the dynamics between the sudden and isentropic limiting cases of performance as a function of work stroke duration. Here, we again investigate the low temperature results, as increasing temperature will, as before, tend to flatten the curve and result in performance that is independent of work stroke duration.
Net work and efficiency, shown in Fig.~\ref{fig:TG_GHD}(a) and (b), respectively, demonstrate convergence to the isentropic limit over the same time window as previously observed. As this convergence was previously shown to be caused by the longitudinal dynamics of the harmonically trapped gas, this was the expected result.

However, we here observe a more significant disagreement between the results for a rapid work stroke duration and the sudden quench results, which are significantly lower, and are thus not shown on the figure. This may again be attributed to the fact that GHD is not applicable on such short timescales, and suggests that there are dynamics on much shorter timescales which GHD is not able to access. However, we note that GHD is known to be accurate for the work stroke durations investigated here, thanks to previous comparison with experiment, and efforts to benchmark the theory, achieved in previous works.
Such rapid work stroke durations are, however, not experimentally relevant, and in general would require a more microscopic theory applicable at finite temperatures, and is beyond the scope of this work.



\section{Conclusions}
In this chapter, we have examined the operation of a quantum many-body Otto engine with unitary work strokes driven by control over the interparticle interaction strength, and purely diffusive coupling between the working fluid and reservoir. This extends previous analysis of this engine cycle, achieved in Ref.~\cite{keller2020feshbach}, and done for the ground state of the weakly interacting gas in the limit of quasistatic work strokes. 
Here, we have made use of the thermodynamic Bethe ansatz and the recently pioneered GHD framework to analyze this engine operating at finite temperature, with arbitrary interaction strengths, and over a breadth of work stroke durations.

In the limit of a sudden quench, we employed formulas for efficiency and net work, derived in Chap.~\ref{Chap:3}, which express these quantities in terms of equilibrium observables, such as Glauber's local second-order correlation function. Here, in combination with a local density approximation, these formulas allow for a detailed analysis of the engine cycle via an approximation for the total integrated correlation function, allowing us to examine the behaviour of the sudden quench engine in the limits of ground state operation, and high temperatures. 

We further derived new results for the quasistatic engine cycle in the ground state of the strongly interacting gas, which is in good agreement with the low temperature results in the near-Tonks-Girardeau regime. Through this, we have observed that the low temperature working fluid universally operates under the largest efficiency for the isentropic interaction-driven engine cycle. However, the dependence of the net work on temperature is nontrivial, being maximized at low temperatures for weak interactions, and high temperatures for strong interactions. This is explained through the dependence of the interaction energy on the temperature in the various regimes that the 1D Bose gas possesses. 

Utilizing the recently developed theory of generalized hydrodynamics, applicable to integrable and near-integrable models, we have explored the nonequilibrium dynamics of the quantum Otto engine cycle. Through this, we show that the transition from sudden quench to isentropic performance occurs within a short window, converging to near its maximum on timescales on the order of that associated with the longitudinal trap. This is explained through the breathing mode dynamics of its longitudinal profile, which further give rise to slight oscillations in the convergence to optimal performance. This demonstrates the utility of the generalized hydrodynamic framework, which is further applicable to study the large-scale dynamics of any quantum or classical model solvable via the thermodyamic Bethe ansatz, making itself a useful tool for studying quantum thermodynamic systems going forward.

Finally, for the sudden quench engine cycle, we have utilized the efficient work, a variation of the `efficient power' criterion first introduced in Ref.~\cite{yilmaz2006new}, in order to evaluate optimal operation conditions in the weakly and strongly interacting regimes. These results demonstrate the unique dependence on temperatures which each regime has, with weak interactions unambiguously achieving better operation under high temperatures, and strong interactions at low temperatures. This is quantified analytically through inspecting the dependence of the net work on both the peak density and Glauber's local second-order correlation function.


%% file: Chapter7/Chapter7.tex


\chapter[Finite-time quantum Otto engine	]{A finite-time quantum Otto engine with tunnel coupled one-dimensional Bose gases	}
\label{Chap:7}	
\pagestyle{headings}


\textit{In this chapter, we undertake a theoretical study of a finite-time quantum Otto engine cycle driven by inter-particle interactions in a weakly interacting one-dimensional (1D) Bose gas in the quasicondensate regime. Utilizing a $c$-field approach, we simulate the entire Otto cycle, i.e. the two work strokes and the two equilibration strokes. More specifically, the interaction-induced work strokes are modelled by treating the working fluid as an isolated quantum many-body system undergoing unitary evolution. The equilibration strokes, on the other hand, are modelled by treating the working fluid as an open quantum system tunnel-coupled to another quasicondensate which acts as either the hot or cold reservoir, albeit of finite size. We find that, unlike a uniform 1D Bose gas, a harmonically trapped quasicondensate cannot operate purely as a \emph{heat} engine; instead, the engine operation is enabled by additional \emph{chemical} work performed on the working fluid, facilitated by the inflow of particles from the hot reservoir. The microscopic treatment of dynamics during equilibration strokes enables us to evaluate the characteristic operational time scales of this Otto thermochemical engine, crucial for characterizing its power output, without any \emph{ad hoc} assumptions about typical thermalization timescales. We analyse the performance and quantify the figures of merit of the proposed Otto thermochemical engine, finding that it offers a favourable trade-off between efficiency and power output, particularly when the interaction-induced work strokes are implemented via a sudden quench. We further demonstrate that in the sudden quench regime, the engine operates with an efficiency close to the near-adiabatic (near maximum efficiency) limit, while concurrently achieving maximum power output. }

\section{Introduction}\label{sec:Introduction}

Quantum heat engines (QHEs) \cite{koch2022making, rossnagel2016single, simmons2023thermodynamic, bouton2021quantum, Nitrogen-vacancy-heat-engine, fogarty2020many, keller2020feshbach, li2018efficient, singh2020optimal} provide a concrete platform for understanding the fundamental laws of thermodynamics in the quantum domain \cite{brandao2015second, masanes2017general}. Their exploration has recently expanded to include interacting many-particle systems \cite{mukherjee2021many, halpern2019quantum, chen2019interaction, watson2024interaction, herrera2023correlation, latune2020collective}, hence offering access to quantum many-body effects such as entanglement, quantum coherence, and correlations. Such quantum effects can be exploited for demonstrating either a form of quantum advantage \cite{jaramillo2016quantum} or a uniquely quantum functionality of QHEs not accessible classically \cite{koch2022making,bouton2021quantum}.

In QHEs that rely on interacting many-body systems as their working fluid, the strength of interparticle interactions provides a tool for engine operation not available in noninteracting systems. For example, it is possible to tune or quench the interaction strength to either do work on, or extract work from, the working fluid, analogous to volumetric compression and expansion work strokes in a conventional Otto engine cycle \cite{chen2019interaction, watson2024interaction, keller2020feshbach,li2018efficient}. Alternatively, one can quench the interaction strength to change the internal energy of the system without exchanging heat through coupling it to a thermal reservoir in a conventional Otto engine. This energy can then be utilized to extract work from the working fluid, as was recently demonstrated in Refs. \cite{simmons2023thermodynamic, koch2022making}.

Recent studies have also focused on optimising the finite time performance of interaction-driven many-body engines \cite{li2018efficient,keller2020feshbach, fogarty2020many,williamson2023many,solfanelli2023qhe}. 
In general, the maximum efficiency for any QHE is attained when the work strokes are executed in the near-adiabatic or quasistatic limit, in accordance with the quantum-adiabatic theorem \cite{abah2019shortcut, hartmann2020many}. This leads to zero power output due to infinitely long engine driving times \cite{li2018efficient, abah2019shortcut, mukherjee2021many}. On the other hand, if the work strokes are rapid, the production of irreversible work due to non-adiabatic excitations results in a significant reduction of efficiency \cite{keller2020feshbach,shiraishi2016universal,abah2019shortcut}. Therefore, a major challenge in quantum thermodynamics is to design QHEs that provide a favourable trade-off between efficiency and power output, meaning they can operate with maximum or near-adiabatic efficiencies while providing non-zero power output in finite time \cite{campbell2017trade, campisi2016power, shiraishi2016universal}.

Recent developments aimed at optimising efficiency in finite-time operations have predominantly focused on employing shortcut to adiabaticity (STA) protocols \cite{guery2019shortcuts,abah2019shortcut, keller2020feshbach}. However, given that STA protocols come with certain drawbacks such as modulation instability \cite{keller2020feshbach, li2018efficient}, additional energetic costs for implementation \cite{calzetta2018not}, and challenges in experimental realization \cite{chen2010transient}, the exploration of alternative and preferably simpler approaches to operate at near-maximum efficiency while maintaining non-zero power output becomes important. Furthermore, when evaluating the power, the dynamics of equilibration of the working fluid with the reservoir during the thermalization strokes are often ignored. This is reasonable when assuming that the reservoir size is infinite and that the thermalization time with such a reservoir is much shorter than the duration of the work strokes \cite{keller2020feshbach, li2018efficient, boubakour2023interaction, li2021shortcut, ccakmak2019spin, rezek2006irreversible}. However, these assumptions may not hold true in laboratory experiments, where the system typically interacts with a finite-sized reservoir, meaning that the characteristic thermalization time of the system with the reservoir becomes important for working out the power output of an engine.

In this chapter, we conduct a theoretical investigation of an experimentally realizable quantum Otto engine driven by the quench of inter-particle interactions in a weakly interacting one-dimensional (1D) Bose gas in the quasicondensate regime. In addition to conducting microscopic simulations of the interaction-driven work strokes of such an engine, which we conduct using a computational $c$-field approach based on the (stochastic) projected Gross-Pitaevskii equation (SPGPE) \cite{Blakie_cfield_2008, blakie2008dynamics, bayocboc2022dynamics, Thomas2021, bayocboc2023frequency, whatisqushock}, we simulate the equilibration strokes of the working fluid with the reservoir. This is done  by treating the working fluid as an open quantum system in thermal and diffusive contact with another, larger quasicondensate serving as the reservoir. Through this microscopic treatment of the equilibration strokes, we evaluate characteristic operational timescales for these strokes which enables us to quantify the power output of the engine for experimentally realistic time scales of the full Otto cycle. From this analysis, we demonstrate that, unlike in the uniform 1D Bose gas investigated in Refs.~\cite{chen2019interaction, watson2024interaction}, operation as a \textit{heat} engine is not possible for the harmonically trapped system. However, allowing for additional chemical work, in the form of particle flow from the hot reservoir to the working fluid, enables engine operation. In this scenario, the engine cycle may instead be considered as an effective \textit{chemical} engine cycle. Such a cycle is shown to possess a favourable trade-off between power output and efficiency when the work strokes are implemented via a sudden quench of the interaction strength. We show that within the sudden quench regime, the engine functions with efficiency nearing the limit of near-adiabatic (near-maximum) efficiency, while attaining maximum power output.

The motivation for choosing the 1D Bose gas as our physical system is threefold. 
First, it is experimentally realisable using ultracold and ultradilute atomic gases confined to highly anisotropic trapping potentials that can be created by using either atom chips 
\cite{Esteve2006,hofferberth2007non,Karen_Yang_2008,Bouchoule2011} or two-dimensional (2D) optical lattices \cite{Greiner2001,Moritz2003,Tolra2004,Kinoshita2004,Kinoshita2005}. In such potentials, all energy scales of the system are much smaller than the transverse excitation energy, and therefore, the dynamics, while frozen in the two transverse dimensions, only occur in the remaining longitudinal dimension \cite{Kruger2010,Armijo2011,Kruger2022}. 
Furthermore, the strength of interatomic interactions in these experiments can be tuned by using confinement induced resonances \cite{Olshanii1998,Haller2009,Haller2010} or a magnetic Feshbach resonance \cite{Chin2010}. Second, in the uniform limit, the ultracold and ultradilute 1D Bose gases can be well characterised by the relatively simple elastic two-body processes in the (low-energy) $s$-wave scattering channel \cite{Olshanii1998}. This, in turn, means that such gases can be described by the integrable 1D Lieb-Liniger model \cite{Lieb-Liniger-I}, which---in the uniform limit---offers exact many-body solutions to equilibrium thermodynamic properties of the system through the thermodynamic Bethe ansatz \cite{yang1969thermodynamics,kheruntsyan2003pair,Karen_Yang_2008,kerr2024analytic}. Such exact solutions can be used for benchmarking approximate analytic approaches or many-body computational techniques (see, e.g., \cite{kheruntsyan2003pair,Drummond2004,kerr2024analytic,Watson_Maxwell}). Third, the $c$-field method of the SPGPE that we use here to model the Otto engine cycles is computationally most efficient in 1D, rather than in 2D and 3D, which aids the breadth and depth of the analysis that can be carried out in different parameter regimes. We note, however, that the Otto engine model proposed and analysed here using the SPGPE can, in principle, be extended to 2D and 3D Bose gases as well \cite{Blakie_cfield_2008, blakie2008dynamics}.

The rest of the chapter is organised as follows: In Section \ref{sec:2}, we introduce our model of the Otto cycle and describe the $c$-field approach that we use to simulate the complete finite-time Otto cycle. Following this, in Section \ref{sec:3}, we focus on the unitary work strokes and identify the timescales governing a sudden quench, an intermediate-time quench, and a near-adiabatic (quasistatic) quench. The knowledge of these timescales will enable us to evaluate the trade-off between efficiency and power when we analyse the finite-time performance of the engine. Next, in Section \ref{sec:4}, we shift our focus to the equilibration strokes with the reservoir. In this Section, we investigate various dynamical scenarios governing the equilibration strokes and evaluate the operational timescales of these strokes, which will help us quantify the power output of the full engine cycle in an experimentally realisable scenario. In Section \ref{sec:5}, we utilize the findings of the previous Sections to evaluate the performance of the proposed interaction-driven chemical Otto engine in its full four stroke cycle. We quantify figures of merit of the engine, such as the power output, efficiency, and the trade-off between power and efficiency, as we increase the duration of the work strokes from the sudden quench regime to the near-adiabatic (quasistatic) regime. Finally, in Section \ref{sec:conclusion}, we summarise our findings and discuss the outlook.

\section{The Otto cycle} \label{sec:2} 
\begin{figure}[t!]
  \centering
  \includegraphics[scale=0.7]{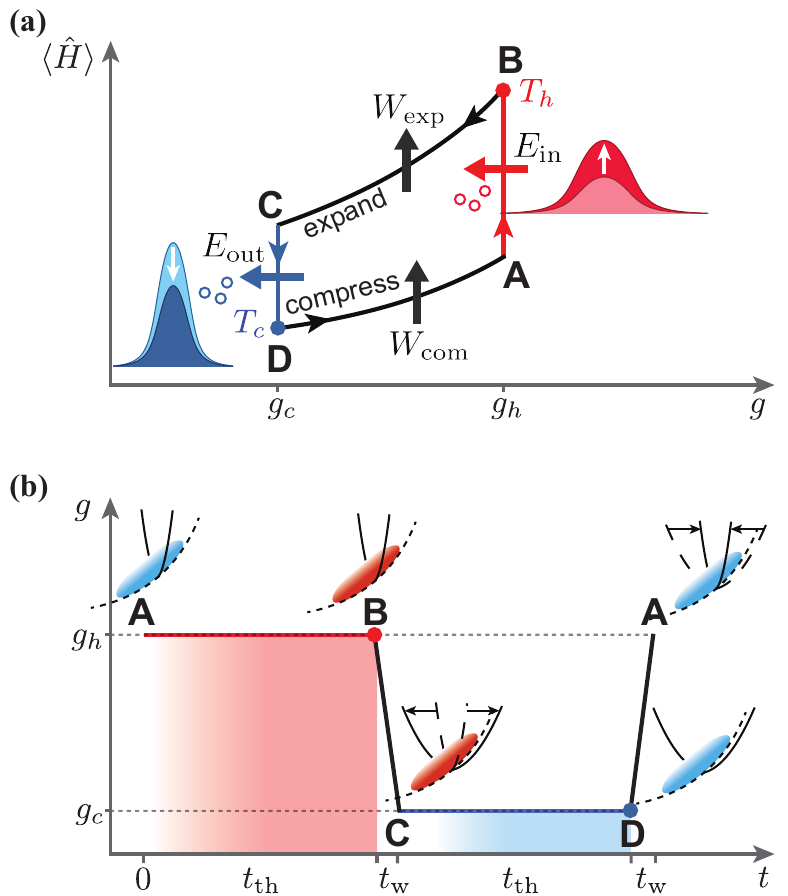}
\caption{
An interaction-driven quantum Otto cycle with a harmonically trapped 1D Bose gas as the working fluid. Panel (a) shows a schematic diagram of the four strokes of the Otto cycle in the working fluid energy $\langle \hat{H}\rangle$ versus interaction strength $g$ parameter space. For further details and notations, see text. Panel (b) illustrates the time sequence of the interaction strength quench (not to scale) during the four strokes of the Otto cycle: two thermalization strokes of duration $t_{\mathrm{th}}$, and two work strokes of duration $t_{\mathrm{w}}$. As $g\simeq 2\hbar \omega_{\perp}a$, the change in $g$ can be achieved via an increase or reduction of the frequency of transverse confinement $\omega_{\perp}$, which then makes the work strokes \textbf{DA} and \textbf{BC} analogous, respectively, to (transverse) compression or expansion strokes of the regular, volumetric Otto cycle. Note that this figure is effectively the same as what was shown in Fig.~\ref{fig:Otto_cycle}, and is repeated here for convenience. }

  \label{fig:Fig1}
\end{figure}

The Otto engine cycle, demonstrated schematically in Fig.~\ref{fig:Fig1}, is widely studied in the field of quantum thermodynamics due to its relative simplicity \cite{Myers2022quantum}, in addition to being the closest model to real-world engine cycles.
It consists of two unitary work strokes alternated with two isochoric equilibration strokes in which the working fluid is coupled to external thermal reservoirs. In particular, the unitary work strokes, denoted \textbf{BC} and \textbf{DA} in Fig.~\ref{fig:Fig1}(a), correspond to volumetric compression and expansion, respectively, and are implemented via an external control parameter over the working fluid. The equilibration strokes \textbf{AB} and \textbf{CD} consist of coupling the working fluid to hot and cold thermal reservoirs, at temperatures $T_h$ and $T_c$, respectively, while maintaining constant volume. In the following subsections, we will describe our model and the implementation of the individual strokes, as well as their combined operation in a full four-stroke cycle of the Otto engine.

\subsection{The Model}
In this chapter, we consider a working fluid consisting of a harmonically trapped ultracold 1D Bose gas in the weakly interacting quasicondensate regime \cite{Petrov_2000_Regimes, Mora-Castin-2003,kheruntsyan2005finite, kheruntsyan2003pair, garrett2013condensation, clade2009observation, Jacqmin_2011_subpoissonian}. The Hamiltonian of this system ($s$), in second-quantized form, is given by 
\begin{equation}\label{eq:hamiltonian}
     \hat{H}_s = \int dz \ \hat{\Psi}_s^{\dag} \Biggl[- \frac{\hbar^2}{2m} \frac{\partial^2}{\partial z^2} \ + \frac{1}{2} m \omega^2 z^2 + \frac{g_s}{2} \hat{\Psi}_s^{\dag} \hat{\Psi}_s    \Biggr] \hat{\Psi}_s,
\end{equation}
where $\omega$ is the longitudinal trapping frequency, $m$ is the particle mass, and $\hat{\Psi}_s^\dagger(z)$ and $\hat{\Psi}_s(z)$ are the bosonic field creation and annihilation operators, respectively. Further, $g_s$ is the strength of repulsive ($g_s>0$) interparticle interactions within the working fluid that can be related to the frequency of transverse confinement, $\omega_\perp$, and the three-dimensional $s$-wave scattering length, $a$, via the relationship $ g_s \!\simeq\! 2 \hbar \omega_{\perp}a$, valid away from confinement-induced resonances \cite{Olshanii1998}.

 Modelling the work strokes of the Otto cycle corresponds to simulating the unitary, real-time dynamics of the working fluid governed by the Hamiltonian~(\ref{eq:hamiltonian}) in response to a change of an external parameter. In this chapter, we study an interaction-driven Otto cycle, enacted through control over the interaction strength $g_s$.
 In practice, the interaction strength can be tuned either by changing the scattering length $a$ using magnetic Feshbach resonance \cite{Chin2010}, or by varying the frequency of the transverse confinement $\omega_\mathrm{\perp}$ \cite{schemmer2018monitoring}, both methods leading to identical results reported here. Tuning the interaction strength by increasing or reducing $\omega_\perp$ can be regarded, respectively, as transverse compression or expansion of the working fluid \cite{watson2024interaction}, offering an analogy to the conventional volumetric Otto cycle even when the interaction strength is changed via a magnetic Feshbach resonance. During compression stroke \textbf{DA}, work, $W_\mathrm{com}>0$, is done on the working fluid by increasing the interaction strength, whereas in expansion stroke \textbf{BC}, the interaction strength is decreased allowing work, $W_\mathrm{exp}<0$, to be done by the working fluid. The net work extracted in one complete cycle is thus $W = W_\mathrm{{com}} + W_\mathrm{{exp}}$, which must be negative, $W<0$, for the cycle to operate as an engine.

To model the equilibration strokes, on the other hand, we consider a coupled system in which the working fluid, described by the Hamiltonian (\ref{eq:hamiltonian}), is coupled to another, larger 1D quasicondensate which serves as the reservoir. The coupling to the hot and cold reservoirs is alternated between the compression and expansion work strokes, but apart from that we assume that the hot and cold reservoirs are described identically, except for their respective thermal equilibrium temperatures. More specifically, for modelling the equilibration strokes, we employ the tunnel-coupled model of two quasicondensates \cite{bayocboc2022dynamics} described by the following Hamiltonian:
\begin{equation}
\label{eqn:tunnelcoupled}
    \hat{H}_\mathrm{coupled} = \hat{H}_{s} + \hat{H}_{h(c)}  - \hbar J \int dz \  [\hat{\Psi}^{\dag}_{s} \hat{\Psi}_{{h(c)}} + \hat{\Psi}^{\dag}_{h(c)} \hat{\Psi}_{s}].
\end{equation}
Here, the subscript $s$ denotes the system or the working fluid, whereas the subscripts $h$ and $c$ denote the hot and cold reservoirs, respectively. The Hamiltonians $\hat{H}_{h(c)}$ for the hot (cold) reservoirs have the same structure as Eq.~(\ref{eq:hamiltonian}), except that the field operators $\hat{\Psi}_s(z)$ are replaced by $\hat{\Psi}_{h(c)}(z)$. The parameter $J$ characterizes the tunnel-coupling strength between the working fluid and the reservoir; it can be determined precisely in experiments by probing their mutual two-point phase-correlation function using matter-wave interferometry \cite{betz2011two}.

\subsection{Numerical method}

To implement the four strokes of the quantum Otto cycle, we utilize the numerical $c$-field method, developed for studying the finite-temperature dynamics of Bose gases \cite{Blakie_cfield_2008, bayocboc2022dynamics, Thomas2021, bayocboc2023frequency, whatisqushock, blakie2008dynamics}. 
This involves decomposing the quantum field operator $\hat{\Psi}_i(z,t)$ into two distinct regions: the classical, or $c$-field region, composed of highly occupied low-energy modes that can be described by a complex field amplitude $\psi^{(\mathbf{C})}_{i}(z,t)$, and the thermal region containing sparsely occupied high-energy modes that act as an effective thermal bath for the $c$-field. We now detail the numerical implementation of the entire finite-time Otto cycle using the $c$-field method.

~

\noindent \textit{Step 1: Unitary compression work stroke,} \textbf{DA}: We prepare the initial finite-temperature thermal equilibrium state of the working fluid at point \textbf{D} of Fig.~\ref{fig:Fig1} using the stochastic projected Gross-Pitaevskii equation (SPGPE) \cite{gardiner2003stochastic,rooney2012stochastic},
\begin{equation*}
    d\psi^{(\mathbf{C})}_{s}(z,t) = \mathcal{P} ^{(\mathbf{C})}\Biggl\{- \frac{i}{\hbar} \mathcal{L}^{(\mathbf{C})}_s \psi^{(\mathbf{C})}_{s}(z,t) dt  \end{equation*}
\begin{equation}
\label{eq:SPGPE}
    \hspace{3.8cm}+\frac{\Gamma}{k_{B}T_{s}}(\mu_{s} - \mathcal{L}^{(\mathbf{C})}_s)\psi^{(\mathbf{C})}_{s}(z,t) dt + dW_{\Gamma}(z,t) \Biggr\},
\end{equation}
where $\psi^{(\mathbf{C})}_{s}(z,t)$ is the classical field of the working fluid or the system $(s)$. In Eq.~(\ref{eq:SPGPE}), the parameters $\mu_s$ and $T_s$ refer to the chemical potential and temperature of the thermal region and determine the total number of particles in the $c$-field region of the working fluid, $\psi^{(\mathbf{C})}_{s}(z,t)$. The operator $\mathcal{P}^{(\mathbf{C})}$ is a projector operator that sets up the boundary between the $c$-field and thermal region, defined by a cut-off energy, $\epsilon_\mathrm{cut}$. The Gross-Pitaevskii operator, $\mathcal{L}^{(\mathbf{C})}_s$, is given by
\begin{equation}
\label{eqn:GPE_operator}
    \mathcal{L}^{(\mathbf{C})}_s = - \frac{\hbar^2}{2m} \frac{\partial^2}{\partial z^2} + \frac{1}{2}m \omega^2 z^2 + g_s |\psi_s^{(\mathbf{C})}(z,t)|^2,
\end{equation}
We note here that we have matched up the strength of interparticle interactions and the temperature of the working fluid to those of the cold reservoir, which is to be considered later (in stroke \textbf{CD}, see below), i.e., we have chosen $g_s = g_c$ and $T_s=T_c$ for the initial thermal equilibrium state of the working fluid.

Finally, the stochastic noise term, $dW_{\Gamma}(z,t)$, in Eq.~(\ref{eq:SPGPE}) corresponds to complex white noise, defined by the correlation \cite{blakie2008dynamics, gardiner2003stochastic, rooney2012stochastic},
\begin{equation}
\label{eqn:noise-correlation}
    \langle dW^*_{\Gamma}(z,t) dW_{\Gamma}(z',t) \rangle  = 2\Gamma \delta(z-z')dt.
\end{equation}
 where $\Gamma$ is the growth rate that characterises the strength of the coupling between the $c$-field and the effective thermal bath, with $\langle \cdot \rangle$ referring to stochastic averaging over a large number of independent stochastic realisations (trajectories). In practice, the growth rate $\Gamma$ may be chosen according to numerical convenience as it does not affect the final thermal equilibrium configuration \cite{blakie2008dynamics, rooney2012stochastic}.

At the end of this preparation stage, the working fluid is at point \textbf{D} in the Otto cycle diagram of Fig.~\ref{fig:Fig1}. To initiate the first (compression) work stroke of the Otto cycle, \textbf{DA}, we assume that the working fluid is isolated from all external reservoirs and its interaction strength is quenched from an initial value, $g_s=g_c$ to the final value $g_h$. We assume that the interaction quench is realised over a finite duration according to a linear protocol,
\begin{equation}\label{eq:g_com}
    g_s(t) = g_c+  \ (g_h-g_c) t / t_\mathrm{w},
\end{equation}
where $t_\mathrm{w}$ is the duration of the quench.
The working fluid now undergoes a unitary evolution, which is modelled by the following projected Gross-Pitaevskii equation (PGPE)
\cite{blakie2005projected,Blakie_cfield_2008}:
\begin{equation}
\label{eqn:PGPE}
    i \hbar \frac{\partial}{\partial t} \psi_\mathrm{s}^{(\mathbf{C})}(z,t) = \mathcal{P} ^{(\mathbf{C})} \Biggl\{ - \frac{\hbar^2}{2m} \frac{\partial^2}{\partial z^2} + \frac{1}{2} m \omega^2z^2 + g_s(t)|\psi_\mathrm{s}^{\mathbf{(C)}}|^2 \Biggr\}.
\end{equation}
Through this stroke, mechanical work $W_\mathrm{com} >0$ is done on the working fluid. The quantity $W_\mathrm{com}$ is computed numerically by evaluating the change in the Hamiltonian energy of the working fluid after completion of the stroke at point \textbf{A}, i.e, $W_\mathrm{com}=\langle \hat{H}_s \rangle_{\mathbf{A}}-\langle \hat{H}_s \rangle_{\mathbf{D}}$.
The duration $t_\mathrm{w}$ in which the compression stroke \textbf{DA} is completed determines the state of the working fluid at the end of the stroke. If this compression stroke is executed via a sudden quench, the working fluid at point \textbf{A} will be in a highly non-equilibrium state with no definable temperature.
In contrast, if the compression stroke is carried out using a slow quasistatic quench, then the working fluid at point \textbf{A} will have a definite temperature, $T_s > T_c$.

~

\noindent\textit{Step 2: Equilibration with hot reservoir,} \textbf{AB}: Upon completion of the unitary compression stroke \textbf{DA} at time $t_{\mathrm{w}}$, we next model the subsequent equilibration stroke \textbf{AB} with the hot reservoir by switching on the tunnel coupling between the working fluid and the hot reservoir. In an experimental setup, the tunnel-coupled system of 1D quasicondensates, or the 1D bosonic Josephson junction, can be realized using a quantum degenerate Bose gas confined in a tunable double well potential in the transverse direction \cite{betz2011two}. The dynamics of the working fluid coupled to the hot reservoir are now modelled using the coupled PGPEs \cite{bayocboc2022dynamics},   
\begin{equation}
\label{eqn:coupled_PGPEs_a}
        i \hbar \frac{\partial}{\partial t} \psi^{(\mathbf{C})}_{\mathrm{s}}(z,t) = \mathcal{P}^{(\mathbf{C})} \Biggl\{ - \frac{\hbar^2}{2m} \frac{\partial^2}{\partial z^2} + \frac{1}{2}m \omega^2 z^2 +  g_s|\psi^{(\mathbf{C})}_{\mathrm{s}}|^2 - \hbar J \psi^{(\mathbf{C})}_{h}(z,t) \Biggr\}, 
         \end{equation}
 \begin{equation}       
    i \hbar \frac{\partial}{\partial t} \psi^{(\mathbf{C})}_{h}(z,t) = \mathcal{P}^{(\mathbf{C})}  \Biggl\{ - \frac{\hbar^2}{2m} \frac{\partial^2}{\partial z^2} + V_h(z)   + g_h|\psi^{(\mathbf{C})}_{h}|^2 - \hbar J \psi^{(\mathbf{C})}_{\mathrm{s}}(z,t) \Biggr\},
    \label{eqn:coupled_PGPEs_b}
    \end{equation}
where, $\Psi^{\mathbf{C}}_h$ is the $c$-field for the hot reservoir $(h)$, which is at a temperature $T_h$. The initial state of the hot reservoir is prepared exactly in the same way as the working fluid, using the SPGPE, Eq.~(\ref{eq:SPGPE}), except with the subscript $s$ replaced by $h$.

\begin{figure}[t!]
    \centering
    \includegraphics[scale=0.7]{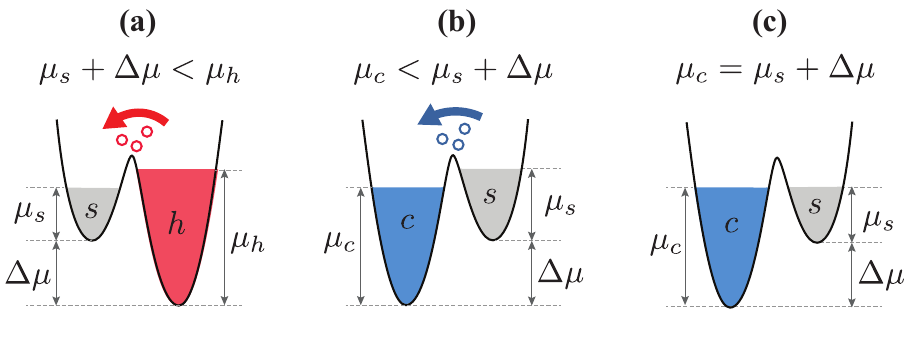}
    \caption{Illustration of three typical chemical potential offset ($\Delta \mu$) settings relevant for modelling the thermalization strokes when the working fluid is tunnel-coupled to a hot ($h$, red) or cold ($c$, blue) reservoir via a transverse double-well potential. In panel (a), the chemical potentials satisfy $\mu_s+\Delta \mu < \mu_h$, resulting in net particle flow from the reservoir to the working fluid, which is the case for stroke \textbf{AB}; panel (b) corresponds to the reverse situation, which is the case for stroke \textbf{CD} with $\mu_s+\Delta \mu > \mu_c$; finally, panel (c) illustrates a balanced situation of $\mu_s+\Delta \mu = \mu_c$, which does not result in any net particle flow between the working fluid and the cold reservoir (with the same being true if the cold reservoir is replaced by the hot one). }
    \label{fig:Chemical_offset}
\end{figure}

To initiate equilibration stroke \textbf{AB}, the tunnel-coupling parameter is quenched from $J=0$ to a constant value, which we chose to be $J=2 \omega$ for definiteness. Moreover, during the equilibration stroke with the hot reservoir, the interaction strength parameter of the working fluid is kept constant at value $g_s = g_h$. The external trapping potential in Eq.~(\ref{eqn:coupled_PGPEs_b})\,(b) is given by, 
\begin{equation}
\label{trap_chemicalshift}
    V_h(z) = \frac{1}{2} m \omega^2z^2 - \Delta \mu,
\end{equation}
where we have introduced a chemical potential offset, $\Delta \mu$, in order to control the net flow of particles from the reservoir to the working fluid if $\mu_s+\Delta \mu < \mu_h$, or vice versa -- from the working fluid to the reservoir -- if we were to chose $\mu_s+\Delta \mu > \mu_h$ (see Fig.~\ref{fig:Chemical_offset} for illustrations). The latter arrangement, with $\mu_s+\Delta \mu > \mu_c$, is the case for the stroke \textbf{CD} (see below).
We note here that, while the choice of $\Delta \mu$ affects the overall \emph{net} flow of particles upon completion of the \textbf{DA} stroke, transient transfer and oscillations of particles between the working fluid and the reservoir is still possible, and is in fact required for establishing eventual mutual thermal equilibrium.

The coupled PGPEs,~ Eqs.~(\ref{eqn:coupled_PGPEs_a}) and (\ref{eqn:coupled_PGPEs_b}), are evolved for a duration of time, which we refer to as thermalization time $t_\mathrm{th}$, during which the working fluid comes into thermal equilibrium with the hot reservoir. In the process, energy, $E_\mathrm{in} = \langle \hat{H}_s \rangle_{\mathbf{B}}-\langle \hat{H}_s \rangle_{\mathbf{A}}>0$, is transferred from the hot reservoir to the working fluid.

~

\noindent \textit{Step 3: Unitary expansion work stroke,} \textbf{BC}: After completion of the equilibration stroke with the hot reservoir, we switch off the tunnel coupling, i.e. set $J=0$, and perform the expansion work stroke by evolving the working fluid according to the PGPE given in Eq.~(\ref{eqn:PGPE}), except that now the interaction strength, $g(t)$ is decreased from the value $g_h$ back to $g_c$ according to the following linear protocol:
\begin{equation}\label{eq:g_exp}
    g(t) = g_{h}-(g_{h}-g_{c}) t  / t_\mathrm{w}.
\end{equation}
During the expansion work stroke, which is completed in the same duration ($t_\mathrm{w}$) as the compression stroke, the working fluid performs work $W_\mathrm{exp} < 0$. Similarly to $W_\mathrm{com}$, the quantity $W_\mathrm{exp}$ is computed numerically by evaluating the change in the Hamiltonian energy of the working fluid after completion of the expansion stroke at point \textbf{B}, i.e, $W_\mathrm{exp}=\langle \hat{H}_s \rangle_{\mathbf{B}}-\langle \hat{H}_s \rangle_{\mathbf{A}}$.

~

\noindent \textit{Step 4: Equilibration with cold reservoir,} \textbf{CD}: On completion of the expansion stroke, we enable tunnel coupling with the cold reservoir and simulate the dynamics of equilibration using the coupled PGPEs given in Eqs.~(\ref{eqn:coupled_PGPEs_a}) and (\ref{eqn:coupled_PGPEs_b}), except that the subscript $h$ is replaced by $c$. During this stroke, energy $E_\mathrm{out} = \langle \hat{H}_s \rangle_{\mathbf{D}}-\langle \hat{H}_s \rangle_{\mathbf{C}}<0$, is being transferred from the working fluid to the cold reservoir, returning the working fluid and the overall Otto cycle to the point of initialization \textbf{D}.
  
~ 

The overall performance of this engine cycle may be quantified through the net work, 
\begin{equation}
W =  W_\mathrm{{com}} + W_\mathrm{{exp}}, 
\end{equation}
which must be negative for the Otto cycle to operate as an engine, efficiency, 
\begin{equation}
\eta = - W/E_\mathrm{in}, 
\end{equation}
and power output, 
\begin{equation}
P = -W/t_\mathrm{tot}, 
\end{equation}
where $t_\mathrm{tot}$ is the total cycle time, given by $t_\mathrm{tot} = 2 (t_\mathrm{w} + t_\mathrm{th})$.

\section{Quantifying the timescales for the unitary work strokes} \label{sec:3}

\begin{figure}[t!]
    \centering
    \includegraphics[scale=0.38]{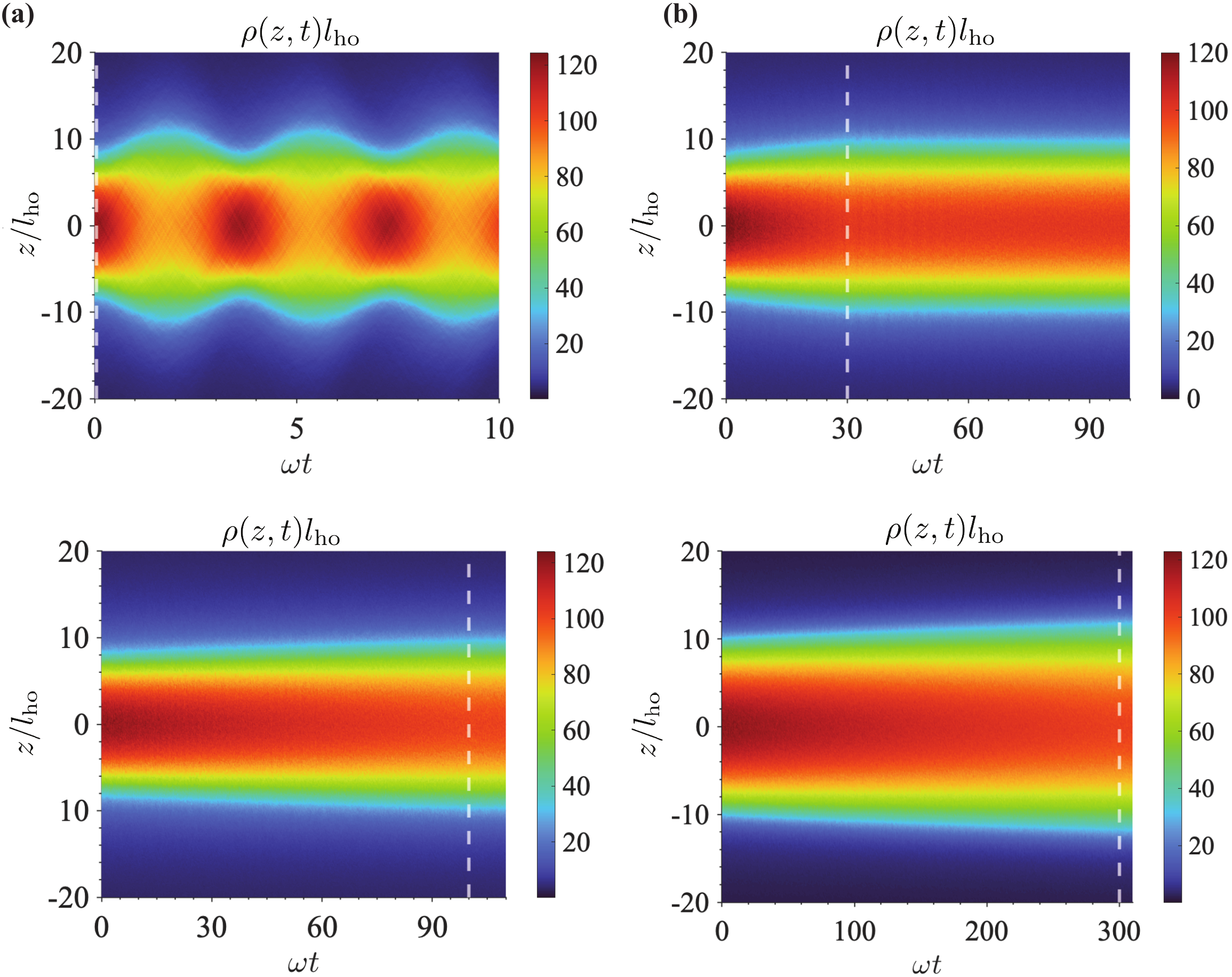}
    \caption{Time-evolution of the real-space density profile $\rho(z,t)$ of a weakly interacting 1D Bose gas, following an interaction strength quench. After preparing the initial state $(t = 0)$, a linear quench of the interaction strength $g_s$ was performed according to Eq.~(\ref{eq:g_com}), from an initial value of $g_c$ to the final value $g_h=1.80g_c$, and the working fluid was evolved according to the 1D PGPE (\ref{eqn:PGPE}). 
    The quench duration (shown as vertical dashed lines) was chosen as: (a) $t_\mathrm{w} = 0.05/\omega$ (sudden quench); (b) $t_\mathrm{w} = 30.0/\omega$ (intermediate); (c) $t_\mathrm{w} = 100/\omega$ (intermediate); and (d) $t_\mathrm{w} = 300/\omega$ (near-adiabatic or quasistatic quench). The dimensionless position $z/l_{\mathrm{ho}}$ is defined using the characteristic harmonic oscillator length along the longitudinal dimension, $l_{\mathrm{ho}} = \sqrt{\hbar/m \omega}$. All results in this chapter are for a gas of $^{87}$Rb atoms with mass $m \simeq 1.44 \times 10^{-25}$ kg and 3D scattering length $a_\mathrm{s} \simeq 5.31\,$nm. The initial thermal equilibrium temperature of the working fluid here is $T_s = 86.3$ nK and the total number of particles is $N_s \simeq 1750$ which was obtained using the chemical potential $\mu_s = 6.62\times10^{-31}$ J. The other relevant physical parameters that were used are as follows: $\omega/2\pi = 20.0$ Hz and $g_c = 1.27 \times 10^{-38}$ J$\cdot$m. The transverse confinement frequency that was used to achieve this value of interaction strength was $\omega_{\perp}/2\pi = 1.81$ kHz (recall that $ g_s \!\simeq\! 2 \hbar \omega_{\perp}a$).} 
     \label{fig:Fig3_Turbo}
\end{figure}

In this Section, we analyze the three distinct time-scales over which the work strokes may be performed via a quench of the interaction strength. These include: (\emph{i}) the sudden quench, where the efficiency is lowest due to the production of maximum irreversible work \cite{abah2019shortcut, rezek2006irreversible}; (\emph{ii}) the near-adiabatic (quasistatic) quench, which corresponds to near-maximal efficiency but minimal power output due to extremely long driving time \cite{born1928beweis}; and (\emph{iii}) the intermediate quench, which lies in between the first two extremes and highlights the trade-off between power and efficiency.

In theory, a sudden quench is often treated as if it were instantaneous \cite{bayocboc2023frequency, bouchoule2016finite, Thomas2021}. However, in practice, a ``sudden'' quench is not truly instantaneous but occurs over a finite duration \cite{schemmer2018monitoring}. This duration must be fast enough to be nearly sudden with respect to the characteristic timescale for longitudinal dynamics, $t_{\parallel}=2\pi/\omega$, yet slow enough (near adiabatic) in relation to the characteristic timescale for transverse dynamics, $t_{\perp}=2\pi/\omega_{\perp}$, so that one avoids exciting any transverse excitations and maintains the 1D nature of the system \cite{schemmer2018monitoring}.

Accordingly, in our simulations, the unitary work strokes in the sudden quench regime are completed in a finite time by performing a linear quench of the interaction strength described in Eqs.~(\ref{eq:g_com}) and (\ref{eq:g_exp}), satisfying the following the condition:
\begin{equation}
\label{eqn:t_sudden}
  t_{\perp} \ll  t_\mathrm{w} \ll t_{\parallel}.
\end{equation}
The quasistatic or near-adiabatic engine cycle, on the other hand, corresponds to completing the work strokes over timescales during which the system remains approximately in thermal equilibrium states. This implies that the work strokes are near-adiabatic relative to both time-scales introduced above, i.e.,
\begin{equation}
\label{eq:t_ad}
     t_{\perp} \ll  t_{\parallel} \ll t_{\mathrm{w}}.
\end{equation}
Finally, to complete the work strokes in the intermediate regime, we follow the following condition:
\begin{equation}
\label{eq:t_inter}
   t_{\perp} \ll t_{\parallel} \sim t_\mathrm{w}.
\end{equation}

In Fig.~\ref{fig:Fig3_Turbo}, we demonstrate the time-evolution of the real-space density profile of the working fluid as we quench the interaction strength to perform the unitary \emph{compression} work stroke \textbf{CD} in time $t_\mathrm{w}$. (The dynamics during the \emph{expansion} strokes are similar and will not be shown). To identify distinct dynamical regimes as a function of the quench duration, we complete the interaction strength quench over various values of $t_\mathrm{w}$ and then continue simulating the unitary post-quench dynamics of the system (instead of immediately implementing the next stroke of the Otto cycle -- the thermalization stroke \textbf{BA}). This is done for diagnostics purposes only -- as to identify the presence, or lack thereof, of any longitudinal excitations in the system -- for the quench to be regarded as sudden or intermediate, as opposed to near-adiabatic (quasistatic) quench.

The vertical dashed lines in Figs.~\ref{fig:Fig3_Turbo}\,(a)--(d) mark the completion of the interaction quench over duration $t_{\mathrm{w}}$. In Fig.~\ref{fig:Fig3_Turbo}\,(a), this duration is too short for the dotted line to be visible; here, we are in the sudden quench regime and can clearly see excitation of breathing mode oscillations \cite{ fang2014quench, tschischik2013breathing,bayocboc2023frequency, schemmer2018monitoring, schmitz2013quantum,bouchoule2016finite} after the interaction quench is completed. 
In contrast, Fig.~\ref{fig:Fig3_Turbo}\,(b) and (c) depict the dynamics after the intermediate quench. Here, post-quench breathing mode oscillations are significantly suppressed, however, we can still observe weak breathing mode excitations during the quench, indicating that these quenches are not yet slow enough to be classified as near-adiabatic or quasistatic.
Finally, Fig.~\ref{fig:Fig3_Turbo} (d) illustrates what we classify as a near-adiabatic (quasistatic) quench. Over these timescales, no observable non-adiabatic excitations are produced. This regime of the Otto engine is expected to result in near-maximum efficiency due to minimal irreversible work produced during the work strokes \cite{rezek2006irreversible, abah2019shortcut, ccakmak2016irreversibility, keller2020feshbach}.

\section{Operational timescales for thermalization strokes under various dynamical scenarios} \label{sec:4}
The power output of an engine cycle is inversely dependent on the total cycle time. This is broken down into a sum of the duration of the unitary work strokes, which are controlled externally, and the equilibration strokes, which are not externally controlled but are often assumed to be fast in comparison to that of the work strokes \cite{keller2020feshbach, boubakour2023interaction, li2018efficient, li2021shortcut, ccakmak2019spin, rezek2006irreversible}. 
Alternative methods in which control over system-reservoir coupling in the equilibration (or thermalization) strokes is utilized to enhance the rate of relaxation have recently been proposed in, e.g., Refs.~\cite{dann2019shortcut,villazon2019swift}. However, to the best of our knowledge there is no currently proposed protocol that provides shortcuts to equilibration in the context of the interacting 1D Bose gas. Though such a prospect is interesting and important for future work, it is beyond the scope of the current chapter.

In this Section, we analyze these equilibration strokes in order to determine the characteristic operational timescale that governs the thermalization of the working fluid. In particular, we explore the effects of various factors, such as the size and temperature of the reservoir, as well as the duration of the prior work stroke which can leave the system in a highly non-equilibrium state after a sudden quench; or in a near-equilibrium state after a quasistatic quench.

\begin{figure}[tp]
    \centering
    \includegraphics[scale=0.38]{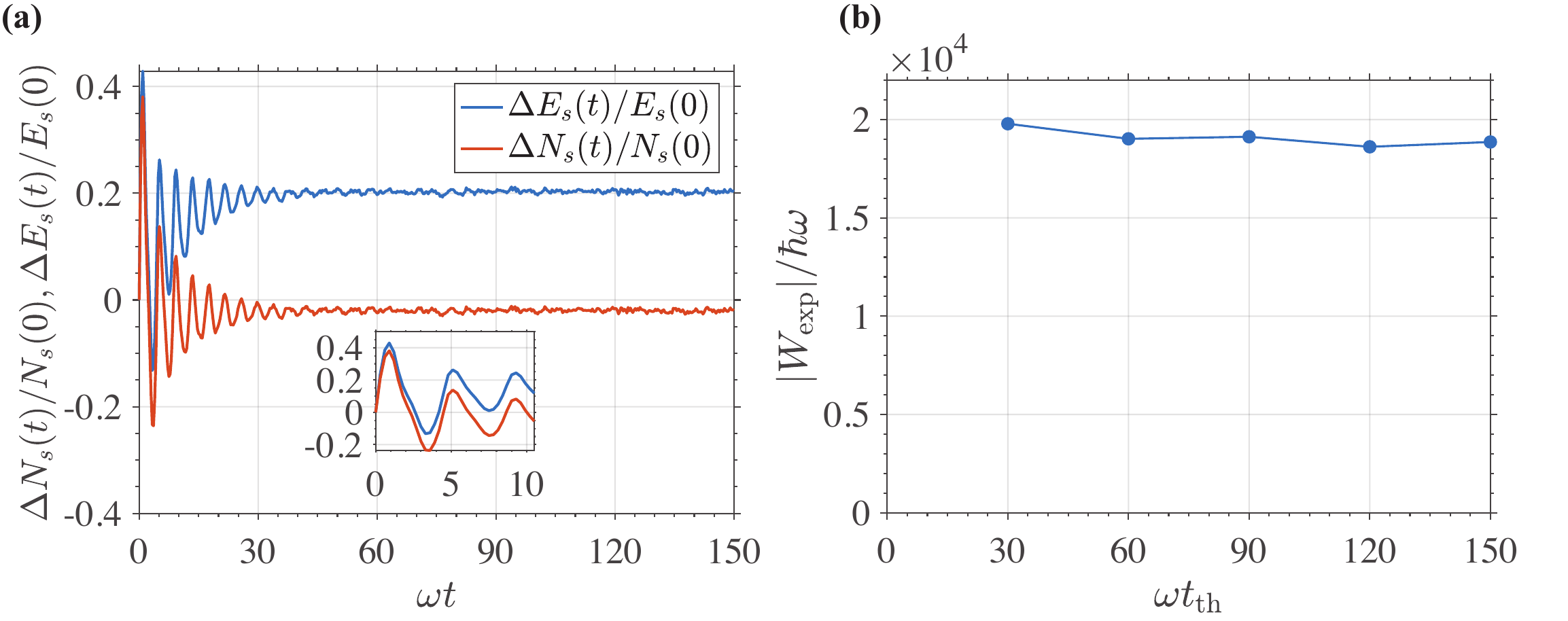}
    \caption{Relative change in energy and particle number, $\Delta E_s(t)/E_s(0)$ and $\Delta N_s(t)/N_s(0)$, of the working fluid versus dimensionless time $\omega t$ during the thermalization stroke \textbf{AB}. (b) Absolute value of the work extracted from the working fluid in the sudden-quench expansion stroke \textbf{BC} as a function of the dimensionless duration of the thermalization stroke \textbf{AB} $\omega t_{th}$. The dimensionless duration of expansion stroke \textbf{BC} is $\omega t_\mathrm{w}=0.05$. All parameter values for the initial state of the system and the ratio of the values of $g_c$ and $g_h$ are the same as in Fig.~\ref{fig:Fig3_Turbo}, except that now (during the expansion stroke) the interaction strength is quenched from $g_h$ back to $g_c$. The hot reservoir is prepared with interaction strength $g_h = 2.29 \times 10^{-38}$ J$\cdot$m and temperature $T_h=258$ nK. The ratio of particle number between the reservoir and the system is $N_h/N_s \simeq 7.31$. The particle number of the reservoir is $N_h \simeq 12800$, which is obtained using the chemical potential $\mu_h=39.7 \times 10^{-31}$ J. The ratio of temperature between the initial states of the system and the hot reservoir is $T_h/T_c =3.00$. The results in this figure correspond to the chemical offset arrangement shown in Fig.~\ref{fig:Chemical_offset}(c), where we have maintained $\Delta N_s (t \gg 1/\omega) \simeq 0$, by using $\mu_h \simeq \mu_s + \Delta \mu$, with $\Delta \mu = 29.5  \times 10^{-31}$ J.} 
   
    \label{fig:Fig3}
\end{figure}

\begin{figure}[tp]
    \centering
    \includegraphics[scale=0.39]{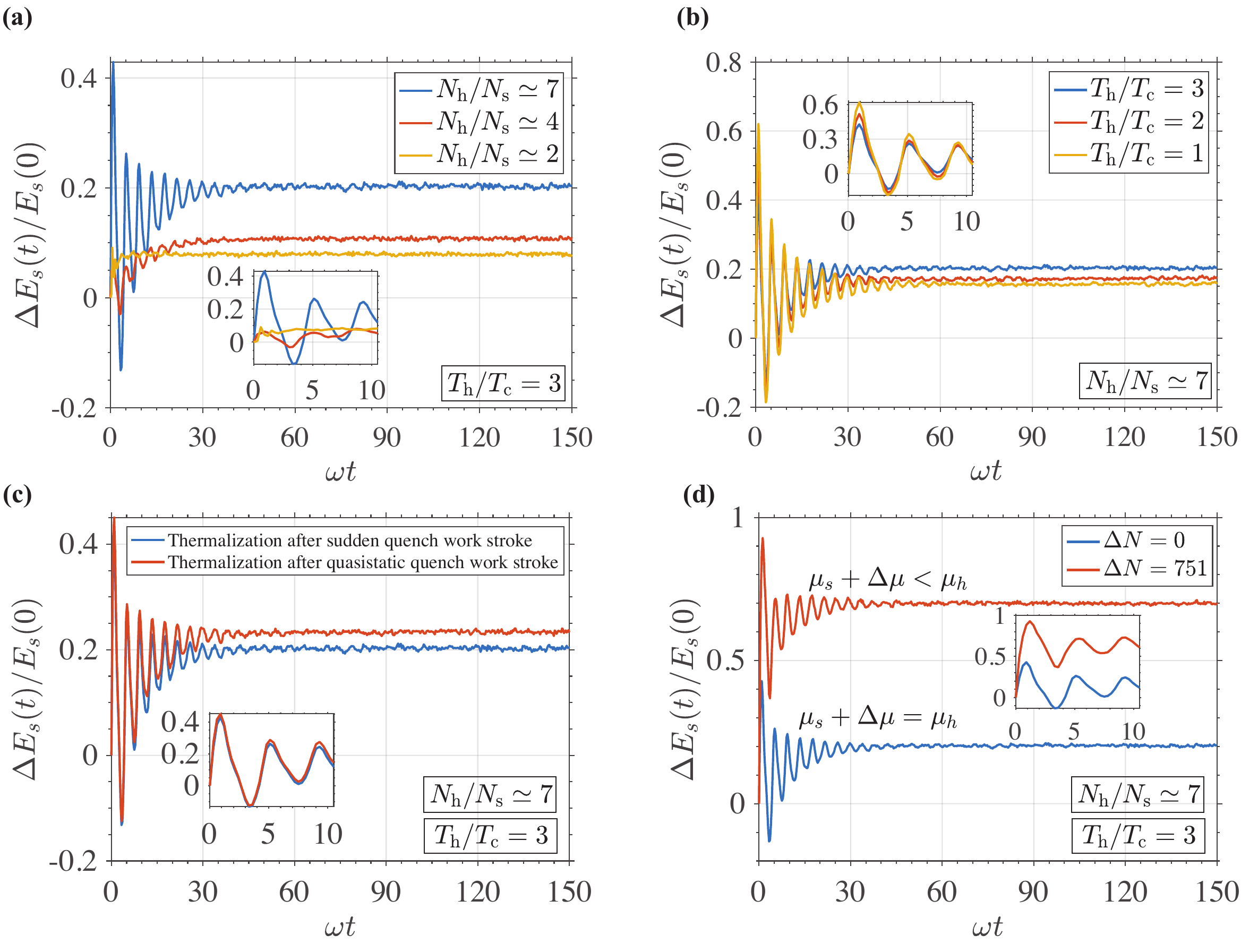}
    \caption{Relative change in energy of the working fluid, $\Delta E_s(t)/E_s(0)$, versus the dimensionless time $\omega t$ during the thermalization stroke \textbf{AB}, for various scenarios: (a) different ratios of particle number, $N_h/N_s \simeq 7$, $N_h/N_s \simeq 4$, and $N_h/N_s \simeq 2$; (b) different temperature ratios, $T_h/T_c = 3$, $T_h/T_c = 2$, and $T_h/T_c = 1$; (c) dynamics of thermalization following a sudden quench in the work stroke ($t_{\mathrm{w}} = 0.05/\omega$) compared to a quasistatic quench ($t_{\mathrm{w}} = 300/\omega$); and (d) dynamics of thermalization in a pure heat engine scenario ($\Delta N = 0$) compared to a chemical engine scenario with a net particle flow of $\Delta N = 751$ into the system from the hot reservoir. In all cases, the initial parameters of the system and the ratio of interaction strength quench for the work stroke are the same as in Fig.~\ref{fig:Fig3_Turbo}. In all panels, the system's initial particle number, $N_s\simeq 1750$ and temperature, $T_s=T_c=86.3$ nK were the same.
      }
    \label{fig:Fig4}
\end{figure}

Specifically, we focus on the thermalization stroke \textbf{AB} that follows immediately after the unitary compression stroke \textbf{DA} studied in the previous Section.
In Fig.~\ref{fig:Fig3}\,(a), we illustrate the time evolution of relative changes in the working fluid's (or system's, for which we use the subscript $s$) energy and particle number, $\Delta E_s(t)/E_s(0) = (E_s(t) - E_s(0))/E_s(0)$ and $\Delta N_s(t)/N_s(0) = (N_s(t) - N_s(0)/N_s(0)$, during this thermalization stroke. Here, the quantities $E_s(t) = \langle \hat{H_s}(t)\rangle$ and $N_s(t) = \int \langle \hat{\Psi}^{\dagger}_s (z,t)\hat{\Psi}_s (z,t) \rangle  dz$
are evaluated by replacing the creation and annihilation field operators by time-evolving stochastic realizations of the $c$-fields, $\psi^{(\mathbf{C})*}(z,t)$ and $\psi^{(\mathbf{C})}(z,t)$, and by replacing quantum mechanical ensemble averages by stochastic averages over a large number (typically 2000) of stochastic realisations.
In the examples shown in Fig.~\ref{fig:Fig3}\,(a), we consider the scenario where the thermalization stroke \textbf{AB} is initiated immediately after executing the work stroke \textbf{DA} via a finite-time sudden quench (as in Fig.~\ref{fig:Fig3_Turbo}(a)); completion of the work stroke at $t=t_\mathrm{w}$, corresponds to the start of the thermalization stroke \textbf{AB}, which we reset to be the zero, $t= 0$, in the horizontal axes labels. 
 We observe that, after an initial rapid exchange of particles, which occurs in the form of damped oscillations, we have maintained---in this example---zero net exchange in the particle number of the working fluid, i.e. $\Delta N_{s}(t\gg 1/\omega) \simeq0$. This is achieved by tailoring the chemical potential offset, $\Delta \mu$, such that we maintain zero net flow of particles, as was illustrated in Fig.~\ref{fig:Chemical_offset}\,(c).
Even though $\Delta N_{s}(t\gg 1/\omega) \simeq0$, we observe a net increase in the energy of the working fluid after a sufficiently long duration of the thermalization stroke, $\Delta E_\mathrm{s}(t\geq 40/\omega)$, which is due to purely temperature imbalance. We point out here that, even though the working fluid may have not come to complete thermal equilibrium at time $t\sim 40/\omega$, the bulk of the energy transfer, which is to be converted into useful work in the next stroke \textbf{BC}, has already taken place by this time.

In Fig.~\ref{fig:Fig3}~(b), we show how the duration of the thermalization stroke \textbf{AB} affects the amount of work, $W_\mathrm{exp} <0$, done by the working fluid during the subsequent expansion stroke \textbf{BC}. More specifically, we show the absolute value of the work, $|W_\mathrm{exp}|$, as a function of the duration, $t_\mathrm{th}$, of the thermalization stroke \textbf{AB} during which the working fluid is kept in contact with the hot reservoir. We observe that after $t_\mathrm{th} \simeq 40/\omega$ the absolute work $|W_\mathrm{exp}|$ shows negligible further change with an increased duration of the thermalization stroke \textbf{AB}. This is consistent with the observation made in Fig.~\ref{fig:Fig3}\,(a) that the bulk of the energy from the hot reservoir, that is being converted into work, has already been transferred to the working fluid by $t_\mathrm{th}\simeq 40/\omega$.

In Figs.~\ref{fig:Fig4}\,(a) and (b) we illustrate, respectively, the effects of the size and temperature of the hot reservoir on the dynamics of the relative change in energy, $\Delta E_s (t)/E_s(0)$, of the working fluid during the thermalization stroke \textbf{AB}. In both these cases, just as we begin the thermalization stroke, the working fluid is in an out-of-equilibrium state after the completion of the compression work stroke via a sudden quench ($t_\mathrm{w}=0.05/\omega$). The completion of the work stroke at $t=t_\mathrm{w}$, corresponds to the start of the thermalization stroke \textbf{AB}, which we reset to be the zero, $t= 0$, in Fig.~\ref{fig:Fig4}, just as we did in Fig.~\ref{fig:Fig3}.

In the examples of Fig.~\ref{fig:Fig4}(a), the thermalization stroke is implemented using three different hot reservoirs, each with a different particle number but an identical temperature, whereas the number of particles in the working fluid is kept constant at $N_s=1750$. The energy transfer from the reservoir to the working fluid again takes place through damped oscillations.
The net increase in energy of the working fluid after a sufficiently long duration of the thermalization stroke, $\Delta E_s(t\!\gg1/\!\omega)$, increases with the size of the hot reservoir, as expected. Furthermore, when using a small reservoir, the amplitude of oscillations, responsible for energy transfer, is smaller than in the case of a larger reservoir, and the oscillations damp faster. 
This means that, when using a smaller reservoir, the thermalization stroke can be terminated at an earlier time, since the bulk of the energy transfer has already occurred during the first few oscillations. 

For the examples of Fig~\ref{fig:Fig4}~(b), we employ the same procedure as in Fig.\ref{fig:Fig4}(a), but in this case we consider three different temperatures of the hot reservoir, and keep its total particle number fixed. In these three scenarios, we observe that the dynamics of the relative change in energy of the working fluid are largely insensitive to the temperature of the reservoir and that the bulk of energy transfer from the reservoir to the working fluid occurs on the same timescale. This suggests that the temperature of the hot reservoir may not have a significant effect on the Otto engine power output. Additionally, we see that the net energy transfer from reservoirs of different temperatures does not vary much, suggesting that these temperatures may not have a significant effect on the efficiency of the engine either. This is consistent with a similar recent finding from Ref.~\cite{estrada2024quantum}, for a study of the conventional (volumetric) Otto cycle with partially condensed, harmonically trapped Bose-Einstein condensate as the working fluid.

In Fig.~\ref{fig:Fig4}\,(c), we analyse the dynamics of energy transfer from the reservoir to the working fluid, depending on whether the working fluid, after the previous work stroke, was left in a highly non-equilibrium state (such as in a sudden quench work stroke) or in a near-equilibrium state (such as after a quasistatic work stroke). Here, the curve corresponding to energy transfer after a sudden quench work stroke is the same as the respective curve from Fig.~\ref{fig:Fig4}\,(a). The curve corresponding to energy transfer after a quasistatic work stroke, on the other hand, is shown in red. As we see, the dynamics of energy transfer in both cases are very similar: thermalization in both cases occurs over similar time scales, and the net increase in energy of the working fluid is approximately the same, with the net energy increase after a quasistatic work stroke being only marginally larger than that after a sudden quench work stroke. This suggests that the rate and magnitude of energy transfer during the thermalization stroke does not strongly depend on the state (highly non-equilibrium versus near equilibrium) of the working fluid. Therefore, executing the work strokes in a quasistatic manner does not lead to faster thermalization nor a more significant increase in energy of the system at the end of the thermalization stroke with the hot reservoir.

Finally, in Fig.~\ref{fig:Fig4}~(d), we demonstrate our model's capability to function as a \textit{chemical} engine. 
We compare the relative increase in energy of the working fluid during the thermalization stroke \textbf{AB} in two scenarios: first, corresponding to a pure \emph{heat} engine, when there is no net flow of particles from the hot reservoir to the working fluid, i.e. $\Delta N_{s}(t\gg 1/\omega) \simeq0$, using the scheme of Fig.~\ref{fig:Chemical_offset}~(c) and repeating the result for $N_{h}/N_s=7$ from Fig.~\ref{fig:Fig3}\,(a) (blue curve); and second, corresponding to a \emph{chemical} engine, where we allow for an additional flow of $\Delta N$ particles from the hot reservoir into the working fluid, using the scheme of Fig.~\ref{fig:Chemical_offset}~(a) (red curve). We observe a significant increase in the energy of the working fluid as we perform additional chemical work via the particle inflow. Thus, increasing the particle inflow from the hot reservoir to the working fluid provides an opportunity to increase the beneficial net work, as we have more energy, $E_\mathrm{in}$, at our disposal to be utilized to perform work, $W_\mathrm{exp}$, in the subsequent expansion stroke.

\section{Performance of the full Otto cycle} \label{sec:5}

In this Section, we combine the analysis of characteristic timescales for the work and thermalization strokes in the interaction-driven Otto cycle to evaluate the overall performance of the proposed Otto engine and the trade-off between its power and efficiency.

\subsection{Impossibility of operating as a heat engine}

\begin{figure}[tp]
    \centering
    \includegraphics[scale=0.4]{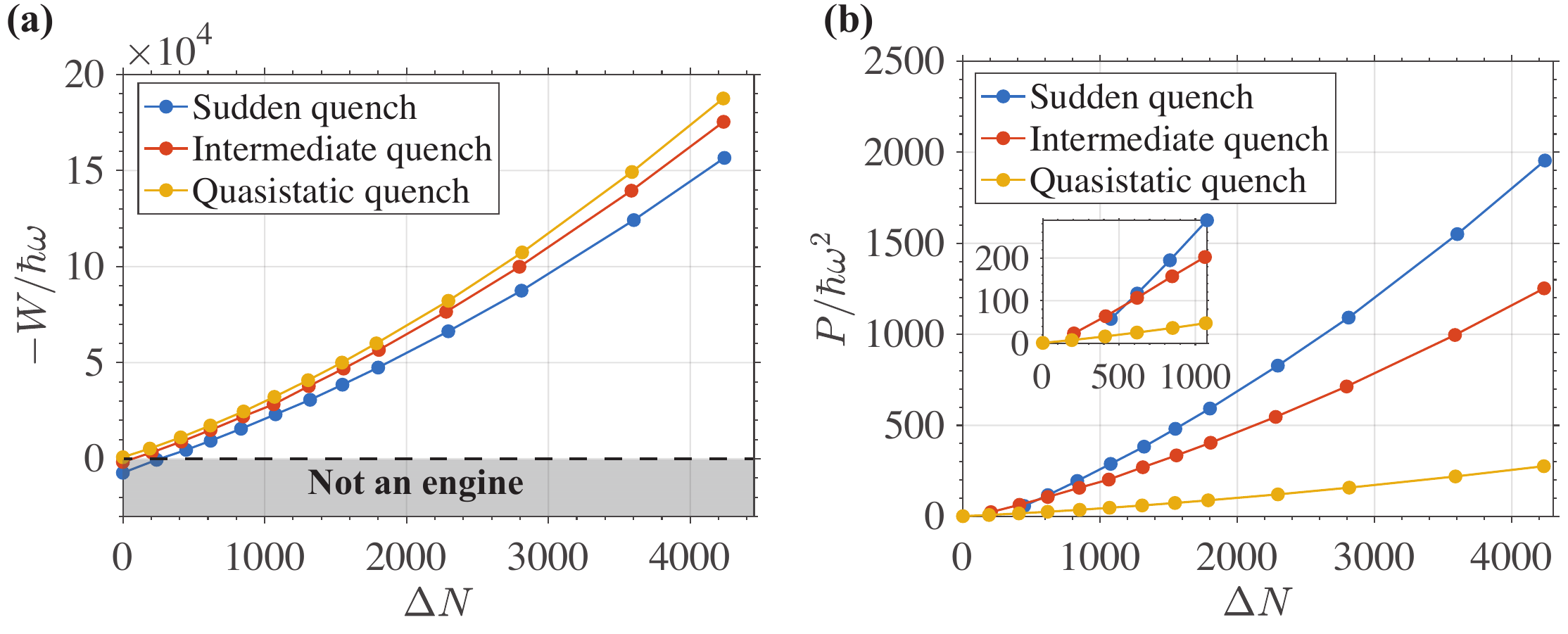}
    \caption{Net work and power output of an interaction-driven Otto engine as a function of the number of particles, $\Delta N$, exchanged with the hot and cold reservoirs during the thermalization strokes for sudden, intermediate, and quasistatic quenches. In panel (a), we show the net work $-W/\hbar \omega$ (in units of $\hbar \omega$), with $-W>0$ corresponding to positive net work done by the working fluid. For the first two data points of the blue curve and the first data point of the red curve, we have $-W<0$, meaning these points do not correspond to performance as an engine. In panel (b), we show the power output, $P/\hbar \omega^2$ (in units of $\hbar\omega^2$) of the Otto engine cycle corresponding to the data in (a). The data points that do not correspond to engine operation have been removed from (b). The duration of the thermalization stroke with each of the reservoirs was fixed at $t_\mathrm{th} = 40/\omega$ in all three cases. All other parameters are the same as in Fig.~\ref{fig:Fig3}. }
    \label{fig:Fig5}
\end{figure}

We first discuss a full interaction-quench Otto cycle with a harmonically trapped 1D Bose gas operating in a pure \emph{heat} engine setup, i.e., when the chemical potential offset $\Delta \mu$ (see Fig.~\ref{fig:Chemical_offset}) is chosen in such a way that there is no net particle flow between the working fluid and the reservoirs. This study was motivated by an attempt to extend the \emph{uniform} 1D Bose gas results of Refs.~\cite{watson2024interaction,chen2019interaction} to a harmonically trapped system, which is easier to realise experimentally. What we found, however, was that a harmonically trapped 1D Bose gas, unlike the uniform system, did not result in heat engine operation regime with large and negative $W<0$: the net work was either positive in the sudden quench scenario (implying that the system gained energy as a result of the cycle, rather than lost energy to a useful work), or was negative, but very small, in the quasistatic quench. This finding is illustrated in Fig.~\ref{fig:Fig5} (a), where see that $-W$ as a function of the number of particles $\Delta N$ is negative (i.e., $W>0$) for $\Delta N=0$, in the sudden-quench Otto cycle. The quantity $-W$ becomes positive (i.e., $W<0$) only at some finite $\Delta N$, in which case we refer to the Otto cycle as a \emph{chemical} engine (see next subsection), rather than \emph{heat} engine.

The main reason hindering the operation of the Otto cycle as a heat engine using a harmonically trapped 1D Bose gas is that the net work is now not only a function of the difference of the local atom-atom correlation functions at the hot and cold thermal equilibrium points \textbf{B} and \textbf{D} of the Otto cycle diagram of Fig.~\ref{fig:Fig1} (for details, see \cite{watson2024interaction}), but it also depends on the inhomogeneity of the density profile and, in particular, on the peak density at these equilibrium points. The overall effect of this is that, while the atom-atom pair correlation has a favourable dependence on the temperature $T$ for the net work done by the fluid to be large and positive, the dependence of the peak density on the temperature is not favourable and it cancels out the positive net work that would be otherwise realisable in a uniform system where the densities in the hot and cold equilibrium points are the same.

\subsection{Operating as an Otto \textit{thermochemical} engine}


Although operating our interaction-driven Otto cycle as a purely heat engine is not feasible using a harmonically trapped Bose gas, we find that one can still operate it as a \emph{thermochemical engine} by performing additional chemical work on the working fluid during the thermalization stroke \textbf{AB}. This additional chemical work is performed via the inflow of particles $\Delta N$ from the hot reservoir to the working fluid, using the chemical potential offset arrangement shown in Fig.~\ref{fig:Chemical_offset}~(a) and demonstrated in Fig.~\ref{fig:Fig4}~(d). The increase in total particle number results in a corresponding increase in the energy of the working fluid, which is available to be converted into mechanical work in the subsequent work stroke \textbf{BC}. After completing this work stroke, we couple the working fluid to the cold reservoir and transfer the same excess number of particles $\Delta N$ to the cold reservoir in the equilibration stroke \textbf{CD}, hence returning the working fluid into the state with the same initial number of particles.

The efficiency, $\eta = -W/E_\mathrm{in}$ of such a thermochemical Otto engine can be calculated by simply evaluating the energy differences at the end of each stroke, as described in Sec.~\ref{sec:2}. Unlike the case of a pure heat engine, the energy, $E_\mathrm{in} = \langle \hat{H}_s \rangle_{\mathbf{B}}-\langle \hat{H}_s \rangle_{\mathbf{A}}>0$, in an Otto thermochemical engine includes a contribution from chemical work and can be expressed as
\begin{equation}
    E_\mathrm{in} = Q_\mathrm{in} + W_\mathrm{chemical}.
\end{equation}
Here, $Q_{\mathrm{in}}$ is the heat taken in by the working during the thermalization stroke with the hot reservoir, \textbf{AB}, whereas $W_\mathrm{chemical}$ represents the additional chemical work done on the working fluid via the transfer of $\Delta N$ particles.
Though calculating the individual contributions, $Q_{\mathrm{in}}$ and $W_{\mathrm{chemical}}$, in the total $E_{\mathrm{in}}$ is a nontrivial task, as the heat and particle transport are intrinsically coupled processes (see, e.g., \cite{brantut2013thermoelectric, husmann2018breakdown}), we emphasise here that the chemical work is included in the overall energetic cost of evaluating the efficiency of the Otto thermochemical engine, as was done in Chapter \ref{Chap:6}, by defining the efficiency via 
$\eta = -W/E_\mathrm{in}$, rather than via $\eta = -W/Q_\mathrm{in}$.

\subsection{Trade-off between power and efficiency}

We finally analyse the overall performance of the finite-time Otto thermochemical engine and evaluate the trade-off between its power and efficiency.
In Fig.~\ref{fig:Fig5}, we show the net work and power output of the Otto thermochemical engine as a function of the number of particles $\Delta N$ exchanged with the reservoir, for three types of work strokes, corresponding to: sudden interaction quench, intermediate quench, and quasistatic quench. We see that, as we increase $\Delta N$, both the net work and power output increase in all three scenarios. Furthermore, as evident in Fig.~\ref{fig:Fig5}\,(a), the maximum work output occurs when the unitary work strokes are implemented through a quasistatic (near-
adiabatic) quench of the interaction strength, as expected. We observe, however, that transitioning from the slowest quasistatic quench to the fastest sudden quench regime results in a
relatively minor loss in work output, despite the work stroke being executed orders
of magnitude faster. This suggests that in the explored engine cycle, non-adiabatic
excitations contribute minimally to irreversible work and therefore optimising the quench protocol via a shortcut to adiabaticity \cite{del2013shortcuts,del2019focus} may not be necessary to operate at near-maximum efficiency \cite{li2018efficient, keller2020feshbach, abah2019shortcut, fogarty2020many}.

In Fig.~\ref{fig:Fig5}~(b), which shows the power, we observe that beyond a specific threshold value of $\Delta N \simeq 530$, the engine driven by a sudden quench of the interaction strength achieves a higher power output compared to the engine driven by a slow quasistatic and an intermediate quench for the work strokes. This increased power output can be attributed to the combination of two factors: first, the total engine driving time, $t_\mathrm{tot}$, in the sudden quench scenario is significantly shorter than the intermediate and the quasistatic quench; second, as demonstrated in Fig.~\ref{fig:Fig5}~(a), in a sudden quench engine, there is a relatively minor loss in net work $-W$, despite the work strokes being executed significantly faster. Consequently, given that power output is defined as $P = -W/t_\mathrm{tot}$, the sudden interaction quench engine significantly outperforms both the quasistatic and intermediate quench engines in terms of power output.

\begin{figure}[tbp]
    \centering
    \includegraphics[scale=0.40]{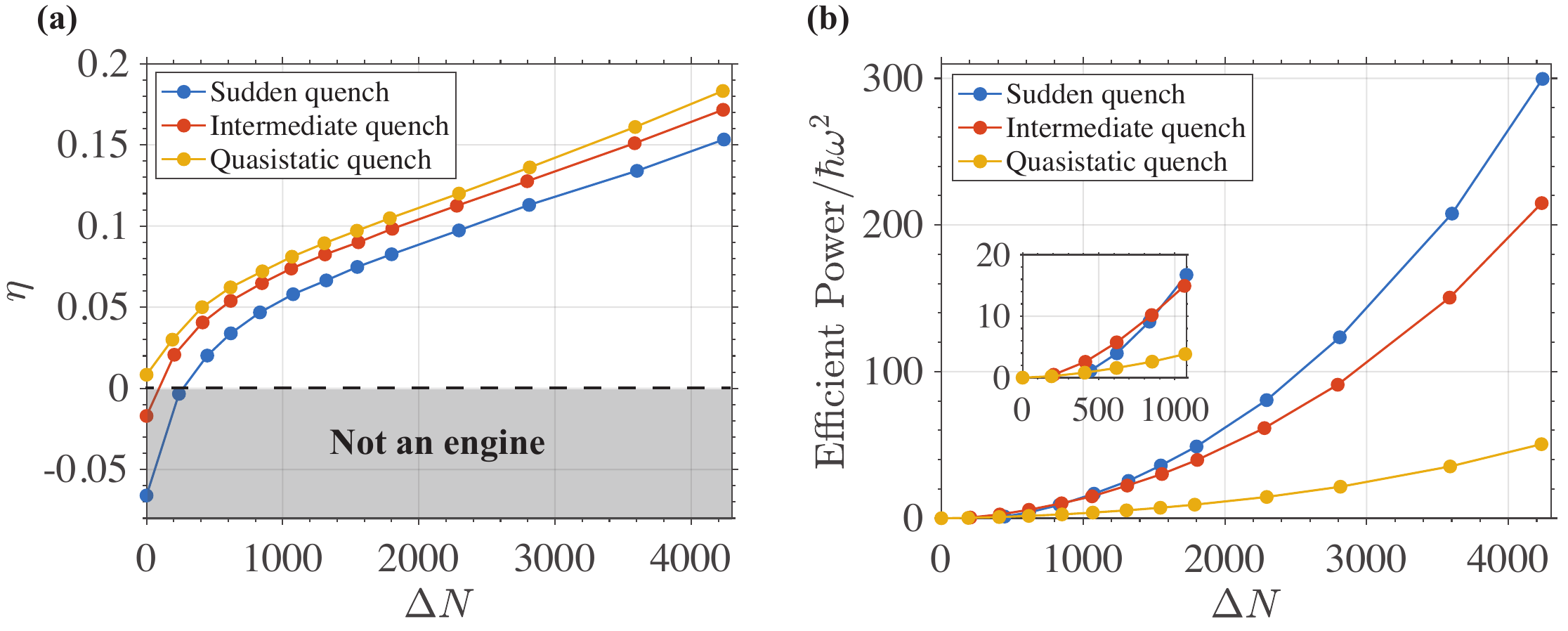}
    \caption{Efficiency, $\eta$, and efficient power as a function of the number of particles $\Delta N$ exchanged with the hot and cold reservoirs during the thermalization strokes for sudden, intermediate, and quasistatic quenches. All initial parameters of the working fluid and the reservoir are the same as in Fig.~\ref{fig:Fig3}. The thermalization time with the reservoir was fixed at $t_\mathrm{th} = 40/\omega$ for all three cases. In panel (a), we show the efficiency $\eta$, with $\eta>0$ corresponding to operation as an engine, i.e., when net work $-W >0$; the first two data points of the blue curve and the first data point of the red curve in (a) do not correspond to engine operation, as $-W < 0$ and hence $\eta<0$. In panel (b), we show the efficient power (see text) of the Otto engine cycle, corresponding to the data in (a). The data points that do not correspond to engine operation have been removed from (b). } 
    \label{fig:Fig6}
\end{figure}

In Fig.~\ref{fig:Fig6}~(a), we plot the efficiency as a function of the number of particles, $\Delta N$, for the thermochemical Otto engine driven by sudden, intermediate, and near-adiabatic (quasistatic) quenches of interaction strength during the work strokes. Consistent with observations in work and power output, an increase in $\Delta N$ leads to enhanced efficiency for all three quench scenarios. Furthermore, we observe that the sudden quench engine operates at efficiencies that are very close to the near-maximum limit achieved by the quasistatic work strokes. This result aligns with the results shown in Fig.~\ref{fig:Fig5}~(a), and means one can operate at near-maximum efficiencies by implementing the simplest finite-time and even sudden work strokes without relying on any optimisation protocol such as the STA to achieve similarly high efficiency \cite{del2013shortcuts, keller2020feshbach,abah2019shortcut, li2018efficient, fogarty2020many}. The results of Figs.~\ref{fig:Fig5}~(a) and \ref{fig:Fig6}~(b) confirm that irreversible work due to non-adiabatic excitations is not significant in this model.

Furthermore, to quantify the trade-off between efficiency and power, we use the parameter ``efficient power'', which was first proposed in Ref.~\cite{yilmaz2006new}. Efficient power is simply a product of the efficiency and power output, and provides a direct relation between the increase in the power output per unit decrease in the efficiency \cite{yilmaz2006new,myers2020bosons}. In Fig.~\ref{fig:Fig6}~(b), we show the efficient power for the engines driven by sudden, intermediate and quasistatic interaction quenches for work strokes as a function of $\Delta N$. We observed that the engine operating in the sudden quench regime provides the maximum efficient power provided a certain threshold value of $\Delta N$ (equal to $\Delta N \simeq 900$ in this example) is crossed.

The main takeaway from the finite-time analysis of the proposed Otto thermochemical engine cycle is the favourable trade-off between efficiency and power output achieved by executing the work strokes through a sudden quench of the interaction strength. Furthermore, as we increase the number of particles, $\Delta N$, exchanged with the reservoirs, we observe a boost in engine performance across all three chosen quench times of the work strokes considered in this study: sudden quench, intermediate, and quasistatic quench. We additionally note that the qualitative conclusions presented above remain valid as long as we are within the weakly interacting regime of the 1D Bose gas, irrespective of the specific values of the parameters chosen.

\section{Conclusions} \label{sec:conclusion}

In this chapter, we simulated a finite-time quantum Otto cycle driven by a quench of atomic interactions of a 1D Bose gas in the weakly interacting quasicondensate regime. Our analysis included a simulation of both the unitary work strokes and the thermalization strokes for the proposed Otto cycle. To simulate the work strokes, we treated the working fluid as an isolated many-body quantum system undergoing unitary evolution, starting from a thermal initial state. 
The thermalization strokes, on the other hand, were simulated by treating the working fluid as an open many-body quantum system coupled to another many-body quantum system serving as the reservoir, both treated microscopically. We identified characteristic operational timescales for these thermalization strokes in experimentally realistic regimes.

The Otto engine's performance was evaluated in three different scenarios corresponding to three typical timescales for the execution of the work strokes: a sudden quench, an intermediate quench, and a slow quasistatic (near-adiabatic) quench. We first found that, contrary to a uniformly trapped Bose gas \cite{watson2024interaction}, a harmonically trapped system does not function as a heat engine. Nonetheless, we have also found that engine operation can be restored by enabling additional chemical work in the form of particle inflow from the hot reservoir to the working fluid. Thus, we have found that a harmonically trapped 1D Bose gas can operate effectively as a thermochemical Otto engine. We have shown that such a thermochemical Otto engine, when operating in the sudden quench regime, achieves an efficiency that is quite close to the near-maximum limit obtained by implementing the work strokes in a quasistatic fashion. Thus, in our proposed engine cycle, it is possible to operate at near-maximum efficiency by executing the simplest finite time quench (linear quench) or even a sudden quench of the interaction strength, without relying on optimization protocols such as the STA. The primary reason for this is the minimal amount of irreversible work generated by non-adiabatic excitations during the finite-time driving of the Hamiltonian to execute the work strokes. Hence, when work strokes are executed through a sudden quench of the interaction strength, we observe a favourable trade-off between efficiency and power output in the thermochemical Otto engine explored here.

The capability of SPGPE to simulate the thermalization strokes of a quantum Otto cycle suggests that future work could focus on tailoring the tunnel-coupling between the working fluid and external reservoirs to facilitate a more rapid onset of thermalization. Such an investigation would be particularly useful for possible future experimental realization of such an engine cycle, shortening the total cycle time and thus increasing power output. Further, one may apply the method developed in this Chapter to alternative quantum thermodynamic cycles, allowing for full simulation of a quantum Carnot or Stirling engine cycle. Finally, the framework of using three tunnel-coupled 1D Bose gases to simulate a full Otto engine cycle could be applied more broadly to the set of quantum thermal devices introduced in Chapter \ref{Chap:3}, these being the refrigerator, heater, and thermal accelerator.


\newpage
\noindent
The work presented in Chapter \ref{Chap:7} was adapted from the accepted publication of Ref.~\cite{nautiyal2024finitetime}, and the contribution of each named author to that work is presented below in Table.~\ref{Tab:Chap7}.

\noindent
\cite{nautiyal2024finitetime} V. V. Nautiyal, \textbf{R. S. Watson}, and K. V. Kheruntsyan, \href{https://arxiv.org/abs/2404.16470}{A finite-time quantum Otto engine with tunnel coupled one-dimensional Bose gases}, accepted by the New Journal of Physics on 25 April 2024.

\begin{table}[h]
	\begin{center}
	\begin{tabular}{|c|l|l|}
		\hline
		Contributor & Statement of contribution & \% \\
		\hline
		V. V. Nautiyal				& writing of text 					& 70\\
															& proof-reading							& 20 \\
															& numerical calculations 		& 100\\
															& preparation of figures 		& 70 \\
															& initial concept						& 20 \\
		\hline
		\textbf{R. S. Watson}				& writing of text 					& 15\\
															& proof-reading							& 30 \\
															& preparation of figures 		& 20 \\
															& initial concept						& 40 \\
		\hline
		K. V. Kheruntsyan								& writing of text 					& 15\\
															& proof-reading							& 50 \\
															& preparation of figures 		& 10 \\
															& initial concept						& 40 \\
		\hline
	\end{tabular}
	\end{center}
 \caption{}\label{Tab:Chap7}
\end{table}

%% file: Conclusion/Conclusion.tex
\chapter[Conclusion and outlook]{Conclusions and outlook}  
\label{Chap:Conclusion}


In this thesis we have investigated various aspects of quantum thermodynamics for integrable and near-integrable systems, with a focus on the one-dimensional (1D) Bose gas.
We have utilized a variety of theoretical and computational tools available for this model, such as analytic expressions for Glauber's local second-order correlation function, and the recently developed theory of generalized hydrodynamic for simulating the large-scale dynamics of integrable and near-integrable models. Our work opens a number of pathways to the study of equilibrium and non-equilibrium thermodynamics in integrable models, and provides insight into the role that correlations may play in the emerging field of quantum thermodynamics.

In Chapter \ref{Chap:3}, we introduced and explored a quantum thermal machine (QTM) realised in a 1D Bose gas in its integrable uniform configuration, and driven by a sudden quench of interparticle interaction strengths. 
For the case of an Otto engine cycle, the net work was expressed in terms of a difference of atom-atom correlations, described by Glauber's local second-order correlation function. Thus, utilising analytic expressions for both this local correlation function and the total system energy at thermal equilibrium, we gave a detailed analysis of the performance of this engine cycle, in terms of net work and efficiency, in three of the parameter regimes of this model. Though we focused on three experimentally accessible regimes for the 1D Bose gas, the analysis may be extended to all other parameter regimes, as each possesses its own analytic expressions for all relevant quantities.

We further made use of the expressions derived for net work and heat transfer with the external reservoirs to investigate operation of this Otto cycle as a variety of alternative thermal devices. Through this, we were able to define the coefficients of performance and limiting cases of operation for the refrigerator, heater, and thermal accelerator.
Notably, though the Otto cycle introduced in this chapter was applied to the uniform 1D Bose gas, it is more broadly applicable to systems with short-range $s$-wave interactions. 
As such, a possible future avenue of study would be in applying the techniques developed to investigate operation of this interaction-driven sudden quench protocol in a variety of other quantum systems. Further, one may extend this method to arbitrary external potentials, which do not fundamentally affect the form of the equations, but could be used to tailor the local correlations. This suggests various possible QTM's, where one could consider quenching the external potential in combination or in sequence with the interaction strength, as was explored for a conventional Otto cycle assisted by tuning interaction strengths in Ref.~\cite{boubakour2023interaction}.

In a practical context, one may consider the question of how the net work produced be such quantum devices may be extracted in order to perform some beneficial process. Though such protocols exist and have been experimentally implemented for single-body engines \cite{lindenfels2019spin}, with additional proposals for similar extraction methods in the context of many-body non-interacting systems \cite{bouton2021quantum}, proposals for work extraction protocols for many-body \emph{interacting} systems are often hard to find. We highlight here the proposal of Chen \textit{et al.} in Ref.~\cite{chen2019interaction} for the adiabatic interaction-driven quantum heat engine for a 1D Bose gas, where they suggest that a 1D spinor Fermi gas, which may be effectively mapped to the Lieb-Liniger model, could couple its additional spin degrees of freedom to an external magnetic field. One further possibility is utilizing a two-component 1D Bose gas, where performing work strokes on a single gas component may enact some measurable beneficial work process on the second component. Though, perhaps the simplest suggestion is that found in Ref.~\cite{simmons2023thermodynamic}, where alternate control over the optical and magnetic degrees of freedom allow for transfer of energy between these in a manner similar to a conventional engine cycle, where the gas generates beneficial work by acting on its container.

Our work in Chapter \ref{Chap:Maxwell} extends our investigation into the role that Glauber's correlation function plays in the thermodynamics of quantum systems. In particular, using the fact that the local second-order correlation function is a thermodynamic quantity for the 1D Bose gas, and may be calculated from the Helmholtz free energy via application of the Hellmann-Feynman theorem, we derive a new Maxwell relation between this correlation function and the entropy of the system. One may utilise this to calculate the entropy of an ultracold atomic system by simply integrating this Maxwell relation. To demonstrate the utility of this in the laboratory---as a new experimental method for deducing the entropy, we perform a numerical experiment using the stochastic projected Gross-Pitaevskii equation (SPGPE), which is a classical $c$-field method applicable to finite temperature gases in the weakly interacting, quasicondensate regime. We calculate the system entropy, which is often highly nontrivial to accomplish in experiment, of a uniform 1D Bose gas from local correlations, which are typically much more accessible.

The relation derived in Chapter \ref{Chap:Maxwell} is of general importance to experimental investigation of ultracold atomic gases with short range $s$-wave interactions, such as the Yang-Gaudin model \cite{PhysRevLett.19.1312,gaudin1967systeme,guan_YangGaudin}, or even the attractively interacting 1D Bose gas \cite{PhysRevLett.75.1687,PhysRevA.91.053607,PhysRevB.88.205131}.
Further, this Maxwell relation could be applied in variety of other computational approaches, such as the density-matrix renormalisation group method, which is a flexible computational tool capable of simulating ground state properties for a wide range of quantum models. Finally, we note the breadth of applicability beyond $s$-wave interacting systems, to interacting spin chains such as the transverse field Ising model, where the role of Glauber's second-order correlation function is played by the spin-spin correlation function between the neighbouring sites, as was discussed in the conclusions to Chapter \ref{Chap:Maxwell}.
This highlights the profound role that the calculation and measurement of particle-particle correlation functions may play in the future for determining the thermodynamic properties ultracold atomic gases.

In Chapter \ref{Chap:5} we benchmark the recently developed theory of generalised hydrodynamics (GHD), a numerical tool which is uniquely capable of simulating the large-scale nonequilibrium dynamics of integrable and near-integrable systems. We compare GHD simulation for a handful of nonequilibrium scenarios in a 1D Bose gas against a variety of established alternative theoretical methods. In particular, we focus first on quantum dynamics emanating from the release of a localised density perturbation, which results in a quantum shockwave for the case of a density bump, and a train of grey solitons for a density dip. 
We compare the evolution of GHD in these scenarios against a variety of alternative theoretical tools at zero and finite temperatures, and over the entire range of interaction strengths.
Through this, we define the regions of GHD's applicability, which aligns with the validity of the local density approximation (LDA), introduced in Chapter \ref{Chap:2}.
We further benchmark higher-order Navier-Stokes GHD, which is capable of modelling dynamical thermalization, for the case of a quantum Newton's cradle. We observed excellent agreement when compared against the SPGPE method for the 1D Bose gas, validating the long-time dynamics and relaxation under GHD for the case of a weakly interacting quasicondensate at finite temperature.

As GHD is less than a decade old, there remains a large number of interesting nonequilibrium scenarios in a variety of theoretical models that the theory can be benchmarked against.
Perhaps most importantly, it was noted recently by Malvania \textit{et al.} in Ref.~\cite{malvania2021generalized} that GHD is remarkably effective at describing highly nonequilibrium dynamics even at extremely low atom numbers, where the hydrodynamic assumption is questionable. Further, experimental probes into the rapid onset of the applicability of hydrodynamics, also known as `hydrodynamisation', suggest that GHD may be valid even before local relaxation to a thermal equilibrium state described by the generalized Gibbs ensemble introduced in Chapter \ref{Chap:2}. Such investigations into the effectiveness of GHD to describe phenomena beyond its assumed conditions of applicability serve to highlight the exciting future of this already rapidly developing hydrodynamic theory.

In Chapter \ref{Chap:6}, we further our exploration of the quantum Otto engine cycle introduced in Chapter \ref{Chap:3}, using GHD, which we benchmarked in Chapter \ref{Chap:5}, to perform a detailed exploration of a thermochemical Otto engine cycle with a harmonically trapped 1D Bose gas as a working fluid.
In particular, we applied GHD to study the finite time performance of the interaction-driven engine cycle, utilising the expressions for net work and efficiency derived in Chapter \ref{Chap:3} to contrast this against the idealised limiting cases of sudden and quasistatic operation. To the best of our knowledge, this represents the first application of GHD to the investigation of a quantum thermodynamic device.
As such, our work highlights the utility of GHD for investigating finite time operation of quantum thermal machines in the wide variety of known integrable models. In particular, simulation of experimentally realistic models, such as the Yang-Gaudin, XXZ spin chain, and Hubbard model, in the context of quantum thermodynamics presents an exciting outlook for possible future explorations into novel quantum thermal machines.

Finally, in Chapter \ref{Chap:7}, we simulate the entire interaction-driven quantum Otto engine cycle via the $c$-field SPGPE method with three tunnel-coupled 1D Bose gases in the weakly interacting quasicondensate regime. This extends the work done in Chapter \ref{Chap:6} where the thermalization strokes of the thermodynamic cycle were not simulated, remaining beyond the scope of generalised hydrodynamics. Through this, we found that operation as a heat engine, i.e. where the thermalization strokes consist of no net particle transfer, precluded the generation of beneficial net work. However, if net particle transfer between the tunnel-coupled gases was allowed, the Otto cycle was shown to be capable of operating as a \textit{thermochemical} engine cycle, previously investigated in Chapter \ref{Chap:6}. Analysing the trade-off between net work and efficiency as a function of total cycle time resulted in the sudden quench cycle presenting a favourable trade-off between power and efficiency under large atom number transfer. Possible future extensions to this work include using the SPGPE framework to analyse other thermodynamic cycles, such as the Carnot or Stirling engine cycles, or even investigating operation of the alternative thermal machines, such as the accelerator, heater, and refrigerator, introduced in Chapter \ref{Chap:3}. Further, one may utilize the ability of SPGPE to simulate tunnel coupling between two 1D Bose gases to tailor a time-dependent tunneling rate that may be capable of accelerating relaxation to thermal equilibrium. Such an application would have utility for experimental realisation, as shortening the total cycle time presents the opportunity to increase power output.


%% file: PreliminaryAndBackPages/Back.tex
\backmatter

\normalfont
\cleartooddpage

\pagestyle{empty}

\begin{table}[b!]
\begin{center}
\textit{``All change is a miracle to contemplate; but it is a miracle which is taking place every instant.''}

\end{center}
\begin{flushright}
Henry David Thoreau,\\
Walden

\end{flushright}
\end{table}